\documentclass[letterpaper,10pt,oneside]{book}
\usepackage[latin1]{inputenc}
\usepackage[T1]{fontenc}
\usepackage[scaled]{uarial}
\usepackage{amsmath,amsfonts,amssymb,bbm}
\usepackage{graphicx}
\usepackage{footnote}
\usepackage{float}
\usepackage{subfigure}
\usepackage{setspace}
\DeclareGraphicsExtensions{.bmp,.png,.pdf,.jpg,.eps}
\usepackage{physics}
\usepackage{mathrsfs}
\usepackage{fancyhdr}

\usepackage[colorlinks]{hyperref}
\hypersetup{
    colorlinks=true,
    linkcolor=blue,
    filecolor=magenta,  
    urlcolor=[rgb]{0.00,0.00,0.50},    
    citecolor=red,
    }

\usepackage{url}
\usepackage{amsmath}
\usepackage[english]{babel}

\usepackage{amsmath}
\numberwithin{equation}{section}

\DeclareMathAlphabet\mathbfcal{OMS}{cmsy}{b}{n}
\DeclareMathAlphabet{\boldmathe}{T1}{cmr}{bx}{it}
\newcommand{\mbf}[1]{\boldmathe{#1}}
\newcommand{\mbfgr}[1]{\textit{\mbox{\boldmath$#1$}}}

\def\id{\mathbbm{1}}
\def\ha{\tfrac{1}{2}}
\def\has{\tiny\frac{1}{2}}
\def\vA{\mbf{A}}
\def\vB{\mbf{B}}
\def\vE{\mbf{E}}
\def\vK{\mbf{K}}
\def\vV{\mbf{V}}
\def\vG{\mbf{G}}
\def\vI{\mbf{I}}
\def\vJ{\mbf{J}}
\def\vL{\mbf{L}}
\def\va{\mbf{a}}
\def\vb{\mbf{b}}
\def\ve{\mbf{e}}
\def\vn{\mbf{n}}
\def\vp{\mbf{p}}
\def\vr{\mbf{r}}
\def\vs{\mbf{s}}

\def\vnabla{\mbfgr{\nabla}}
\def\vsigma{\,\mbfgr{\sigma}}

\def\vpi{\mbfgr{\pi}}
\def\hvJ{\,\hat{\hskip-1mm\vJ}}
\def\be{\begin{equation}}
\def\ee{\end{equation}}
\def\cC{\,{\mathcal{C}}}
\def\cQ{{\mathcal{Q}}}

\def\cW{{\mathcal{W}}}

\def\A{\mathbb A}
\def\B{\mathbb B}
\def\Z{\mathbb Z}
\def\N{\mathbb N}
\def\R{\mathbb R}

\def\P{\mathbb P}

\def\I{\mathbb{I}}

\usepackage{todonotes}

\usepackage{natbib}

\usepackage[letterpaper]{geometry}
\let\OLDthebibliography=\thebibliography
\def\thebibliography#1{\OLDthebibliography{#1}%
\addcontentsline{toc}{chapter}{\refname}}

\usepackage{fancyhdr}

\begin{document}

\vskip0.5cm
\begin{titlepage}
\begin{minipage}[c]{12cm}
\centering
{\bfseries\Large UNIVERSIDAD DE SANTIAGO DE CHILE \par}
{\bfseries  \large FACULTAD  DE CIENCIAS \par}
{\bfseries \large Departamento de F\'isica \par}
\end{minipage}
\begin{minipage}[c]{3cm} 
\centering
\begin{figure}[H]
 \begin{center}
\includegraphics[scale=1]{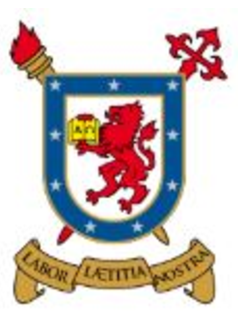}
\end{center}
\end{figure}
\end{minipage}

\vspace{3.5cm}
\centering
{\bfseries\large Hidden symmetries and nonlinear\\ (super)algebras \par}
\vspace{1.5cm}
{\bfseries Luis Abel Inzunza Melo \par}
\vspace{2cm}
\begin{flushright}
\begin{minipage}[t]{7cm} 
{\bfseries  PhD Thesis in Science, mentioning Physics \par}
\vspace{1.5cm}
{\bfseries Thesis Supervisor:\\ Mikhail S. Plyushchay \par}
\end{minipage}
\end{flushright}

\vspace{3cm}
{\bfseries  Santiago - Chile \par}
{\bfseries  2020 \par}

\end{titlepage}

\newpage
\thispagestyle{empty}

${}$
\vskip 20.5cm
\begin{flushright}
{\bfseries   Luis Abel Inzunza Melo, 2020 \par}
Reconocimiento-No Comercial 4.0 Internacional 
\end{flushright}

\chapter*{Abstract}

\rfoot[e1]{\thepage}

\fancypagestyle{plain}{
\fancyfoot[R]{\thepage}
}

\pagestyle{fancy}

\pagestyle{fancy}
The relevance of hidden symmetries is explored at the level of classical and quantum mechanics in a variety of physical systems related to conformal and superconformal invariance. Hidden symmetries, that correspond to nonlinear in momenta integrals of motion, generally lead to nonlinear algebras.

First, analyzing the $\mathfrak{sl}(2,\mathbb{R})$ symmetry, it is concluded that both the asymptotically free (at infinity) and the harmonically confined models are two different forms of dynamics described by the same symmetry algebra. A mapping between these two dynamics is constructed, and its applications are studied in one-, two- and three-dimensional systems.

Second, 
rational extensions of the conformal mechanics model of de Alfaro, Fubini and Furlan (AFF) are derived 
by employing the generalized Darboux transformation. In general, the obtained systems have an almost equidistant spectrum with some gaps inside, and their spectral properties imply the presence of hidden symmetries. The supersymmetric extensions of the AFF model are also studied, and the origin of the hidden bosonized superconformal symmetry of the quantum harmonic oscillator is established.

Finally, a three-dimensional generalization of the AFF system is considered. 
The model describes a particle with electric charge $e$ in Dirac monopole background of magnetic charge $g$, and subjected to the central potential 
$\frac{m\omega^2}{2}r^{2}+\frac{\alpha}{2mr^2}$. When $ \alpha=(eg)^2 $, 
the classical trajectories are periodic for arbitrary initial conditions and at the quantum level, 
the spectrum acquires a peculiar degeneration. These characteristics are described by hidden symmetries, 
which can be obtained from the model without harmonic term by means of the mentioned mapping. 
A complementary spin-orbit coupling term gives rise to a supersymmetric extension of the system, characterized by superconformal symmetry. The spectrum-generating operators of the new model are shown to be nonlocal.

\vskip 1cm

\textcolor{blue}{\underline{\emph{Keywords:}}}
Hidden symmetries; (Super-)Conformal symmetry; de Alfaro, Fubini and Furlan model; 
Harmonic oscillator; 
Supersymmetric quantum mechanics; Rationally extended systems;
Darboux duality;
Klein four-group; 
Dirac monopole.

\pagestyle{fancy}
\renewcommand\headrulewidth{0pt}
\lhead{}\chead{}\rhead{}\cfoot{}
\rfoot{\vspace*{0\baselineskip}\thepage}

\pagenumbering{roman}

\chapter*{Dedicatory}
${}$
\vspace{2cm}
\begin{flushright}
\emph{``... y la realidad plausible cae de pronto sobre mi...\\
me incorporo a medias en\'ergico, convencido, humano\\
y voy a escribir estos versos para convencernos de lo contrario...''}

\vspace{0.5 cm}
\'Albaro de Campos (Fernando Pessoa), \emph{La Tabaquer\'ia}. Extracto. 

\vspace{2cm}
Dedicado a quienes me apoyaron (y soportaron) durante todo el proceso.

\end{flushright}

\chapter*{Acknowledgements}
The work on this Thesis was mainly supported by
CONICYT (now ANID) scholarship  21170053. 
 I would also like to thank 
the University of Santiago  for the sponsorship 
of research projects  FCI-PM-02 
and USA 1899,
and the FONDECYT Project 1190842,  in which I participated.

I am especially grateful for all the support and help 
provided by  my Thesis supervisor,
 Professor Mikhail S. Plyushchay, who has been with me at all times. 
In the same vein, I would like to thank Professor 
Andreas Wipf for his warm hospitality and help 
during my stay at the Friedrich-Schiller University, Jena, 
Germany, in 2019.

Finally, I want to thank my family and friends for their emotional support. 
They are always in my thoughts.

\chapter*{Notations}
Here we summarize some common notations used in the  manuscript.  In this Thesis we use $\hbar=c=1$. \\

\underline{\textit{Geometry}}:\\
$g_{\mu\nu}$ and $\eta_{\mu\nu}$: The  general  metric tensor and Minkowski metric tensor.\\
$x_\mu=g_{\mu\nu}x^{\nu}=\sum_{\nu}g_{\mu\nu}x^{\nu}\,$ and 
$g_{\mu\nu}x^{\mu}x^{\nu}=\sum_{\mu,\nu}g_{\mu\nu}x^{\mu}x^{\nu}\,$: 
The Einstein summation convention.\\
$\zeta^\mu$: A Killing vector component.\\
$A\wedge B$ and $d$: The exterior product and the exterior derivative, respectively.\\
$ \pounds_{X}{T}$: The Lie derivative of a tensor field 
$T$ along the flow of the vector field $X$.\\
$i_{X}\omega\equiv
\omega(X,\underbrace{\ldots\ldots\ldots}_{r-1\text{ entries}})$: 
The contraction between a vector field and a differential $r$-form $\omega$,\\ 
${}\qquad\qquad\qquad\qquad\qquad\,\,$ which, in turns,
is a differential $(r-1)$-form.\\

\underline{\textit{Classical mechanics}}:\\
$\mathcal{M}$: The configuration space.\\
$T\mathcal{M}_q$: The tangent space at $q \in \mathcal{M}$.\\ 
$T_*\mathcal{M}_q$: The cotangent space at $q \in \mathcal{M}$.\\ 
$T\mathcal{M}$: The tangent bundle.\\ 
$T_*\mathcal{M}$: The cotangent bundle.\\ 
$q^i$ and  $\dot{q}^{i}=\frac{dq^i}{dt}$: The generalized coordinates on $\mathcal{M}$ and its velocities.\\
$\mathcal{L}$ , $p_i=\frac{\partial\mathcal{L}}{\partial \dot{q}^{i}}$ and $H$: The Lagrangian, 
the canonical momenta and the Hamiltonian.\\
$\omega=dq_i\wedge dp^i$: The symplectic two-form. \\

\underline{\textit{Supersymmetric quantum mechanics}}: \\
$H$: The quantum Hamiltonian.\\
$L$: A dimensionless quantum Hamiltonian.\\
$\psi_*$, $\widetilde{\psi}_*$: Two linearly independent eigenstates of $L$, with eigenvalue $\lambda_*$\,.\\
$W(\underbrace{\ldots\ldots\ldots}_{n \text{ entries}})$: The generalized Wronskian of $n$ functions.\\
$\breve{L}$: A dimensionless supersymmetric partner of $L$.\\
$A^\pm$: The first order mutually conjugate intertwining operators.\\
$\A_{n}^\pm$: The higher order mutually conjugate intertwining operator.\\
$\Omega_*(x)\,,\breve{\Omega}_*(x)$: The Jordan states constructed by means of $\psi_*$ and $\widetilde{\psi}_*\,,$
respectively.\\
$\mathcal{H}:$ A matrix-valued super-Hamiltonian operator.\\
$\mathcal{Q}_a:$ A Supercharge.\\
$\mathcal{N}$: The number of supercharges.\\
Pauli matrices: 
$
\sigma_1=\left(\begin{array}{cc}
0 & 1\\
1 & 0
\end{array}\right)\,,\qquad 
\sigma_2=\left(\begin{array}{cc}
0 & -i\\
i & 0
\end{array}\right)\,,\qquad 
\sigma_3=\left(\begin{array}{cc}
1 & 0\\
0 & -1
\end{array}\right)\,.
$\\
$\Pi_\pm=\frac{1}{2}(1\pm \sigma_3)$: Projectors to $\sigma_3$ subspaces. \\

\underline{\textit{Conformal mechanics}}:\\
$H$, $D$, and $K$:  Generators of the $\mathfrak{so}(2,1)$ algebra.\\
$\mathcal{J}$ and $\mathcal{J}_\pm$:  Generators of the $\mathfrak{sl}(2,\R)$ algebra.\\
$H_\nu:$ The Hamiltonian of an asymptotically free conformal invariant system.\\
$\mathscr{H}_\nu$ and $\mathcal{C}_\nu^\pm$ The Hamiltonian of de Alfaro, Fubini and Furlan model and its ladder operators.\\
$\mathfrak{S}$: The conformal bridge transformation operator.\\

\underline{\textit{Rationally extended systems}}:\\
$\Delta_\pm$: The positive-negative Darboux scheme.\\
$A_{(\pm)}^\pm$: The self-conjugate intertwining operators of the positive-negative Darboux scheme.\\
$L_{(\pm)}$: The rationally extended system associated with the positive-negative Darboux scheme.\\
$\mathcal{A}^\pm\,,$ $\mathcal{B}^\pm\,,$ and $\mathcal{C}^\pm\,$: The spectrum-generating ladder operators 
of the ABC-type.\\
$\mathfrak{A}_i^\pm\,,$ $\mathfrak{B}_i^\pm\,,$ and $\mathfrak{C}_i^\pm\,$: 
The extended families of  ladder operators 
of the ABC-type.\\
$\mathfrak{S}_z^\pm\,,$ : 
The extended families of intertwining operators.\\
$\mathcal{U}_{0,z}^{(2\theta(z)-1)},$ and $\mathcal{I}_{N,z}^{(1-2\theta(z-N))}$ : 
The extended subsets of generators of a nonlinear superalgebra.\\

\underline{\textit{Three-dimensional conformal mechanics in a monopole background}}:\\
$\nu=(eg)^2$: Here $e$ and $g$ are the particle's electric charge  
and the monopole's magnetic  charges,
 respectively.\\
$\alpha:$ The coupling of the  conformal mechanics potential.\\
$\vI_1$, $\vI_2$, $\va$ and 
$\va^\dagger$: Dynamical integrals for the case
$\alpha=\nu^2$. \\ 
$\vJ$: The Poincar\'e vector integral.\\
$T^{(ij)}, T^{[ij]}$: Symmetric and anti-symmetric tensor integrals.\\

\underline{\textit{A charge-monopole superconformal model}}\\
$\vK=\vJ+\frac{1}{2}\vsigma$: The total angular momentum.\\
$k=j\pm 1/2$: The eigenvalue of $\vK^2$.\\
$\pm \omega \vsigma\cdot \vJ$: The spin-orbit coupling.\\ 
$\Theta$, $\Theta^\dagger$, 
$\Xi$ and  $\Xi^\dagger$: Scalar intertwining operators.\\
$\mathcal{H}$ and 
$\breve{\mathcal{H}}:$ Pauli type supersymmetric Hamiltonians in exact and spontaneously broken phase.\\
$\mathcal{Q}$, $\mathcal{Q}^\dagger$, 
$\mathcal{W}$, $\mathcal{W}^\dagger$: Nilpotent fermionic operators. \\
$\mathcal{R}$,
$\mathcal{G}$ and $\mathcal{G}^\dagger$: The 
$R-$symmetry generators and the lowering and rising supersymmetric ladder operators. \\
$\mathcal{P}_\pm$: Projectors onto subspaces with fixed $k$. \\
$\mathcal{B}$ and $\mathcal{F}$: Generic bosonic and fermionic three-dimensional generators.\\
$\mathscr{B}$ and $\mathscr{F}$: Generic bosonic and fermionic one-dimensional generators.\\

\newpage 


\rfoot[e1]{}

\fancypagestyle{plain}{
\fancyfoot[R]{}
}

\pagestyle{fancy}

\tableofcontents 

\cleardoublepage 
\listoftables 

\cleardoublepage
\listoffigures 

\chapter*{Introduction}

\addcontentsline{toc}{chapter}{Introduction}

\setcounter{page}{1}
\pagenumbering{arabic}


\rfoot[e1]{\thepage}

\fancypagestyle{plain}{
\fancyfoot[R]{\thepage}
}

\pagestyle{fancy}

Symmetries play a very important role in 
the construction of  the fundamental theories that we have in 
physics nowadays.
Examples of that are the  general relativity and   
the Standard Model of particle physics, just to name a few. 
In this Thesis, we study hidden symmetries that control nontrivial aspects of 
classical dynamics, 
as well as spectral peculiarities in  quantum and supersymmetric quantum mechanics models.

From a classical mechanics perspective,  Noether's theorem reveals that behind the 
invariance of
 action under a symmetry transformation,  
 there exists a conservation law. In general,
the principle of least action  assumes the existence of a Lagrangian
$\mathcal{L}$, which in mechanics depends on the generalized coordinates and  its velocities. 
Geometrically, these coordinates belong to a configuration space $\mathcal{M}$, which points are usually denoted by $q$,
and their associated velocities are  vectors that live on
 the tangent space  
$T\mathcal{M}_{q}$ at $q$ , which in turns,
are generated by the action of a particular  tangent vector
field.  Then, 
naturally the 
Lagrangian is a function  on 
the tangent bundle   $T\mathcal{M}=\cup_{q\in \mathcal{M} }T\mathcal{M}_{q}$
 of 
$\mathcal{M}$
\textcolor{red}{[\cite{Nakahara,Sundermayer}]}. 
In this framework, 
symmetry
is a one-parametric
 transformation
generated by some conserved
  vector field.
To compare transformations associated with
 two different vector fields, say $X=X^\mu\frac{\partial}{\partial q^\mu}$
 and  $Y=Y^\mu\frac{\partial}{\partial q^\mu}$,
we compute the Lie derivative\footnote{The Lie derivative 
evaluates the change of a tensor field 
(including scalar functions, vector fields and one-forms), 
along the flow defined by another vector field \textcolor{red}{[\cite{Nakahara}]}.}
of $Y$ along the flow of $X$, denoted by
 $ \pounds_{X}Y$, and it is not difficult to show that 
 this operation reduces to the usual commutator between 
 two vector fields $[X,Y]\in T\mathcal{M}$.
This gives rise to a Lie algebra 
of vector fields on  $T\mathcal{M}$
 \textcolor{red}{[\cite{Nakahara}]}.

On the other hand,  when we go 
to the Hamiltonian formalism, the dynamical variables considered now
are
the generalized coordinates and their canonical momenta $p_i=\frac{\partial \mathcal{L}}{\partial \dot{q}_i}$. 
One can show that under a general change of coordinates, $p_i$ transform as the components of a vector 
in the cotangent space $T_*\mathcal{M}_{q}$ at $q$  \textcolor{red}{[\cite{Nakahara}]}.
Then
the phase space is naturally identified with 
 the cotangent bundle 
$T_*\mathcal{M}=\cup_{q\in \mathcal{M} }T_*\mathcal{M}_{q}$
 with 
local coordinates $(q^i,p_i)$ on it 
\textcolor{red}{[\cite{Arnold, Nakahara, Sundermayer}]}.
Here, the 
 symplectic form
$\omega=dq^i\wedge dp_i$
encodes the Poisson bracket structure. Namely,  
with
a given function $F=F(q,p)$  
on  the
 phase space, 
 a 
 Hamiltonian vector field,
$$X_F=
\frac{\partial F}{\partial p^i}\frac{\partial}{\partial q_i}
-\frac{\partial F}{\partial q^i}\frac{\partial}{\partial p_i}\,,$$ 
is associated, 
such that the contraction  $i_{X_F}\omega\equiv
\omega(X_F,.)=dF$.
 For two 
Hamiltonian vector fields $X_F$ and $X_G$,
it follows then that 
$ \pounds_{X_F}X_G=X_{\{F,G\}}$ and 
$ \pounds_{X_F}G=\{G,F\}$.
 If $ F $ 
is identified as the Hamiltonian of the system, then 
the
last relation corresponds to the equation of motion for $ G $
\textcolor{red}{[\cite{Arnold,Sundermayer}]}.  In this formalism, a  symmetry transformation is a  
flow produced by a Hamiltonian vector field whose generating function in phase space is conserved in time.

As  it is known, the Lie algebra mentioned above corresponds to a more abstract concept. A Lie group 
is a smooth manifold with an additional group structure, and 
 any Lie group gives rise to a Lie algebra, which is its tangent space at the identity \textcolor{red}{[\cite{Nakahara,Gilmore}]}. 
When a  group ``acts'' on some target space
 (that could be the same group manifold), 
an explicit form of its elements is required. 
This  
leads us to the representation theory. 
In  Hamiltonian classical mechanics,
 the target space 
is 
 $T_*\mathcal{M}$, 
the
 Lie algebra generators are identified 
with the Hamiltonian vector fields, and the group action corresponds to Hamiltonian flows.  
In the case of 
quantum  theory, we look, 
in accordance with
the celebrated  Wigner theorem \textcolor{red}{[\cite{Wigner0,Wigner,Waimberg1}]},
for irreducible 
unitary representations of the quantum symmetry group of the system, 
and 
target space 
is the Hilbert space generated by   
eigenstates of the quantum Hamiltonian operator.
In fact, the ``algebraic'' approach claims that the entire Hilbert space can be generated by the
 action of the symmetry operators on an arbitrary solution of the corresponding
 Schr\"odinger equation, i.e., the spectrum of the system is explained by symmetry.

Symmetries are intrinsic properties of the geometry that characterizes a given manifold.  
Suppose we have a space-time manifold with 
a metric structure 
 $ds^2=g_{\mu \nu}dx^{\mu}dx^{\nu}$.
 If $ds^2$ is invariant under a certain change of coordinates, we have an ``isometry'', which in accordance with 
 the discussion above, is generated by a particular vector 
 field, called Killing vector field  \textcolor{red}{[\cite{Nakahara}]}.
 We can ask for 
mechanical systems 
that 
respect the isometries  of the  space-time
where they live,  that
gives rise to
important physical consequences.  
For example,  the construction of an action principle in Minkowski space 
that is invariant under the Poincar\'e 
 group  transformations 
$x^\mu\rightarrow y^{\mu}=\Lambda^{\mu}_\nu x^\nu+a^\mu$, 
where $\Lambda^{\mu}_\nu $ are the Lorentz transformations,
is just the same as to 
impose
 the relativity postulates.
In this way,  Poincar\'e invariant 
quantum field theories involve in their description  field operators which provide  certain
representations of this symmetry group \textcolor{red}{[\cite{Waimberg2,Sundermayer}]}.  
The isometry condition for infinitesimal transformations  $x^\mu \rightarrow  x^\mu+\zeta^\mu$
corresponds to the Killing equation 
$$\frac{\partial g_{\mu\nu}}{\partial x^\lambda}\zeta^\lambda+
g_{\mu\lambda} \frac{\partial\zeta^\lambda}{\partial x^\nu}+
g_{\lambda\nu} \frac{\partial\zeta^\lambda}{\partial x^\mu}=0\,,$$
and for Poincar\'e  transformations in Minkowski space 
its solutions are given by 
$\zeta^\mu=a^\mu+\omega^{\mu\nu} x_\nu$,
where $\omega^{\mu \nu}$ is an antisymmetric matrix. 
To obtain the corresponding Killing vector fields
 we use the fact 
that  Poincar\'e transformations admit the unitary representation  
$\exp(i(a^\mu T_\mu-\frac{1}{2}\omega^{\mu\nu}M_{\mu\nu}))$, 
where $ T_\mu$ and $M_{\mu\nu}$ are our candidates for 
translations and Lorentz transformations 
generators,
respectively. To identify them we should compare
$y^{\mu}=x^\mu+\zeta^\mu$ with 
$$
\exp(i(a^\nu T_\nu-\frac{1}{2}\omega^{\alpha\beta}M_{\alpha\beta}))x^\mu \approx
x^\mu+i(a^\nu T_\nu-\frac{1}{2}\omega^{\alpha\beta}M_{\alpha\beta})x^\mu\,,
$$
which implies that 
$T_\mu=i\partial_\mu$ and $ M_{\mu\nu}=i(x_\mu\partial_{\nu}-x_\nu\partial_{\mu})+\Sigma_{\mu\nu}\,.$
Here $ \Sigma_{\mu \nu} $ are operators that do not act on the coordinates,
 but their representations tell us about the spin
 of the  corresponding 
 fields.

The notion of Killing vectors is generalized to the so-called conformal Killing vectors, 
which are related to 
 the 
 coordinate changes so 
 that $ ds^2 \rightarrow \Omega(x) ds^2 $, 
where $\Omega(x)$ is the conformal factor.
Such
transformations correspond, particularly,  to 
dilatations $x^\mu\rightarrow c x^\mu$ 
and special conformal transformations $x^\mu\rightarrow (x^\mu-b^\mu x^2)/(1-2b^\nu x_\nu+b^2x^2)$
  \textcolor{red}{[\cite{Francesco}]}. 

Conformal symmetry, as well as conformal field theories, 
have made an huge contribution on different aspects of physics, such as condensed matter, 
electrodynamics, and gravity, 
just to mention a few examples \textcolor{red}{[\cite{Ginsparg,Jackiw}]}.
The two-dimensional case is special in
 this context.
  Indeed, consider
the change of coordinates 
 $$x^1\rightarrow x^1+f_1(x^1,x^2)\,,\qquad x^2\rightarrow x^2+f_2(x^1,x^2)\,,$$
in flat space. This transformation can be shown to be of the conformal type
 if and only if $ f_1 (x ^ 1, x ^ 2) $ and $ f_2 (x ^ 1, x ^ 2) $ 
satisfy the Cauchy-Riemann equations, i.e., they are the real and imaginary parts of a 
holomorphic function. In the case of infinitesimal 
transformations,  however, we can be less restrictive.  
To see this better, it is natural to 
take the complex 
coordinate
$z=x^{1}+ix^{2}$, together with its 
 complex conjugate $\bar{z}$, and consider 
the infinitesimal transformation 
$z\rightarrow z+\varepsilon(z)$, 
where $\varepsilon(z)$ is assumed to be a meromorphic function which admits a Laurent expansion around $z=0$.
In this situation 
a (primary) field $ \phi (z, \bar{z}) $ 
infinitesimally transforms as
$\delta \phi=-(\varepsilon \partial_z+\bar{\varepsilon} \partial_{\bar{z}})\phi$,
 from where 
 we identify 
 the symmetry generators 
$l_n= -z^{n + 1}\partial_z $ and $ \bar{l}_n=-\bar{z}^{n + 1} \partial_{\bar{z }} $, with 
$ n \in \Z $. They produce a direct sum of two copies of the 
infinite-dimensional
Witt algebra, 
while the global conformal group that maps the complex plane onto itself 
is obtained from the subalgebra
$\mathfrak{sl}(2,\mathbb{C})=\mathfrak{sl}(2,\R)\oplus\mathfrak{sl}(2,\R)$, 
which, in
turn,
is generated by
$\{l_0,\, \bar{l}_0 ,\, l _\pm,\,\bar{l} _\pm\}$, \textcolor{red}{[\cite{Francesco}]}. 
Using these properties one can introduce 
a conformal field theory 
that does not even need a specific action principle.
 This corresponds to 
the so-called conformal bootstrap \textcolor{red}{[\cite{Polyakov1}]}. This type of theories, 
``minimal models'', as they are often called \textcolor{red}{[\cite{Polyakov2, Francesco}]}, 
appears in the study of critical points in the
second-order phase transition phenomena, 
and their main advantage is the calculation of the correlation functions of 2 and 3 points,
only by symmetry arguments.
On the other hand, conformal theories in higher dimensions became popular 
after Maldacena's famous article \textcolor{red}{[\cite{Maldacena}]}, where a duality between a gravity
theory in AdS (type $ IIB $ string theory in $ AdS_5 \cross S^5 $) and a conformal field theory in the boundary 
($ \mathcal{N} = 4 $ 
supersymmetric
Yang-Mills) was shown.
This AdS/CFT correspondence along with holographic techniques have 
found applications not only in black holes physics but also in other areas such as QCD 
\textcolor{red}{[\cite{Ammon,App1,Brod2}]}.

Beyond the Standard Model it has been postulated supersymmetry,    
 based on transformations
that relate
 bosons and fermions 
\textcolor{red}{[\cite{Waimberg3}]}. 
These models refer to an action principle defined in the 
``super-space'', 
which is a place where bosonic and fermion quantities (described by Grassmann's variables) live together. 
To overcome the Coleman-Mandula theorem: 
``space-time and internal symmetries cannot be combined in any but a trivial way'', 
see \textcolor{red}{[\cite{Pelc}]}, the concept of symmetry is generalized to 
a $ \Z_2 $-graded 
algebra, 
or superalgebra, which is characterized by the supercommutator $ [ [A, B]] $,
\begin{itemize}
\item $ [[A,B]]=[A,B]$ if $A$ and $B$ are bosonic generators,
\item $ [[A,B]]=[A,B]$ if 
one generator is bonosic while another is fermionic,
\item $ [[A,B]]=\{A,B\}=AB+BA$ if both generators are fermionic.
\end{itemize}
To discriminate between bosonic and fermionic objects it is necessary 
to introduce a grading operator $ \Gamma $, $\Gamma^2=1$,  that commutes with all bosonic generators and anti-commutes 
with fermionic
 ones. The conserved 
 quantities that generate
 the supersymmetric transformations are called supercharges 
and are the fermionic operators. 
For the study of supersymmetry
outside the framework of 
quantum field theory, 
the concepts of pseudo-classical mechanics \textcolor{red}{[\cite{psedoclasical2,psedoclasical1,Casalbuoni}]}
and
its
 quantum version, supersymmetric quantum mechanics 
\textcolor{red}{[\cite{Witten1, Witten2,Cooper}]},
were introduced.
The latter has become an invaluable tool
in the study of solvable potentials, and is closely related to the theory of  
integrable classical field systems
 and their solitonic and finite-gap type solutions \textcolor{red}{[\cite{MatSal}]}. 
 Details of this formalism are presented in the next chapter.

At this point it is clear that symmetries govern physics,
 and in this context, the notion of hidden symmetries becomes relevant \textcolor{red}{[\cite{Cariglia}]}.
To explain it, let us consider again classical mechanics.
If,  regardless of the initial conditions, it happens that the nature of the trajectories in some system is 
``special '' (in a geometric sense), 
this 
should indicate on the presence of the hidden symmetries. Form the perspective of symmetry 
transformations, these objects  mix the coordinate and velocity (momenta) variables in Lagrangian (Hamiltonian) formalism.
 At the quantum level, 
 hidden
 symmetries can explain 
peculiar
 properties of the physical spectrum,
 such as a degeneration.
Take, for example, the case of the Kepler-Coulomb problem, where  
we know that the system is invariant under rotations and that the particle 
trajectories, being conical sections, lie 
in the plane orthogonal to the angular momentum vector. 
We also know that the geometric properties are determined by the energy and the angular momentum itself,
 but there is one more special property,
 the orientation of the trajectory,
 which is given by the so-called Laplace-Runge-Lentz vector 
 to be the second-order  in canonical momenta quantity. 
 This vector
  integral is also
  relevant at a quantum level because
 it explains the ``accidental'' degeneration in the spectrum of the 
 hydrogen atom model \textcolor{red}{[\cite{Pauli}]}.
From now on,  the nonlinear  
  in canonical momenta  integrals of motion 
different from Hamiltonian,
like the mentioned Laplace-Runge-Lentz vector,
 will be called hidden symmetries. To study the geometric interpretation of these objects, which are usually related to Killing tensors and conformal Killing tensors 
\textcolor{red}{[\cite{Cariglia}]}, a good approach corresponds to the Eisenhart-Duval lift \textcolor{red}{[\cite{Cariglia2}]}, 
the
 procedure by which classical trajectories are identified 
 with the null geodesics of a non-trivial  geometry  with two extra dimensions. 
Some other well known 
examples where these objects play a key role 
are the three-dimensional isotropic
harmonic oscillator \textcolor{red}{[\cite{Jauch,Frad}]}, 
the anisotropic harmonic oscillator \textcolor{red}{[\cite{Bonatsos,deBoer}]},
the Higgs oscillator \textcolor{red}{[\cite{Zhedanov, Evnin}]}, nonlinear supersymmetry 
\textcolor{red}{[\cite{Plyunonlinear}]}
and a charged particle in a monopole background \textcolor{red}{[\cite{PlyWipf,InzPlyWipf2}]}. 

The hidden symmetries satisfy nonlinear algebras in the general case. 
The first examples of nonlinear algebras introduced in field theory literature 
 were the infinite $W$ algebras  \textcolor{red}{[\cite{Zamolodchikov}]},
 which are necessary to study the nature of the infinite-dimensional groups that appear in two-dimensional 
conformal models.
The listed  above   systems
 are examples of elementary models whose 
associated integrals of motions satisfy finite $W$ algebras, which in turns, have played a relevant role in understanding of their infinite counterpart \textcolor{red}{[\cite{deBoer}]}.

In the particular case of 
 one-dimensional quantum mechanics, the
  supersymmetric algorithm allows us to build families of solvable potentials that have spectral peculiarities, perfectly encoded in hidden symmetries.
 A good example of this are the rational deformations of the harmonic oscillator, 
characterized by a potential of the form $ x^2-2 \ln (W (x))'' $, where $W (x)$ is a regular polynomial 
on the real line
\textcolor{red}{[\cite{Krein,Adler}]}.
Systems of this nature find importance in the field of exceptional orthogonal polynomials, see 
for example
\textcolor{red}{[\cite{Dubov,Quesne2012,Gomez2}]}. 
 The corresponding spectrum  of  
 this kind of systems is divided into $ g $ subsets of equidistant energy levels, isolated from each other. 
The first $(g-1)$ subsets, or bands, have a finite number of levels, while the last band has
infinite number of  equidistant discrete levels.
In \textcolor{red}{[\cite{CarPly}]}, the spectrum-generating ladder operators for these systems were built, and they turned out to be higher order symmetry operators.

This Thesis reviews in a self-contained manner
the results obtained  within
the framework of a 
three-years research project,  
in which
we address the following  problems: 
\newpage
\underline{\emph{a) Connection between different mechanical systems through symmetries}} \\
The 
$\mathfrak{so}(2,1)$ conformal algebra 
$$
[D,H]=iH\,,\qquad
[D,K]=-iK\,,\qquad
[K,H]=2iD\,,
$$
describes different quantum systems with continuous spectrum, that is, $ H $ could represent the Hamiltonian of a free particle, Calogero models, 
monopole-charge system, etc. This algebra is isomorphic to 
the $ \mathfrak{sl} (2, \R) $
algebra,
$$
[\mathcal{J}_0,\mathcal{J}_\pm]=\pm \mathcal{J}_\pm\,,\qquad
[\mathcal{J}_-,\mathcal{J}_+]=2 \mathcal{J}_0\,,
$$
where $ \mathcal {J}_0 $ is a compact generator that represents the Hamiltonian of a confined system, such as the harmonic oscillator.
We address the problem of establishing a mapping between 
these two forms of dynamics associated with conformal algebra. Such a transformation 
would be useful, particuarly, 
for mapping conserved quantities that are easier to identify for one system than for the other.

\underline{\emph{b) Hidden and  bosonized supersymmetry}}\\
In quantum mechanics, the reflection operator $ \mathcal{R} $ is defined by $ \mathcal{R} x = -x \mathcal{R} $ and 
$ \mathcal{R} p = -p \mathcal {R} $.
If we choose the supersymmetric grading operator $\Gamma$ to be
  $ \mathcal{R} $, we can construct 
  bosonized  supersymmetric systems 
\textcolor{red}{[\cite{Boson1}; $\quad$ \cite{PlyPara};$\quad$ \cite{Gamboa2}; $\quad$ \cite{CorNiePly}; 
\cite{Boson2, Boson3, jakubsky}]}
 which do not employ fermionic  
   degrees of freedom. 
We focus on the origin of the hidden 
bosonized superconformal symmetry of the harmonic oscillator in one dimension 
\textcolor{red}{[\cite{Hiden1,BalSchBar,CarPly2,Hiden3}]}, 
that is, we build an unconventional supersymmetric system that, after   nonlocal   transformation of the 
Foldy-Wouthuysen type
  and a dimensional reduction \textcolor{red}{[\cite{jakubsky}]}, produces the 
    superalgebra we are looking for.

\underline{\emph{c) Hidden symmetries in rationally extended conformal mechanics}}\\
 The simplest conformal 
invariant system that one can construct is 
$$
S=\int_{t_1}^{t_2}\left(\frac{1}{2}\dot{q}^2+\frac{g}{2 q^2}\right) dt\,,\qquad q>0\,,
$$
where $g$ is a dimensionless constant that should be 
 non-negative
   in classical mechanics and $g\geq -1/4$
at   the quantum level. 
This model does not have a well-defined invariant ground state and to
eliminate this deficiency, 
de Alfaro, Fubini, and Furlan used a particular coordinate 
 and time change to transform
the latter action 
into
$$
S=\int_{\tau _1}^{\tau_2}\left(\frac{1}{2}\dot{y}^2+\frac{g}{2 y^2}+\frac{\omega^2}{2}y^2 \right)d\tau\,,\qquad x>0\,.
$$
The corresponding 
Hamiltonian is compact and has a well-defined ground state at the quantum level, see \textcolor{red}{[\cite{AFF}]}.
This system, called the de Alfaro, Fubini and Furlan model 
(AFF), and its 
supersymmetric extensions  \textcolor{red}{[\cite{SCM1,SCM2,SCM3,SCM4,SCM5}]}
have attracted   a great
attention over the years in a  variety of fields such as
 particles dynamics
 in black hole backgrounds 
 \textcolor{red}{[\cite{BlackHold1};$\,\,$ \cite{BlackHold2}; $\,\,$\cite{BlackHold3}; $\,\,$\cite{BlackHold5};$\quad$ \cite{Galaj}]},
 $\quad$ cosmology $\quad$\textcolor{red}{[\cite{DGH};$\quad$ \cite{PioWal}]},$\qquad$
 nonrelativistic $\qquad\,$ AdS/CFT $\qquad\,$ correspondence $\qquad\,$ \textcolor{red}{[\cite{GAdS1};$\,\,$\cite{GAdS2, BarFue, Jack}]}, 
QCD confinement problem
 \textcolor{red}{[\cite{App1,Brod2}]},
  physics of Bose-Einstein condensates  \textcolor{red}{[\cite{App2,App3}]} and anyon statistics \textcolor{red}{[\cite{leinaas1,leinaas2, mackenzie}]}.
We apply the generalized Darboux-Crum-Krein-Adler transformation (DCKA)
 \textcolor{red}{[\cite{Moutard1,Moutard2,Darboux,Crum,Krein,Adler,MatSal}]} to the AFF model to
construct
  rational deformations of this system. The objective is to follow the approach
 given in \textcolor{red}{[\cite{CarPly}]} to find the ladder operators that generate spectrum of these systems.

\underline{\emph{d) Hidden symmetries in three-dimensional conformal mechanics}}\\
Consider a charged particle moving in a magnetic field generated by a Dirac monopole, i.e., 
in a monopole background \textcolor{red}{[\cite{Sakurai}]}, which is also 
  subject
to a central potential of the form
$ V (\vr) = \frac {\alpha}{2m \vr ^ 2} $. In 
\textcolor{red}{[\cite{PlyWipf}]}
it   had
already been shown that the system has hidden symmetries when $ \alpha = (eg)^2 $, where 
$ e $ and $ g $ are the 
particle's electric
   charge and the monopole   magnetic charge, respectively.
It was also shown that the system allows   an
$ \mathcal{N} = 4 $ supersymmetric extension. 
We investigate the possibility of obtaining hidden integrals of motion when the central potential is changed 
for
$ V (\vr) = \frac{\alpha}{2m \vr ^ 2} 
+\frac {m \omega \vr^2}{ 2} $, and we look for possible supersymmetric extensions. 
The results obtained from problem {\emph{a)} are used to investigate this problem.

The
results of investigation
  of the  listed problems
  were 
reported in the articles \textcolor{red}{[\cite{CarInzPly,InzPly1,InzPly2,InzPly3,InzPlyWipf1,InzPlyWipf2}]}.
  
 The
subsequent  main part  of the   Thesis
is organized as follows.
 In Chap. \ref{ChSUSY} we review the supersymmetric quantum 
mechanics formalism as well as the generalized 
Darboux transformations
and   their 
confluent extensions.  
In Chap. \ref{ChConformal} we revisit the one-dimensional
conformal mechanics model  of
de Alfaro, Fubini and Furlan \textcolor{red}{[\cite{AFF}]}, as well as
its $\mathcal{N}=2$ supersymmetric extension, leading us  to  
the $\mathfrak{osp}(2,2)$ superconformal symmetry.  
In Chap. \ref{ChBridge}, based on \textcolor{red}{[\cite{InzPlyWipf1}]},
we consider 
the conformal bridge transformation and 
its applications to models in one and two dimensions.  In Chap. \ref{ChHiddenboson}, 
we explain the origin of the
 hidden bosonic superconformal symmetry of the harmonic oscillator 
 \textcolor{red}{[\cite{InzPly1}]}. 
In Chap. \ref{ChRQHO} we 
review the results of ref.  \textcolor{red}{[\cite{CarInzPly}]}, where 
 rational extensions of the conformal mechanics model characterized by the  
 potential $ \frac{m(m+1)}{x^2}$ 
 with 
  $m=1,2,\ldots$, 
 as well as its spectrum-generating ladder operators are constructed. 
 In Chap. \ref{ChNonLinearSUSY}, following \textcolor{red}{[\cite{InzPly2}]},
we consider 
supersymmetric
 extensions of the 
rationally deformed system of 
 Chap. \ref{ChRQHO}, 
 as well as its complete spectrum-generating nonlinear superalgebra. 
In Chap. \ref{ChKlein} we 
exploit 
  a discrete Klein four-group  symmetry 
  of the 
  Schr\"odinger
   equation for the AFF model  to 
 generalize the construction of 
rationally
   extended systems  
 and the spectrum-generating ladder 
operator sets
  for the case in which 
integer parameter $m$ is replaced 
    by a
    real number 
$\nu\geq-1/2$
\textcolor{red}{[\cite{InzPly3}]}. 
In this case, the confluent Darboux 
transformations appear naturally. 
Chap. \ref{Chapmono1} and \ref{Chapmono2} 
  are devoted to investigation, 
    in the  light  of hidden symmetries,
   of the
   conformal mechanics in a monopole background
   as well as its  supersymmetric extension, 
   which is characterized by a three-dimensional realization 
   of the $\mathfrak{osp}(2,2)$ superconformal symmetry \textcolor{red}{[\cite{InzPlyWipf2}]}.  
   The Thesis ends with its Conclusion and Outlook. In Appendix, some technical 
   details are collected.

%

\chapter{Supersymmetric quantum mechanics }
\label{ChSUSY}

The application of supersymmetric ideas in nonrelativistic quantum mechanics
has given us a better understanding of the problem of solvable potentials and 
its associated hidden symmetries.  
In this context, the main technique is the factorization method \textcolor{red}{[\cite{Infeld,Cooper}]}, which  relates 
a particular quantum mechanical system with another one (the so-called superpartner).
In the one-dimensional case, the formalism of construction of such  operators (starting from a 
well known quantum system) receives 
the name of Darboux-Crum-Krein-Adler  transformation  \textcolor{red}{[\cite{Moutard1,Moutard2,Darboux,Crum,Krein,Adler,MatSal}]}.
An algorithmic procedure involves a given number of
eigenstates of the original system, typically called   
``seed states'', and in its confluent extension 
Jordan states are also considered \textcolor{red}{[\cite{Jordan2,Jordan1,Jordan3}]}. 
In this chapter we revisit these methods.

Generalization $\,\,$to $\,\,$higher $\,\,$spatial $\,\,$dimensions $\,\,$can $\,\,$be $\,\,$reformulated 
$\,\,$in $\,\,$different $\,\,$ways, $\,\,$see$\,\,$
\textcolor{red}{[\cite{HSUSY1,HSUSY2,HSUSY3,HSUSY4,HSUSY5,HSUSY6}]}.
In this Thesis, we just consider the approach of a
given Dirac Hamiltonian, whose square produces a supersymmetric Hamiltonian operator
\textcolor{red}{[\cite{Cooper}]}.

\section{The one-dimensional case}

In one-dimensional systems, 
the factorization method consists in introducing intertwining operators 
of the form 
\begin{equation}
A=\sqrt{\frac{\hbar^2}{2m}}\frac{d}{dx}+W(x)\,,\qquad
A^\dagger=-\sqrt{\frac{\hbar^2}{2m}}\frac{d}{dx}+W(x)\,,
\end{equation} 
which satisfy
\begin{eqnarray}
\label{susy1}
&
AA^\dagger=H_+\,,\qquad
A^\dagger A=H_-\,,\qquad
H_\pm=-\frac{\hbar^2}{2m}\frac{d^2}{dx^2}+W(x)^2\pm \frac{\hbar}{\sqrt{2m}}W(x)'\,.
\end{eqnarray}
Here $W(x)$ is called  superpotential and $H_{\pm}$ are the superpartner systems.
Now, let us assume we know a function $\psi_*$
such that $H_-\psi_*=0$. This defines a nonlinear Riccati type equation for $W$
\begin{equation}
W(x)^2-\frac{\hbar}{\sqrt{2m}}\frac{dW}{dx}=u(x)\,,\qquad u=(x)= \frac{\hbar^2}{2m}\frac{1}{\psi_*}\frac{d^2}{dx^2}\psi_*\,,
\end{equation}
and a particular solution of which is  
\begin{equation}
\label{intert1}
W(x)=-\frac{\hbar}{\sqrt{2m}}\frac{\psi_{*}'}{\psi_{*}}=-\frac{\hbar}{\sqrt{2m}}\ln(\psi_*)'\quad
\Rightarrow\quad
A=\frac{\hbar}{\sqrt{2m}}\left(\frac{d}{dx}-\ln(\psi_*)'\right)=
\frac{\hbar}{\sqrt{2m}}\psi_*\frac{d}{dx}\frac{1}{\psi_{*}}\,,
\end{equation} 
which in turns implies $A\psi_{*}=0$. 
This result allows us to conclude the following: For a given well 
known physical system we can select one of the two linear independent 
(formal) zero
 energy solutions to recognize the associated superpotential 
and use it to construct a new quantum mechanical system given by $H_+$. 
From the first equation in (\ref{intert1}) 
it can be concluded that 
$\psi_*$ must not have zeros 
in the domain of $H_-$ to obtain a \emph{priori},
 a regular superpotential and a \emph{posteriori}, a well defined 
superpartner in the same domain.   
If the selected state does not fulfill  this condition we call the resulting system 
as a ``virtual system'' that makes no physical sense\footnote{Such virtual systems are useful in the context of higher order supersymmetry, see for example  
\textcolor{red}{[\cite{Arancibia,Plyushchay,CarInzPly}]}.
}.
One typically refers to $\psi_*$ as a seed state.

On the other hand, the action of operator 
$A$ on other eigenfunctions of $H_-$ produces eigenstates of $H_+$.
 To show this
statement we use Eq. (\ref{susy1}) to deduce the intertwining relations 
\begin{equation}
\label{Intertwiningrelation0}
AH_-=H_+A\,,\qquad
A^\dagger H_+=H_-A^\dagger\,.
\end{equation}
 Then, if $\psi_\lambda$ is an eigenstate of $H_-$ with eigenvalue $\lambda$ we get 
 \begin{equation}
 H_-\psi_{\lambda}=\lambda\psi_{\lambda}\qquad\Rightarrow\qquad
 H_+\,(A\psi_{\lambda})=\lambda\,(A\psi_{\lambda})\,.
 \end{equation}
  Of course these  relations also work  for the second linearly independent solution of the form 
 \begin{equation}\label{tildepsi}
\widetilde{\psi}_{\lambda}=\psi_{\lambda} \int^x\frac{d\zeta}{(\psi_{\lambda}(\zeta))^2}\,,
\end{equation}
which together with $\psi_\lambda$ satisfies $W(\psi_\lambda,\widetilde{\psi_\lambda})=1$, where $W(.,.)$ is the Wronskian of two functions. 

It is not difficult to show that operator $A^\dagger$ annihilates the state
  $A \widetilde{\psi_*}=1/\psi_*$ 
which is one of the zero eigenvalue solutions of $H_+$\footnote{Note that with this method we only obtain one 
of the two linear independent solutions since $A\psi_*=0$. To obtain the second linear independent solution we should 
extend the transformation by applying it to  Jordan states.}. Knowing this, 
one can say something about the spectrum of the latter Hamiltonian
in correspondence with the behavior of the seed state.
First, acting on physical states of $H_-$, operator $A$ produces physical states of $H_+$ and second,
if $\psi_*$ is a physical state, then the spectrum of $H_+$ does not have this energy level. On the other hand,
if the seed state is nonphysical, two things could happen: 1) $1/\psi_*$ is normalizable and
system $H_+$ possesses an extra level and 2)  $1/\psi_{*}$ is nonphysical and both systems are isospectral.

Finally, suppose we have a given number of differential operators
 denoted by $ I_i $, each of them of a certain differential order $ d_i $,
which together with $H_-$
span a symmetry algebra. In this context, it is not difficult to show the relation 
$[H_+,AI_iA^\dagger]=A[H_-,I_i]A^\dagger$, which means that when operator $I_i$ is the
  integral
of motion of $H_-$, then $A(I_i)A^\dagger$ (of differential 
order $d_i+2$) is the integral for $H_+$ and the
system is described (in the general case) by a certain nonlinear deformed algebra. 
In conclusion, the method not only serves to map states but also to 
 obtain hidden integrals of motion of the generated system.
This procedure is known as ``Darboux-dressing''.

We have the complete picture to extend our superpartner systems to supersymmetric quantum mechanics. 
We use our Hamiltonians and  intertwiners 
to construct the $2\cross 2$ matrix  operators 
\begin{equation}
\mathcal{H}=\left(\begin{array}{cc}
H_+ & 0\\
0 & H_-
\end{array}\right)\,,\qquad
\mathcal{Q}_1=\left(\begin{array}{cc}
0 & A\\
A^\dagger & 0
\end{array}\right)\,,\qquad
\mathcal{Q}_2=i\sigma_3\mathcal{Q}_1\,,
\end{equation}   
 which satisfy the $\mathcal{N}=2$\footnote{Here $\mathcal{N}$ indicates the number of true fermionic integrals.} Poincar\'e superalgebra
\begin{equation}
\label{Poincare0}
[\mathcal{H},\mathcal{Q}_{a}]=0\,,\qquad
\{\mathcal{Q}_a,\mathcal{Q}_{b}\}=2\delta_{ab}\mathcal{H}\,,
\end{equation}
 with $\Z_2$ grading operator $\Gamma=\sigma_3$. In the case in which the state
 $\psi_*$ (or $1/\psi_*$) is the physical
 ground state of $H_-$ ($H_+$),
  then the spinor $(0,\, \quad \psi_*)^t$ (or $(1/\psi_* ,\,  \quad 0)^t$) is the supersymmetric 
  invariant ground state 
 of $\mathcal{H}$.  Otherwise  supersymmetry is spontaneously broken.  
 
The method described in this paragraph is called the Darboux transformation
 and is the first step in an iterative process. 
In the next step we can produce a third new Hamiltonian by taking a seed 
state from $H_+$ and so on. The final form of the method after a several number of steps 
 is called the ``Darboux-Crum-Krein-Adler  transformation''
 (DCKA transformation for short) whose details 
 are explored in the following section.    

\section{DCKA transformation}
\label{Chap1Darbux}
Let us start with the equation 
\be
\label{Sch}
L\psi_\lambda=\lambda\psi_\lambda\,,\qquad
L=
-\frac{d^2}{dx^2}+V(x)\,,
\ee
corresponding to the eigenvalue problem of 
a Schr\"odinger type operator $L$. 
In this paragraph
we treat Eq. (\ref{Sch}) as a formal second order differential equation
on some interval $(a,b)$. Consider now a set of solutions $\psi_{k}$
corresponding to eigenvalues $\lambda_{k}$,
$k=1,\ldots,n$. We use them as seed states 
for our DCKA transformation  and  generate the 
new   Schr\"odinger  operator
\be
\label{Dar}
\breve{L}\Psi_{\lambda}=
\lambda\Psi_\lambda\,,\qquad
\breve{L}=
-\frac{d^2}{dx^2}+V(x)-2\frac{d^2}{dx^2}
\ln W(\psi_{1},\ldots,\psi_{n})\,.
\ee
If the set of the seed states is chosen in such a way 
that the generalized Wronskian of $n$ functions 
\be
W(f_1(x),\ldots,f_n(x))=\text{det}\left(\frac{d f_{i}(x)}{dx^{j-1}}\right)\,,\qquad i,j=1\ldots,n\,,
\ee 
takes nonzero values on $(a,b)$,
then the potential of the generated system will also be nonsingular 
there.
 In general case,  
solutions of (\ref{Dar}) are obtained 
from  solutions of Eq. (\ref{Sch}) as follows
\begin{equation}
\label{Darstates}
\Psi_{\lambda}=\frac{W(\psi_{1},\ldots,\psi_{n},\psi_{\lambda})}{W(\psi_{1},\ldots,\psi_{n})}=\A_{n}\psi_{\lambda}\,,
\end{equation}
where $\A_{n}$ is the  differential operator of
order $n$  defined recursively as 
\be
\label{generic-inter}
\A_{n}=A_n\ldots A_1\,,\qquad A_k=\A_{k-1}\psi_k\frac{d}{dx}\left(\frac{1}{\A_{k-1}\psi_k}\right),
\qquad k=1,\ldots,n,\qquad \A_0=1\,.
\ee
Note that this operator is the natural generalization of 
 (\ref{intert1})
with $\hbar/\sqrt{2m}=1$ and 
by the construction,  $\ker\A_n=\text{span}\{\psi_{1},\ldots,\psi_{n}\}$.
Operator $\A_n$ and its Hermitian conjugate $\A_n^\dagger$ intertwine the operators 
$L$ and $\breve{L}$,
\be
\label{inter-gen}
\A_nL=\breve{L}\A_n\,,\qquad \A_n^\dagger \breve{L}=L\A_n^\dagger\,,
\ee
and satisfy relations 
\be
\label{poly1}
\A_n^\dagger\A_n=\prod_{k=1}^{n}(L-\lambda_k)\,,\qquad
\A_n\A_n^\dagger=\prod_{k=1}^{n}(\breve{L}-\lambda_k)\,.
\ee
From the first equation in (\ref{poly1}) one can find that
$\ker\A_n^\dagger=\text{span}\{\A_n\widetilde{\psi}_{1},\ldots,\A_n\widetilde{\psi}_{n}\}$.
Similarly to (\ref{Darstates}), $\A_{n}^\dagger \Psi_{\lambda}=\psi_{\lambda}$
for  $\Psi_\lambda\notin \text{ker}\,\A_{n}^\dagger$,
and 
$$
\A_{n}^\dagger \widetilde{(\A_n\widetilde{\psi}_k)}=\psi_k\in \text{ker}\,\A_n\,.
$$

Following the same approach as in the previous section,
we can also use  
 the pair $L$ and $\breve{L}$ and their corresponding intertwining operators 
to construct an $\mathcal{N}=2$ superextended  system 
described by the $2\times 2$ matrix Hamiltonian  and the supercharges
given by
\be\label{Hlambda*}
\mathcal{H}=
\left(
\begin{array}{cc}
H_1\equiv  \breve{L}-\lambda_*&    0 \\
0 & H_0\equiv  L-\lambda_*     
\end{array}
\right),\qquad
\mathcal{Q}_1=
\left(
\begin{array}{cc}
  0&    \A_n \\
\A_n^\dagger &  0     
\end{array}
\right),
\qquad 
\mathcal{Q}_2=i\sigma_3\mathcal{Q}_1\,,
\ee
where $\lambda_{*}$ is a constant associated
 with the energy levels of the seed states.
  These generators produce 
the (anti)commutation relations 
\be\label{N2susy}
[\mathcal{H},\mathcal{Q}_a]=0\,, \quad
\{\mathcal{Q}_a,\mathcal{Q}_b\}=2\delta_{ab}P_n(\mathcal{H}+\lambda_*)\,,
\ee
which for $n=1$ correspond to an $\mathcal{N}=2$ Poincar\'e
supersymmetry and for $n=2,3,\ldots$, we have a nonlinear deformation of the latter supersymmetry 
(here $P_{n}(\eta)$ represents a polynomial of order $n$ in $\eta$). Examples of this kind of systems 
will be the main focus in Chap. \ref{ChNonLinearSUSY}.

The iterative nature of DCKA transformation allows 
us to derive some useful Wronskian identities 
for a given set of eigenstates. They are shown  
in Appendix \ref{ApenWI}.

\section{Jordan states and confluent Darboux transformation}
 Jordan states correspond to functions that are annihilated by a certain polynomial of the 
Schr\"odinger operator $L$ \textcolor{red}{[\cite{Jordan1}]}.
They  were used, for 
example, in the construction 
of  isospectral  
$\quad$deformations $\quad\,$of $\quad\,$ the$\quad\,$ harmonic$\quad\,$ oscillator$\quad\,$ \textcolor{red}{[\cite{Car2,CarPly}; \cite{InzPly1}]},
and also they can be used to construct solutions 
of the KdV equation \textcolor{red}{[\cite{Correa2016,JM1}]}.
These Jordan states will play a key role throughout 
this manuscript. This time we will focus our attention on 
building solutions of the fourth order differential equation
$(L-\lambda_*)^2\chi_*=0$. 

Let us take an  eigenstate $\psi_*$ with eigenvalue $\lambda_*$ 
as a seed state of the  Darboux transformation. The corresponding intertwining operators 
are
\be
A_{\psi_*}=\psi_*\frac{d}{dx}\left(\frac{1}{\psi_*}\right)\,,\qquad
A_{\psi_*}^\dagger=-\frac{1}{\psi_*}\frac{d}{dx}\psi_*\,.
\ee  
According to Eq.  (\ref{poly1}),  their product gives us the shifted Schr\"odinger operator 
$A_{\psi_*}^\dagger A_{\psi_*}=L-\lambda_*$, whose kernel is spanned by
the linear independent states $\psi_*$ and $\widetilde{\psi}_*$. 
The problem of constructing Jordan states reduces then  to solving 
equations 
\be\label{Omega12}
A_{\psi_*}^\dagger A_{\psi_*}\Omega_*=(L-\lambda_*)\Omega_*=\psi_*\,,\qquad
A_{\psi_*}^\dagger A_{\psi_*}\breve{\Omega}_*=(L-\lambda_*)\breve{\Omega}_*=\widetilde{\psi}_*\,.
\ee
Their solutions are given, up to a linear combination of
 $\psi_*$ and  $\widetilde{\psi}_*$,  by particular solutions of respective
  inhomogeneous 
 equations, 
\begin{eqnarray}
\label{omega1}
\Omega_*=\psi_*\int_{a}^{x}\frac{d\zeta}{\psi_*^2(\zeta)}\int_{\zeta}^{b}\psi_*^2(\eta)d\eta
\,,\qquad
\breve{\Omega}_*=\psi_*\int_{a}^{x}\frac{d\zeta}{\psi_*^2(\zeta)}\int_{\zeta}^{b}\psi_*
(\eta)\widetilde{\psi}_*(\eta)d\eta\,.
\end{eqnarray}
Here the  integration limits  are chosen coherently 
with  the region where  the operator $L$  is defined,
and we have the relations
\begin{eqnarray}
\label{JorWronskian}
W(\psi_*,\Omega_*)=\int_{x}^{b}\psi_{*}^2d\zeta\,,\qquad
W(\psi_*,\breve{\Omega}_*)=\int_{x}^{b}\psi_{*}\widetilde{\psi}_*d\zeta\,,
\end{eqnarray}
which will be useful to produce nonsingular confluent 
Darboux transformations.

Let us  inspect  now the role of Jordan states (\ref{omega1})  in  DCKA transformation 
generated by a set of the seed states $\{\psi_n\}$.     
The intertwining operator (\ref{generic-inter}) and 
Eqs. (\ref{inter-gen}) and (\ref{Omega12}) give us the relations
\be
\label{Dar-jor}
\A_n\psi_{*}=(\breve{L}-\lambda_{*})\A_n\Omega_{*}\,,\qquad
\A_n\widetilde{\psi}_{*}=(\breve{L}-\lambda_{*})\A_n\breve{\Omega}_{*}\,.
\ee
If the state $\psi_*$ (or $\widetilde{\psi}_*$) is annihilated by 
$\A_n$, i.e., if the set of the seed states $\{\psi_n\}$ includes  $\psi_*$ (or $\widetilde{\psi}_*$),
the function  $\A_n\Omega_{*}$ (or $\A_n\breve{\Omega}_{*}$) 
will be an eigenstate of $\breve{L}$ with eigenvalue $\lambda_*$
which is available to produce another Darboux transformation 
if we consider  $\breve{L}$ as an intermediate system.   
Otherwise,  
the indicated  function is a Jordan state of 
$\breve{L}$, and in correspondence  with (\ref{omega1}) we have 
\begin{eqnarray}
\label{omega2}
&\A_n\Omega_*=(\A_n\psi_*)\int_{a}^{x}\frac{d\zeta}{(\A_n\psi_*)^2(\zeta)}\int_{\zeta}^{b}(\A_n\psi_*)^2(\eta)d\eta
\,,\qquad&\\&
\A_n\breve{\Omega}_*=(\A_n\psi_*)\int_{a}^{x}\frac{d\zeta}{(\A_n\psi_*)^2(\zeta)}\int_{\zeta}^{b}(\A_n\psi_*)
(\eta)\widetilde{\A_n\psi_*}(\eta)d\eta\,&
\end{eqnarray}
up to a linear combination with $\A_n\psi_*$ and  $\widetilde{\A_n\psi_*}$.

Having in mind that Jordan states appear 
naturally  in the confluent generalized Darboux transformations \textcolor{red}{[\cite{Jordan1}]}, 
one can consider directly a generalized Darboux transformation 
based on the  following set of the seed states\,:
 $(\psi_1,\Omega_1,\ldots,\psi_n,\Omega_n)$.
 This generates a Darboux-transformed  
 system which we denote by 
 $\widehat{L}_{[2n]}$. The  intertwining operator 
$\A_{2n}^{\Omega}$ as a differential operator of order $2n$ is built 
according to the same rule (\ref{generic-inter}),
but with the inclusion  of Jordan states into the set
of generating functions. 
By the construction, this operator annihilates the chosen $2n$ seed  states,
and  one can show that 
\be
\label{Polly2}
(\A_{2n}^\Omega)^\dagger\A_{2n}^\Omega=\prod_{i=1}^{n}(L-\lambda_i)^2\,,\qquad
\A_{2n}^\Omega(\A_{2n}^\Omega)^\dagger=\prod_{i}^{n}(\widehat{L}_{[2n]}-\lambda_i)^2\,.
\ee
This, in particular, means that 
$\ker(\A_{2n}^\Omega)^\dagger=\text{span}\{\A_{2n}^\Omega\widetilde{\psi}_{1},
\A_{2n}^\Omega\breve{\Omega}_{1},\ldots,\A_{2n}^{\Omega}\widetilde{\psi}_{n},\A_{2n}^{\Omega}\breve{\Omega}_{n}\}$.

\section{A three-dimensional example}
\label{SecSUSY3d}
Unlike the one-dimensional case, three-dimensional supersymmetric quantum mechanics does not have a unique generalization.
Here, following \textcolor{red}{[\cite{Cooper}]}, 
we begin with  a charged massless Dirac particle in
a four-dimensional 
 Euclidian space. Assuming the presence of an 
external electromagnetic field, 
the Dirac's equation takes the form ($\hbar=e=c=1$)
\begin{equation}
\label{DiracEq}
\gamma^\mu P_\mu\Psi=0\,, \qquad
P_\mu= -i\partial_\mu+ A_\mu\,,\qquad\mu=0,1,2,3\,,
\end{equation}
where $A_\mu$ is the associated $U(1)$ gauge potential, 
the metric is just $\delta_{\mu\nu}$ and 
$\gamma^\mu$ are the Euclidean gamma matrices
\begin{eqnarray}
\gamma_{i}=\left(\begin{array}{cc}
0 & -i\sigma_i\\
i\sigma_i & 0 
\end{array}\right)\,,\qquad
\gamma_{0}=\left(\begin{array}{cc}
0 & 1 \\
1 & 0 
\end{array}\right)\,, \qquad \Gamma=\gamma_5=i\gamma_0\gamma_1\gamma_2\gamma_3\gamma_4=
\left(\begin{array}{cc}
1 & 0\\
0 & -1 
\end{array}\right)\,. 
\end{eqnarray}
Assuming that the gauge field does not depend on $t$,
 we can look for stationary solutions of the form  $\Psi=e^{i\lambda t}\Phi(\vr)$.  
Expanding equation (\ref{DiracEq}) in  terms 
of this ansatz
we get 
\begin{eqnarray}
\left(
\begin{array}{cc}
0 & \lambda\\
\lambda & 0 
\end{array}\right)\Phi=
\mathcal{Q}_1\Phi\,,\qquad
\mathcal{Q}_1=\left(
\begin{array}{cc}
0 & \vsigma\cdot(\vnabla+i\vA)-A_0\\
-\vsigma\cdot(\vnabla+i\vA)-A_0 & 0 
\end{array}\right) \,.
\end{eqnarray}
By applying $\mathcal{Q}_1$ from the left,
the Schr\"odinger equation  
$
\lambda^2\Phi=\mathcal{H}\Phi\,$ is obtained,  where 
\begin{eqnarray}
\label{Pauli1}
\mathcal{H}=(\vp+\vA)^2+A_0^2 +\Pi_+\vsigma\cdot (\vE+\vB)+\Pi_-\vsigma\cdot(\vE-\vB)\,,\qquad \Pi_\pm=\frac{1}{2}(1\pm \Gamma)\,, 
\end{eqnarray}
is a Pauli Hamiltonian operator with $\vE=-\vnabla A_0$ and  $\vB=\vnabla \cross \vA$. 
The operator $\mathcal{Q}_1$, together with $\mathcal{Q}_2=i\Gamma\mathcal{Q}_1$ and $\mathcal{H}$ produce a 
three-dimensional realization of the $\mathcal{N}=2$
 Poincar\'e supersymmetry with grading operator $\Gamma$.  
Furthermore, in the dual case $\vE=\vB$ (antidual case 
$\vE=-\vB$) the system possesses the nontrivial bosonic integral 
of motion $\mbfgr{\mathcal{S}}^-=\Pi_-\vsigma$ 
($\mbfgr{\mathcal{S}}^+=\Pi_+\vsigma$), 
and the commutation relations 
$[\mathcal{S}_i^-,\mathcal{Q}_a]$ 
($[\mathcal{S}_i^+,\mathcal{Q}_a]$) with $a=1,2$, produce other 3  pairs of supercharges. 

The system described by (\ref{Pauli1}) has been studied properly 
in \textcolor{red}{[\cite{HSUSY3}]} where authors show that dual and anti-dual cases are the only 
ones that admit extensions of the Poincar\'e supersymmetry. 
In \textcolor{red}{[\cite{PlyWipf}]}, the case of dual and anti-dual dyon (where the magnetic fields 
is due to a Dirac magnetic monopole) was considered, and it was shown that 
the system possesses the exceptional $\mathcal{N}=4$ superconformal algebra $D(1,2;\alpha)$ \textcolor{red}{[\cite{HSUSY2}]}.

\section{Remarks}

The tools considered in this chapter are going to be 
our principal methods for the rest of this Thesis. 
When we study one-dimensional potentials, 
our initial system for the DCKA transformation will always 
be a Newton-Hooke conformal invariant  particle 
\textcolor{red}{[\cite{NH1,NH2,NH3,NH4}]}, the properties 
of which 
are described in  Chaps. \ref{ChConformal} and \ref{ChBridge}. Our principal target is to study the 
hidden symmetries of the nontrivial resulting systems 
and their supersymmetric extensions. This is the main content 
of Chaps. \ref{ChHiddenboson}-\ref{ChKlein}.

In Chap. \ref{Chapmono1} and \ref{Chapmono2} we study a three-dimensional 
generalization of the system introduced in Chap. \ref{ChConformal}, as well as
its supersymmetric extensions. 
The resulting system will have  superconformal symmetry  
that can be reinterpret in accordance with Sec. \ref{SecSUSY3d}, but with 
a nontrivial gauge potential.    


\chapter{ One-dimensional conformal mechanics }
\label{ChConformal}

As it was noted in the introduction, 
conformal invariance appears as a natural extension of the Poincar\'e symmetry of space-time, 
and involves the set of transformations that perform the change 
$g_{\mu \nu}dx^{\mu}dx^{\nu}\rightarrow\Omega(x)g_{\mu \nu}dx^{\mu}dx^{\nu}$, where 
 $g_{\mu\nu}$ is the metric tensor 
and $\Omega(x)$ the conformal factor \textcolor{red}{[\cite{Francesco,Sundermayer}]}. The transformations that 
make this job (preservation of angles) 
are the space-time dilatations  and the special conformal transformations. 
Some examples of space-time 
manifolds 
that allow this extension are the flat space (Minkowski),
together with  de Sitter (dS) and Anti de Sitter (AdS) spaces \textcolor{red}{[\cite{Francesco,Nakahara}]}.

 The $\mathfrak{so}(2,1)$ conformal algebra is given by
\begin{equation}
\label{so(2,1)preambulo}
[D,H]=iH\,,\qquad
[D,K]=-iK\,,\qquad
[K,H]=2iD\,,
\end{equation}
being $H$, $D$ and $K$ the generators of time translations, dilatations and special conformal transformations, 
for details we recommend \textcolor{red}{[\cite{SCM5}]}.
Taking the linear combinations 
\begin{equation}
\label{Jintermsofso2}
\mathcal{J}_0=\frac{1}{2}(\alpha^{-1} H+\alpha K)\,,\qquad
\mathcal{J}_1=\frac{1}{2}( \alpha^{-1} H- \alpha K)\,,\qquad
\mathcal{J}_2=D\,,\
\end{equation}
where $\alpha$ is a constant that compensates the dimensions of $K$ and $H$, 
we obtain the Lorentz algebra in $(2+1)$-dimensional 
 Minkowski space, with metric $\eta^{\mu\nu}=\text{diag}(-1,1,1)$,
given  by 
\begin{equation}
[\mathcal{J}_{\mu},\mathcal{J}_{\nu}]=-i\epsilon_{\mu\nu\rho}\mathcal{J}^{\rho}\,,\qquad \epsilon^{012}=1\,,
\end{equation} 
which, in turn, is isomorphic to the $\mathfrak{sl}(2\,,\R)$ algebra, \textcolor{red}{[\cite{Mikisl2R}]},
\begin{equation}
\label{sl(2R)preambulo}
[\mathcal{J}_0,\mathcal{J}_\pm]=\pm \mathcal{J}_\pm\,,\qquad
[\mathcal{J}_-,\mathcal{J}_+]=2 \mathcal{J}_0\,,\qquad \mathcal{J}_\pm=\mathcal{J}_1\pm i\mathcal{J}_2=
\frac{1}{2\alpha}( H- \alpha^2 K \pm i2\alpha D) \,.
\end{equation}  
This algebra has the automorphisms
$
\mathcal{J}_0\rightarrow \mathcal{J}_0\,,$
$\mathcal{J}_\pm\rightarrow -\mathcal{J}_\pm\,,$ and 
$\mathcal{J}_0\rightarrow -\mathcal{J}_0\,,
\mathcal{J}_\pm\rightarrow -\mathcal{J}_\mp\,,$ and the Casimir element is given by
\begin{equation}
\label{CasimirInvariant}
\mathscr{F}=-\mathcal{J}_\mu \mathcal{J}^\mu=\mathcal{J}_0^2-\frac{1}{2}(\mathcal{J}_+\mathcal{J}_-+
\mathcal{J}_-\mathcal{J}_+)=KH-D^2\,.
\end{equation}

One of the objectives of this Thesis is to study models that have
both this symmetry and some supersymmetric extensions of it. 
We also study possible nonlinear extensions of 
(super)conformal algebra, performed in terms of hidden symmetries.

This chapter is devoted to the analysis of classical and quantum conformal mechanical models.
 In Sec. \ref{SecAFF} we review the theory behind the de Alfaro, Fubini, and Furlan (AFF) model, 
presented in \textcolor{red}{[\cite{AFF}]}, that looks for a well-defined one-dimensional quantum system with a 
conformal invariant ground state. In Sec. \ref{SecOSP22Conformal} we use the tools developed
 in Chap. \ref{ChSUSY} to construct the  $\mathfrak{osp} (2,2)$ supersymmetric extension  
of the AFF model.

 
\section{The de Alfaro, Fubini and Furlan model}
\label{SecAFF}

Consider  the one-dimensional system given by the action, 
\textcolor{red}{[\cite{AFF}]},
\be
\label{conformalaction}
I[q]=\int \mathcal{L}(q,\dot{q}) dt\,, \quad
\mathcal{L}=\frac{1}{2}\left(\dot{q}^2-\frac{g}{q^2}\right)\,,
\ee
where 
$q$ takes values on the positive real line and
 has dimension $[q]=[\sqrt{t}]$, besides 
  $g$ is a dimensionless coupling constant 
which classically is assumed to be positive 
to avoid the ``problem of fall to  the center''.
This action could represent, for example, a
 Calogero model of two particles, but omitting the degree of 
freedom of the center of mass
\textcolor{red}{[\cite{Calogero1,Calogero2}]}.

On can show that the action (\ref{conformalaction})
is invariant under  time translations $t\rightarrow t+\alpha t$, 
space-time dilatations 
\begin{equation}
x\rightarrow e^{\frac{\beta}{2}}x\,,\qquad
t\rightarrow e^{\beta}t\,,
\end{equation}
and the spacial conformal transformations 
\begin{equation}
x\rightarrow \frac{x}{1-\gamma t}\,,\qquad
t\rightarrow \frac{t}{1-\gamma t}\,,
\end{equation}
where $\alpha$, $\beta$ and $\gamma$ are parameters of 
the corresponding transformations. 
This symmetry is  generated by the
 Hamiltonian $H_g$, 
  the dilatations 
 generator $D$, and generator of special conformal transformations  
 $K$,
 \begin{equation}
 \label{conformalgenerators}
 H_g=\frac{1}{2}(p^2+\frac{g}{q^2})\,,\qquad
 D=\frac{1}{4}(qp+pq) -H_gt\,,\qquad
 K=\frac{1}{2}q^2-2Dt-H_gt^2\,,
\end{equation}      
 where $p=\dot{q}$. 
 These are the integrals of motion that satisfy the equation 
 of the form
 $\frac{d}{dt}{A}=\frac{\partial A}{\partial t} + \{A,H\}=0$
 where $\{,\}$ denotes Poisson brackets. We often call objects 
 of this type as ``dynamical integrals'', and in this case they obey
 the classical version of
 $\mathfrak{so}(2,1)$ algebra 
\begin{equation}
\label{so(2,1) cap 1}
\{D,H_g\}=H_g\,,\qquad
\{D,K\}=-K\,,\qquad
\{H_g,K\}=-2D\,,
\end{equation}
and the
Casimir invariant (\ref{CasimirInvariant}) takes the value 
$\mathscr{F}=\frac{1}{4}g$. The 
last relation in (\ref{conformalgenerators}) 
gives us the  solution of the corresponding  Euler-Lagrange 
equation
derived from 
(\ref{conformalaction}), 
\be\label{q(t)}
q(t)=\sqrt{2(at^2+2bt+c)}=\sqrt{2\left(a(t+\frac{b}{a})^2+\frac{\mathscr{F}}{a^2}\right)}\,,
\ee
where real-valued  constants $a$, $b$ and $c$
 correspond to the  
values of the integrals  $H_g$, $D$ and $K$, respectively (for a
given initial configuration).

Note that in the case of $ g = 0 $, $H_g$
 takes the form of an object that looks like
 the Hamiltonian of a free particles, 
but is defined in the restricted domain
$\R^+$. The notable difference between this system and the free particle $H_{f}$, 
which lives in $ \R $, is that the latter has two additional integrals of motion, 
namely the momentum $p$ and the Galileo boost generator $\chi=\tilde{q}-pt$, 
with $\tilde{q}\in \R$.
They produce Heisenberg algebra and together with the generators 
$D_{f}=\frac{\chi P}{2}$ and $K_{f}=\frac{\chi^2}{2}$, 
leading to  the Schr\"odinger symmetry \textcolor{red}{[\cite{Niedfree,Duval,Henkel,GAdS1,Aizawa}]},
\begin{eqnarray}
&
\label{so(2,1) free}
\{D_{f},H_{f}\}=H_{f}\,,\qquad
\{D_{f},K_{f}\}=-K_{f}\,,\qquad
\{H_{f},K_{f}\}=-2D_{f}\,,&\\&
\label{Schr free cap 1}
\{\chi,p\}=1\,,\qquad
\{H_{f},p\}=\{K_{f},\chi\}=0\,\qquad
\{H_{f},\chi\}=-p\,,\qquad
\{K_{f},p\}=\chi\,,\qquad &\\&
\{D_{f},\chi\}=-\frac{1}{2}\chi\,,\qquad
\label{Schr free cap 1 2}
\{D_{f},p\}=\frac{1}{2}p\,.&
\end{eqnarray}
The model (\ref{conformalaction}) has a problem at the quantum level, as we explain below:
In the Schr\"odinger picture, the quantum version of the generators (\ref{conformalgenerators})
are given by ($\hbar=1$)
\begin{equation}
\label{Qso(2,1)}
H_\nu=\frac{1}{2}\left(-\frac{d^2}{dq^2}+\frac{\nu(\nu+1)}{q^2}\right)\,,\qquad
D=\frac{1}{4i}\left(q\frac{d}{dq}+\frac{d}{dq}q\right)\,,\qquad
K=\frac{q^2}{2}\,,
\end{equation}
where we have parameterized $g$ as 
$\nu(\nu+1)$. 
Obviously, the operators $ D $ and $ K $ are not integrals of motion,
 however they can be promoted to dynamical integrals, in the sense of the Heisenberg equation
$\frac{d O}{dt}=\frac{\partial O }{\partial t}-i[O,H_\nu ] $, by means of the unitary transformation 
\be
\label{recipe}
 O \rightarrow {}_{H}O=e^{-iH_\nu t } O  e^{iH_\nu t}\,.
\ee
This is the general recipe for moving from the Schr\"odinger picture 
 to the Heisenberg picture, and in this last framework
  the operators $ D $ and $ K $ are changed for
${}_HD=D-H_\nu t$ and 
${}_HK=K-2D t-H_\nu t^2$, respectively. 

The
Hamiltonian  $H_\nu$ is self-adjoint 
for the cases in which $\nu\geq 0$ and admits self-adjoint 
extensions for  $\nu\geq -1/2$, \textcolor{red}{[\cite{Landau,kirsten}]}. 
In these cases, $H_\nu$ has a 
 continuous spectrum $E=\kappa^2/2$, 
with $\kappa \in \R$,  in the domain  
$\{ \psi\in L^2((0,\infty),dq)\vert \psi(0^+)=0\}$ 
and the physical eigenstates are given by 
\begin{equation}
\label{states calogero}
\psi_\nu (q;\kappa)=\sqrt{q}J_{\nu+\frac{1}{2}}(\kappa q)\,,
\end{equation} 
where $J_{\alpha}(\zeta)$ is the Bessel function of the first kind
\begin{equation}
J_\alpha(\zeta)=
\sum_{n=0}^{\infty}\frac{(-1)^{n}}
{n!\Gamma(n+\alpha+1)}\zeta^{2n+\alpha}\,.
\end{equation}

From here it is not difficult to show that the state 
$ e^{i \alpha D} \psi_\nu (x; \kappa) $ corresponds to the energy $ e ^ {\alpha} E $,
 which implies that the only scale-invariant solutions are those with  zero energy eigenvalue, 
which in this case are given by the nonphysical solutions $ q^{\nu + 1} $ and 
$ q^{-\nu} $, the first of which is not bounded at infinity and the second
diverges when $ q = 0 $. This means that conformal symmetry is spontaneously broken at
the quantum level.

To$\,$ find$\,$ a$\,$ conformal $\,$invariant$\,$ model$\,$ with$\,$ a$\,$ well-defined$\,$ ground$\,$ state,$\,$
 the$\,$ proposal$\,$ in$\,$ \textcolor{red}{[\cite{AFF}]}
 is to consider the following 
  change of the variables at the classical level
  \begin{equation}
  \label{trans}
  y(t)=\frac{q(t)}{\sqrt{u+vt+wt^2}}\,, \qquad d\tau=\frac{dt}{u+vt+wt^2}\,,
  \end{equation}
where $u>0$, $v$ and $w>0$ are real
constants with dimensions $[u]=1$, $[v]=1/t$ and $[w]=1/t^2$,
and $y>0$. 
This is in fact related to a change of coordinates in an  AdS${}_2  $ space,
 where $t$ is not a good global coordinate, in contrast to $ \tau $
  \textcolor{red}{[\cite{BlackHold2}]}. 
Under the transformation  (\ref{trans}),
action 
 (\ref{conformalaction}) takes the form  
 \begin{eqnarray}
\label{conformalaction2}
&\int\mathcal{L}(y,y')d\tau
+
\frac{1}{4}\int d\tau\frac{d}{d\tau}[(v+2wt(\tau))q^2(t(\tau))]=
I[y]+I_{surface}\,,\,\,&
\end{eqnarray} 
 where $\mathcal{L}(y,y')=\frac{1}{2}(y'^2-\frac{g}{y^2}-\omega^2y^2)\,,$
 $y'=\frac{dy}{d\tau}$, and $\omega^2=(4wu-v^2)/4$. 
 {}Action  $I[y]=\int \mathcal{L}d\tau$ is 
 the so-called de Alfaro, Fubini and Furlan (AFF)  model,
from where 
  we obtain the new time translation generator  
\begin{eqnarray}
\label{mostgeneralH}
\mathscr{H}_g=\frac{1}{2}\left(p^2+\frac{g}{y^2}+\omega^2 y^2\right)\,,\qquad p=y'\,.
\end{eqnarray}
 The evolution 
parameter $\tau=\frac{1}{\omega}\text{acrtan}(\frac{v+2wt}{2\omega})$ 
varies in the 
finite interval 
$(-\frac{\pi}{2\omega},\frac{\pi}{2\omega})$,
and new Hamiltonian (\ref{mostgeneralH}) is conjugate to this 
good global time coordinate.
As $\omega$  is a dimensionful  parameter, $[\omega]=[1/t]$,
(\ref{mostgeneralH}) breaks the manifest scale invariance
of the original system  (\ref{conformalaction}), 
and via such a basic mechanism the mass and length scales 
are introduced in holographic QCD (often referred to as ``AdS/QCD'')
\textcolor{red}{[\cite{App1,Brod2}]}.

In spite of the introduced scale, 
 the action of the new system is conformal invariant 
as we will see now.
The dilatation generator $\mathscr{D}$ and 
the conformal transformation generator $\mathscr{K}$ associated with the 
action 
$I[y]$ are given by the
explicitly depending on time $\tau$ integrals 
\begin{eqnarray}
\label{NHgenD}
&\mathscr{D}=\frac{1}{2}\left(yp\cos(2\omega \tau)+\left(\omega y^2-
\mathscr{H}_g{\omega}^{-1}\right)\sin(2\omega \tau)\right)\,,&\\
&\mathscr{K}= \cos(2\omega \tau)\frac{y^2}{2}+\frac{\mathscr{H}_g}{\omega^2}\sin^2(\omega\tau)
-\frac{\sin(2\omega\tau)}{2\omega}yp\,,\label{NHgenH}&
\end{eqnarray}
which 
generate  the Newton-Hooke symmetry, \textcolor{red}{[\cite{NH1,NH2,NH3,NH4}]},
\begin{equation}\label{NHalg}
\{\mathscr{H}_g,\mathscr{D}\}=-(\mathscr{H}_g-2\omega^2 \mathscr{K})\,,\qquad
\{\mathscr{H}_g,\mathscr{K}\}=-2\mathscr{D}\,,\qquad
\{\mathscr{D},\mathscr{K}\}=-\mathscr{K}\,,
\end{equation}
whose Casimir invariant is $\mathscr{F}=\mathscr{K}\mathscr{H}_g-\mathscr{D}^2-\omega^2\mathscr{K}^2=g/4$.
Using Eqs. (\ref{NHgenD}) and (\ref{NHgenH}),
one can find solution to the equation of motion of the 
system (\ref{conformalaction2}),

\be\label{ytau}
y^2(\tau)=\frac{2}{\omega^2}(a\sin^2(\omega\tau)+\omega b \sin(2\omega\tau)+\omega^2c\cos(2\omega \tau))\,,
\ee
where $a>0$, $b$ and $c>0$ are constants 
corresponding to the values of the integrals
 $\mathscr{H}_g$, $\mathscr{D}$ and $\mathscr{K}$, respectively, and  
obeying the relation 
$ac-b^2-\omega^2c^2=g/4$. {}From the explicit form of the solution
we see that it is periodic with the period $T=\pi/\omega$ not depending on the value 
of the coupling constant\footnote{System
given by Hamiltonian (\ref{mostgeneralH}) is an isoperiodic deformation 
of the half-harmonic oscillator of frequency $\omega$ \textcolor{red}{[\cite{Aso}]}.} $g$.
The finite interval in which the  evolution parameter $\tau$
varies  corresponds to the period of the motion of the 
system (\ref{conformalaction2}), and one can consider
$\tau$ as the compact evolution parameter that
takes values on the closed interval 
$[-\frac{\pi}{2\omega},\frac{\pi}{2\omega}]$
with identified ends.

As in the previous case, if one sets $ g = 0 $, $ \mathscr{H}_g $
 is formally reduced to the Hamiltonian of the harmonic oscillator, 
however the object is defined in $ \R ^ + $ (we will call it  as \emph{half} harmonic oscillator).
If  we extend for this case the domain to the entire real line,  i.e., 
we exchange $y\rightarrow \tilde{y}\in \R$, the resulting system $\mathscr{H}_{\text{os}}$ has 
the additional 
dynamical integrals
\begin{equation}
\label{chigamapegama}
\chi_\omega=\tilde{y}\cos(\omega \tau)-\frac{p}{\omega}\sin(\omega \tau)\,, \qquad
P_\omega=\omega \tilde{y}\sin(\omega \tau)+p\cos(\omega \tau)\,.
\end{equation}
They are identified as the initial conditions of the oscillatory motion and in terms of them,
 the Hamiltonian is read as $\mathscr{H}_{\text{os}}=\frac{1}{2}(P_\omega^2+\omega^2\chi_\omega^2)$.
The generators (\ref{chigamapegama}), 
together with generators  $\mathscr{H}_{\text{os}}$, 
$\mathscr{D}_{\text{os}}=\frac{\chi_\omega P_\omega}{2}$ and
$\mathscr{K}_{\text{os}}=\frac{1}{2}\chi_{\omega}^2$ 
 produce the following Poisson brackets relations 
\begin{eqnarray}
\label{Schr free os cap 1}
&
\{\mathscr{H}_{\text{os}},\mathscr{D}_{\text{os}}\}=
-(\mathscr{H}_{\text{os}}-2\omega^2 \mathscr{K}_{\text{os}})\,,\quad
\{\mathscr{H}_{\text{os}},\mathscr{K}_{\text{os}}\}=-2\mathscr{D}_{\text{os}}\,,\quad
\{\mathscr{D}_{\text{os}},\mathscr{K}_{\text{os}}\}=-\mathscr{K}_{\text{os}}\,,
&\\&
\{\chi_\omega ,P_\omega\}=1\,,\qquad
\{\mathscr{H}_{\text{os}}, P_\omega\}=\omega^2\chi_\omega\,,\qquad
\{\mathscr{H}_{\text{os}}, \chi_\omega\}=P_\omega\,\qquad
\{\mathscr{K}_{\text{os}},\chi_\omega\}=0\,,&\\&
\{\mathscr{K}_{\text{os}},P_\omega\}=\chi_\omega\,, \qquad
\{\mathscr{D}_{\text{os}},\chi_\omega\}=-\frac{1}{2}\chi_\omega\,,\qquad
\{\mathscr{D}_{\text{os}},P_\omega\}=\frac{1}{2}P_\omega\,.&
\end{eqnarray}
Note that if instead to take $\mathscr{H}_{\text{os}}$ we consider 
$\hat{\mathscr{H}}=\mathscr{H}_{\text{os}}-\omega^2\mathscr{K}_{\text{os}}=\frac{1}{2}P_{\omega}^2$
one gets  the algebraic relations 
(\ref{so(2,1) free}), (\ref{Schr free cap 1})  and (\ref{Schr free cap 1 2}),
which mean that  generators 
$\{\mathscr{H}_{\text{os}}\,, \mathscr{D}_{\text{os}}\,,\mathscr{K}_{\text{os}}\,,\chi_\omega\,,P_\omega
\}$ are just  another  basis for the Schr\"odinger symmetry. In fact by taking the limit 
$\omega\rightarrow 0$ we recover the
free particle generators. 

According to  \textcolor{red}{[\cite{Dirac}]}, 
starting from a given symmetry algebra, one can freely designate a particular generator or a linear combination of generators as
 Hamiltonian, leading to different forms of dynamics.
 This terminology was introduced in the context of special relativity,
 however, the two models discussed above are good examples in
 nonrelativistic mechanics.

At the quantum level, the AFF Hamiltonian 
takes the form 
\begin{eqnarray}
\label{Lg}
\mathscr{H}_{\nu}=\frac{1}{2}\left(-\frac{d^2}{dy^2}+\omega^2y^2+\frac{g(\nu)}{y^2}\right)\,,\qquad
g(\nu)=\nu(\nu+1)\,,
\end{eqnarray}
which as well as $H_\nu$ in (\ref{Qso(2,1)}), 
has a bounded spectrum restricted from below 
in the domain 
$\{ \psi\in L^2((0,\infty),dy)\vert \psi(0^+)=0\}$
for $\nu\geq-1/2$ \textcolor{red}{[\cite{Falomir1,Falomir2}]}. 
 The normalized eigenstates of the system and its 
 respective energy values 
 are given by 
 \begin{equation}
 \label{AFF states}
 \psi_{\nu,n}(y)=\sqrt{\frac{2n!\omega^{\nu+\frac{3}{2}}}{\Gamma(n+\nu+\frac{3}{2})}}\,\,y^{\nu+1}L_{n}^{(\nu+\frac{1}{2})}(\omega y^2)e^{-\frac{\omega y^2}{2}}\,,\qquad
 E_{\nu,n}=\omega(2n+\nu+\frac{3}{2})\,,
\end{equation}  
where 
\begin{equation}
\label{Laguerre}
\mathcal{L}_n^{(\alpha)}(\eta)=
\sum_{j=0}^{n}\frac{\Gamma(n+\alpha+1)}{\Gamma(j+\alpha+1)}\frac{(-\eta)^{j}}{j!(n-j)!}\,,
\end{equation} 
are the generalized Laguerre Polynomials.  Note that 
$g$ in (\ref{Lg}) vanishes for $\nu=0$ and for $\nu=-1$ (where we have some problems with boundary conditions)
and for both cases $\mathscr{H}_{\nu}$ looks like an harmonic oscillator Hamiltonian. Indeed, the well 
known relations 
\be
\label{hermiteLaguerre}
H_{2n+1}(\eta)=  (-1)^{n}2^{2n+1}L_{n}^{(1/2)}(\eta^2)\,,\qquad 
H_{2n}(\eta) =  (-1)^{n}2^{2n}L_{n}^{(-1/2)}(\eta^2)\,,
\ee 
where functions $H_{n}(\eta)$ are the Hermite polynomials, 
show us that in the first case eigenfunctions (\ref{AFF states}) become the odd eigenstates of the harmonic oscillator (vanishing  at the origin),
and in the second case, they take the form of the even eigenstates of the latter mentioned system (which do not vanish at $x=0$, thereby 
 violating  
the imposed boundary conditions). 

Instead to do a direct quantization of generators 
$\mathscr{K}$ and $\mathscr{D}$, 
it is worth it to consider complex combinations of them.
In particular, in the Schr\"odinger picture  we construct 
 \begin{eqnarray}
 \label{Cpm}
&\mathcal{C}_\nu^\pm=
\mathscr{H}_\nu-2\omega^2\mathscr{K}\pm 2i\omega \mathscr{D}=\left(\mathscr{H}_\nu-\omega^2y^2\pm 2\omega (y\frac{d}{dy}+\frac{1}{2})
\right)\,, &
\end{eqnarray}
which, together with $\mathscr{H}_\nu$, produce the commutator 
relations 
 \begin{equation}
 \label{sl2RAFF}
 [\mathscr{H}_{\nu},\mathcal{C}_\nu^\pm]=\pm 2\omega\mathcal{C}_\nu^\pm\,,\qquad
 [\mathcal{C}_\nu^-,\mathcal{C}_\nu^+]=4\omega\mathscr{H}_\nu\,,
 \end{equation}
 and by using the identification
  $\mathscr{H}_\nu=2\omega \mathcal{J}_0$ and $\mathcal{C}_\nu=2\omega \mathcal{J}_\pm$,
 we recognize the $\mathfrak{sl}(2,\R)$ algebra
 (\ref{sl(2R)preambulo}). 
  On the  Hilbert space
of the AFF system, the  states (\ref{AFF states})
 correspond to 
an infinite-dimensional 
unitary  irreducible representation of the 
$\mathfrak{sl}(2,\R)$ algebra of the discrete type series $\mathcal{D}^+_\alpha$ 
with $\alpha=\frac{1}{2}\nu+\frac{3}{4}$,
 and the Casimir operator takes the value
$\mathscr{F}_\nu=\mathcal{J}^\mu\mathcal{J}_\mu=-\alpha(\alpha-1)=
\frac{3}{16}-\frac{1}{4}\nu(\nu+1)$,  \textcolor{red}{[\cite{Mikisl2R}]}.

As operators $\mathcal{C}_\nu^\pm $ are not integrals of motion, when we go to 
the Heisenberg picture, it is necessary to  replace operators $\mathcal{C}^\pm$
by the dynamical integrals ${}_H\mathcal{C}^\pm=e^{\mp i 2\omega t} \mathcal{C}^\pm$. 

 Relations  (\ref{sl2RAFF}) clearly show us that 
 $\mathcal{C}_\nu^\pm$ are ladder operators which 
 change the  energy in $\pm 2\omega$. Their action 
 can be computed  by means of  the corresponding recurrence 
relations of Laguerre polynomials,
\begin{eqnarray}
& \nonumber y\frac{d}{dy}L^\alpha_n(y)-yL^\alpha_n(y)+\alpha L_n^\alpha=
(n+1)L_{n+1}^{\alpha-1}\,,\qquad \frac{d}{dy}L_n^\alpha(y)-L_n^\alpha(y)=-L_n^{\alpha+1}(y)\,,
\nonumber &\\
&\frac{d}{dy}L^{\alpha}_{n}(y)=-L_{n-1}^{\alpha+1}(y)\,,\qquad 
y\frac{d}{dy}L_n^\alpha(y)+\alpha L_n^\alpha(y)=
(n+\alpha)L_n^{\alpha-1}(y)\,.\label{recurrence-relations}&
\end{eqnarray}
Using this we get 
\begin{eqnarray}
\label{Conpsi}
&
\mathcal{C}_\nu^-\psi_{\nu,n}=2\omega\sqrt{n(n+\nu+\frac{1}{2})}\,\psi_{\nu,n-1}\,,&\\&
\mathcal{C}_\nu^+\psi_{\nu,n}=2\omega\sqrt{(n+1)(n+\nu+\frac{3}{2})}\,\psi_{\nu,n+1}\,,&
\end{eqnarray}
from where we see that the lowering operator 
$\mathcal{C}_\nu^-$ annihilates the ground state of the system. 
In next section we will show that these operators 
have their own origin in supersymmetric quantum mechanics.

\section{The $\mathfrak{osp}(2|2)$ superconformal symmetry}
\label{SecOSP22Conformal}

The aim of this section is to construct an $\mathcal{N}=2$
super-extension of the AFF model 
 (\ref{Lg}). To this end, we apply the method 
 introduced in Chap. \ref{ChSUSY}.
 
For the construction let us use the ground state 
$\psi_{\nu,0}\propto y^{\nu+1}e^{-\omega y^2/2}$ as a seed state
for the first order Darboux transformation.
 The associated intertwining operators
are   
\begin{equation}
\label{Anu}
A_{\nu}^-=\frac{1}{\sqrt{2}}\left(\frac{d}{dy}+\omega y-\frac{\nu+1}{y}\right)\,,\qquad
A_{\nu}^+=(A_{\nu}^-)^\dagger\,,
\end{equation}
which produce 
\begin{equation}
\label{producA}
A_\nu^+A_{\nu}^-=\mathscr{H}_{\nu}-\omega (\nu+\frac{3}{2}):=H_-\,,\qquad
A_\nu^-A_{\nu}^+=\mathscr{H}_{\nu+1}-\omega(\nu+\frac{1}{2}):=H_+\,,\end{equation}
and intertwining relations take the form (\ref{Intertwiningrelation0}).
Using the recurrence relations that Laguerre 
polynomials satisfy (\ref{recurrence-relations}),
 one
gets the explicit action of $A_\nu^\pm$ on
 eigenstates (\ref{AFF states}), 
\begin{equation}
\label{Aonpsi}
A_\nu^-\psi_{\nu,n}=-\sqrt{2n\omega}\,\psi_{\nu+1,n-1} \,,\qquad
A_{\nu}^+\psi_{\nu+1,n-1}=-\sqrt{2n\omega }\,\psi_{\nu,n}\,.
\end{equation}
With the help of  (\ref{Anu}) we can construct 
the matrix generators  
\begin{eqnarray}
\label{Poincare1}&
\mathcal{H}_\nu^{e}=\left(\begin{array}{cc}
\mathscr{H}_{\nu+1}-\omega (\nu+1/2) & 0 \\
0 & \mathscr{H}_{\nu}-\omega(\nu+3/2)
\end{array}\right)\,, &\\&
\mathcal{Q}_\nu^{1}=\left(\begin{array}{cc}
0 & A_\nu^-\\
A_\nu^+ & 0
\end{array}\right)\,,\quad
\mathcal{Q}_\nu^{2}=i\Gamma \mathcal{Q}_\nu^{1},&
\end{eqnarray}
where $\Gamma=\sigma_3$ is the $\Z_2$ grading operator. These generators produce the Poincar\'e supersymmetry 
(\ref{Poincare0}).
Operator $ \mathcal{H}_\nu^e \, $ has the spectrum 
$ 2 \omega n $, $n=0,1,\ldots,$ and the unique ground state 
$ (0 ,\, \, \psi_{\nu, 0})^{t} $ 
is annihilated by all generators in (\ref{Poincare1}), therefore supersymmetry is in the exact phase.

On the other hand, the system (\ref{Lg})
 possesses the nonphysical
solutions $\psi_{\nu,n}^-=\psi_{\nu,n}(iy)$ of the 
eigenvalues $-E_{n,\nu}$\footnote{
The stationary  Schr\"odinger  equation $\mathscr{H}_\nu \psi_{\nu,n} = E \psi_{\nu,n}$ has a discrete symmetry group, and the transformation defined as $ y \rightarrow iy$ and $ E_{\nu, n} \rightarrow -E _ {\nu,n} $ is an element of this group, see Chap. \ref{ChKlein}. The nonphysical eigenstates produced by the action of the mentioned group can be used in the Darboux transformations, resulting in new solvable systems.}. 
Then, instead of the ground state we could select the function 
$ \psi_{\nu, 0}^- \propto y^{\nu + 1}e^{\omega y^2/2} $
 as a seed state. The resulting intertwining operators are
\begin{equation}
\label{Bnu}
B_\nu^{-}= \frac{1}{\sqrt{2}}\left(\frac{d}{dy}-\omega y-\frac{\nu+1}{y}\right)\,,\qquad
B_\nu^{+}= (B_\nu^-)^\dagger \,. 
\end{equation}
Their products give us 
\begin{equation}
\label{ProductB}
B_\nu^+B_\nu^-=\mathscr{H}_\nu+\omega (\nu+\frac{3}{2})=H_-+\omega(2\nu+3)\,,\qquad
B_\nu^-B_\nu^+=\mathscr{H}_{\nu+1}+\omega(\nu+\frac{1}{2})=H_++\omega(2\nu+1)\,,
\end{equation}
and in terms of $H_\pm$ the intertwining relations take the form 
\begin{equation}
B_\nu^-H_-=(H_+-2\omega)B_\nu^-\,,\qquad
B_\nu^+H_+=(H_-+2\omega)B_\nu^+\,.
\end{equation} 
Coherently with this,
the action of operators $B_\nu^\pm$ on the eigenstates is 
\begin{equation}
\label{Bonpsi}
B_{\nu}^-\psi_{\nu,n}=-\sqrt{(2n+2\nu+3)\omega}\,
\psi_{\nu+1,n}
\,,\qquad
B_{\nu}^+\psi_{n,\nu+1}=-\sqrt{(2n+2\nu+3)\omega}\,
\psi_{\nu,n}\,.
\end{equation}
Just like we did with 
$ A_\nu^\pm $, we can also use $ B_\nu^\pm $ 
to build other matrix operators
\begin{eqnarray}
\label{Rnu}
&
\mathcal{H}_{\nu}^{b}=\left(\begin{array}{cc}
\mathscr{H}_{\nu+1}+\omega(\nu+1/2) & 0 \\
0              & \mathscr{H}_{\nu}+\omega (\nu+3/2)
\end{array}\right) \,,&\\&
\label{Snu}
\mathcal{S}_\nu^{1}=\left(\begin{array}{cc}
0 & B_\nu^-\\
B_\nu^+ & 0
\end{array}\right)\,,\qquad
\mathcal{S}_\nu^{2}=i\Gamma \mathcal{S}_\nu^{1}\,,&
\end{eqnarray}
which again will satisfy the $\mathcal{N}=2$ Poincar\'e 
supersymmetry, but now, in the spontaneously broken 
phase\footnote{The seed state $\psi_{\nu,0}^-$ is nonphysical and $1/\psi_{\nu,0}^-$ does not satisfy the boundary condition at the origin. }; 
the spectrum of $\mathcal{H}_\nu^{b}$ is $\omega(2n+2\nu+3)$, $n=0,1,\ldots,$ and 
there is no  physical eigenstate which is 
simultaneously annihilated by both odd operators 
$\mathcal{S}_\nu ^{a}$. On the other hand one can reinterpret 
the object $\mathcal{H}_\nu^{b}$ as a linear combination 
of   $\mathcal{H}_\nu^{e}$ and the nontrivial integral
\be
\mathcal{R}_\nu= \frac{1}{2\omega}(\mathcal{H}_\nu^{e}- \mathcal{H}_\nu^{b})
=
\frac{1}{2}\sigma_3 -(\nu+1)\,,
\ee
that plays the role of what will become
an $R$ symmetry generator.  

Now, remember that the system (\ref{Lg}) 
 has the two second
order ladder operators (\ref{Cpm}). Namely, they are constructed from 
$A_\nu^\pm$ and $B_\nu^\pm$ as follows  
\begin{eqnarray}&
\label{FactorCnu+1}
B_\nu^-A_\nu^+=\mathcal{C}_{\nu+1}^+\,,\qquad
A_\nu^-B_\nu^+=\mathcal{C}_{\nu+1}^-\,,
&\\&
A_\nu^+B_\nu^-=\mathcal{C}_{\nu}^+\,,\qquad
B_\nu^+A_\nu^-=\mathcal{C}_{\nu}^- \,.
\label{FactorCnu}&
\end{eqnarray}
By using this structure together with the Eqs. 
(\ref{Bonpsi}) and (\ref{Aonpsi}) it is easy to check the relations
 (\ref{Conpsi}). 
Also, by means of the Eqs. (\ref{producA}) and (\ref{ProductB}),
in addition with  the intertwining relations corresponding to 
$ A_\nu^\pm $ and 
$B_\nu^\pm $, it is easy to derive the $ \mathfrak{sl}(2,\R) $ algebra (\ref{sl2RAFF}).

Returning to the  matrix operators subject, the relations (\ref{FactorCnu+1})-(\ref{FactorCnu})
show us that  the  
anti-commutator between generators  $\mathcal{S}_\nu^{a}$
and 
$\mathcal{Q}_\nu^{a}$ produces the even operators 
\begin{equation}
\label{Gnu}
\mathcal{G}_\nu^\pm=\left(\begin{array}{cc}
\mathcal{C}_{\nu+1}^\pm & 0 \\
0  & \mathcal{C}_{\nu}^\pm 
\end{array}\right)\,,
\end{equation} 
which are  the corresponding  super-extensions
of the ladder operators of  systems 
$\mathcal{H}_\nu^{e}$ and $\mathcal{H}_\nu^{b}$. 
Then, all together the  generators $\{\mathcal{H}_\nu^{e}\,,\mathcal{G}_\nu^\pm\,,\mathcal{R}_\nu\,,\mathcal{Q}_\nu^{a}\,,\mathcal{S}_\nu^{a}\} $ 
satisfy  the superalgebraic relations
\begin{eqnarray}\label{HRQ0}
&[\mathcal{H}_\nu^{e},\mathcal{R}_\nu]=[\mathcal{H}_\nu^{e},\mathcal{Q}_\nu^a]=0\,,&\\
\label{evencommutation}
&[\mathcal{H}_\nu^{e},\mathcal{G}_\nu^{\pm}]=\pm2\omega \mathcal{G}_\nu^{\pm}\,, \qquad 
[\mathcal{G}_\nu^{-},\mathcal{G}_\nu^{+}]=4\omega\left(\mathcal{H}^{e}_\nu-\omega^2\mathcal{R}_\nu\right)\,,&\\
\label{evenodd}
&[\mathcal{H}_\nu^{e},\mathcal{S}_\nu^a]=-2 i \omega \epsilon^{ab}\mathcal{S}_\nu^b\,,\qquad
[\mathcal{R}_\nu,\mathcal{Q}_\nu^a]=-i\epsilon^{ab}\mathcal{Q}_\nu^b\,,
\qquad
[\mathcal{R}_\nu,\mathcal{S}_\nu^a]=-i\epsilon^{ab}\mathcal{S}^b_\nu\,,&\\
\label{fq1}
&[\mathcal{G}_\nu^-,\mathcal{Q}_\nu^a]=\omega (\mathcal{S}_\nu^a+i\epsilon^{ab}\mathcal{S}_\nu^b), \qquad 
[\mathcal{G}_\nu^+,\mathcal{Q}_\nu^a]=-\omega (\mathcal{S}_\nu^a-i\epsilon^{ab}\mathcal{S}_\nu^b)\,,&\\
\label{fq3}
&[\mathcal{G}_\nu^-,\mathcal{S}_\nu^a]=\omega (\mathcal{Q}_\nu^a-i\epsilon^{ab}\mathcal{Q}_\nu^b)\,, \qquad 
[\mathcal{G}_\nu^+,\mathcal{S}_\nu^a]=-\omega (\mathcal{Q}_\nu^a+i\epsilon^{ab}\mathcal{Q}_\nu^b)\,,&\\
\label{anti1}
&\{ \mathcal{Q}_\nu^a,\mathcal{Q}_\nu^b\}=2\delta^{ab}\mathcal{H}_\nu^{e}\,, \qquad 
\{ \mathcal{S}_\nu^a,\mathcal{S}_\nu^b\}=2\delta^{ab}(\mathcal{H}_\nu^{e} -2\omega \mathcal{R}_\nu)\,,&\\
\label{anti2}
&\{\mathcal{Q}^a_\nu,\mathcal{S}^b_\nu\}=\delta^{ab}(\mathcal{G}_\nu^{+}+\mathcal{G}_\nu^-)+
i\epsilon^{ab}(\mathcal{G}_\nu^+-\mathcal{G}_\nu^-)\,.\label{QSGG}
\end{eqnarray}
From here we realize that operators  $\mathcal{G}^\pm$
and $\mathcal{S}_\nu^a$ are not integrals of motion, and in the Heisenberg picture we have instead the dynamical integrals 
  $ {}_{H}\mathcal{G}^\pm=  e^{\mp 2\omega t}\mathcal{G}^\pm$
and $ {}_{H}\mathcal{S}_\nu^a =e^{-i\sigma_3 \omega t}\mathcal{S}_\nu^a$.

 Superalgebra $\,$  (\ref{HRQ0})-(\ref{QSGG})$\,$ 
$\,$is$\,$ identified$\,$ with$\,$ the$\,$ $\mathfrak{osp}(2|2)$
 $\,$superconformal$\,$ symmetry$\,$ 
 $\,$\textcolor{red}{[\cite{InzPly1,InzPly2,InzPly3}]}, and 
has the automorphism $f=f^{-1}$
given by the  transformations 
$\mathcal{H}_{\nu}^{e}\rightarrow \mathcal{H}_{\nu}^{e}-4\mathcal{R}_{\nu}=\mathcal{H}_{\nu}^b$,
$\mathcal{R}_{\nu}\rightarrow -\mathcal{R}_\nu$, 
$\mathcal{G}_{\nu}^\pm\rightarrow \mathcal{G}_{\nu}^{\pm}$,
$\mathcal{Q}_\nu^1\rightarrow -\mathcal{S}_{\nu}^{1}$, 
$\mathcal{Q}_\nu^2\rightarrow \mathcal{S}_{\nu}^{2}$,
$\mathcal{S}_\nu^1\rightarrow -\mathcal{Q}_{\nu}^{1}$
$\mathcal{S}_\nu^2\rightarrow \mathcal{Q}_{\nu}^{2}$.
Transformation $f$ shows us what would happen 
with the superalgebra if we
 had chosen $\mathcal{H}_\nu^{b} $ instead of
 $ \mathcal{H}_\nu^{e}$ as our time translation generator. 

For future applications, we present the superalgebraic
structure in terms of nilpotent fermionic operators
\begin{eqnarray}
\mathcal{Q}_\nu =\left(\begin{array}{cc}
0 & A_\nu \\
0 & 0
\end{array}\right)\,,\qquad
\mathcal{W}_\nu =\left(\begin{array}{cc}
0 & 0\\
B_\nu^+ & 0
\end{array}\right)\,,
\end{eqnarray}
and its Hermitian counterpart, as follows, 
\begin{eqnarray}
\label{Ospnil1}
&[\mathcal{H}_\nu^{e},\mathcal{G}_\nu^{\pm}]=\pm 2\omega\mathcal{G}_\nu^{\pm}\,, \qquad 
[\mathcal{G}_\nu^{-},\mathcal{G}_\nu^{+}]=4\omega(\mathcal{H}^{e}_\nu-\omega^2\mathcal{R}_\nu)\,,&\\
&[\mathcal{H}_\nu^{e},\mathcal{W}_\nu]=- 2\omega\mathcal{W}_\nu\,,\qquad
[\mathcal{R}_\nu,\mathcal{Q}_\nu]=\mathcal{Q}_\nu\,,\qquad
[\mathcal{R}_\nu,\mathcal{W}_\nu]= -\mathcal{W}_\nu\,,\qquad
&\\&
\{\mathcal{Q}_\nu,\mathcal{Q}_\nu^\dagger\}=\mathcal{H}_\nu^{e}\,,\qquad
\{\mathcal{W}_\nu,\mathcal{W}_\nu^\dagger\}=\mathcal{H}_\nu^{e}-2\omega \mathcal{R}_\nu\,, &\\&
\{\mathcal{Q}_\nu,\mathcal{S}_\nu\}=\mathcal{G}_\nu^- \,,\qquad
[\mathcal{G}_\nu^-,\mathcal{Q}_\nu^\dagger]= 2 \omega \mathcal{W}_\nu \,,\qquad
[\mathcal{G}_\nu^-,\mathcal{W}_\nu^\dagger]= 2\omega \mathcal{Q}_\nu \,,
\label{Ospnilf}
\end{eqnarray}
in addition with corresponding Hermitian conjugate relations. 
In this base, we have the automorphism 
$\mathcal{H}_\nu^{e}\rightarrow\mathcal{H}_\nu^{b}\,,$
$\mathcal{G}_\nu^{\pm}\rightarrow\mathcal{G}_\nu^{\pm}\,,$
$\mathcal{R}_\nu\rightarrow -\mathcal{R}_\nu\,,$
 $\mathcal{Q}_\nu^a\leftrightarrow \mathcal{S}_\nu^a\,.$

As was for the bosinic case, one can use this structure as an 
approach to the study of the super-harmonic oscillator 
system,
whose corresponding $\mathfrak{osp}(2|2)$ generators are 
\be
\label{ospos}
\{\mathcal{H}_{os}\,,\mathcal{G}^\pm\,,\mathcal{R}\,,\mathcal{Q}^{a}\,,\mathcal{S}^{a}\} =\{\mathcal{H}_\nu^{e}\,,\mathcal{G}_\nu^\pm\,,\mathcal{R}_\nu\,,\mathcal{Q}_\nu^{a}\,,\mathcal{S}_\nu^{a}\} |_{\nu=-1, y\rightarrow \tilde{y}}\,,
\ee
where $\tilde{y}\in \R$.
The super-Hamiltonian  $\mathcal{H}_{os}=\text{diag}(\mathscr{H}_{os}+\omega,\mathscr{H}_{os}-\omega)$
is a composition of two copies of an harmonic oscillator Hamiltonian, 
displaced from each other. On the other hand, from the perspective of the Darboux transformation, 
the seed states used to construct the fermionic operators 
$\mathcal{Q}^{a}$ and $\mathcal{S}^{a}$ are
 $\psi_{0}(\tilde{y}) \propto e^{-\tilde{y}^2/2}$ and 
$\psi_{0}(i\tilde{y})\propto  e^{\tilde{y}^2/2}$ respectively, see \textcolor{red}{[\cite{InzPly1}]}, and 
as a consequence, 
 both resulting systems $\mathcal {H}_{os} $ and 
$ \mathcal{H}_{os} -4 \mathcal{R}_0 $ have the exact Poincar\'e supersymmetry, 
 in contrast to the AFF case, since 
$ \psi_{0} (i \tilde{y})^{- 1} \propto \psi_{0} (\tilde{y}) $.
Finally, 
the intertwining operators are reduced to the usual harmonic oscillator ladder operators,
\be
A^\pm  |_{\nu=-1, y\rightarrow \tilde{y}}=a^\pm\,,\qquad 
B^\pm |_{\nu=-1, y\rightarrow \tilde{y}}=-a^\mp\,,\qquad a^\pm =\frac{1}{\sqrt{2}}\left(\omega \tilde{y}\mp \frac{d}{d\tilde{y}}  \right)\,.
\ee
A radical difference with the super-extended AFF model is that for
the super-harmonic oscillator system we can also build the 
additional operators 
\be
\label{Fysigma}
\mathcal{F}^\pm=\left(\begin{array}{cc}
a^\pm & 0\\
0 & a^\pm 
\end{array}\right)\,,\qquad 
\Sigma_{1}=\frac{1}{2}\sigma_{1}\,,\qquad 
\Sigma_{2}=-\frac{1}{2}\sigma_{2}\,,
\ee
that  supplement the $\mathfrak{osp}(2|2)$ superalgebra 
with the (anti)-commutation relations 

\begin{eqnarray}
\label{qho3}
&[\mathcal{H}_{os},\mathcal{F}^{\pm}]=\pm \omega \mathcal{F}^{\pm}\,,\qquad
[\mathcal{F}^\mp,\mathcal{G}^{\pm}]=\mp \omega \mathcal{F}^{\pm}\,,\qquad
[\mathcal{F}^-,\mathcal{F}^{+}]=\omega \mathbb{I}\,,&\\
&\{\Sigma_{a},\Sigma_b\}=\frac{1}{2}\delta_{ab}\mathbb{I}\,,\qquad
 [\mathcal{H}_{os},\Sigma_a]=-i\omega \epsilon_{ab}\Sigma_{b} \,,\qquad
 [\mathcal{R},\Sigma_a]=i\epsilon_{ab}\Sigma_{b}\,,&\\
&\{\Sigma_{a},\mathcal{Q}_b\}=\frac{1}{2}[\delta_{ab}(\mathcal{F}^{+}+\mathcal{F}^{-})
-i\epsilon_{ab}(\mathcal{F}^{+}-\mathcal{F}^{-})]\,,&\\
&\{\Sigma_{a},\mathcal{S}_b\}=\frac{1}{2}[ \delta_{ab}(\mathcal{F}^{+}+\mathcal{F}^{-})+
i\epsilon_{ab}(\mathcal{F}^{+}-\mathcal{F}^{-})]\,,&\\
&[\mathcal{F}^-,\mathcal{Q}_a]= \omega(\Sigma_a+i\epsilon_{ab}\Sigma_b)\, ,\qquad 
[\mathcal{F}^+,\mathcal{Q}_a]=- \omega(\Sigma_a-i\epsilon_{ab}\Sigma_b)\,,&\\
\label{qho4}
&[\mathcal{F}^-,\mathcal{S}_a]=\omega(\Sigma_a-i\epsilon_{ab}\Sigma_b) \,,\qquad 
[\mathcal{F}^+,\mathcal{S}_a]=- \omega(\Sigma_a+i\epsilon_{ab}\Sigma_b)\,,&\\
\label{SigGJ0}
&[\Sigma_{a},\mathcal{F}^\pm]=[\Sigma_{a},\mathcal{G}^\pm]=0\,.&
\end{eqnarray}
Again, operators (\ref{Fysigma}) do not commute with $\mathcal{H}_{\text{os}}$
so in the Heisenberg picture we will have the dynamical integrals 
${}_H\mathcal{F}^\pm= e^{\mp i \omega t},\mathcal{F}^\pm$  and 
${}_H\Sigma^\pm= e^{-i \sigma_3 \omega t}\Sigma^\pm$ .

Note that generators  $\{\mathcal{F}^\pm, \I,\mathfrak{S}_{a}\}$
produce an ideal sub-supergebra, which we identify with the natural super-extension of Heisenberg's symmetry. 
In fact, the superalgebraic structure generated by (\ref{ospos}), along with the Eqs.
 (\ref{qho3})-(\ref{SigGJ0}) is a semi-direct sum of this super-Heisenberg symmetry 
and the superalgebra $\mathfrak{osp}(2|2) $, corresponding to 
an  $ \mathcal{N} = 2 $ super-extension of the Schr\"odinger symmetry 
\textcolor{red}{[\cite{beckers1, beckers2, InzPly1}]} .

\section{The zero frequency limit}
\label{Sec0omegalimit}

In this paragraph we take the limit $ \omega \rightarrow 0 $ 
in supersymmetric generators introduced  in last section,
getting new  $ \mathcal{N} = 2 $ super-extended  systems. 
We start with the supersymmetric AFF model generators,
but now we consider the basis
\begin{eqnarray}
&
\hat{\mathcal{D}}_\nu=\frac{i}{4\omega}(\mathcal{G}_\nu^--\mathcal{G}_\nu^+)\,,\qquad 
\hat{\mathcal{K}}_\nu= \frac{1}{4\omega^2} (\mathcal{H}_\nu^{e}-\mathcal{G}_\nu^--\mathcal{G}_\nu^-)\,,\qquad 
\hat{\mathcal{H}}_\nu=\frac{1}{2}(\mathcal{H}_\nu^{e}+\mathcal{H}_\nu^{b})-\omega^2 \hat{\mathcal{K}}_\nu\,,
&\\&
\xi_\nu^a= \frac{1}{2}\epsilon^{ab}(\mathcal{Q}_\nu^a-\mathcal{S}_\nu^a)\,,\qquad 
\mathcal{\pi}_\nu^a= \frac{1}{2\omega }\epsilon^{ab}(\mathcal{Q}_\nu^a+\mathcal{S}_\nu^a)\,,\qquad 
\mathcal{Z}_\nu= \frac{1}{2}\mathcal{R}_\nu\,.&
\end{eqnarray}
The generators defined in this way satisfy 
\begin{eqnarray}
&
[\hat{\mathcal{D}}_\nu,\hat{\mathcal{H}}_\nu]=i\hat{\mathcal{H}}_\nu\,,\qquad
[\hat{\mathcal{D}}_\nu,K_\nu]=-i\hat{\mathcal{K}}_\nu\,,\qquad
[\hat{\mathcal{H}}_\nu,\hat{\mathcal{D}}_\nu]=-2\hat{\mathcal{D}}_\nu\,,&\\&
\{\zeta_\nu^{a},\zeta_\nu^{b}\}=2\hat{\mathcal{K}}_\nu\delta^{ab}\,,\qquad 
\{\pi_\nu^{a},\pi_\nu^{b}\}=2\hat{\mathcal{H}}_{\nu}\delta^{ab}\,,\qquad 
\{\zeta_\nu^{a},\zeta_\nu^{b}\}=2\hat{\mathcal{D}}_\nu\delta^{ab}+2\epsilon^{ab}\mathcal{Z}_\nu\,,
&\\&\label{superS3}
[\hat{\mathcal{D}}_\nu,\pi_\nu^a]=\frac{i}{2}\pi_\nu^a\,,\quad 
[\hat{\mathcal{D}}_\nu,\xi_\nu^a]=-\frac{i}{2}\xi_\nu^a\,,\quad 
[\mathcal{Z}_\nu,\pi_\nu^a]=-\frac{i}{2}\epsilon_{ab}\pi_\nu^b\,,\quad 
[\mathcal{Z}_\nu,\xi_\nu^a]=-\frac{i}{2}\epsilon_{ab}\xi_\nu^b\,,
&\\&
\label{superS3+}
[\hat{\mathcal{H}}_\nu,\xi_\nu^a]=-i\pi_\nu^a\,,\qquad
[\hat{\mathcal{K}}_\nu,\pi_a]=i\xi_\nu^a\,.
\end{eqnarray}
This is the usual way in which the superalgebra $ \mathfrak{osp} (2,2) $
 is presented for supersymmetric extensions of the conformal model
 (\ref{conformalaction}) at the quantum level \textcolor{red}{[\cite{Leiva,SCM5}]}.
 So it is not a surprise that at the zero frequency limit we get
\begin{eqnarray}
&
\hat{\mathcal{H}}_\nu|_{\omega=0}=\frac{1}{2}(p^2+\frac{\nu^2}{y^2})\mathbb{I}+\frac{\nu}{2y^2}\sigma_3\,,&\\&
\hat{\mathcal{D}}_\nu|_{\omega=0}= \frac{1}{4i}\left(y\frac{d}{dy}+\frac{d}{dy}y\right)\mathbb{I}:=\mathcal{D}\,,\qquad 
\hat{\mathcal{K}}_\nu|_{\omega=0}= \frac{y^2}{2}\mathbb{I}:=\mathcal{K}\,, &\\&
\xi_\nu^a |_{\omega=0}=\frac{y}{\sqrt{2}}\sigma_a,,\qquad
\pi_\nu^a|_{\omega=0}=\frac{1}{\sqrt{2}} \left(
p\sigma_a
-\frac{\nu+1}{y}\epsilon_{ab}\sigma_b\right)\,,
&
\end{eqnarray}
where $\mathbb{I}=\text{diag}(1,1)$ and $p=-i\frac{d}{dy}$. 

We can repeat this procedure for the super-Schr\"odinger
 symmetry, which we have derived 
for the super-harmonic oscillator system. In this case the generators 
\be
\label{freeparticlesupergen}
\{\hat{\mathcal{H}}_\nu,\hat{\mathcal{D}}|_{\omega=0}, \hat{\mathcal{K}}|_{\omega=0} \,,
\mathcal{Z}_\nu\,,\xi_\nu^a|_{\omega=0}\,,\pi_\nu^a |_{\omega=0}\}|_{\nu=-1, y\rightarrow \tilde{y}}=
\{\mathcal{H}_0, \mathcal{D},\mathcal{K},\mathcal{Z}\,,
\xi_a\,,\pi_a\}
\ee
  reflect the superconformal symmetry of the super-extended free particle,
 which in turn includes the additional integrals
$\Sigma_1=\frac{1}{2}\sigma_1$,  
$\Sigma_2=-\frac{1}{2}\sigma_2$ and 
\begin{equation}
\mathcal{P}=\frac{i}{2}(\mathcal{F}^+-\mathcal{F}^+)|_{\omega=0,y\rightarrow \tilde{y}}= -\frac{i}{\sqrt{2}}\frac{d}{d\tilde{y}} \mathbb{I} \,,\qquad 
\mathcal{X}=\frac{1}{2\omega}(\mathcal{F}^++\mathcal{F}^+)|_{\omega=0,y\rightarrow \tilde{y}}=
\frac{\tilde{y}}{\sqrt{2}} \mathbb{I} \,.
\end{equation}
Together, generators  
$\{\mathcal{H}_0, \mathcal{D},\mathcal{K},\mathcal{X}\,,\mathcal{P}\,,
\xi_a\,,\pi_a\,,\Sigma_a\}$ produce the super-Schr\"odinger symmetry, now for the super-free particle 
system \textcolor{red}{[\cite{Aizawa,InzPly1}]},  
\begin{eqnarray}
\label{superS1}
&
[\mathcal{D},\mathcal{H}_0]=i\mathcal{H}_0\,,\qquad 
[\mathcal{D},\mathcal{K}]=-i\mathcal{K}\,,\qquad 
[\mathcal{K},\mathcal{H}_0]=2i\mathcal{D}\,,\qquad
[\mathcal{X},\mathcal{P}]=\frac{1}{2}i\mathbb{I}\,,
&\\&\label{superS2}
[\mathcal{H}_0,\mathcal{X}]=-i\mathcal{P}\,,\qquad 
[\mathcal{K},\mathcal{P}]=i\mathcal{X},\qquad 
[\mathcal{D},\mathcal{P}]=\frac{i}{2}\mathcal{P}\,,\qquad
[\mathcal{D},\mathcal{X}]=-\frac{i}{2}\mathcal{X}\,,
&\\&\label{superSH3}
[\mathcal{D},\pi_a]=\frac{i}{2}\pi_a\,,\quad 
[\mathcal{D},\xi_a]=-\frac{i}{2}\xi_a\,,\quad 
[\mathcal{Z},\pi_a]=-\frac{i}{2}\epsilon_{ab}\pi_b\,,\quad 
[\mathcal{Z},\xi_a]=-\frac{i}{2}\epsilon_{ab}\xi_b\,,
&\\&\label{superSH3+}
[\mathcal{H}_0,\xi_a]=-i\pi_a\,,\qquad
[\mathcal{K},\pi_a]=i\xi_a\,,
&\\&
\label{superS4}
[\mathcal{Z},\Sigma_a]=\frac{i}{2}\epsilon_{ab}\Sigma_b\,,\qquad
[\mathcal{P},\pi_a]=-i\Sigma_a\,,\qquad 
[\mathcal{X},\xi_a]=i\Sigma_a\,,
&\\&\label{superS5}
\{\Sigma_a,\pi_{b}\}=\delta_{ab}\mathcal{P}\,, \qquad 
\{\Sigma_a,\xi_{b}\}=\delta_{ab}\mathcal{X}\,,\qquad
\{\Sigma_a,\Sigma_b\}=\frac{1}{2}\delta_{ab}\mathbb{I}\,,
&\\&\label{superS6}
\{\pi_a,\pi_b\}=2\delta_{ab}\mathcal{H}_0\,,\qquad
\{\xi_a,\xi_b\}=2\delta_{ab}\mathcal{K}\,,\qquad
\{\pi_a,\xi_b\}=2\delta_{ab}\mathcal{D}+2\epsilon_{ab}\mathcal{Z}\,.
\end{eqnarray}

\section{Remarks}
In this chapter we have considered one-dimensional
 conformal and an $\mathcal{N}=2$ super-conformal mechanical models.
In the bosonic case, there are many models that share the same conformal symmetry 
and some examples are the charged particle in a Dirac monopole background, 
Landau problem, rational Calogero models of $ N $ particles, 
geodesic  motion in extreme black holes, the free particle and the harmonic oscillator in $ d $ dimensions, to name a few. In particular, some systems in various dimensions are especially rich thanks to the presence of conformal symmetry. Such is the case of the rational Calogero model, which is not only integrable, but also super-integrable, 
see \textcolor{red}{[\cite{CorLefPly}]} and references therein.
On the other hand, higher extensions of superconformal models  are also a regular topic in scientific literature 
\textcolor{red}{[\cite{SCM1,SCM2,SCM3,SCM4,SCM5}]}.

In Sec \ref{SecAFF} we have emphasized that the models
 (\ref{conformalaction}) and (\ref{conformalaction2}) 
represent two different forms of dynamics associated with conformal algebra.
 In the next chapter we will show that there is a non-unitary mapping between both models.
 We call it  the conformal bridge transformation,
 and it might be useful to obtain hidden symmetries for higher 
dimensional conformal invariant models.


\chapter{The conformal bridge}
\label{ChBridge}


As we highlighted in the previous chapter, the 
conformal invariant systems with or without a harmonic potential are just
two different dynamical phases of the same algebraic structure. However, 
there seems to be no direct relationship at the eigenstate level because 
one of the Hamiltonians is a non-compact generator, in contrast to the another Hamiltonian
(the harmonically trapped one), which is compact. 
The objective of this chapter is to show that there is a non-unitary
 transformation that effectively maps one quantum mechanical system to the other but in an unorthodox way.
To do so, let us start with algebra (\ref{so(2,1)preambulo}) without specifying 
a particular form of the generators. Then we construct the operators
\begin{equation}
\label{Generalbridge}
\mathfrak{S}=e^{-\alpha K} e^{\frac{H}{2\alpha}}
e^{i\ln(2)D}\,,\qquad
\mathfrak{S}^{-1}= e^{-i\ln(2)D}e^{-\frac{H}{2\alpha}}e^{\alpha K}\,,
\end{equation}
which from now on we will call as ``conformal bridge'', because by means of the Baker-Campbell-Hausdorff formula 
\begin{equation}
e^{A}Be^{-A}=B+[A,B]+\frac{1}{2!}[A,[A,B]]+\ldots\,,
\end{equation}
one can show  that 
\begin{eqnarray}
\label{ConfBrid chapter 2}
&
\mathfrak{S}(H)\mathfrak{S}^{-1}=\alpha \mathcal{J}_-\,, \qquad
\mathfrak{S}(D)\mathfrak{S}^{-1}=-i\mathcal{J}_0\,, \qquad
\mathfrak{S}(K)\mathfrak{S}^{-1}=-\frac{1}{\alpha}\mathcal{J}_+\,.
&
\end{eqnarray}
Here, $\mathcal{J}_0$ and $\mathcal{J}_\pm$ correspond to the generators 
of the $\mathfrak{sl}(2,\R)$ algebra given in 
(\ref{sl(2R)preambulo}).  
Note that the transformed generators in (\ref{ConfBrid chapter 2}) 
still satisfy the $\mathfrak{so}(2,1)$
symmetry, i.e., the transformation is an automorphism 
of the algebra.

 Anyway, as we showed in the previous chapter, for the  one-dimensional case
$H$ could represent  the Hamiltonian of a  free particle or that of the model (\ref{conformalaction}) and  
 on the other hand, $\mathcal{J}_0$ could be  the Hamiltonian of 
 an harmonic oscillator or that of the  AFF model. 
Therefore, the conformal bridge transformation  produces a mapping between 
these two forms of dynamics as follows: the formal eigenstates of  $ -iD $ are transformed into
those of 
$\mathcal {J} _0 $, and on the other hand, eigenstates
of $H$ are mapped to eigenstates of the lowering 
operator $\mathcal{J}_-$, which are 
in turns coherent states for $\mathcal{J}_0$. Of course, these statements
 remain true for any other higher-dimensional representation of the generators.
 In the following sections we explore the  scope  of this transformation
 with examples in one and two dimensions.

The content of this chapter is based on \textcolor{red}{[\cite{InzPlyWipf2}]}.
Here we only consider the basic elements and important results
related with quantum mechanics examples, even though construction
 can be extended to the classical level, as we briefly discus in Sec. \ref{Remarks3}.

\section{Free particle/ harmonic oscillator conformal bridge}

Let us identify $\alpha$ with $\omega$
and $H$, $K$ and $D$ with the free particle conformal symmetry 
generators in the Schr\"odinger picture ($\hbar=m=1$),
\begin{equation}
\label{free}
H=-\frac{1}{2}\frac{d^2}{dx^2}:=H_0\,,\qquad
D=\frac{1}{2i}\left(\frac{d}{dx}+\frac{1}{2}\right)\,,\qquad
K=\frac{x^2}{2}\,.
\end{equation}
Then, the conformal bridge takes the form 
\begin{equation}
\label{U0KH+}
\mathfrak{S}=\exp(-\omega\frac{x^2}{2})\exp(-\frac{1}{4\omega}\frac{d^2}{dx^2})\exp(
\ln\sqrt{2}\,\left(x\frac{d}{dx}+\frac{1}{2}\right))\,,
\end{equation}
besides $\mathcal{J}_0$ and $\mathcal{J}_\pm$ are the symmetry generators of 
the harmonic oscillator,
\begin{equation}
\label{harmonicoscillatorgen}
2\omega \mathcal{J}_0=\frac{1}{2}\left(-\frac{d^2}{dx^2}+\omega^2x^2\right):=H_{os}\,,\qquad
2\omega \mathcal{J}_\pm=-(a^\pm)^2\,,\qquad
a^\pm=\frac{1}{\sqrt{2}}\left(\omega x\mp \frac{d}{dx}\right)\,.
\end{equation}
As we saw in the previous chapter,
these operators are well defined for $x\in \R$, 
and there are many more symmetries for these systems. In the case 
of the free particle we have the momentum 
operator $p=-i\frac{d}{dx}$ 
and the Galilean boost, which 
in the Schr\"odinger picture at $t=0$ is just $x$.
These objects are connected with the 
Heisenberg generators 
$a^\pm$, appearing in (\ref{harmonicoscillatorgen}),  via 
the conformal bridge 
as follow 
\begin{equation}
\label{Heisenbergmap}
\mathfrak{S} (x) \mathfrak{S}^{-1}=\frac{1}{\omega}a^+ \,,\qquad
\mathfrak{S}(p) \mathfrak{S}^{-1}=-ia^-\,,
\end{equation}  
and, therefore,  the transformation 
is also an automorphism of the Schr\"odinger symmetry. 

For the sake of simplicity,  we set $\omega=1$
along the rest of this chapter. 

The relation (the inverse Weierstrass transformation of a monomial)
\begin{equation}
\label{the wellknown relation}
e^{-\frac{1}{4}\frac{d^2}{dx^2}}x^{n}=2^{-n}H_n(x)\,,
\end{equation}
where $H_n(x)$ are the Hermite polynomials, implies that 
\begin{equation}
\psi_{n}(x)=\frac{1}{\sqrt{\pi^{1/2}n!}}H_{n}(x)e^{-\frac{x^2}{2}}=(2\pi)^{\frac{1}{4}}
\sqrt{2^{n}n!}\,\,
\mathfrak{S}\left(\frac{x}{\sqrt{2}}\right)^{n}\,,
\end{equation}
which are the eigenstates of $H_{os}$ with eigenvalues 
$E_{n}=\omega(n+\frac{1}{2})$.

With the exception of $1$ and $x$, 
which are annihilated by  $H_0$, the functions 
$x^n$ are not solutions of the free particle Schr\"odinger 
equation. They are in fact the rank $n$ Jordan
 states of the zero energy, 
as  the relations 
\begin{eqnarray}
&
(H)^{j}x^{2l}=\left(-\frac{1}{2}\right)^{2l}\frac{(2l)!(2l-1)!}{(2l-j)!(2l-1-j)!}x^{2(l-j)}\,,&\\&
(H)^{j}x^{2l+1}=\left(-\frac{1}{2}\right)^{2l+1}\frac{(2l)!(2l+1)!}{(2l-j+1)!(2l-j)!}x^{2(l-j)+1}\,,
\end{eqnarray}
 valid for $j=0,\ldots,l$, show us.  If we set $j=l$ in  these formulas, a subsequent application
of   $H$ from the left 
produces 0 on the right hand side of the equation, and for  this reason 
the nonlocal operator  
in (\ref{the wellknown relation})
 produces 
a polynomial of order $n$. Also  these Jordan states satisfy 
the equation $2iDx^{n}=(n+1/2)x^{n}$, so
 it is not really a surprise that after applying $ \mathfrak{S} $, 
one obtains the harmonic oscillator eigenstates.

Now let us put our attention to the plane waves. We know 
that the functions $e^{i\frac{\kappa}
{\sqrt{2}} x}$ are eigenstates 
of the free particle with energy $E=\kappa^2$, then 
the application of 
$\mathfrak{S}$ produces
\begin{equation}
\mathfrak{S}e^{i\frac{\kappa}
{\sqrt{2}} x}=2^{\frac{1}{4}} \exp(-\frac{x^2}{2}+\frac{\kappa^2}{4}+i\kappa x)=(2\pi)^{\frac{1}{4}}
\sum_{n=0}^{\infty}\sqrt{\frac{2^{n}}{n!}}\,\,(ik)^n\psi_{n}(x):=\psi_{CS}(x,\kappa)\,.
\end{equation}   
These funtions are eigenstates of $a^-$ and $(a^-)^2$, 
and up to a normalization factor, they are coherent states of 
$H_{os}$, 
\textcolor{red}{[\cite{Schrodinger,Klauder,Gazeau}]}.
 In fact, by applying 
the evolution operator $U=e^{iH_{os}t}$ one gets 
$\psi_{CS}(x,\kappa,t)=e^{\frac{i t}{2}}\psi_{CS}(x,\kappa e^{it})$ which 
is a solution of the harmonic oscillator time-dependent Schr\"odinger equation.  
To obtain the over-complete set of coherent states, an analytical continuation in $\kappa$
must be done, allowing complex values.

From these results one can  formulate  a general recipe:

\begin{itemize}
\item  Under the conformal bridge transformation, the formal states of the operator 
$2iD$, that are also the rank $n$ Jordan state of zero energy, are mapped to
normalizable 
eigenstates of $J_0$. 

\item Eigenstates of the Hamiltonian $H$ are transformed
 into
coherent states of the system $J_0$, 
which are eigenstates of $J_-$  and 
 conserve their form with time evolution. To have the overcomplete set,
negative energy solutions (complex $\kappa$) should be also considered in this map. 

\item The conformal bridge also serves to map other symmetries
from one system to another, as was  the case for
generators of the Heisenberg algebra (\ref{Heisenbergmap}).
\end{itemize}

In \textcolor{red}{[\cite{InzPlyWipf2}]}, it is shown how to obtain the squeezed states
by applying the conformal bridge  to Gaussian packets, and 
there is also an interesting discussion about the relation of this transformation and the 
Stone-von Newman
theorem \textcolor{red}{[\cite{Tak}]}. Nevertheless, 
we prefer to not dwell with  these details here.

\section{Conformal bridge and the AFF model}
\label{conformal mechanics bridge}
Let us now set up $H$ as
the Hamiltonian operator of the two-body Calogero model 
with omitted  center of mass degree of freedom 
\begin{eqnarray}\label{Hnudef}
H=\frac{1}{2}\left(-\frac{d^2}{dx^2}+\frac{\nu(\nu+1)}{x^2}\right):=H_\nu\,,
\end{eqnarray}
besides $D$ and $K$ take the same form given in (\ref{free}). 
With this choice, the conformal bridge 
is also labeled  by $\nu$,
\begin{equation}
\mathfrak{S}_\nu=\exp(-\frac{x^2}{2})
\exp(\frac{1}{4}\left(-\frac{d^2}{dx^2}+
\frac{\nu(\nu+1)}{x^2}\right))
\exp(
\ln\sqrt{2}\,\left(x\frac{d}{dx}+\frac{1}{2}\right))\,,
\end{equation}
and operators $J_0$, $J_\pm$ are 
now 
\begin{equation}
2\omega J_0=\frac{1}{2}\left(-\frac{d^2}{dx^2}+\frac{\nu(\nu+1)}{x^2}+x^2 \right):=\mathcal{H}_\nu\,,\qquad
2\omega J_\pm=-(a^\pm)^2+\frac{\nu(\nu+1)}{2x^2}:=\mathcal{C}_\nu^\pm\,.
\end{equation}

Following the recipe described above, we look 
for the zero energy solutions and its Jordan states, 
then 
consider the set of functions
  $x^{\nu+1+2n}$,  $n=0,1,\ldots$.
The function with $n=0$ represents a formal, diverging at infinity,
 eigenstate of the differential operator
$H_\nu$ with $\nu\geq -1/2$ of  eigenvalue $E=0$.
For  $n\geq 1$  this functions  are the Jordan states of rank $n$ 
corresponding to the same eigenvalue of $H_\nu$.
The functions $x^{\nu+1+2n}$ are  at the same time 
eigenstates of the operator $2i D$ with eigenvalues 
$\nu+2n+3/2$. 
The Jordan states with $n\geq 1$ satisfy the relations
\begin{eqnarray}
\label{j-th_element1+}
(H_{\nu})^{j}x^{\nu+1+2n}=
\frac{(-2)^{j}\Gamma(n+1)}{\Gamma(n+1-j)}\frac{\Gamma(n+\nu+3/2)}{\Gamma(n+\nu+3/2-j)}x^{\nu+1+2(n-j)}\,,
\quad
j=0,1,\ldots,n\,,
\end{eqnarray}
which can be proved by  induction.
Eq. (\ref{j-th_element1+}) extends to the case $j=n+1$ giving
$(H_{\nu})^{n+1}x^{\nu+1+2n}=0$ due to  appearing of a  simple pole in the 
denominator.
\vskip0.1cm 
Using relation (\ref{j-th_element1+}) one can compute the conformal bridge transformation in functions
$x^{2n +\nu+1}$, which gives  
\begin{eqnarray}
\mathfrak{S}_\nu \left( 
\frac{x}{\sqrt{2}}
\right)^{\nu+1+2n}=2^{\frac{-1}{4}}(-1)^{n}\sqrt{n!\Gamma(n+\nu+3/2)}\,\,\psi_{\nu,n}(x)\,,
\end{eqnarray}
 where eigenstates $\psi_{\nu,n}(x)$ correspond to
 (\ref{AFF states}) (with $\omega=1$ and $y=x$).

On the other hand, application of the operator
$\mathfrak{S}_\nu$ to the eigenstates 
(\ref{states calogero}) (with $x=q$)
of the system $H_\nu$ 
gives
\begin{eqnarray}
\label{coherent0}
&\mathfrak{S}_{\nu}\psi_{\kappa,\nu}(\frac{1}{\sqrt{2}}x)=2^{\frac{1}{4}}e^{-\frac{1}{2}x^2+\frac{1}{4}\kappa^2}
\sqrt{x}J_{\nu+1/2}(\kappa x):=\phi_\nu(x,\kappa)\,.&
\end{eqnarray} 
 These are the coherent states of the AFF model \textcolor{red}{[\cite{Perelomov}]},
which satisfy 
\begin{equation}
\mathcal{J}_-\phi_\nu(x,\kappa)
=-\frac{1}{4}\kappa^2\phi_\nu(x,\kappa)\,.
\end{equation}
By allowing  the $\kappa>0$ to become a  complex parameter $z$,
coherent states can be constructed with  complex eigenvalues
of the operator $\mathcal{J}_-$.
Application of the evolution operator $e^{-it H_\nu}$ 
to these states gives the time-dependent coherent states
\begin{eqnarray}
\phi_{\nu}(x,z,t)=2^{1/4}\sqrt{x}\mathcal{J}_{\nu+1/2}(z(t) x)e^{-x^2/2+z^2(t)/4-it}\,,
\end{eqnarray} 
where $z(t)=z e^{-it}$.
In the case of $\nu=0$, these time-dependent coherent states of 
the AFF
model are the odd Schr\"odinger cat states of the
quantum harmonic oscillator \textcolor{red}{[\cite{Dodonov}]}, 
\begin{equation}
\phi_{0}(x,z,t)\propto e^{-\frac{x^2}{2}+\frac{z^2(t)}{4}-\frac{it}{2}}\sin(z(t)x)\,.
\end{equation}

\section{The conformal bridge and Landau problem}

The generalization of the conformal bridge between   free particle and harmonic oscillator 
to the $d-$dimensional case is straightforward; since the problem is separable in 
Cartesian coordinates, the conformal  bridge operator is just
 $\mathfrak{S}(\vr)=\mathfrak{S}(x_1)\ldots\mathfrak{S}(x_d)$. 
Each $\mathfrak{S}(x_i)$ touch only the objects constructed in terms  
of $x_i$ and $p_i=-\frac{d}{dx_i}$, leaving invariant the other coordinates.
On the other hand, as both systems posses the  $\mathfrak{so}(d)$ symmetry, 
the generalized angular momentum tensor $M_{ij}=x_ip_j-x_jp_i$ remains intact
after the similarity  transformation.

On the other hand, a nontrivial relation between two-dimensional free particle, whose
conformal symmetry generators are 
\begin{eqnarray}
\label{2dimensionConf}
H=\frac{1}{2}(p_x^2+p_y^2)\,,\qquad 
D=\frac{1}{2}(xp_x+yp_y +1)\,,\qquad
K=\frac{1}{2}(x^2+y^2)\,,
\end{eqnarray}
 and
 the Landau problem in the symmetric gauge, can be established by means of the 
two-dimensional conformal bridge operator
\begin{equation}\label{Sxy}
\mathfrak{S}(x,y)=\mathfrak{S}(x)
\mathfrak{S}(y)\,,
\end{equation}
with $\mathfrak{S}(x)$ 
and $\mathfrak{S}(y)$ of the form  (\ref{U0KH+}). This is the subject of this section.

Consider now the Landau problem for a scalar particle on $\R^2$.
In the symmetric gauge
$\vec{A}=\frac{1}{2}B(-q_2,q_1)$, 
the Hamiltonian operator (in units $c=m=\hbar=1$) is given by  
\begin{equation}
\label{Landau}
H_{\text{L}}=\frac{1}{2}\vec{\Pi}^{2},\qquad
\Pi_j=-i\frac{\partial}{\partial q_j}-eA_j\,, \qquad
[\Pi_1,\Pi_2]=ieB\,.
\end{equation}
Assuming $\omega_c=eB>0$, this operator can be  factorized as 
\begin{eqnarray}
&H_{\text{L}}=\omega_c(\mathcal{A}^+\mathcal{A}^-+\frac{1}{2})\,, &\\&
\label{Landaulad}
\mathcal{A}^\pm=\frac{1}{\sqrt{2\omega_c}}(\Pi_1\mp
 i\Pi_2)\,,\qquad [\mathcal{A}^-,
\mathcal{A}^+]= 1\,.&
\end{eqnarray}
Setting  $\omega_c=2$, we can identify $q_i$ with dimensionless variables  $q_1=x$,
$q_2=y$. Then  we present $\mathcal{A}^\pm$ as linear combinations of the 
usually defined ladder operators $a_x^\pm$ and $a_y^\pm$
 (the shape of which corresponds to the third equation in (\ref{harmonicoscillatorgen})),
 in terms of which we also 
define the  operators $\mathcal{B}^\pm$,
\begin{equation}
\mathcal{A}^\pm=\frac{1}{\sqrt{2}}
(a_y^\pm \pm ia_x^\pm)\,,\qquad
\mathcal{B}^\pm=\frac{1}{\sqrt{2}}(a_y^\pm \mp ia_x^\pm)\,.
\end{equation}
The operators  $\mathcal{B}^\pm$  satisfy relation 
$[\mathcal{B}^-,\mathcal{B}^+]=1$, 
and commute with $\mathcal{A}^\pm$. They are integrals of motion, 
and their  non-commuting Hermitian linear  combinations 
$\mathcal{B}^++\mathcal{B}^-$
and  $i(\mathcal{B}^+-\mathcal{B}^-)$
are identified with the coordinates of the center of the cyclotron motion.
 In terms of the ladder operators 
$a_x^\pm$, $a_y^\pm$ the Hamiltonian $H_{\text{L}}$ takes the
form of a linear combination of the Hamiltonian of the isotropic oscillator $H_{\text{iso}}$ and 
angular momentum operator $M$,
\begin{equation}\label{HLM3iso}
H_{\text{L}}=H_{\text{iso}} -M\,,\qquad
H_{\text{iso}}=a_x^+a_x^-+a_y^+a_y^-+1\,,\qquad
M=xp_x-yp_y=-i(a_x^+a_y^--a_y^+a_y^- )\,.
\end{equation}
On the other hand, $H_{\text{iso}}$ and $M$
are presented in terms of $\mathcal{A}^\pm$ and 
$\mathcal{B}^\pm$
as follows,
\begin{equation}\label{M3Hiso}
M=
\mathcal{B}^+\mathcal{B}^--
\mathcal{A}^+\mathcal{A}^-\,,\qquad
H_{\text{iso}}=\mathcal{B}^+\mathcal{B}^-+
\mathcal{A}^+\mathcal{A}^-+1\,,
\end{equation}
and we have the commutation relations
$
[M,\mathcal{B}^\pm]=\pm\mathcal{B}^\pm,$
$
[M,\mathcal{A}^\pm]=\mp\mathcal{A}^\pm.$ 
By taking into account the invariance of the angular momentum under
similarity transformation, we find that its linear combination with the dilatation operator 
is transformed into the Hamiltonian of the Landau problem,
\be
\mathfrak{S}(x,y)(2iD-M)\mathfrak{S}^{-1}(x,y)=H_{\text{L}}\,.
\ee

Let us now introduce complex coordinate in the plane,
\begin{equation}
w=\frac{1}{\sqrt{2}}(y+ix)\,, \quad \text{and}\qquad \bar{w}=\frac{1}{\sqrt{2}}(y-ix)\,.
\end{equation}
The elements of conformal algebra and angular momentum 
operator take then the form 
\begin{equation}
\label{conformalcomplex}
H=-\frac{\partial^2}{\partial w \partial \bar{w}}\,,\quad
D=\frac{1}{2i}\left(w\frac{\partial}{\partial w}+\bar{w}\frac{\partial}{\partial \bar{w}}+1\right)\,,\quad
K=w\bar{w}\,,\quad
M=\bar{w}\frac{\partial}{\partial \bar{w}}-w\frac{\partial}{\partial w}\,, 
\end{equation}
and  we find  that the operator (\ref{Sxy}) generates the  
similarity transformations 
\begin{eqnarray}
&
\mathfrak{S}(x,y)w\mathfrak{S}^{-1}(x,y)=\mathcal{A}^+\,, 
\qquad
\mathfrak{S}(x,y)\left(\frac{\partial}{\partial w}\right)
\mathfrak{S}^{-1}(x,y)=\mathcal{A}^-\,,
&\label{wASig}\\
&
\mathfrak{S}(x,y)\bar{w}\mathfrak{S}^{-1}(x,y)=\mathcal{B}^+ \,,
\qquad
\mathfrak{S}(x,y)\left(\frac{\partial}{
\partial \bar{w}}\right)\mathfrak{S}^{-1}(x,y)=\mathcal{B}^-\,,
&\label{wBSig}\\
&
\mathfrak{S}(x,y)\left(w\frac{\partial} {\partial w}\right)\mathfrak{S}^{-1}(x,y)=
\mathcal{A}^+\mathcal{A}^-\,,\qquad
\mathfrak{S}(x,y)\left(\bar{w}\frac {\partial} {\partial \bar{w}}\right)\mathfrak{S}^{-1}(x,y)
=\mathcal{B}^+\mathcal{B}^-\,.
&\label{wdwSig}
\end{eqnarray}
Observe that each pair of relations in (\ref{wASig}) and  (\ref{wBSig}) 
has a form similar as the  one-dimensional transformation (\ref{Heisenbergmap}),
where, however, the coordinate and momentum are Hermitian operators.

Simultaneous eigenstates of the operators $w\frac{\partial }{\partial w}$ and 
$\bar{w}\frac{\partial }{\partial\bar{w}}$,
which satisfy the relations 
$w\frac{\partial}{\partial w}\phi_{n,m}=n\phi_{n,m}$ 
and 
$\bar{w}\frac{\partial }{\partial\bar{w}}\phi_{n,m}=m\phi_{n,m}$
with $n,m=0,1,\ldots$, 
are
\begin{equation}
\phi_{n,m}(x,y)= w^n(\bar{w})^{m}=2^{-(n+m)/2}\sum_{k=0}^{n}
\sum_{l=0}^{m}{n\choose k}{m\choose l}(i)^{n-m+l-k}y^{k+l}x^{n+m-k-l}\,, 
\end{equation}
where the binomial theorem has been used.
Employing Eq. (\ref{conformalcomplex}) we find that 
\begin{eqnarray}
&
M\phi_{n,m}=(m-n)\phi_{n,m}\,,\qquad
2iD\phi_{n,m}=(n+m+1)\phi_{n,m}\,,&\\&
\label{KyHenphi}
K\phi_{n,m}=\phi_{n+1,m+1}\,,\qquad
H\phi_{n,m}=-nm\phi_{n-1,m-1}\,.
&
\end{eqnarray}
The last equality shows  
that  $\phi_{0,m}$ and $\phi_{n,0}$ are the 
zero energy eigenstates of the two-dimensional free particle, 
while the $\phi_{n,m}$ with $n,m>0$ are the  Jordan states
corresponding to the same zero energy value.
Application of the  operator $\mathfrak{S}(x,y)$ 
to these functions yields
\begin{equation}
\label{landaustates}
\mathfrak{S}(x,y)\phi_{n,m}(x,y)=2^{2(n+m)+\frac{1}{2}}e^{-\frac{(x^2+y^2)}{2}}H_{n,m}(y,x)
=\psi_{n,m}(x,y)\,,
\end{equation}
where 
\begin{equation}
H_{n,m}(y,x)=2^{-(n+m)}\sum_{k=0}^{n}
\sum_{l=0}^{m}{n\choose k}{m\choose l}(i)^{n-m+l-k}H_{k+l}(y)H_{n+m-k-l}(x)\,,
\end{equation}
are the complex Hermite polynomials, see \textcolor{red}{[\cite{Hermite}]}. 
These functions are  
eigenstates of the operators $H_{\text{L}}$, $M$
and $H_{\text{iso}}$, 
\begin{eqnarray}
&
H_{\text{L}}\psi_{n,m}=(n+\frac{1}{2})\psi_{n,m}\,,\qquad 
M\psi_{n,m}=(m-n)\psi_{n,m}\,,&\\&
H_{\text{iso}}\psi_{n,m}=(n+m+1)\psi_{n,m}\,,&
\end{eqnarray}
and we note that $\psi_{n,n}$ is rotational invariant.

Eqs. (\ref{wASig}), (\ref{wBSig}),  and (\ref{KyHenphi}) show that 
the operators $\mathcal{A}^\pm$ and $\mathcal{B}^{\pm}$ act as the ladder operators 
for the  indexes  $n$ and $m$, respectively, 
while the operators $\hat{\mathcal{J}}_\pm=-\frac{1}{2}((a_x^\pm)^2+(a_y^\pm)^2)$,
increase or decrease simultaneously  $n$ and $m$ by one.

Application 
of the operator  $\mathfrak{S}(x,y)$ to exponential functions of the most general 
form   $e^{\alpha w+\beta \bar{w}}$ with $\alpha,\beta \in \mathbb{C}$
 gives  here, similarly to the one-dimensional case,  the coherent states
of the Landau problem as well of the isotropic harmonic oscillator,
\begin{eqnarray}
\label{coherentLandau}
\begin{array}{lcl}
\psi_{\text{L}}(x,y,\alpha,\beta)&=& \mathfrak{S}(x,y)
e^{\frac{1}{\sqrt{2}}((\alpha+\beta)y +i(\alpha-\beta)x)}=\sqrt{2}
e^{-\frac{(x^2+y^2)}{2}+(\alpha+\beta)y +i(\alpha-\beta)x-\alpha\beta}\\
&=& \sum_{n=0}^{\infty}\sum_{l=0}^{n}\frac{1}{n!}{n \choose l}\alpha^{l}\beta^{n-l}\psi_{l,n-l}(x,y)\,.
\end{array}
\end{eqnarray}
Applying to them, in particular,  the evolution operator $e^{-itH_{\text{L}}}$,
we  obtain the
time dependent solution to the Landau problem, 
\begin{equation}
\psi_{\text{L}}(x,y,\alpha,\beta,t)=e^{-\frac{i t}{2}}\psi_{\text{L}}(x,y,\alpha e^{-it},\beta)\,, 
\end{equation} 
whereas under rotations these states transform as 
\begin{equation}
e^{i\varphi M}\psi_{\text{L}}(x,y,\alpha,\beta)=\psi_{\text{L}}(x,y,\alpha e^{-i \varphi},\beta e^{i \varphi})\,.
\end{equation}
As the function  $e^{\alpha w+\beta \bar{w}}$
is a common eigenstate of the differential operators
$\frac{\partial}{\partial w}$ and  $\frac{\partial}{\partial \bar{w}}$
with eigenvalues $\alpha$ and $\beta$, respectively, then our 
transformation yields 
\begin{equation}
\mathcal{A}^-\psi_{\text{L}}(x,y,\alpha,\beta)=\alpha\psi_{\text{L}}(x,y,\alpha,\beta)\,,\qquad
\mathcal{B}^-\psi_{\text{L}}(x,y,\alpha,\beta)=\beta\psi_{\text{L}}(x,y,\alpha,\beta)\,, 
\end{equation}
that provides another explanation  why the wave functions 
(\ref{coherentLandau})  are 
the coherent states  for the planar harmonic oscillator as well as
for the Landau problem.

\section{Remarks}
\label{Remarks3}
Note that if we apply $\mathfrak{S}$ from the right to the equations in
(\ref{ConfBrid chapter 2}),
 we get intertwining relations  of the form 
\begin{eqnarray}
&
\mathfrak{S}H=\alpha \mathcal{J}_-\mathfrak{S}\,, \qquad
\mathfrak{S}D=-i\mathcal{J}_0\mathfrak{S}\,, \qquad
\mathfrak{S}K=-\frac{1}{\alpha}\mathcal{J}_+\mathfrak{S}\,,
&
\end{eqnarray}
which are very similar to the usual intertwining relations of the supersymmetric 
quantum mechanics, however, here the ``intertwining operators'' are nonlocal  and non-unitary  
operators described by an infinite series of powers of second derivatives. This is the reason why
non-normalizable functions are mapped to bound states and vice-versa. 

One can go further and try to obtain a classical version of the conformal bridge by using 
``Hamiltonian flows'' of the form 
\begin{equation}\label{TransCan}
f(\alpha)=\exp(\alpha F)\star f:=f+\sum_{n=1}^\infty 
\frac{\alpha^n}{n!}\{F,\{\ldots,\{F,f\underbrace{\}\ldots\}\}}_{n}=:T_F(\alpha)(f)\,.
\end{equation}
where $F$ represents a symmetry generator, $\alpha$ is a transformation parameter  
and $f=f(q,p)$
corresponds to a  function on phase space. The composed transformation 
\begin{equation}\label{Tabg0}
T_{\beta\alpha\gamma}:=T_K(\beta)\circ T_{H_0}(\alpha) \circ T_D(\gamma)=
 T_K(\beta)\circ T_D(\gamma) \circ T_{H_0}(2\alpha)\,,
 \end{equation}
with the election 
\begin{equation}\label{abgfix}
\alpha=\frac{i}{2}\,,\qquad
 \beta=-i\,,\qquad \gamma=-\ln 2\,,
\end{equation}
is the classical analog, the operator (\ref{Generalbridge}) 
(generators should be fixed at $t=0$).  This is a complex 
canonical transformation, so one should expect that there is some relation
  with $\mathcal{PT}$ symmetry \textcolor{red}{[\cite{PT1,PT2,PT3}]}. Actually,
in the case of the classical bridge between free particle and harmonic oscillator,
  the function $T_{iD}(\tau)(x)$, i.e, the ``imaginary'' flux of $x$ due to $D$, 
is the one that is mapped to a complex combination 
of  position and momentum of the harmonic oscillator.
Besides,  the transformation of the free particle trajectory does not have a clear 
 interpretation. 

Finally it is worth emphasizing that Hamiltonians 
of the form $xp$ have found application in mathematics, namely,  in the study of Riemann hypothesis, see \textcolor{red}{[\cite{Connes,Berry,Regniers,Sierra,Bender2017}]}.


 \chapter{Hidden bosonized superconformal symmetry}
 \label{ChHiddenboson}

It is well known that the one-dimensional quantum harmonic oscillator system 
is characterized 
by a bosonized superconformal symmetry  \textcolor{red}{[\cite{Hiden1,BalSchBar,CarPly2,Hiden3}]}, however,
 the origin of this symmetry had not been clarified, 
until the article \textcolor{red}{[\cite{InzPly1}]} appeared, and this chapters 
summarize the main results of that work.
We show that this supersymmetry can be derived 
by applying a nonlocal transformation (of the nature of a
Foldy-Wouthuysen transformation)  to 
 a particular  super-extended  system.
The latter system itself can not be obtained directly 
from a given superpotential, i.e., is outside of the Darboux transformation scheme,
 however its corresponding generators are, in fact, linear combinations of 
 the  $\mathfrak{osp}(2|2)$ symmetry generators that the 
super-harmonic oscillator  system possesses. They were
introduced in Chap \ref{ChConformal}, Sec. \ref{SecOSP22Conformal}.
The mentioned system can also be obtained by taking a certain limit in an
 isospectral deformation of the harmonic oscillator, produced with a confluent
 Darboux transformation.

\section{Dimensionless generators}
So far we have turned our attention to Hamiltonian operators of the form
\be
\label{Hdimensionless}
H=\frac{1}{2}\left(-\frac{d^2}{dy^2}+ V(y) \right)\,,\qquad [y]=\sqrt{t}\,,
\ee
where $ V (y) $ is the potential of the harmonic oscillator or that of the AFF model.
However, when we are working with the DCKA transformation, it is worth using dimensionless operators. 
For this reason,
 we consider the change of variables $x=\sqrt{\omega} y$, in term of which 
the Hamiltonian (\ref{Hdimensionless}) takes the form 
$H=\frac{\omega}{2}L$, where depending on the situation we are looking at, the operator
$ L $ as well as its eigenstates and its spectrum could be
\be
\label{Lqhosc}
L_{\text{os}}=-\frac{d^2}{dx^2}+x^2\,,\qquad 
\psi_{n}(x)=\frac{H_{n}(x)e^{-\frac{x^2}{2}}}{\sqrt{\pi^{1/2}n!}}\,,\qquad E_n=2n+1\,, 
\ee
or 
\begin{eqnarray}
&\label{AFFless}
L_\nu=-\frac{d^2}{dx^2}+x^2+\frac{\nu(\nu+1)}{x^2}\,,
&\\&
 \psi_{\nu,n}(x)=\sqrt{\frac{2n!}{\Gamma(n+\nu+\frac{3}{2})}}\,\,x^{\nu+1}L_{n}^{(\nu+\frac{1}{2})}( x^2)e^{-\frac{ x^2}{2}}\,,\qquad
E_{\nu,n}=4n+2\nu+3\,. 
\end{eqnarray}
It is also convenient to redefine the first order ladder operators of 
the harmonic oscillator as 
\begin{equation}
\label{Laderdimensionless1}
a^{\pm}=\mp\frac{d}{dx}+x,\qquad [a^+,a^-]=2\,,\qquad  [L_{\text{os}}, a^\pm]=\pm 2a^\pm\,.
\end{equation}
and the same for the second order ladder operators of the AFF system 
which are now given by 
\begin{eqnarray}
&\label{Ladderdimensionless}
\mathcal{C}_\nu^{\pm}=-(a^\pm)^2+\frac{\nu(\nu+1)}{x^2}\,,&\\&
[L_\nu,\mathcal{C}_\nu^{\pm}]=\pm \Delta E \mathcal{C}_\nu^{\pm}\,,\qquad
[\mathcal{C}_\nu^{-},\mathcal{C}_\nu^{+}] =8L_{\nu}\,,\qquad \Delta E=4\,.
&
\end{eqnarray}
In the Heisenberg picture operators $a^\pm$ and $\mathcal{C}_\nu^{\pm}$ 
are respectively replaced by ${}_Ha^{\pm}= e^{\mp 2it} a^{\pm}$ and 
 ${}_H  \mathcal{C}^{\pm}= e^{\mp 4it}  \mathcal{C}^{\pm}$, which will be 
 dynamical integrals of motion for the corresponding systems.

\section{Hidden superconformal symmetry of the quantum harmonic oscillator}\label{SecHiden}
In this paragraph we show how the aforementioned superconformal 
symmetry appears for the
one-dimensional
 bosonic harmonic oscillator system, the Hamiltonian
of which is given by (\ref{Lqhosc}). 

As the ladder operators (\ref{Laderdimensionless1}) anticommute with  reflection operator
$\mathcal{R}$ defined by  $\mathcal{R}^2=1$, 
$\mathcal{R}x=-x\mathcal{R}$, and their
anti-commutator produces
$\{a^+,a^-\}=2L_{\text{os}}$,  it is clear that if one set 
 $\mathcal{R}$ as the $\Z_2$-grading operator, 
then:
\begin{itemize}
\item   $a^\pm$ are identified as odd, fermionic generators,
\item $L_{\text{os}}$ and quadratic  operators
$(a^{\pm})^2$
are identified as even, bosonic generators since
$[\mathcal{R},L_{\text{os}}]=[\mathcal{R},(a^\pm)^2]=0$.

\end{itemize}

Consider now the dynamical integrals of motion
\begin{equation}\label{localGen}
J_0=\frac{1}{4}L_{\text{os}},\qquad 
J_\pm=\frac{1}{4}e^{\mp4it}(a^{\pm})^2\,,\qquad \alpha_{\pm}=\frac{1}{4}e^{\mp i2t}a^{\pm}\,.
\end{equation}
They produce the (anti)commutator relations 
\begin{eqnarray}
\label{sl2}&
[J_0,J_\pm]=\pm J_\pm,\qquad [J_-,J_+]=2J_0\,,
&\\&
\label{osp(21.1)}
\{\alpha_{+},\alpha_-\}=\frac{1}{2}J_0\,,
\qquad\{\alpha_{\pm},\alpha_\pm\}=\frac{1}{2}J_\pm\,,
&\\&
\label{osp(21.3)}
[J_0,\alpha_{\pm}]=\pm\frac{1}{2}\alpha_{\pm},\qquad
[J_\pm,\alpha_{\mp}]=\mp\alpha_{\pm}\,.&
\end{eqnarray}
The superalgebra (\ref{sl2}), (\ref{osp(21.1)}),  (\ref{osp(21.3)}) 
describes the hidden superconformal 
$\mathfrak{osp}(1|2)$ symmetry of the quantum harmonic 
oscillator \textcolor{red}{[\cite{Hiden1,BalSchBar}]}.
The set of even integrals  $J_0$, $J_\pm$ generates
the $\mathfrak{sl}(2,\R)$ subalgebra (\ref{sl2}), and  relations
(\ref{osp(21.3)}) 
mean that fermionic generators $\alpha_\pm$
form  a spin-$1/2$ representation  of this Lie subalgebra.
One can extend this superalgebra  by introducing the
fermionic operators
 \begin{equation}\label{beta}
\beta_{\pm}=i\mathcal{R}\alpha_{\pm}\,.
\end{equation}
which give rise to the additional super-algebraic relations  
\begin{eqnarray}
\label{beta1}
&
[J_0,\beta_{\pm}]=\pm\frac{1}{2}\beta_{\pm}\,,\qquad
[J_{\pm},\beta_{\mp}]=\mp\beta_\pm\,,
&\\&
\qquad\{\beta_{\pm},\beta_{\pm}\}=\frac{1}{2}J_{\pm}\,,\qquad \{\beta_{+},\beta_{-}\}=\frac{1}{2}J_0\,,
\qquad
\{\alpha_{\pm},\beta_{\mp}\}=\mp\frac{i}{2}Z\,,
&\\&
[Z,\alpha_{\pm}]=\frac{i}{2}\beta_{\pm}\,,
\qquad [Z,\beta_{\pm}]=-\frac{i}{2}\alpha_\pm\,,\label{betaf}
&
\end{eqnarray}
where
\begin{equation}
Z=-\frac{1}{4}\mathcal{R}\,.
\end{equation} 
However this extension is nonlocal since $\mathcal{R}$ can be presented as 
$\mathcal{R}=\sin (\frac{\pi}{2}L_{\text{os}})$.

We will show soon that  superalgebra given by
Eqs.  (\ref{sl2})-(\ref{osp(21.3)})  and (\ref{beta1})-(\ref{betaf}) 
is just another basis for the $\mathfrak{osp}(2|2)$ superconformal algebra
presented in  Chap. \ref{ChConformal}, Sec \ref{SecOSP22Conformal}.

\section{Extended system with super-Schr\"odinger symmetry and 
nonlocal Foldy-Wouthuysen transformation}\label{Sec3Hiden}
The approach with nonlocal Foldy-Wouthuysen transformation and a subsequent reduction was used to 
clarify the origin of the hidden bosonized supersymmetry (that is outside the conformal symmetry) in 
\textcolor{red}{[\cite{Gamboa2,jakubsky}]}, and 
in this section we demonstrate that the bosonized superconformal 
symmetry introduced above  can be ``extracted'' from
the symmetry generators of the
extended quantum harmonic oscillator system
described by the matrix Hamiltonian
\be\label{Hextbr}
\mathcal{H}=
\left(
\begin{array}{cc}
L_{\text{os}} &  0 \\
 0&   L_{\text{os}}     
\end{array}
\right).
\ee
It is natural to identify the diagonal matrix 
$\Gamma=\sigma_3$ as a $\Z_2$-grading 
operator, implying that Hamiltonian (\ref{Hextbr})  is an 
even generator, besides the anti-diagonal integrals
$\sigma_a$, $a=1,2$,  can be considered as odd supercharges. 
The peculiarity of the system (\ref{Hextbr}) is that
these  supercharges  anticommute not 
for Hamiltonian but for central element, 
$\{\sigma_a,\sigma_b\}=2\delta_{ab}\mathbb{I}$,
$\mathbb{I}=\text{diag}\,(1,1)$. On the other hand, 
all the energy levels of the extended system $\mathcal{H}$ 
(including the lowest $ E_0 = 1> 0 $)
 are doubly degenerate.
Furthermore, neither the supercharges nor the Hamiltonian can annihilate any 
eigenstate 
or linear combination of them, so the system is in 
the  spontaneously broken supersymmetric phase.
Additionally, one can  also construct the dynamical integrals 
 \begin{eqnarray}
&\label{J+-t}
\qquad \mathcal{J}_\pm=\frac{1}{4}e^{\mp i4t}\left(
\begin{array}{cc}
(a^{\pm})^2 &  0 \\
 0&     (a^{\pm})^2   
\end{array}
\right)=\left(
\begin{array}{cc}
J_\pm &  0 \\
 0&   J_\pm      
\end{array}
\right),
&\\&
\label{C+-t}
\mathcal{F}_\pm=\frac{1}{4}e^{\mp i2t}\left(
\begin{array}{cc}
  a^{\pm}& 0  \\
 0 &   a^{\pm}
\end{array}
\right)=\left(
\begin{array}{cc}
  \alpha_\pm& 0  \\
 0 &   \alpha_\pm
\end{array}
\right),
&\\&
\mathcal{Q}_\pm=\frac{1}{4}e^{\mp i2t}
\left(
\begin{array}{cc}
  0&   a^\pm  \\
 a^\pm&   0     
\end{array}
\right)=
\left(
\begin{array}{cc}
  0&   \alpha_\pm  \\
 \alpha_\pm &   0     
\end{array}
\right),
\qquad \mathcal{S}_\pm=i\sigma_3\mathcal{Q}_\pm.
\label{J+-t2}&
\end{eqnarray}
Diagonal operators $\mathcal{J}_\pm$ and 
$\mathcal{F}_\pm$ are identified 
here as even generators, and 
antidiagonal dynamical integrals $\mathcal{Q}_\pm$
and $\mathcal{S}_\pm$ are odd.  All
these generators produce the superalgebra\,:
\begin{eqnarray}
&
\label{supmat1}
[\mathcal{J}_0,\mathcal{J}_\pm]=
\pm \mathcal{J}_\pm\,,
\qquad [\mathcal{J}_-,
\mathcal{J}_+]=2\mathcal{J}_0\,,
&\\&
\label{supmat2}
[\mathcal{J}_0,\mathcal{F}_\pm]=\pm\frac{1}{2}\mathcal{F}_\pm\,,
\qquad
[\mathcal{J}_\pm,\mathcal{F}_\mp]=\mp\mathcal{F}_
\pm\,,
\qquad
[\mathcal{F}_-,\mathcal{F}_+]=\frac{1}{2}\mathcal{I}\,,
&\\&
\label{supmat2+}
[\mathcal{J}_0,\mathcal{Q}_\pm]=\pm\frac{1}{2}\mathcal{Q}_\pm\,,\quad
[\mathcal{J}_0,\mathcal{S}_\pm]=\pm\frac{1}{2}\mathcal{S}_\pm\,,\quad
[\mathcal{J}_\pm,\mathcal{Q}_\mp]=\mp \mathcal{Q}_\pm\,,\quad
[\mathcal{J}_\pm,\mathcal{S}_\mp]=\mp \mathcal{S}_\pm\,,
&\\&
\label{supmat3}
\{\Sigma_a,\Sigma_b\}=2\delta_{ab}\,\mathcal{I}\,,
\qquad 
\{\Sigma_1,\mathcal{Q}_\pm\}=\mathcal{F}_\pm\,,
\qquad
\{\Sigma_2,\mathcal{S}_\pm\}=\mathcal{F}_\pm\,,
&\\&
\label{supmat4}
\{\mathcal{Q}_\pm,\mathcal{Q}_\pm\}=\frac{1}{2}\mathcal{J}_\pm\,,
\quad
\{\mathcal{Q}_+,\mathcal{Q}_-\}=\frac{1}{2}\mathcal{J}_0\,,\quad
\{\mathcal{S}_\pm,\mathcal{S}_\pm\}=\frac{1}{2}\mathcal{J}_\pm\,,
\quad
\{\mathcal{S}_+,\mathcal{S}_-\}=\frac{1}{2}\mathcal{J}_0\,,
&\\&
\label{supmat5}
\{\mathcal{Q}_+,\mathcal{S}_-\}=-\frac{i}{2}\mathcal{Z}\,,
\qquad
\{\mathcal{Q}_-,\mathcal{S}_+\}=\frac{i}{2}\mathcal{Z}\,,
&\\&
\label{supmat6+}
[\mathcal{Z},\Sigma_a]=\frac{i}{2}\epsilon_{ab}\Sigma_b\,,\qquad
[\mathcal{Z},\mathcal{Q}_{\pm}]=\frac{i}{2}\mathcal{S}_{\pm}\,,
\qquad [\mathcal{Z},\mathcal{S}_{\pm}]=-\frac{i}{2}\mathcal{Q}_\pm\,,
&\\&
\label{supmat6}
[\mathcal{F}_\pm,\mathcal{Q}_\mp]=\mp\frac{1}{4} \Sigma_1,\qquad
[\mathcal{F}_\pm,\mathcal{S}_\mp]=\mp\frac{1}{4} \Sigma_2\,,
&
\end{eqnarray}
where 
\begin{eqnarray}\label{mathcalJ0}
&
\mathcal{J}_0=\frac{1}{4}\mathcal{H}=
\left(
\begin{array}{cc}
J_0 &  0 \\
 0&   J_0      
\end{array}
\right)
\,,&\\&
\label{Sigmadef}
\Sigma_1=\frac{1}{2}\sigma_1\,,\qquad 
\Sigma_2=-\frac{1}{2}\sigma_2\,, \qquad
\mathcal{Z}=-\frac{1}{4}\sigma_3\,,
\qquad 
\mathcal{I}=\frac{1}{4}\mathbb{I}\,.&
\end{eqnarray} 
The not shown  (anti)commutators between generators are equal to zero.
The system presented here cannot be obtained by the usual Darboux transformation procedure, 
since it is not possible to find a superpotential, so that the potentials relative to the superpartners are exactly $x^2$,
see 
\textcolor{red}{[\cite{InzPly1}]}. However, when considering  the base change
\begin{eqnarray}
&
\mathcal{H}_{os}=2(\mathcal{J}_0-\mathcal{Z}) \,,\qquad 
\mathcal{G}^\pm=-2\mathcal{J}_\pm \,,  &\\&
\label{Qa osi}
\mathcal{Q}_1=2\sqrt{2}\left( Re(\mathcal{Q}^-)+Im(\mathcal{S}^-)\right)\,,\qquad 
\mathcal{Q}_2=2\sqrt{2}\left(Re(\mathcal{S}^-)-Im(\mathcal{Q}^-)\right)\,,
&\\&
\label{Qb osi}
\mathcal{S}_1= 2\sqrt{2}\left(Re(\mathcal{Q}^-)- Im(\mathcal{S}^-)\right)\,,\qquad 
 \mathcal{S}_2= 2\sqrt{2}\left( Re(\mathcal{S}^-)+ Im(\mathcal{Q}^-)\right)\,,
&
\end{eqnarray}
and identifying  $2\mathcal{Z}$ with the generator of the $R$ symmetry, 
one realizes that the generators defined in this way satisfy 
  the $\mathfrak{osp}(2|2)$ superconformal algebra  (\ref{HRQ0})-(\ref{QSGG}).
Actually, the generators (\ref{Qa osi}) - (\ref{Qb osi}) match with $\mathcal{Q} _a $ 
and $ \mathcal{S}_a $ in (\ref{ospos}) when $ t = 0 $\footnote{putting $ \omega = 1 $ and therefore $ y = x $.}, 
and in addition, the operators $ \Sigma_a $ and $ \mathcal {F}_\pm $ are, up to a proportionality factor,
the generators of Heisenberg's superextended symmetry. 
With this information at hand we identify 
(\ref{supmat1})-(\ref{supmat6}) as another expression for super-Schr\"odinger symmetry.

By comparing with what we have in the previous section, 
it is obvious that the matrix 
integrals $\mathcal{J}_0$, $\mathcal{J}_\pm$, $\mathcal{Z}$,
$\mathcal{Q}_\pm$, $\mathcal{S}_\pm$
of the extended system (\ref{Hextbr})
are analogous to the corresponding 
integrals $J_0$, $J_\pm$, $Z$,
$\alpha_\pm$, $\beta_\pm$
of the quantum harmonic oscillator.
Because of the extension, the nonlocal integrals 
$Z$ and $\beta_\pm$ of the system 
(\ref{Lqhosc}) are changed here for
the corresponding local matrix integrals 
$\mathcal{Z}$ and $\mathcal{S}_\pm$.
The anti-commutator of additional fermionic integrals $\Sigma_a$ 
 with $\Sigma_b$ generates
a central charge $\mathcal{I}$, and 
 via the anti-commutators 
with odd dynamical integrals $\mathcal{Q}_\pm$ and
$\mathcal{S}_\pm$ they produce additional 
bosonic integrals $\mathcal{F}_\pm$,
see Eq. (\ref{supmat3}). 

{}The comparison of the symmetries and generators 
of the systems  (\ref{Hextbr}) and (\ref{Lqhosc}) 
indicates  that the local $\mathfrak{osp}(1|2)$ 
and nonlocal $\mathfrak{osp}(2|2)$ hidden superconformal
symmetries of the quantum harmonic oscillator 
can be obtained by a certain projection (reduction)
of the local symmetries of the matrix system (\ref{Hextbr}).
To find the exact relation between these two systems and 
their symmetries, 
we apply to the extended system a unitary transformation 
$\mathcal{O}\mapsto \widetilde{\mathcal{O}}=U\mathcal{O}U^\dagger$
generated by the nonlocal matrix operator
\begin{equation}\label{Utrans}
U=U^\dagger=U^{-1}=\frac{1}{2}
\left(
\begin{array}{cc}
 1+\mathcal{R} &  1-\mathcal{R}  \\
  1-\mathcal{R}&   1+\mathcal{R}      
\end{array}
\right).
\end{equation}
 Under this 
  transformation,  
   the central element $\mathcal{I}$ and 
 generators of the   $\mathfrak{sl}(2,\R)$ subalgebra, $\mathcal{J}_0$ and 
 $\mathcal{J}_\pm$,  do not change, 
 while other generators take the following form\,:
\begin{equation}
\widetilde{Z}=\frac{1}{4}\left(
\begin{array}{cc}
  -\mathcal{R}& 0    \\
 0&   \mathcal{R}     
\end{array}
\right)\,,
\end{equation} 

\begin{equation}
\widetilde{\mathcal{Q}_\pm}=
\left(
\begin{array}{cc}
  \alpha_\pm& 0    \\
 0&   \alpha_\pm     
\end{array}
\right),\qquad
\widetilde{\mathcal{S}_\pm}=
\left(
\begin{array}{cc}
  i\mathcal{R}\alpha_\pm& 0    \\
 0&   -i\mathcal{R}\alpha_\pm     
\end{array}
\right)
=\left(
\begin{array}{cc}
  \beta_\pm& 0    \\
 0&   -\beta_\pm     
\end{array}
\right),
\end{equation}
\be
\widetilde{\Sigma_1}=\frac{1}{2}\sigma_1\,,\qquad
\widetilde{\Sigma_2}=-\frac{1}{2}\sigma_2\mathcal{R}\,,
\qquad
\widetilde{\mathcal{F}_\pm}=\sigma_1\alpha_\pm \,.
\ee
{}Note that the transformation 
diagonalizes the dynamical odd integrals 
$\mathcal{Q}_\pm$ and 
$\mathcal{S}_\pm$ which initially
have had the anti-diagonal form.
Therefore, the transformation is of the same nature as a 
Foldy-Wouthuysen$\,\,$ 
transformation$\,\,\,\,$ for$\,\,\,\,$  a$\,\,\,\,$  Dirac$\,\,\,\,$  particle$\,\,\,\,$  in$\,\,\,\,$  
external$\,\,\,\,$  electromagnetic$\,\,\,\,$ field$\,\,\,\,$ 
\textcolor{red}{[\cite{FW}]}. 
On the other hand, the transformed even, $\widetilde{\mathcal{Z}}$, and odd, 
$\widetilde{\mathcal{S}_\pm}$, generators of the
super-extended Schr\"odinger symmetry of the system (\ref{Hextbr}) 
take a  nonlocal form.
We can reduce (or, in other words, project) the transformed system 
and its symmetries 
to the proper subspace of eigenvalue $+1$ of the matrix
$\sigma_3$ which corresponds,
according to Eq. (\ref{mathcalJ0}), 
to the single (non-extended) quantum harmonic oscillator system.
In this procedure (which can be done using projector  
 $\Pi_+=\frac{1}{2}(1+\sigma_3)$) we looses operators 
  $\widetilde{\mathcal{F}_\pm}$ and $\widetilde{\Sigma_{b}}$
   because they are anti-diagonal, but on the other hand, we retrieve 
all the generators of the bosonized superconformal symmetry given in the previous section.

\section{Two-step isospectral  Darboux chain}\label{Sec5}

As we have indicated previously, the extended system (\ref{Hextbr}) 
cannot be produced by means of the usual supersymmetric algorithm
 based on some superpotential $ W (x) $.
In this section we will show that an option to generate this system 
is through a two-step confluent Darboux transformation:
 The extended system obtained in this way will have a set of true and dynamical  integrals
of motion,  and after the application of a certain limit, 
these integrals will give us the generators of the super-extended  Schr\"odinger 
symmetry related to  (\ref{Hextbr}).

Consider the functions  $\psi_0(x)$, which is the normalized ground state of 
 (\ref{Lqhosc}),
and $\chi_0(x)$, given by
\begin{equation}
\chi_0(x;\mu)=\mu\widetilde{\psi_0(x)}+\Omega_0\,,
\end{equation}
where  $\Omega_0$ is a Jordan state of energy $E=1$, whose form corresponds to 
(\ref{omega1}),
 and $\mu$ is a real constant. By construction  $\chi_0$ 
satisfy $H_-^2\chi_0=0$ with $H_-=a^+ a^-=L_{\text{os}}-1$, and the application of $a^-$ on it gives us 
\begin{equation}\label{varphi0}
\varphi_{-0}(x;\mu)=a^{-}\chi_0(x;\mu)=\frac{\mu+I_0(x)}{\psi_0(x)}=\mu\psi_{-0}(x)+\widetilde{\psi_{-0}(x)}\,,\qquad 
I_0(x)=\int_{-\infty}^x (\psi_0(t))^2dt\,,
\end{equation}
where $\psi_{-0}(x)=e^{x^2/2}$ is a nonphysical eigenstate of $L_{\text{os}}$
with negative energy $E=-1$ and 
$\widetilde{\psi_{-0}(x)}$ is its corresponding linear independent partner constructed  according to (\ref{tildepsi}).

If we choose the value of parameter $\mu$ in 
one of the infinite intervals
$(-\infty,-1)$ or $(0,\infty)$ for which  $\varphi_{-0}(x;\mu)$
is a nodeless on a real line function being 
a nonphysical 
eigenstate of $H_+=a^-a^+$ of zero eigenvalue,
 $H_+\varphi_{-0}(x;\mu)=0$, then we can use it as a seed state 
 for a new Darboux transformation 
which produces the first order differential operators
\be
A^-_\mu=\varphi_{-0}(x;\mu)\frac{d}{dx}\frac{1}{\varphi_{-0}(x;\mu)}=
 \frac{d}{dx}+W(x;\mu)\, , \qquad 
A^+_\mu= (A^-_\mu)^\dagger\,,
\ee
where
\be
W(x;\mu)=-(\ln \varphi_{-0}(x;\mu))'=- x -\frac{\psi_0(x)}{\varphi_{-0}(x;\mu)}\,.
\ee
These operators factorize 
$H_+$ and 
\be\label{H-muWron}
H_\mu=H_++2W'=H_--2\left(\ln (I_0(x)+ \mu)\right)''\,,
\ee
$A^+_\mu A^-_\mu=H_+$, $A^-_\mu A^+_\mu=H_\mu$,
and intertwine them, 
$A^-_\mu H_+=H_\mu A^-_\mu$, 
$A^+_\mu H_\mu=H_+ A^+_\mu$.
Considering the second order differential operators
given by a composition of the first order Darboux generators,
 \be\label{secondA}
 \A^-_\mu=A^-_\mu a^-\,,\qquad
 \A^+_\mu=a^+A^+_\mu\,,
 \ee
we find that they intertwine the Hamiltonian operators
$H_-$ and $H_\mu$, 
\be\label{A2inter}
\A^-_\mu H_-=H_\mu \A^-_\mu\,,\qquad
\A^+_\mu H_\mu=H_- \A^+_\mu\,,
\ee
and also satisfy relations $\A^+_\mu\A^-_\mu=(H_-)^2$,
$\A^-_\mu\A^+_\mu=(H_\mu)^2$.
By construction,
\be \ker\,(\A^-_\mu)=\text{span}\,\{\psi_0(x),\chi_0(x;\mu)\}\,.
\ee
The Darboux-deformed oscillator system described by the 
Hamiltonian operator $H_\mu$ is \emph{completely isospectral}
to the system $H_-$. Its eigenstates 
with eigenvalues $E=2n$, $n=1,2\ldots$, are obtained
by the  mapping 
  $\A^-_\mu$\,:
$\psi_n(x)\mapsto \psi_n(x;\mu)=\A^-_\mu\psi_n(x)$,
$H_\mu\psi_n(x;\mu)=2n\psi_n(x;\mu)$.
The (not normalized) ground state of zero energy of the system
$H_\mu$ is described by wave function 
$\psi_0(x;\mu)=\frac{1}{\varphi_{-0}(x;\mu)}$,
where $\varphi_{-0}(x;\mu)$ corresponds to  (\ref{varphi0}).

Thus, we obtained the completely 
isospectral pair  $H_-$ and $H_\mu$, 
from which 
we compose the extended system described by the matrix 
Hamiltonian operator
\begin{equation}\label{calHmu}
\mathcal{H}_\mu=\left(
\begin{array}{cc}
H_\mu & 0   \\
 0 & H_-    
\end{array}\right).
\end{equation}

On the other hand,
 $A_\mu^-$ and $A^+_\mu$ intertwine 
$H_+=H_-+2$ and $H_\mu$, which implies 

\be\label{Amuinter}
A^-_\mu H_-=(H_\mu-2) A^-_\mu\,,\qquad 
A^+_\mu (H_\mu-2)=H_- A^+_\mu\,.
\ee
 {}For this system we have in fact three Darboux schemes\,: 
\begin{itemize}
\item Scheme $(\psi_0(x),\chi_0(x;\mu))$ which provides us with the intertwining operator $\A^\pm_\mu$.
\item Scheme  $(\varphi_{-0}(x;\mu))$, the intertwining operators of which are  $A^\pm_\mu$ .
\item Scheme $(\psi_0(x),\psi_1(x), a^+\chi_0(x;\mu))$, which gives us the third order intertwining operators 
$\mathcal{A}^-_\mu=A^-_\mu(a^-)^2=\A^-_\mu a^-$
  and $\mathcal{A}^+_\mu=(\mathcal{A}^-_\mu)^\dagger$, that satisfy   $\mathcal{A}^-_\mu H_-=(H_\mu+2)\mathcal{A}^-_\mu$,
    $\mathcal{A}^+_\mu (H_\mu+2)=H_-\mathcal{A}^+_\mu$.
\end{itemize}
 Using the intertwining operators of these three Darboux schemes,
 we construct the odd integrals 
\begin{equation}
\mathcal{Q}_{\mu 1}=\left(
\begin{array}{cc}
0 & \A^- _\mu  \\
 \A^+_\mu & 0  
\end{array}\right),\quad 
\mathcal{Q}_{\mu 2}=i\sigma_3\mathcal{Q}_{\mu 1}\,,
\quad
\mathcal{S}_{\mu 1}=\left(
\begin{array}{cc}
0 & A^-_\mu   \\
 A^+_\mu & 0  
\end{array}\right),\quad 
\mathcal{S}_{\mu 2}=i\sigma_3\mathcal{S}_{\mu 1}\,,
\end{equation}
\begin{equation}
\mathcal{L}_{\mu 1}=\left(
\begin{array}{cc}
0 & \mathcal{A}^-_\mu   \\
 \mathcal{A}^+_\mu& 0  
\end{array}\right),\qquad
 \mathcal{L}_{\mu 2}=i\sigma_3\mathcal{L}_{\mu 1}\,.
\end{equation}
and by means of the relations
\begin{eqnarray}
&
\A^-_\mu A^+_\mu=A^-_\mu a^-A^+_\mu\,,\qquad 
A^+_\mu \A^-_\mu=(H_-+2) a^-\,,&\\&
\mathcal{A}^-_\mu A^+_\mu=A^-_\mu (a^-) A^+_\mu\,,\qquad 
A^+_\mu \mathcal{A}^-_\mu=(H_-+2)(a^-)^2\,,&
\end{eqnarray}
we also construct diagonal (even) operators
\begin{equation}
\mathcal{F}_{\mu-}=\left(
\begin{array}{cc}
A_\mu^-a^-A_\mu^+ & 0   \\
 0 & (H_-+2)a^-
\end{array}\right),\qquad
\mathcal{J}_{\mu-}=\left(
\begin{array}{cc}
A_\mu^-(a^-)^2A_\mu^+ & 0   \\
 0 & (H_-+2)(a^-)^2
\end{array}\right),
\end{equation}
and Hermitian conjugate operators $\mathcal{F}_{\mu+}$ and 
 $\mathcal{J}_{\mu+}$.
 With respect to the Hamiltonian $\mathcal{H}_\mu$,
  the only pair of time-independent 
  integrals are the supercharges $\mathcal{Q}_{\mu a}$, $a=1,2$. 
To obtain dynamical integrals one should unitary transform other operators
with   $U(t)=\exp\left(i\mathcal{H}_\mu t\right)$\,. 

The generators considered here produce 
  a kind of  a nonlinear deformation of the super-Schr\"odinger 
  symmetry. We are not interested here in explicit form
  of such a nonlinear superalgebra, but just note that
  when $\mu\rightarrow \pm \infty$, we have
   $(\ln(I(x) + \mu))' \rightarrow 0$.
   As a result, in any of the  two  limits
   the Hamiltonian $H_\mu$ transforms into $H_-$,
   and the matrix Hamiltonian transforms into extended Hamiltonian
   (\ref{Hextbr}) shifted for the minus unit matrix\,: 
   $\mathcal{H}_\mu\rightarrow \mathcal{H} -\mathbb{I}$.
   In this limit we also have 
$A^\pm_\mu\rightarrow -a^\mp$, and 
find that the constructed operators 
transform as follows\,:
\begin{eqnarray}
\mathcal{Q}_{\mu 1}\rightarrow -(\mathcal{H}-1)\sigma_1\,,&&\quad
\mathcal{Q}_{\mu 2}\rightarrow (\mathcal{H}-1)\sigma_2\,,\\
\mathcal{S}_{\mu a}\rightarrow-\breve{\mathcal{S}}_a\,,&&\quad
\mathcal{L}_{\mu a}\rightarrow-(\mathcal{H}-2+\sigma_3)
\widehat{\mathcal{Q}}_a\,,\\
\mathcal{F}_{\mu -}\rightarrow (\mathcal{H}-\sigma_3)\mathcal{F}_-\,,&&
\quad 
\mathcal{F}_{\mu +}\rightarrow \mathcal{F}_+(\mathcal{H}-\sigma_3)\,,\\
\mathcal{J}_{\mu -}\rightarrow (\mathcal{H}-\sigma_3)\mathcal{J}_-\,,&&
\quad 
\mathcal{J}_{\mu +}\rightarrow \mathcal{J}_+(\mathcal{H}-\sigma_3)\,.
\end{eqnarray}
In such a way we reproduce all the corresponding integrals
of the system  (\ref{Hextbr}) that generate the super-extended 
Schr\"odinger symmetry 
lying behind the hidden superconformal symmetries
$\mathfrak{osp}(1|2)$ and $\mathfrak{osp}(2|2)$
of a single quantum harmonic oscillator.

The isospectral deformation $V_\mu(x)$  of the 
harmonic oscillator potential is illustrated by 
Figure \ref{HidenFig1}, while Figure \ref{HidenFig2} illustrates the action 
of the intertwining operators $\A^\pm_\mu$ and 
$\mathcal{A}^\pm_\mu$.

\begin{figure}[h]
\begin{center}
\includegraphics[scale=0.7]{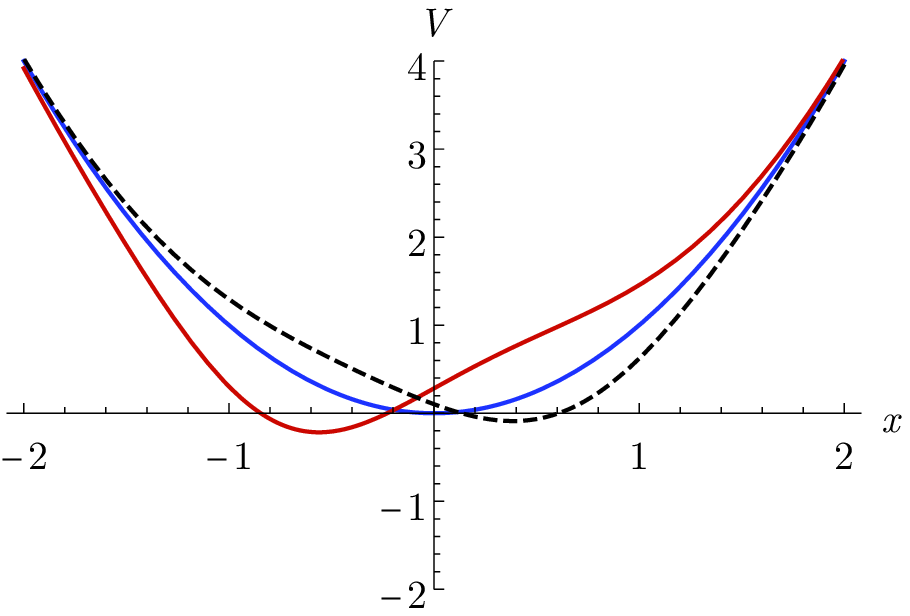}
\includegraphics[scale=0.7]{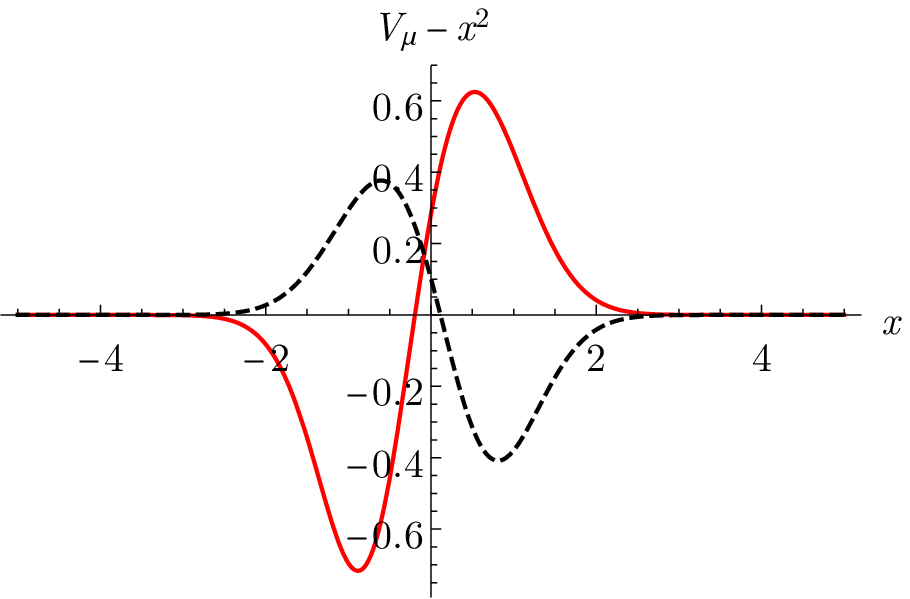}
\caption[Behavior of the potential, Sec. 5.4 ]{\small{ On the left: Isospectrally deformed potential $V_\mu$ 
at $\mu=1$  and $\mu=-3$ is shown by continuous red and dashed black
lines, respectively. 
On the right: The difference $V_\mu(x)-x^2$  given by the last term in Eq. (\ref{H-muWron}) 
is shown for the same values $\mu=1$  and $\mu=-3$.
With increasing value of modulus 
of the deformation parameter $\mu$ 
the amplitudes of minimum and maximum of  the difference $V_\mu(x)-x^2$ decrease, 
and in both  limits $\mu \rightarrow\pm \infty$
the deformed potential $V_\mu(x)$ transforms into harmonic potential $V=x^2$
shown on the left by continuous blue line.
}}
\label{HidenFig1}
\end{center}
\end{figure}
\begin{figure}[h]
 \begin{center}
\includegraphics[scale=0.3]{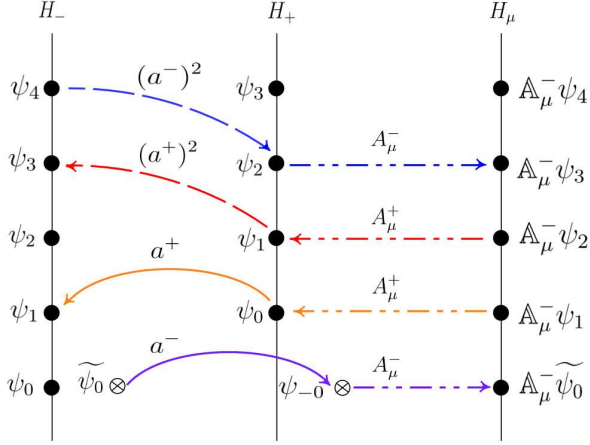}
\caption[Intertwining operators behavior, Sec. 5.4 ]{\small{Mapping of eigenstates of the systems $H_-$ and $H_\mu$  
by intertwining operators 
$\A^\pm_\mu$ and $\mathcal{A}^\pm_\mu$ 
via eigenstates of intermediate system $H_+$.
The ground state $\A^-_\mu\widetilde{\psi_0 }$
 of 
$H_\mu$ is obtained by applying  $\A^-_\mu$ to
nonphysical eigenstate $\widetilde{\psi_0 }$
of $H_-$. It also can be generated  by a not shown here  action of  $\mathcal{A}^-_\mu$
on nonphysical  eigenstate $\widetilde{\psi_1}$ of $H_-$ via
nonphysical eigenstate $\psi_{-0}$ of $H_+$. }}
\label{HidenFig2}
\end{center}
\end{figure}

In conclusion of this section we note 
 that the Hamiltonian (\ref{calHmu}) 
and the second order  intertwining operators 
$\A^\pm_\mu$ can be presented 
in alternative  form which corresponds to the anomaly-free 
scheme of quantization of  classical  systems with second-order
supersymmetry \textcolor{red}{[\cite{Plyushchay}]}.
For this we introduce the quasi-amplitude \textcolor{red}{[\cite{Brezh}]}
\be
\Xi(x)=\sqrt{\psi_{-0}(x)\varphi_{-0}(x;\mu)}\,.
\ee
It is a square root of the product of two nonphysical eigenstates  of eigenvalue 
$-1$ of the quantum harmonic oscillator $L_{\text{os}}$.
 The rescaled 
function $\Xi(x)/\sqrt{\mu}$ transforms in the limit $\mu\rightarrow \pm \infty$
 into the 
nonphysical eigenstate $\psi_{-0}$.
This function satisfies Ermakov-Pinney  equation 
\textcolor{red}{[\cite{ermakov1,ermakov2,ermakov3,ermakov4}]}
\be 
-\Xi''+(x^2+1)\Xi=\frac{1}{4\Xi^3}\,.
\ee
In terms of quasi-amplitude,  
the first order differential operators
\be\label{Axi}
A^-_\Xi=\Xi(x)\frac{d}{dx}\frac{1}{\Xi(x)}=\frac{d}{dx}-x-\mathcal{W}(x)\,,
\qquad
A^+_\Xi=(A_\Xi)^\dagger\,,\qquad 
\ee
can be defined,
where 
\be\label{calWsup}
\mathcal{W}(x)=\frac{1}{2\Xi^2(x)}=
\frac{1}{2}\left(\ln (I_0(x)+\mu)\right)'\,.
\ee
Then the Hamiltonian $H_\mu$  and the intertwining operator $\A^-_\mu$
can be presented in the form
\be
\mathcal{H}_\mu=A^-_\Xi A^+_\Xi +\mathcal{W}^2 -  2\mathcal{W}'\sigma_3\,,
\qquad 
\A^-_\mu=-(A^-_\Xi-\mathcal{W})(A^+_\Xi+\mathcal{W})\,.
\ee
Function $\mathcal{W}(x)$ in the anomaly-free scheme of quantization
plays a role of superpotential for corresponding$\,\,\,$  classical $\,\,\,$system $\,\,\,$ with$\,\,\,$
second $\,\,\,$order $\,\,\,$supersymmetry,$\,\,\,$ \textcolor{red}{[\cite{PlyPara,KliPly,Plyushchay}]}.

\section{Remarks}

Along with the harmonic oscillator, there are many bosonic systems that have
 hidden bosonized supersymmetry and the idea of the Foldy-Wouthuysen transformation is not new, see
\textcolor{red}{[\cite{Boson1}; \cite{PlyPara}; \cite{Gamboa2}; \cite{Boson2}; \cite{Boson3}; \cite{jakubsky}]}. 
In fact, one can use the transformation (\ref{Utrans}) in the generators of the 
super-extended free particle given in Chap. \ref{ChConformal}, Sec. \Ref{Sec0omegalimit}, to obtain the hidden superconformal symmetry of the   bosonic  free particle.

The exotic feature here is the supersymmetric system from where we get the
 bosonic superalgebra which, in principle, does not correspond to the 
Darboux transformation scheme.
However, it is possible to obtain such a system  starting from the classical level:
Consider a classical system described by
a Hamiltonian
\be\label{Comm1}
H=p^2+W^2 +W'[\theta^+,\theta^-]
\ee
 with superpotential 
$W(x)=\sqrt{x^2+c^2}$, where $c>0$ is a constant, besides 
$\theta^+$ and  $\theta^-=(\theta^+)^*$  are Grassmann variables 
with a nonzero Poisson bracket 
$\{\theta^+,\theta^-\}_{{}_{PB}}=-i$, that after quantization 
are realized as the creation-annihilation fermionic 
operators  $\theta^\pm\rightarrow \sigma_\pm=\frac{1}{2}(\sigma_1\pm i\sigma_2)$.
 A direct quantum analog of this system is a composition 
 of two isospecral systems and is in the phase of spontaneously broken supersymmetry,
 with nonsingular superpartner potentials $V_\pm=x^2+c^2\pm x/\sqrt{x^2+c^2}$.
 The spectrum of subsystems is different from that 
 of the quantum harmonic  oscillator.
On the other hand, if before the quantization we realize a canonical 
transformation 
$x\rightarrow X=x+ N \partial G(x,p)/\partial p$,
$p\rightarrow P=p - N\partial G(x,p)/\partial x$, 
$\theta^\pm\rightarrow \Theta^\pm=e^{\pm iG(x,p)}\theta^\pm$,
where $N=\theta^+\theta^-$ and 
$G=\frac{1}{2}\arcsin \big((p^2-x^2-c^2)/(p^2+x^2+c^2)\big)$ \textcolor{red}{[\cite{KliPly,InzPly1}]},
we obtain the canonically equivalent form of the 
Hamiltonian $H=P^2+X^2 +c^2$. In the canonically transformed system, the new 
classical Grassmann variables $\Theta^\pm$ completely decouple and are the odd 
integrals
of motion with Poisson bracket $\{\Theta^+,\Theta^-\}_{{}_{PB}}=-i$. 
The quantization of the canonically transformed system 
gives us exactly the  extended quantum system (\ref{Hextbr}) shifted just for the additive 
constant $c^2$.

Another possibility is a ``naive'' application of the comformal bridge. To do so, 
let us start by setting the super-Schr\"odinger symmetry generators for the super-free particle
system,
\begin{eqnarray}
&
\mathcal{H}= -\frac{1}{2}p^2\mathbb{I}\,,\qquad 
\mathcal{D}= \frac{1}{4}\{ x, p\} \mathbb{I}\,,\qquad 
\mathcal{K}= \frac{x^2}{2} \mathbb{I}\,, \qquad \mathcal{Z}=-\frac{\sigma_3}{4}\,,&\\&
\mathcal{P}=p\mathbb{I}\,,\qquad 
\mathcal{X}=x\mathbb{I}\,,\qquad 
\Sigma_1=\sigma_1\,,\qquad 
\Sigma_2=-\sigma_2\,,
\qquad 
\pi_{a}=p\Sigma_{a}\,\qquad 
\xi_{a}=x\Sigma_{a}\,,
&
\end{eqnarray}
where $p=i\frac{d}{dx}$. The conformal bridge transformation produces 
\begin{eqnarray}
&
\mathfrak{S}\mathcal{H} \mathfrak{S}^{-1}=-\frac{(a^-)^{2}}{2}\mathbb{I}\,,\qquad 
\mathfrak{S}\mathcal{D} \mathfrak{S}^{-1}=-\frac{i}{4}L_{\text{os}}\mathbb{I}\,,\qquad 
\mathfrak{S}\mathcal{K}\mathfrak{S}^{-1}=\frac{(a^+)^{2}}{2}\mathbb{I} &\\&
\mathfrak{S}\mathcal{Z} \mathfrak{S}^{-1}=\mathcal{Z}\,,\qquad 
\mathfrak{S}\mathcal{X} \mathfrak{S}^{-1}=\frac{a^+}{\sqrt{2}}\mathbb{I} \,,\qquad 
\mathfrak{S}\mathcal{P}\mathfrak{S}^{-1}=-i\frac{a^-}{\sqrt{2}}\mathbb{I} \,,
&\\&
\mathfrak{S}\Sigma_a\mathfrak{S}^{-1}=\Sigma_a \,,\qquad 
\mathfrak{S}\xi_a \mathfrak{S}^{-1}=\frac{a^+}{\sqrt{2}}\Sigma_a \,,\qquad 
\mathfrak{S}\pi_a\mathfrak{S}^{-1}=-i\frac{a^-}{\sqrt{2}}\Sigma_a \,,
\end{eqnarray}
that up to a complex  proportionality constant, they match with 
the generators presented in Sec. \ref{Sec3Hiden} at $ t = 0 $. 
The conformal bridge works fine in this case because $\mathcal{D} $
 and its transformed version are matrix generators  
  containing two copies of the same differential operator. 
In the general case, both superpartners are different from each other and the transformation fails.



\chapter{Rationally extended conformal mechanics}
\label{ChRQHO}
As we  have shown in Chap. \ref{ChSUSY}, Sec. \ref{Chap1Darbux},  
DCKA transformation allows us to construct 
new quantum systems starting from a 
well known original one. In this context, the systems that appear due to these transformations
 applied to the harmonic oscillator are the  rationally  extended harmonic oscillators, that is,
 a harmonic potential plus a regular rational function of $ x $, and to obtain a well-defined system, 
we have to follow some rules for selecting the set of seed states for transformation.
The selection rule that gives us a regular potential 
 is known 
 as the Krein-Adler theorem,  \textcolor{red}{[\cite{Krein,Adler,Dubov,Quesne2012,Gomez2}]}. 
In the research carried out in the article \textcolor{red}{[\cite{CarInzPly}]}, 
we found new selection rules to construct completely isospectral rational 
extensions for the AFF model with integer coupling constant $ m (m + 1) $, where 
$m=1,2,\ldots$, 
as well as deformations with gaps in their spectrum.
We also learned how to construct the spectrum-generating ladder operators
of these deformed systems  
by using what we call Darboux dualities. 
The content presented in this chapter is a summary of the results obtained in 
\textcolor{red}{[\cite{CarInzPly}]}, an article that in turn 
was inspired by previous research on rational deformations of the harmonic oscillator \textcolor{red}{[\cite{CarPly}]}. 

Before to start, 
let us explain what a Darboux duality is with a simple 
example:
consider the half-harmonic oscillator Hamiltonian $L_0$\footnote{ 
This Hamiltonian is formally
(\ref{Lqhosc}), but defined 
in the domain $\{ \psi\in L^2((0,\infty),dy)\vert \psi(0^+)=0\}$.
The  physical states are  the odd
eigenstates of the  harmonic oscillator system.}. 
When the first $m$ physical states are considered 
as seed states for the DCKA transformation, it is not difficult to show that the 
resulting system is the AFF model $L_m$, defined in (\ref{AFFless}),
 shifted by the constant $-2m$. 
Now, by performing the transformation $ x \rightarrow ix $ in the physical 
eigenstates, we produce new nonphysical solutions, 
and when the first $m$ functions obtained in this way are taken as seed states
for the DCKA transformation, 
the resulting system 
is again  $L_m$ but now shifted by the positive constant $2m$. 
So both Darboux transformation schemes generate essentially the same quantum system,
 and in this sense we call them as dual Darboux schemes. 
The intertwining  operators of both dual schemes are independent of each 
other and it can be shown that operators constructed by means of products of these 
intertwiners are equivalent to powers of $\mathfrak{sl}(2,\R)$ generators \textcolor{red}{[\cite{CarInzPly}]}. 

Here we study rational extended systems built on the basis of the half-harmonic oscillator, 
and for simplicity, we present the following notation to refer to the physical and nonphysical 
eigenstates of the quantum harmonic oscillator system (from now on  QHO),
\begin{equation}
\label{notation}
n\equiv \psi_n(x),\qquad -n\equiv \psi_{n}^-=\psi_n(ix)\,,\qquad \widetilde{n}\equiv \widetilde{\psi_n}\,,
\qquad \widetilde{-n}\equiv \widetilde{\psi_{n}^-}.
\end{equation} 
\section{Generation of rationally extended systems}
\label{rationallyextededinteger}
Rational deformations (extensions) 
 of the QHO  system are constructed following the 
Krein-Adler theorem  \textcolor{red}{[\cite{Krein,Adler}]}, 
which ensures that the Wronskian of the seed states (or henceforth Darboux scheme)
  $(n_1,n_1+1,\ldots,n_\ell,n_\ell+1)$, 
where  the numbers $n_j\in \N $, $j=1,\ldots,\ell$, indicate   
 the chosen seed states, see notations  (\ref{notation}), does not have zeros on the real axis.
The corresponding DCKA transformation produces 
\begin{equation}
\label{DQHO}
L_{(n_1,n_1+1,\ldots,n_\ell,n_\ell+1)}=L +4\ell +
\frac{F(x)}{Q(x)},
\end{equation} 
where  $F(x)$ and  $Q(x)$ are even polynomials, 
with $Q(x)$  taking positive values
on real line and having degree higher by two
of degree of $F(x)$.  According with Chap. \ref{ChSUSY}, 
the spectrum of the system (\ref{DQHO}) is 
almost isospectral to the QHO spectrum:
there are missing energy levels or gaps,  related to the energy levels 
corresponding to seed states.

\vskip0.1cm

On the other hand, deformations of the AFF model $L_m$
 can be obtained from the half-harmonic oscillator by considering the scheme
$(n_1,n_1+1,\ldots,n_\ell,n_\ell+1,2k_1+1,\ldots, 2k_m+1)$, 
where even indexes inside the set  
$n_1,n_1+1,\ldots,n_\ell,n_\ell+1$ represent nonphysical eigenstates 
of $L_{0}$ and 
$k_i$, $i=1,\ldots,m$,   are identified as  
$m$ odd states which were not considered in the first set  of $2n_\ell$ states.
The  Hamiltonian operator
\begin{equation}
\label{REIO}
L_{(n_1,n_1+1,\ldots,n_
\ell,n_\ell+1,2k_1+1,\ldots, 2k_m+1)}=L_{m}+
2m +4\ell + \frac{\widetilde{F(x)}}{\widetilde{Q(x)}},
\end{equation} 
appears as a final result of the DCKA transformation,
where polynomials $\widetilde{F(x)}$ and  $\widetilde{Q(x)}$  
have the properties similar to those of $F(x)$ and $Q(x)$ in 
(\ref{DQHO}).
Note that in this way we can only construct deformations of 
$L_m$. Rational deformations of $L_\nu$, with arbitrary 
values for parameter $\nu$, cannot be connected with the harmonic oscillator
as we did here, and the 
issue about their construction 
is discussed properly in Chap. \ref{ChKlein}.
In general such a 
system has gaps in its spectrum. 
If, however,  the set
$n_1,n_1+1,\ldots,n_
\ell,n_\ell+1,2k_1+1,\ldots, 2k_m+1$ contains all
the $\ell+m$ odd indexes from $1$ to $2k_m+1$,
the generated deformed AFF system 
will have  no gaps in its spectrum  
and  we obtain a system completely isospectral
to $L_{0}+4\ell+2m$. Such completely isospectral
(gapless)  deformations 
in the QHO case are only possible if we include Jordan states in the construction.
\vskip0.1cm

The mirror diagram method developed and used in \textcolor{red}{[\cite{CarInzPly}]} is a
technique such that a dual scheme with nonphysical ``negative'' eigenstates
  (\ref{notation}), 
is derived from a ``positive'' scheme with physical states of $L_{os}$
and, vice-versa.
This can be done by using the algorithmic 
procedure described  in Appendix \ref{MirrorHarmonic} and  the final 
picture is the following: 
\begin{itemize}
\item  For a given positive scheme
$\Delta_+\equiv (l_1^+,\ldots,l_{n_+}^+)$, where 
$l_i^+$ with  $i=1,\ldots, n_+$, 
one gets the negative scheme 
 $\Delta_-=(-\check{0},\ldots,-\check{n}_i^-= l_i^+-l_{n_+}^+,
\ldots,-l_{n_+}^+)$, where $-\check{n}_i^-$,  
means that the corresponding number $-n_i^-$ is omitted
from  the set  $\Delta_-$.
\item  If we have instead  the negative scheme 
$\Delta_-\equiv (-l_{1}^-,\ldots,-l_{n_-}^-)$,  where   $-l_{j}^-$  with 
$j=1,\ldots, n_-$,  one obtains the positive scheme 
 $\Delta_+=(\check{0}, \ldots,\check{n}^+_{j}=l_{n_-}^- -l_{j}^-,\ldots,l_{n_-}^-)$, where symbols 
$\check{n}_j^+$ represent  again the 
states missing from  the list of the chosen seed states.  
\end{itemize}
Obviously, Darboux scheme 
must be constructed in such a way that the generated Hamiltonian 
is a non-singular operator,
that is, by means of the rules discussed above. Then, 
having two dual schemes on hand,  the relation
\be\label{genDual}
e^{-n_+x^2/2}W(\Delta_-)=e^{n_-x^2/2}
W(\Delta_+)\,,
\ee
is valid modulo a multiplicative constant. From here  one 
can see that the Hamiltonians of dual schemes satisfy 
\begin{equation}
\label{L+L-}
L_{(+)}-L_{(-)}=2N\,,\qquad N\equiv n_+ +n_-=l_{n^+}^++1=l_{n^-}^-+1\,,
\end{equation}
where  $L_{(+)}$ and
 $ L_{(-)}$ correspond to 
 \be
\label{L(+-)form}
L_{(\pm)}=
-\frac{d^2}{dx^2}+V(x)-2\frac{d^2}{dx^2}
\ln W(\Delta_\pm)\,.
\ee
  On the other hand,
the intertwining operators $\A_{n_+}^-$ and 
$\A_{n_-}^-$ that correspond to each scheme are constructed following 
(\ref{generic-inter}), however, we prefer to use the more generic notation 
\be
\A_{n_+}^-=A_{(+)}^-\,,\qquad 
\A_{n_-}^-=A_{(-)}^-\,,\qquad 
A_{(+)}^+=(A_{(+)}^-)^\dagger\,,\qquad 
A_{(-)}^+=(A_{(-)}^-)^\dagger\,.
\ee 
By means of the negative scheme we do not eliminate any energy level from the spectrum, 
but instead energy levels can be introduced, but not obligatorily, in its lower part.
In the particular special case  
of completely isospectral deformations of the (shifted) 
$L_m$ systems,
all $m$ seed states composing negative scheme  
are  nonphysical odd eigenstates of $L_0$, 
and the transformation does not introduce any additional energy level.   

The construction of the mirror diagram can be better understood with the following 
example:
Consider the illustration in figure \ref{defAFFFigure1}. In the upper line 
we have represented the first eleven physical eigenstates of the harmonic oscillator by circles, 
where the black ones are the seed states of the positive scheme $ (1,4,5,10,11) $,  which produces a system 
of the type (\ref{REIO}).  In a similar way, the first eleven nonphysical states with negative energy 
are indicated by the circles in the bottom line, and the marked ones are the seed 
states of the corresponding dual negative scheme. 
In general, when considering a scheme of the form $(\ldots,N)$, in the 
upper line we ordered from left to right all the physical state from $\psi_{0}$ to $\psi_N$, besides 
in the bottom line we set from right  to left all the states between $\psi_{0}^-$ to   $\psi_N^-$.
 After marking the 
states of  the positive scheme, the construction of the negative scheme is by means of a sort of an 
``anti-reflection'' transformation with respect to an imaginary line in the center,
 that is parallel to the other two lines.
The construction of the positive scheme from the negative one is analogous.

\begin{figure}[H]
\begin{center}
\includegraphics[scale=0.16]{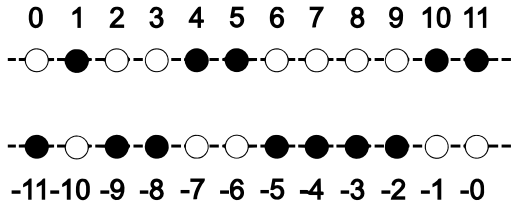} 
\caption[Mirror diagram, Sec. 6.1 ]{\small{A mirror diagram example.}} 
\label{defAFFFigure1}
\end{center} 
\end{figure}
\vskip-0.5cm
This construction seems to be related with the Maya diagram formalism, for a review see
\cite{gomez2019}. 
However, our technique is completely based on the existence of the first-order ladder 
operators and their relationship with the Darboux transformation
(this is the key to its generalization for the AFF model in Chap. \ref{ChKlein}), 
besides for the Maya diagrams it is important 
to study the proprieties of an additional structure called the pseudo-Wronskian, 
which we do not introduce in our work.

\section{Spectrum-generating ladder operators: completely isospectral case}
\label{SecIsocase}

In this section we explore the possibilities of constructing spectrum-generating 
ladder operators for rationally extended isospectral systems. 
We start with the simplest example and then expand on the ideas for the general case.

Consider the simplest 
deformed AFF  system generated via the 
Darboux transformation
based on the nonphysical eigenstate 
$\psi_{3}^{-}=(2x^3+3x)e^{x^2/2}$
of the half-harmonic oscillator $L_{0}$. 
The resulting Hamiltonian takes the form 
\begin{eqnarray}
\label{reio3}
L_{(-)} 
:=L_1 -2 +8\frac{2 x^2-3}{(2 x^2+3)^2}\,.
\end{eqnarray}
By the method of the mirror diagram, 
we find that up to a constant shift,
the system can be generated alternatively by
the  DCKA 
transformation based on the set  $(1,2,3)$\footnote{ The state $\psi_{2}$ is not a physical state of
 the half-harmonic oscillator $L_{0}$. } ,
\be L_{(+)}:=
L_{(-)}+8\,.\ee

The intertwiners of the negative scheme are 
\be\label{A+-(-3)}
A_{(-)}^-=\psi^{-}_3\frac{d}{dx}\frac{1}{\psi^{-}_3}
=\frac{d}{dx}-x -\frac{1}{x}-\frac{4x}{2x^2+3}\,,\qquad
A_{(-)}^+=(A_{(-)}^-)^\dagger\,. 
\ee
They provide us 
the factorization relations 
$A_{(-)}^{+}A_{(-)}^{-}=L_0+7\,,$ 
$A_{(-)}^{-}A_{(-)}^{+}=L_{(-)}+7=L_{(+)}-1$. 
In correspondence with them, 
$A_{(-)}^{-}$ intertwines the Hamiltonian operators 
$L_{0}$ and $L_{(-)}$,
\be
A_{(-)}^{-}L_{0}=L_{(-)}A_{(-)}^{-}=(L_{(+)}-2\Delta E)A_{(-)}^{-}\,, \qquad
\Delta E= 4\,,
\ee
and the intertwining relation for $A_{(-)}^{+}$ is obtained 
by Hermitian conjugation.  

The systems $L_{0}$ and $L_{(+)}$ are also intertwined 
by the third order operators $A^{\pm}_{(+)}$, where 
the operator $A^{-}_{(+)}$ is uniquely 
 specified  by 
 its kernel\,: $\ker A^{-}_{(+)}=\text{span}\,\{\psi_1,\psi_2,\psi_3\}$. 
We have the intertwining relation 
$A^{-}_{(+)}L_{0}=L_{(+)}A^{-}_{(+)}=(L_{(-)}+8)A^{-}_{(+)}$, 
and the conjugate relation for $A^{+}_{(+)}$.

To construct ladder operators for this deformed 
system we can ``Darboux dress''
the ladder operators of the half-harmonic oscillator 
which are nothing else than $(a^\pm)^2$.
The first pairs of operators produced in this way are 
\begin{equation}
\label{reio7}
\mathcal{A}^{\pm}=A_{(-)}^{-}(a^{\pm})^{2}A_{(-)}^{+}\,.
\end{equation}  
These  operators together with the Hamiltonian $L_{(-)}$
generate a nonlinear deformation of the conformal symmetry 
given by the commutation relations
\be
[L_{(-)},\mathcal{A}^{\pm}]=\pm 4  
\mathcal{A}^{\pm}, \qquad
[\mathcal{A}^-,\mathcal{A}^+]=16\left(L_{(-)}+3\right)\left(L_{(-)}+7\right)\left(L_{(-)}+{1}/{2}\right).
\ee

The roots of the  fourth order polynomial 
in the relation 
\be
\mathcal{A}^{+}\mathcal{A}^{-}= 
(L_{(-)}+7)(L_{(-)}+3)(L_{(-)}-1)(L_{(-)}-3)\,,
\ee
 correspond to  eigenstates of
$L_{(-)}$,  which belong to the 
kernel of the lowering  operator, 

\be
\ker\mathcal{A}^{-}=\text{span}\,\{ A_{(-)}^{-}
\widetilde{\psi_{3}^{-}},\,A_{(-)}^{-}\psi_1^{-}\,,
A_{(-)}^{-}\psi_0,\,
A_{(-)}^{-}\psi_1 
\}.
\ee
The last state $\,A_{(-)}^{-}\psi_1=A_{(+)}^-\psi_5$ describes here the ground state
of $L_{(-)}$ of eigenvalue  $E=3$, and other states are nonphysical. 

On the other hand,
the roots in the product 
\be
\mathcal{A}^{-}\mathcal{A}^{+}=(L_{(-)}+11)(L_{(-)}+7)(L_{(-)}+3)(L_{(-)}+1)\,,
\ee
correspond to 
eigenvalues of the eigenstates of $L_{(-)}$
which appear in the kernel of the raising ladder operator,
\be
\ker\mathcal{A}^{+}=\text{span}\{ A_{(-)}^{-}\psi_5^{-},\,
A_{(-)}^{-}\widetilde{\psi_{3}^{-}},\, 
A_{(-)}^{-}\psi_1^{-},\,A_{(-)}^{-}\psi_0^{-}\}\,.
\ee 
All the states in this kernel are nonphysical.
In correspondence with the described properties of 
the ladder operators (\ref{reio7}) they
are the spectrum-generating ladder operators for the
system $L_{(-)}$\,:
acting by them on any physical eigenstate of $L_{(-)}$,
we can generate any other physical eigenstate.
The kernels of the ladder operators 
contain here the same nonphysical
 eigenstate $A_{(-)}^{-}\widetilde{\psi_{3}^{-}}=
A_{(-)}^{-}\psi_1^{-}$. Below we shall see that in the
case of \textbf{non-isospectral} rational deformations of the
AFF system the kernels of analogs 
of such lowering and raising ladder operators contain 
some common physical eigenstates, see for example 
Figure \ref{defAFF1Figure2} in next section.

In a similar way, one can construct 
the ladder operators for  $L_{(-)}$  
via Darboux-dressing of   $(a^\pm)^2$ 
by the third order intertwining operators,
\be
\mathcal{B}^{\pm}=A^{-}_{(+)}(a^{\pm})^2 A^{+}_{(+)}\,,\qquad 
[L_{(-3)},\mathcal{B}^{\pm}]=\pm 4\mathcal{B}^{\pm}\,.
\ee
However, these  differential operators of order $8$ are not independent 
and reduce to the fourth order ladder operators 
(\ref{reio7}) multiplied by the second order polynomials in 
the Hamiltonian,
\be
\mathcal{B}^{-}=\mathcal{A}^{-}(L_{(-)}+1)(L_{(-)}+5)\qquad
\text{and} \qquad
\mathcal{B}^{+}=(\mathcal{B}^{-})^\dagger\,. 
\ee

 As the first and third
order operators 
$A_{(-)}^{\pm} $ and $A_{(+)}^{\pm}$
intertwine the half-harmonic oscillator with the
system $L_{(-)}$ with a nonzero relative 
shift,
we can 
construct 
yet another pair of the ladder operators
for the quantum system $L_{(-)}$,
 \begin{eqnarray}\label{C+-iso}
&
 \mathcal{C}^{-}=A_{(+)}^{-}A_{(-)}^{+}\,, 
 \qquad
 \mathcal{C}^{+}=A_{(-)}^{-}A_{(+)}^{+} \,,
&\\&
 [L_{(-)},\mathcal{C}^{\pm}]=\pm 8\,\mathcal{C}^{\pm}\,,\qquad
 [\mathcal{C}^-,\mathcal{C}^+]=32\left(L^3_{(-)}+6L^2_{(-)}-L_{(-)}+30\right)\,.
&
\end{eqnarray}
The kernel of the lowering ladder operator  
 is
 \be
\ker\mathcal{C}^{-}=\text{span}\,\{(\psi_{(-)}^{-})^{-1},\,A_{(-)}^{-}\psi_1,\,A_{(-)}^{-}\psi_2,\,A_{(-)}^{-}\psi_3
\}\,.
\ee
Here
 $A_{(-)}^{-}\psi_1=A_{(+)}^-\psi_5$ and $A_{(-)}^-\psi_3=A_{(+)}^-\psi_7$
  are the ground  and the first exited states of $L_{(-)}$. 
On the other hand, 
 all the states in the kernel of 
the raising ladder operator are nonphysical\,:
\be
\ker \mathcal{C}^{+}=\text{span}\,\{A_{(-)}^{-}\psi_7^{-},\,A_{(-)}^{-}\psi_2^{-},\,A_{(-)}^{-}
\psi_1^{-},\,A_{(-)}^{-}\psi_0^{-}\}\,.
\ee
 As a result, the space of states
 of  $L_{(-)}$ is separated into  two subspaces, on each of which
 the ladder operators $\mathcal{C}^{+}$ and $\mathcal{C}^{-}$ 
 act irreducibly. One subspace is spanned  by
 the even eigenstates and the 
another subspace corresponds 
to the odd eigenstates.
  The ladder operators 
 $\mathcal{C}^{\pm}$,  unlike $\mathcal{A}^{\pm}$,
 are therefore not spectrum-generating operators  for the system 
 $L_{(-)}$. 
 {}Notice that from the point of view of the basic properties 
 of the ladder operators  $\mathcal{C}^{\pm}$,
 they are similar to the operators 
 $(a^\pm)^4$ 
 in the case of the half-harmonic oscillator
 $L_0$.
 The essential difference here, however,   is that
 the ladder operators
 $\mathcal{C}^{\pm}$ are independent from 
  the spectrum-generating
 ladder operators  $\mathcal{A}^{\pm}$
 and have the same  differential order equal to four.
 We shall see  
 that for \textbf{non-isospectral}
 rational extensions of the AFF systems 
 the direct analogs of the operators  $\mathcal{C}^{\pm}$
 will constitute an inseparable part of the set
 of the spectrum-generating operators.

\vskip0.1cm

The described properties of this  particular example 
are extended  for the general case of 
isospectral deformations and can be summarized as follows.
 No matter what set of the $m$ odd nonphysical eigenstates of 
 the quantum harmonic oscillator 
 we select,  the lower order ladder operators
 $\mathcal{A}^\pm$ obtained by Darboux-dressing 
 of the ladder operators of the half-harmonic oscillator 
 are spectrum-generating operators 
 for  the rationally deformed AFF system.
 They commute for a polynomial of order $2m+1$ 
 in the corresponding  Hamiltonian with which they 
 produce a  deformation of  the conformal $\mathfrak{sl}(2,\R)$ symmetry 
 of the type of $W$-algebra  \textcolor{red}{[\cite{deBoer}]}.
 Other spectrum-generating ladder operators,
 which can be constructed on the basis of other 
 DCKA schemes via the Darbox-dressing procedure, 
 act on physical states 
 in the same way as the operators $\mathcal{A}^\pm$
 of order $2(m+1)$, and
 are equal to them modulo the multiplicative 
 factor in the form of the polynomial in the  
 Hamiltonian operator of the system.
The ladder operators $\mathcal{C}^{\pm}$
constructed by ``gluing'' intertwining operators 
of the two dual  schemes are not spectrum-generating.
Particularly, for the isospectral deformation of the 
system $L_{l_m+1}$
based on the  set of the 
seed states $(-(2l_1+1),-(2l_2+1),\ldots,
-(2l_m+1))$ with $0\leq l_1<l_2<\ldots<l_m$, $l_m\geq 1$, 
the operator ${\mathcal{C}}^-$
annihilates the lowest $l_m+1$ states in the spectrum of the system.

\section{Spectrum-generating ladder operators: non-isospectral case}
\label{sectionSG}
As in the previous section, here we explore the construction of spectrum-generating 
 ladder operators  for non-isospectral deformations 
 of the AFF system through a particular example, and then generalize the ideas.

Let us  start with Darboux's positive scheme
$ (1,4,5,10,11) $ that we have already used as
example to explain the mirror diagram technique
in Sec. \ref{rationallyextededinteger}.
There we had already obtained the negative scheme
which is $ (- 2, -3, -4, -5, -8, -9, -11) $.

After performing the DCKA transformation using the positive scheme, 
we obtain the Hamiltonian operator
\begin{equation}\label{(1,4,5,10,11)}
L_{(+)}:=-\frac{d^2}{dx^2}+x^2-2(\ln W(1,4,5,10,11))''\,,
\end{equation}
where
\begin{eqnarray}
\begin{array}{ll}
W(1,4,5,10,11)\propto & x e^{-\frac{5}{2}x^2}
(467775+4x^2(155925-93555x^2+8x^4(62370
-21945x^2+\\
& +4x^4(735+1145x^2-504x^4+358x^6-88x^8+8x^{10})))) 
\end{array}\,.
\end{eqnarray}
The graph of the resulting 
potential and the quantum spectrum of the system
(\ref{(1,4,5,10,11)}) are shown on Figure  \ref{defAFF1Figure2}.
\begin{figure}[hbt]
\begin{center}
\includegraphics[scale=0.25]{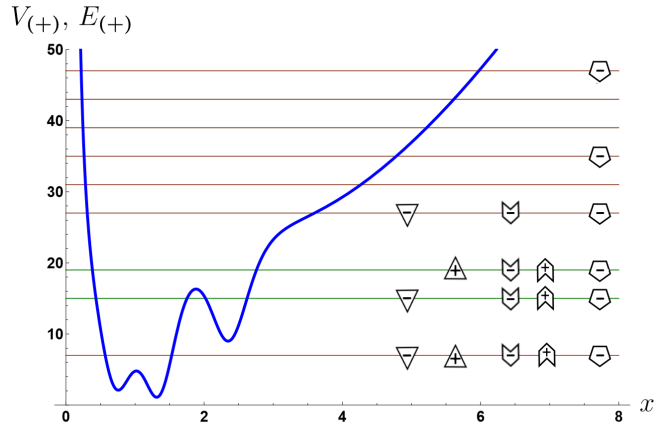}
\caption[Behavior of the potential and physical states 
annihilated by the spectrum-generating ladder operators, Sec. 6.3 ]{\small{Potential of the system (\ref{(1,4,5,10,11)}).
The energy levels of the corresponding physical states
annihilated by ladder operators $\mathcal{B}^-$,
$\mathcal{B}^+$, $\mathcal{A}^-$, $\mathcal{A}^+$, and $\mathcal{C}^-$ are indicated 
from left to right.} }
\label{defAFF1Figure2}
\end{center} 
\end{figure}
\vskip-0.6cm
The potential has three local minima and the system supports
three separated states in its spectrum which are organized in two 
``valence bands''  of one and two states.
On the other hand, the dual scheme  
produces the same Hamiltonian operator but shifted by  a constant,
$L_{(+)}-L_{(-)}=6\Delta E=24$.  
The fact that the mutual shift  of both Hamiltonians is proportional to the 
difference of two consecutive energy levels  in the spectrum  of the AFF model  
allows us to use below exactly the same  rule  for
the construction of the ladder
 operators of the type $\mathcal{C}^{\pm}$
 as in the previous section.  
As we shall see,  the number of physical  
states annihilated by the lowering operator
$\mathcal{C}^-$ in this case is equal exactly to six.    
Later, we also shall see that in some cases  
of the  rational gapped deformations of the AFF  systems, 
the mutual shift 
of the corresponding Hamiltonian operators 
can be equal to the half-integer 
multiple of $\Delta E$, and then 
the procedure for the 
construction of the ladder operators of the 
type $\mathcal{C}^{\pm}$
will require some modification.

 In the DCKA construction  of the Hamiltonian operator $L_{(+)}$, 
 the energy levels corresponding 
 to the physical seed eigenstates of the half-harmonic oscillator $L_{0}$
 were removed from the spectrum  producing two gaps.
 In the (up to a shifted constant) equivalent  system $L_{(-)}$ based on nonphysical seed eigenstates
 of $L_{0}$, the  energy levels were added under the lowest energy 
of the ground  state of $L_{0}$.
 The intertwining  operators  associated with the positive scheme $A_{(+)}^\pm $ 
  have differential order five, while the operators $ A_{(-)}^\pm $, 
obtained from the negative scheme, have differential order  eleven.

The three lowest physical states of the system (\ref{(1,4,5,10,11)})
which correspond to the three separated energy levels 
can be presented in two equivalent forms
\begin{eqnarray}
\phi_0= A_{(-)}^{-}\widetilde{\psi_8^-}=A^-_{(+)}\psi_3\,,\qquad 
\phi_1= A_{(-)}^{-}\widetilde{\psi_4^-}=A^-_{(+)}\psi_7\,, \qquad 
 \phi_2= A_{(-)}^{-}\widetilde{\psi_2^-}=A^-_{(+)}\psi_9\,,
\end{eqnarray}
 where equalities are modulo a nonzero constant multiplier.
We have  here the intertwining relations
\be
A^-_{(+)}L_0=L_{(+)}A^-_{(+)}=(L_{(-)}+24)A^-_{(+)}\,, \qquad 
A_{(-)}^{-}L_0=L_{(-)}A_{(-)}^{-}=(L_{(+)}-24)A_{(-)}^{-}, 
\ee
and the  conjugate relations for $A^+_{(+)}$ and $A_{(-)}^{+}$.

Let us turn now to the construction of the ladder
operators for the system under consideration.
Like in the isospectral case, 
here we have two ways to realize 
Darboux-dressing of the ladder operators 
 $-\mathcal{C}^\pm_0=(a^{\pm})^{2}$.
Using $A^\pm_{(+)}$ for this purpose ,
 we obtain the  operators of order twelve:
\begin{equation}\label{BcalAladder}
\mathcal{B}^{\pm}=A^-_{(+)}(a^{\pm})^{2}A^+_{(+)}\,,\qquad 
[L_{(-)},\mathcal{B}^{\pm}]=\pm\Delta E\mathcal{B}^{\pm}\,.
\end{equation}
The kernel of $\mathcal{B}^{-}$ contains three physical states  
$\phi_0$,  $\phi_1$ and 
$\phi_3= A_{(-)}^{-}\psi_1=A_{(+)}^{-}\psi_{13}$ among other 
9 nonphysical solutions with negative energy.  
They correspond to the ground state, the lowest states in the
isolated 
``valence band'', and the first state in the equidistant part of the spectrum, see Figure \ref{defAFF1Figure2}.
On the other hand $\mathcal{B}^+$ annihilates $\phi_0$, the upper state in the 
valance band  $\phi_2$ and other 10 nonphysical states. Then, 
due to the incapacity of these operators to connect 
 the isolates states with 
the equidistant part of the spectrum, it is obvious that  $\mathcal{B}^\pm$ are not spectrum-generating.

We also can construct 
ladder operators by
using  $A_{(-)}^{\pm}$ instead,
\begin{equation}\label{AaADarboux}
 \mathcal{A}^{\pm}=A_{(-)}^{-}(a^{\pm})^{2}A_{(-)}^{+}\,,
\qquad  
[L_{(+)},\mathcal{A}^{\pm}]=\pm\Delta E \mathcal{A}^{\pm\,}\,.
\end{equation}
These  are also not spectrum-generating operartors
because the leap they make does not allow  to overcome the gaps.  
Operator $\mathcal{A}^{+}$ detects all the 
 states in both  separated valence bands by annihilating them. In addition
 to the indicated physical states,
 the lowering operator $\mathcal{A}^{-}$ also annihilates  the lowest state in the 
 half-infinite equidistant part of the spectrum.

Therefore, the essential difference of the non-isospectral rational deformations 
of the AFF model  from their  isospectral rational extensions is that
there is no pair of spectrum-generating ladder operators constructed 
 via the Darboux-dressing procedure. 
This situation is similar to that in the rationally extended QHO 
systems \textcolor{red}{[\cite{CarPly}]}.

We now construct 
 the ladder operators $\mathcal{C}^\pm$ 
  by ``gluing'' 
 the intertwining operators of different types. As in the case of the isospectral deformations,
 they also will not be the spectrum-generating operators, but
 together with any pair of the ladder operators ${\mathcal{B}^{\pm}}$, or $\mathcal{A}^\pm$
 they will form a spectrum-generating set.
So, let us consider  
\begin{equation}\label{C+-AAdefin}
\mathcal{C}^{-}=A_{(-)}^-A_{(+)}^{+} \,,
\qquad 
\mathcal{C}^{+}=A_{(+)}^{-}A_{(-)}^+\,,
\qquad
[L_{(-)},\mathcal{C}^{\pm}]=\pm 6\Delta E\mathcal{C}^{\pm}\,.
\end{equation}
They are independent
from the ladder operators constructed via the Darboux-dressing procedure,
and their commutator $[\mathcal{C}^-, \mathcal{C}^+]$ is a 
certain polynomial of order $11$ in the Hamiltonian $L_{(-)}$.
The  operators $\mathcal{C}^{\pm}$
 divide the Hilbert  space of  the system 
into  six infinite subsets on which they act irreducibly: 
The $\mathcal{C}^{-}$ transforms a physical 
eigenstate into another  physical eigenstate
by making it skip six levels below and annihilates the first six eigenstates of the spectrum.
The  operator $\mathcal{C}^{+}$ does not annihilate any physical state here and skip the energy of an arbitrary 
state in to six levels above.
Therefore they connect the separated states with the equidistant 
part of the spectrum.

As a result, the pair  $\mathcal{C}^{\pm}$ 
together with any pair of the ladder operators, 
$\mathcal{B}^\pm$ or $\mathcal{A}^\pm$ are the spectrum-generating set.
Figure \ref{Figure3} illustrates the action of the ladder operators and show  how  we can 
use them to obtain a particular state, starting  from an arbitrary one.

\begin{figure}[h]
\begin{center}
\includegraphics[scale=0.40]{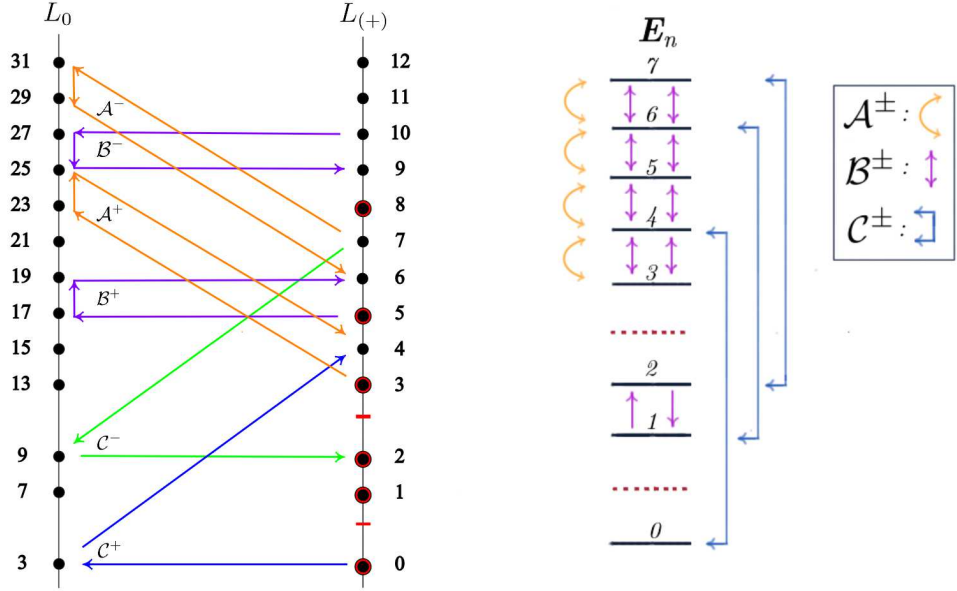}
\end{center} 
\caption[Spectrum-generating ladder operators behavior, Sec.  6.3]{{\label{Figure3}}
\small{On the left: The numbers on the left correspond to 
the indices of the physical eigenstates $\psi_{2l+1}$ of the half-harmonic oscillator 
that are mapped ``horizontally'' by  operator $\A^-_{(+)}$ into  
eigenstates $\Psi_n$ of the system (\ref{(1,4,5,10,11)}). 
Lines show the action 
of the ladder operators coherently with their structure 
(\ref{AaADarboux}), (\ref{BcalAladder}) and 
(\ref{C+-AAdefin}). The marked set of the states $0,\, 1,\, 2,\, 3,\, 5,\,8 $ 
on the right  corresponds to six eigenstates of $L_{(+)}$ annihilated by $\mathcal{C}^-$.
On the right: Horizontal lines correspond to the  energy levels of $L_{(+)}$. 
Upward and downward arrows  represent the action of the rising and lowering ladder operators, 
respectively. As it is shown in the figure on the right, 
following the appropriate  paths, any eigenstate can be transformed into any other eigenstate
by applying subsequently the corresponding ladder operators.}   }
\end{figure}

All the described picture is generalized directly in the case when
the index of the last seed state used in the corresponding
DCKA transformation is odd. 
Then the corresponding scheme based on physical eigenstates 
of $L_0$ is of the form
$(\ldots,2l_m,2l_m+1)$, and the dual scheme
is  $(\ldots,-(2l_m+1))$.
Following the same notation as we used 
in the particular examples, the Hamiltonian operators generated in these  two 
dual schemes are shifted by the distance equal to the separation $\Delta E=4$
of energy levels in the equidistant part of the spectrum times
integer number $l_m+1$\,: 
$L_{(+)}-L_{(-)}=4l_m+4$, see (\ref{L+L-}), and the picture is the following: 
\begin{itemize}
\item Operators $\mathcal{A}^\pm=\A^-_{(-)}(a^\pm)^2\A^+_{(-)}$ 
are of differential order  $2n_-+2$. 
Rising and lowering operators 
of this kind annihilate all the states in the isolated valence bands, in the sense of a group 
of energy levels separated by a gap from the equidistant part of the spectrum.
 They act as regular ladder operators in the equidistant part of the spectrum. 
\item  Operators $\mathcal{B}^\pm= \A^-_{(+)}(a^\pm)^2\A^+_{(+)}$ are 
of differential order  $2n_++2$.  $\mathcal{B}^-$ annihilates all the lowest 
states in each valence band and the lowest state in the 
equidistant part of the spectrum. 
The raising operator $\mathcal{B}^+$ annihilates all the highest 
states in each valence band.  They act in the same way as $\mathcal{A}^\pm$
 in the equidistant part of the spectrum.

\item Operators  $\mathcal{C}^\pm$ of the form (\ref{C+-AAdefin}) 
have a differential order
$n_-+n_+=2l_m+2$, and
their commutation with 
 Hamiltonian produces: 
\begin{equation}\label{LCpm(lm+1)}
[L_{(-)}, \mathcal{C}^\pm]=\pm (l_m+1)\Delta E\mathcal{C}^\pm\,.
\end{equation}
 Lowering operator  $\mathcal{C}^-$ annihilates  $l_m+1$  physical states, where we find 
all of the isolated states and some exited states of the equidistant part. 
Rising operator  $\mathcal{C}^+$ does not annihilate any physical state. 

\end{itemize}

\vskip0.1cm

When we have the schemes $(\ldots,2l_m-1,2l_m)\sim (\dots,-2l_m)$
generating a  gapped rational extension 
of some AFF system, the corresponding Hamiltonian operators
associated with them are shifted mutually for the distance 
$L_{(+)}-L_{(-)}=4l_m+2=(l_m+\frac{1}{2})\Delta E$, 
that is equal to the half-integer multiple of the energy spacing 
in the equidistant part of the spectrum and
in the valence bands with more than one state.
In this case the procedure related to the construction 
of the ladder operators $\mathcal{A}^\pm$ and 
$\mathcal{B}^\pm$ 
and their properties are similar to those 
in the systems generated by the schemes 
$(\ldots,2l_m,2l_m+1)\sim (\dots,-(2l_m+1))$.
However, the situation with the construction of
the ladder operators  of the type $\mathcal{C}^\pm$
in this case is essentially different.
We still can construct the operators $\mathcal{C}^\pm$
of the form (\ref{C+-AAdefin}).
Such operators will be of odd differential order 
$2l_m+1$, and their commutation relations  with any of the
 Hamiltonian operators $L_{(+)}$ and  $L_{(-)}$ 
 will be  of the form $[L, \mathcal{C}^\pm]=\pm (4l_m+2)\mathcal{C}^\pm$.
 This means that these operators acting on physical 
 eigenstates of $L$  will produce nonphysical eigenstates
 excepting the case when the lowering 
 operator $\mathcal{C}^-$ acts on the states 
 from its kernel.  
  The square of these operators 
 will not have the indicated deficiency and will form together 
 with the ladder operators $\mathcal{A}^\pm$ or 
 $\mathcal{B}^\pm$  the set of the
 spectrum-generating 
 operators. This picture can be compared with 
  the case of the half-harmonic oscillator $L_0$, where 
  the first order differential operators 
$a^\pm$ will have the properties similar to those of the 
described operators $\mathcal{C}^\pm$. 
In this case we can however modify slightly the construction
of the ladder operators of the $\mathcal{C}^\pm$ type 
by taking 
\be\label{Cpmnew}
\widetilde{\mathcal{C}}^-=A_{(-)}^-(a^-)A_{(+)}^+\,,\qquad
\widetilde{\mathcal{C}}^+=A_{(+)}^-(a^+)A_{(-)}^+\,.
\ee 
These ladder operators satisfy the commutation
relations 
$[L_{(\pm)},\widetilde{\mathcal{C}}^\pm]=4(l_m+1)\widetilde{\mathcal{C}}^\pm$,
and transform a particular  physical states into other physical states with different 
energy.

To conclude  this section, let us summarize
the structure of the nonlinearly deformed conformal symmetry
algebras generated by different pairs of the corresponding  ladder operators
and Hamiltonians of the rationally deformed 
conformal mechanics systems.
The commutators of the ladder operators 
$\mathcal{A}^\pm$, $\mathcal{B}^\pm$ and $\mathcal{C}^\pm$
with Hamiltonian operators are  given, respectively, 
by Eqs. (\ref{AaADarboux}), (\ref{BcalAladder}) and (\ref{C+-AAdefin})
with  $\Delta E=4$. The commutation relations   of the form
(\ref{AaADarboux}) also are valid for the case of the isospectral deformations
discussed in the previous section.
To write down the commutation relations between 
raising and lowering operators of the same type in general case, 
let us introduce the polynomial functions
\be
P_{n_+}(x)=\Pi_{k=1}^{n_+}(x-2n_k-1)\,,
\qquad R_{n_-}(x)=\Pi_{l=1}^{n_-}(x+2n_l+1),
\ee
where $n_k>0$ are the indices  of the corresponding seed states in the positive 
scheme and $-n_l<0$ are the indices 
of the seed states in the negative scheme. With this notation,
we have the relations $A_{(+)}^+A_{(+)}^-=P_{n_+}(L_0)$, 
$A_{(+)}^-A_{(+)}^+=P_{n_+}(L_{(+)})=P_{n_+}(L_{(-)}+2(n_-+n_+))$, 
and $A_{(-)}^+A_{(-)}^-=R_{n_-}(L_0)$,  $A_{(-)}^-A_{(-)}^+=R_{n_-}(L_{(-)})$. 
Then we obtain
\begin{eqnarray}\label{AARH}
&
[\mathcal{A}^-,\mathcal{A}^+]=(x+1)(x+3) R_{n_-}(x)R_{n_-}(x+4) 
\big\vert_{x=L_{(-)}}^{L_{(-)}-4}\,,
\label{BBPH}&\\&
[\mathcal{B}^-,\mathcal{B}^+]=(x+1)(x+3)P_{n_+}(x+4)P_{n_+}(x) 
\big\vert_{x=L_{(-)}+2N}^{x=L_{(-)}+2N-4}\,,
\label{CCPRH}&\\&
[\mathcal{C}^-,\mathcal{C}^+]=R_{n_-}(x) P_{n_+}(x)\big\vert_{x=L_{(-)}}^{x=L_{(-)}+2N}\,,
\end{eqnarray}
where $N=n_-+n_+$, and relation 
(\ref{AARH}) also is valid in the case of isospectral deformations.
In the case of the non-isospectral deformations given by the dual schemes
$(\ldots,2l_m-1,2l_m)\sim (\dots,-2l_m)$, the corresponding modified operators
(\ref{Cpmnew})  satisfy the commutation relation
\begin{equation}\label{tilCCPRH}
[\widetilde{\mathcal{C}}^-,
\widetilde{\mathcal{C}}^+]=(x+1)R_{n_-}(x) P_{n_+}(x+2)\big\vert_{x=L_{(-)}}^{x=L_{(-)}+2N-2}\,.
\end{equation}

Thus, 
in any rational deformation of the  conformal mechanics model we considered, 
each pair of the conjugate ladder operators of the types 
$\mathcal{A}^\pm$, $\mathcal{B}^\pm$ or $\mathcal{C}^\pm$ 
generates a nonlinear deformation of the conformal $\mathfrak{sl}(2,\R)$ symmetry.
The commutation relations between ladder operators of different types of the form 
$[\mathcal{A}^\pm, \mathcal{C}^\pm]$, etc.  is considered in next chapter, 
and their taking into account gives rise naturally to different nonlinearly extended  versions 
of the superconformal $\mathfrak{osp}(2|2)$ symmetry  \textcolor{red}{[\cite{InzPly2}]}.

\section{Remarks}

The construction of the spectrum-generating ladder operators 
can also be explored by using intertwining operators between 
the final rational extended model and some intermediate system 
in the Darboux chain. This possibility was explored in
\textcolor{red}{[\cite{CarInzPly}]}. Anyway, the final conclusion 
 of this is that one always has a triad of pairs of ladder operators $\mathcal{A}^\pm$, 
$\mathcal{B}^\pm$ and $\mathcal{C}^\pm$
which behaves as described above. The only difference here is the number 
of nonphysical states that appear in the corresponding kernels.

An unresolved question for us is if there is any relationship
between rationally  extended systems and other systems of quantum mechanics, such as the 
conformal model (\ref{conformalaction}) or a $ \mathcal{PT} $ deformation  
of it \textcolor{red}{[\cite{JM1,JM2,plyushchay2020exotic}]}, we are thinking of something like the conformal bridge. 
It can be speculated that if such a relationship exists,
 it would be useful in applications related to integrable systems of infinite
 degrees of freedom,  since $\mathcal{PT}$ symmetric  systems have opened new 
branches in the search for solitonic solutions for the KdV equation  and
 other integrable models \textcolor{red}{[\cite{Correa2016,JM2,Cen}]}. 

In the next chapter we continue with rationally extended AFF models characterized
by integer coupling constants as well as extended QHO systems,
 but now from the perspective of supersymmetric quantum mechanics.

 
\chapter{Nonlinear supersymmetries in rationally extended systems}
\label{ChNonLinearSUSY}

We now turn to the study of the extensions and deformations of the
 superconformal and super-Schr\"odinger symmetries 
that appear in the $\mathcal {N} = 2$ super-extended systems
described by the superpartners 
($L_\text{os}$, $L_{def}$) and ($L_{0}$,  $L_{m,def}$).
Here $L_{def}$ and    $L_{m,def}$ correspond
to rational deformations of the QHO system
and the AFF model with integer 
values of the parameter $\nu=m$, $m\in \N$, respectively. 
As we have seen in the last chapter, the rational deformations of the 
QHO system and the AFF model are characterized, in the general case,
 by a finite number of  missing energy levels, or gaps, 
  in their spectra, and the description of such systems 
  requires more than a couple of spectrum-generating operators.
It is because of this expansion of the sets of ladder operators,
 whose differential order exceeds two, that nonlinearly deformed 
superconformal and super-Schr\"odinger structures appear. 
This chapter, based on the article \textcolor{red}{[\cite{InzPly2}]}, is devoted to the description of the
 complete sets of generators of the indicated symmetries. 
At this point, we will again take advantage of the Darboux duality 
property of the QHO system.


\section{Basic intertwining operators}
 According to \textcolor{red}{[\cite{CarPly,CarInzPly}]}, with each of the dual schemes it is necessary
 first to associate two basic pairs of the intertwining operators. Here, we discuss general properties of such operators
without taking care of the concrete nature of the system built by the DCKA transformation. 
On the way, however, some important  distinctions between  rational deformations of
the AFF  model and harmonic  oscillator  have to be taken into account, and for this reason, 
it is convenient to speak of  two classes of the systems.  
We distinguish between them by introducing the class index $c$,
where $c = 1$ and $c=2$ will correspond to deformed harmonic oscillator 
and  deformed AFF conformal mechanics model, respectively.  

As already established in the previous chapter, we will denote the Hamiltonian produced by the positive 
 scheme $\Delta_+$ (negative scheme  $\Delta_-$)
by  $L_{(+)}$ ($L_{(-)}$), and the corresponding intertwining operators 
by 
$A_{(+)}^-$ and $(A_{(+)}^-)^\dagger \equiv  A_{(+)}^+$ 
( $A_{(-)}^-$ and $(A_{(-)}^-)^\dagger \equiv  A_{(-)}^+$), see Sec. (\Ref{rationallyextededinteger}).
 These operators satisfy the relations
\begin{eqnarray}
\label{L+L-N}
&L_{(+)}-L_{(-)}=2N\,,\qquad N=n_++n_-\,,  &\\&
\label{inter0}
A_{(+)}^-L=L_{(+)}A_{(+)}^-\,, \qquad A_{(-)}^-L=L_{(-)}A_{(-)}^-\,,&
\end{eqnarray} 
and  the corresponding 
Hermitian conjugate  relations for
$A_{(+)}^+$ and $A_{(-)}^+$. Here $ L $ could be $ L_\text{os} $ or $ L_{0} $, depending on the class index $ c $ of the rationally deformed system $ L _ {(\pm)} $ that we want to study.
Applying operator identities 
  (\ref{inter0}) to an arbitrary physical or nonphysical (formal)  eigenstate
 $\varphi_n$ of $L$ different from any seed state of the positive scheme 
 and using Eq. (\ref{L+L-}), one  can derive the equality
\be 
\label{relation-operators}
  A_{(-)}^-\varphi_n=A_{(+)}^-\varphi_{n+N}\,,
\ee
to be valid modulo a multiplicative constant. 
As a result, 
both operators acting on the same state of the harmonic oscillator produce 
different states
of the new system.
We have seen this behavior before in last chapter, Sec. \ref{sectionSG}. 
The Hermitian conjugate operators  $A_{(-)}^+$ and  $A_{(+)}^+$ do a similar job but  
in the opposite direction. 
Eq.  (\ref{relation-operators}) suggests that some 
peculiarities 
 should be taken into account  for class 2 systems\,: 
 the infinite potential barrier  at $x=0$
 assumes that  
 physical states of $L_{0}$ and 
 $L_{(\pm)}$ systems
 are described by odd wave functions.
Then,
in order  for $ A_{(+)}^-$to  transform 
physical states of $L_{0}$ into  physical states of  $L_{(\pm)}$, 
we must take $n + N$  to be odd  for odd $n$
in (\ref{relation-operators}). This
means that 
$A_{(-)}^-$ transforms physical states into physical 
 only if $N$ is even.
In the case of odd $N$, it is necessary to take 
$A_{(-)}^-a^-$ or $A_{(-)}^-a^+$ as a physical  intertwining operator. 
It is convenient to take into account this peculiarity
by denoting 
the remainder of the division $N/c$ by 
$r(N,c)$\,: it takes value $1$ 
in the class $c=2$ of the systems with odd $N$ and equals zero
in all other cases. 

The products of the described 
 intertwining operators are of the form (\ref{poly1}), 
 and  
 for further analysis it 
is useful to write down  them explicitly: 
\begin{eqnarray}
\label{A-A-A+A+Poly}
&A_{(\pm)}^{+}A_{(\pm)}^{-}=P_{n_\pm}(L)\,,\qquad A_{(\pm)}^{-}A_{(\pm)}^{+}=P_{n_\pm}(L_{(\pm)})\,,&\\
\label{polyA}
&P_{n_+}(\eta)\equiv \prod_{k=1}^{n_+}(\eta-2l_k^+-1)\,, \qquad 
P_{n_-}(\eta)\equiv \prod_{k=1}^{n_-}(\eta+2l_k^-+1)\,.&
\end{eqnarray} 
Here $l_k^+$ are indexes of physical states with eigenvalues $2l_k^++1$
 in 
the set $\Delta_+$,
and $-l^-_k$ correspond to nonphysical states with eigenvalues $-2l^-_k-1$ 
in 
the negative scheme $\Delta_-$.
 In the same 
 vein, it is useful to write
\begin{eqnarray}
\label{ak}
&(a^+)^k(a^-)^k=T_k(L_0),\qquad (a^-)^k(a^+)^k=T_k(L_0+2k)\,,&\\
\label{Tk}
&T_{k}(\eta)\equiv \prod_{s=1}^{k}(\eta-2s+1)\,, \qquad T_{k}(\eta+2k)\equiv \prod_{s=1}^{k}(\eta+2s-1)\,.&
\end{eqnarray}
We also have the operator identities 
  \begin{eqnarray}
\label{ide}
(a^-)^{N}=(-1)^{n_-}A_{(-)}^+A_{(+)}^-\,,
\qquad f(L_{(-)})A_{(+)}^-(a^+)^{n_-}=(-1)^{n_-}h(L_{(-)})A_{(-)}^-(a^-)^{n_+}\,,
\end{eqnarray}
and their Hermitian conjugate versions, 
where $f(\eta)$ and $h(\eta)$ are polynomials whose explicit structure 
is given in Appendix \ref{show}. 
In one-gap deformations of the harmonic oscillator and gapless deformations 
of $L_1$ these polynomials reduce to $1$.


\section{Extended sets of ladder and intertwining operators}
\label{interladder}
Actually, instead of three types of ladder operators, we have a
 total of three families of operators
\begin{eqnarray}
\label{genlad}
&\mathfrak{A}_{k}^\pm\equiv A_{(-)}^-(a^\pm)^k A_{(-)}^+\,,\qquad \mathfrak{B}_{ k}^\pm\equiv A_{(+)}^-(a^\pm)^k A_{(+)}^+\,,
&\\
&\mathfrak{C}_{N\pm k'}^-\equiv A_{(+)}^-(a^\mp)^{k'}A_{(-)}^+ \,,\qquad
\mathfrak{C}_{N\pm k'}^+\equiv (\mathfrak{C}_{N\pm k'}^-)^\dagger\,,&\label{genlad+}
\end{eqnarray}
where, formally, $k$ can take any nonnegative integer value  and $k'$  is such that 
$N-k'\geq 0$, otherwise operators (\ref{genlad+})  reduce to $\mathfrak{A}_k^\pm$, 
\textcolor{red}{[\cite{InzPly2}]}. Due to relations (\ref{A-A-A+A+Poly})-(\ref{ide}) one concludes that 
at $k=0$ and $N-k'=0$ all these operators are reduced to certain polynomials in 
 $L_{(\pm)}$.
These objects are generated by taking the commutator relations between
 two arbitrary representatives of the spectrum generator set described in 
the previous chapter, and behave like powers of the ladder operator in the QHO system. 
Calculations with these operators are discussed in detail in Appendix~\ref{apen-red}, so
 this chapter contains only the main results.

Independently of the class of the system, or on whether 
the operators are physical or not, 
the three families  
$\mathfrak{D}_{\rho,j}^\pm=( \mathfrak{A}_{j}^\pm, \mathfrak{B}_{j}^\pm,\mathfrak{C}_{j}^\pm )$,
$\rho=1,2,3$,  $j=1,2,\ldots$, satisfy the commutation relations of the form 
\begin{equation}
\label{sl2rh}
[L_{(\pm)},\mathfrak{D}_{\rho,j}^\pm]=\pm2j \mathfrak{D}_{\rho,j}^\pm \,,
\qquad [\mathfrak{D}_{\rho,j}^-,\mathfrak{D}_{\rho,j}^+]=\mathcal{P}_{\rho,j}(L_{(-)})\,, 
\end{equation}
where  $\mathcal{P}_{\rho,j}(L_{(-)})$ 
is a certain polynomial of the corresponding Hamiltonian operator of the system,
whose order  
of polynomial   is equal to differential order 
of $\mathfrak{D}_{\rho,j}^\pm$ minus one,
see
Appendix  \ref{apen-red}.
Algebra
(\ref{sl2rh}) can be considered as a deformation of
 $\mathfrak{sl}(2,\R)$\,,
 \textcolor{red}{[\cite{JM2}]}.
 \vskip0.1cm
Of all the operators that can be built, our objective is to discriminate against those that are physical and cannot be written as products of lower order elements, belonging to others or to the same family. Having this in mind, 
we have the following assertion related to the three families:
\begin{itemize}
\item From (\ref{sl2rh}) one concludes that $2j\propto \Delta E=2c$.  
Then, for $\mathfrak{A}$ and $\mathfrak{B}$ families, 
the physical operators are those 
whose  index is $j=lc$ with $l \in \N$, while
for $\mathfrak{C}$ family  index should be $j=N+r(N,c)+cs$, where $s$ is 
integer such that $j>0$. 
\item For isospectral deformations of the AFF model, the 
spectrum-generating 
set is given by any pair of the conjugate operators $\mathfrak{A}^{\pm}_{2}$, 
$\mathfrak{B}^{\pm}_{2}$, 
or $\mathfrak{C}^{\pm}_{2}$. 
\item Due to Eq.  (\ref{ide}) one realizes that the basic operators in the general case are 
\begin{eqnarray}
\label{ladgen}
\left\{
\begin{array}{cc}
\mathfrak{A}_{k}^\pm\,, & 0<k<N\,,\\
\mathfrak{B}_{k}^\pm\,, & 0<k<N\,,\\
\mathfrak{C}_{k}^\pm\,, & 0<k<2N+r(N,c)\,,\\
\end{array}
\right.
\end{eqnarray}
\item For  one-gap deformations of the  harmonic 
oscillator, the set of basic ladder operators can be reduced further
to the set
\begin{eqnarray}
\label{basicsubsetonegap}
\left\{
\begin{array}{cc}
\mathfrak{A}_{k}^\pm\,, & 0<k<n_+\,,\\
\mathfrak{B}_{k}^\pm\,, & 0<k<n_-\,,\\
\mathfrak{C}_{k}^\pm\,, & M<k<n_+\,,\\
\end{array}
\right. \qquad 
M=\left\{\begin{array}{ccc}
\max\,(n_-,n_+) & \text{if} & n_-\neq n_+\,,\\
N/2 & \text{if} & n_-=n_+\,,
\end{array}
\right.
\end{eqnarray}
where the relations
$\mathfrak{A}_{n_+}^{\pm}=(-1)^{n_-}\mathfrak{C}_{n_+}^{\pm}$ and 
 $\mathfrak{B}_{n_-}^{\pm}=(-1)^{n_-}\mathfrak{C}_{n_-}^{\pm}$ were taken into account.
\end{itemize}
As is obvious from their explicit form, any of the basic elements belonging to one of the three families
 of ladder operators can be constructed by ``gluing'' two different intertwining 
 operators associated with an alternative DCKA transformation, which are of the form $ A_{(\pm) } a ^ {\pm} $
and $ A_{(\pm)} a^ {\mp} $, so their number should also be reduced. Indeed,
for general deformations  
only  the 
operators
\begin{eqnarray}
\label{genA}
\left\{
\begin{array}{cc}
A_{(\pm)}^-(a^\pm)^{n}\,, & 0\leq n<N\,,\\
A_{(\pm)}^-(a^\mp)^{n}\,,& 0<n<N+r(N,c)\,,
\end{array}
\right.
\end{eqnarray}
and their Hermitian conjugate counterparts 
can be considered as basic, see Appendix \ref{apen-red}.
One can note 
that the total number of the basic intertwining operators $\#_f=2[(4N-2+r(N,c))/c]$ 
is greater than the number of the basic ladder operators 
$\#_{lad} = 2[(4N-3+r(N,c))/c]$ 
which can be constructed  with their help. 
 In particular case of gapless deformations of the AFF model, 
 the indicated set of Darboux generators  can be reduced to 
 those which produce, by  `gluing' procedure, one conjugate pair of the 
 spectrum-generating 
 ladder operators of the form
$\mathfrak{D}_{2,\rho}^\pm$.

For $c=1$ one-gap systems, identity
(\ref{ide}) allows us to reduce further the set of the basic intertwining 
operators, which, together with corresponding Hermitian conjugate ones, 
 is  given by any of the two options,  
\begin{equation}
\label{frakS}
\mathfrak{S}_{z}\equiv \left\{ \begin{array}{lcc}
              A_{(-)}^-{(a^+)}^{|z|}\,,  &  -N<z\leq 0\,,\\\vspace{-0.4cm}
             \\ A_{(-)}^-{(a^-)}^{z}\,,  &  0< z \leq n_+\,,\\\vspace{-0.4cm}
             \\ A_{(+)}^-{(a^+)}^{N-z} \,,  & n_+ < z \leq N\,,\\\vspace{-0.4cm}
             \\ A_{(+)}^-{(a^-)}^{N-z}\,,   & N<z<2N\,, \\
             \end{array}
   \right.\\\qquad \text{or}\qquad \mathfrak{S}_{z}^{'}\equiv \mathfrak{S}_{N-z}\,,
\end{equation}
 see Appendix \ref{apen-red}. 
Here we have reserved  $z=0$ and $z=N$ values for index $z$ to the dual schemes
intertwining operators: in the first choice, $\mathfrak{S}_{0}=A_{(-)}^-$ and
$\mathfrak{S}_{N}=A_{(+)}^-$, and  
for the second choice we have $\mathfrak{S}_{0}'=A_{(+)}^-$ 
and $\mathfrak{S}_{N}'=A_{(-)}^-$. Written in this way, these operators satisfy the  
intertwining relations $
 \mathfrak{S}_{z}L=(L_{(-)}+2z)\mathfrak{S}_{z}$ or $\mathfrak{S}_{z}'L=(L_{(+)}-2z)\mathfrak{S}_{z}'$,
and their Hermitian conjugate versions. 
Then,  to study supersymmetry, we have to choose   either positive or negative scheme
to define the  $\mathcal{N}=2$ super-extended Hamiltonian.
We take $\mathfrak{S}_{z}$ if we work with a negative scheme, 
and $\mathfrak{S}_{z}'$ if positive scheme is chosen for the construction of  super-extension.


\section{Supersymmetric extensions}
\label{susyextension}
For each of the two dual schemes, one can construct an $\mathcal{N}=2$ super-extended
Hamiltonian operator 
following the recipe given in Chap. \ref{ChSUSY}, equation (\ref{Hlambda*}). 
The task is to choose appropriately $H_1=\breve{L}-\lambda^*$
and $H_0=L-\lambda^*$.
We put $\breve{L}=L_{(+)}$ and $\lambda^*=\lambda_+= 2l_1^++1$
for positive scheme, and choose  $\breve{L}=L_{(-)}$ and $\lambda^*=\lambda_-= -2l_1^--1$
for negative scheme.
For  both options, we set $L=L_\text{os}$ if we are dealing with a rational extension of 
harmonic oscillator, and $L=L_{0}$ if we  work with a deformation of the AFF model.
 We name the matrix Hamiltonian associated with  negative scheme as $\mathcal{H}$, and
denote by $\mathcal{H}'$ the Hamiltonian  of positive scheme. 
The  spectrum of  these systems can be  found using 
 the properties of  the corresponding  intertwining operators described in
 Sec. \ref{Dar}, see also refs.  \textcolor{red}{[\cite{CarPly,CarInzPly}]}.
The two  Hamiltonians are connected by relation 
 $\mathcal{H}-\mathcal{H}'=-N(1+\sigma_3)-\lambda_-+\lambda_+$,
 and  $\sigma_3$ plays a role of 
 the $\mathcal{R}$ symmetry
generator  for both  super-extended systems.
In this subsection we finally construct the corresponding 
spectrum-generating superalgebra for $\mathcal{H}$  and  $\mathcal{H'}$. 
The resulting structures are based on the physical operators $\mathfrak{D}_{\rho,j}^\pm$.
As we shall see, 
the supersymmetric versions of the  $c=1$ systems
are described by a nonlinearly extended super-Schr\"odinger symmetry with 
bosonic  generators  to be differential operators of even and odd orders, while in the case 
of the $c=2$ systems we obtain
nonlinearly  extended superconformal symmetry in which bosonic generators 
are of even order only.  
\vskip0.1cm

We construct a pair of fermionic operators on the basis 
of each intertwining operator from the set (\ref{genA}) and 
their Hermitian conjugate counterparts.  
Let us consider first 
the extended nonlinear super-Schr\"odinger symmetry of a one-gap deformed 
harmonic oscillator,
and then we generalize the picture.  
If we choose the negative scheme, then we use $\mathfrak{S}_z$ defined in (\ref{frakS}) to construct  
the set of operators 
\begin{eqnarray}
\label{gencharge}
\mathcal{Q}_1^{z}=
\left(
\begin{array}{cc}
  0&  \mathfrak{S}_z  \\
 \mathfrak{S}_z^\dagger &  0     
\end{array}
\right)\,,
\qquad 
\mathcal{Q}_2^{z}= i\sigma_3\mathcal{Q}_1^{z}\,, \qquad -N<z<2N\,.
\end{eqnarray}
They satisfy  the (anti)-commutation relations 
\begin{equation}
\label{SUSY}
[\mathcal{H},\mathcal{Q}_a^z]=2iz\epsilon_{ab}\mathcal{Q}_{b}^{z}\,,
\qquad \{\mathcal{Q}_a^z,\mathcal{Q}_b^z\}=2\delta_{ab}\mathbb{P}_{z}(\mathcal{H},\sigma_3)\,,\qquad
[\Sigma,\mathcal{Q}_{a}^z]=-i\epsilon_{ab}\mathcal{Q}_{b}^{z}\,,
\end{equation}
where $\Sigma= \frac{1}{2}\sigma_3$ and 
$\mathbb{P}_z$ are some polynomials whose structure is described in 
Appendix \ref{apen-comm}. 
For the choice of the positive scheme to fix extended Hamiltonian, according to 
(\ref{frakS}), the  corresponding fermionic operators are given by 
$\mathcal{Q}_1^{'z}\equiv \mathcal{Q}_1^{N-z}$. They satisfy relations 
of the same form (\ref{SUSY}) but with replacement 
$\mathcal{H}\rightarrow \mathcal{H}'$, $\Sigma=\frac{1}{2}\sigma_3 \rightarrow \Sigma'=-\frac{1}{2}\sigma_3$, 
$\mathbb{P}_{z}(\mathcal{H},\sigma_3)\rightarrow 
\mathbb{P}_{z}'(\mathcal{H}',\sigma_3)=\mathbb{P}_{N-z}(\mathcal{H}'-N(1+\sigma_3)-\lambda_-+\lambda_+,\sigma_3)$,
$\mathcal{Q}_{1}^{z}\rightarrow \mathcal{Q}_{2}^{'z}$ and
$\mathcal{Q}_{2}^{z}\rightarrow \mathcal{Q}_{1}^{'z}$. 
The  fermionic operators $\mathcal{Q}_a^{0}$ (or $\mathcal{Q}_a^{'0}$) 
are the  supercharges 
of the (nonlinear in general case) $\mathcal{N}=2$ Poincar\'e supersymmetry,
which are 
integrals of motion of the system $\mathcal{H}$ 
(or $\mathcal{H}'$), and 
 $\mathbb{P}_{0}=
P_{n_-}(\mathcal{H} + \lambda_{-})\,$
 (or $\mathbb{P}_{0}=P_{n_+}(\mathcal{H}' +
\lambda_{+})$)  with polynomials   $P_{n_\pm}$ defined in (\ref{polyA}).
The operators $\mathcal{Q}_a^{'0}$ are analogous here to supercharges 
in $Q_\nu^{a}$ in the linear case, see Chap. \ref{ChConformal}.
On the other hand, we have here the  
fermionic operators $\mathcal{Q}_a^{'N}$ as  analogs 
of dynamical integrals $\mathcal{S}^a_\nu$ there. 
We recall that in the  simple linear case considered in section  \ref{SecOSP22Conformal}, 
the interchange between positive and negative  schemes
corresponds to the  automorphism of superconformal algebra,
and this observation will be helpful for us  
for the analysis of the nonlinearly extended
super-Schr\"odinger
structures.
Here, actually, 
each of the  $(\#_f-2)/2$ pairs of fermionic operators distinct from supercharges
provides a possible dynamical extension of the super-Poincar\'e symmetry.
As we will see,  all of them are necessary to obtain a closed
nonlinear 
spectrum-generating 
superalgebra of the super-extended system.

To construct any extension of the deformed Poincar\'e supersymmetry, 
we calculate $\{\mathcal{Q}_{a}^{0},\mathcal{Q}_{a}^{z} \}$,  in the negative scheme, 
or $\{\mathcal{Q}_{a}^{'0},\mathcal{Q}_{a}^{'z} \}$ in the positive one.
In the first case we have
\be 
\label{Cn+kQN}
\{\mathcal{Q}_a^{0},\mathcal{Q}_{b}^{z}\}=\delta_{ab}(\mathcal{G}_{-z}^{(2\theta(z)-1)}+ 
\mathcal{G}_{+z}^{(2\theta(z)-1)})+i\epsilon_{ab}(\mathcal{G}_{-z}^{(2\theta(z)-1)}- \mathcal{G}_{+z}^{(2\theta(z)-1)})\,,
\ee
where $z\in (-N,0)\cup(0,2N)$, $\theta(z)=1\, (0)$ for $z>0\, (z<0)$, 
 and  $\mathcal{G}^{(2\theta(z)-1)}_{\pm z}$
are given by 
\be
\label{superC}
\mathcal{G}_{+z}^{(2\theta(z)-1)}=
\left(
\begin{array}{cc}
 \mathfrak{S}_{0}(\mathfrak{S}_{z})^{\dagger} &  0  \\
 0&     (\mathfrak{S}_{z})^{\dagger}\mathfrak{S}_{0}   
\end{array}
\right)\,,\qquad
\mathcal{G}_{-z}^{(2\theta(z)-1)}=(\mathcal{G}_{+z}^{(2\theta(z)-1)})^\dagger\,.
\qquad 
\ee 
Following  definition (\ref{frakS}), one finds  directly that 
 $\mathfrak{S}_{0}(\mathfrak{S}_{z})^{\dagger}$ is equal to $\mathfrak{A}_{|z|}^-$ when $-N<z<0$, 
 while  for $0<z\leq n_+$, this operator is equal to $\mathfrak{A}_{z}^+$,  
and takes the form of  $\mathfrak{C}_{z}^+$ for $n_+<z<2N$.
The operators $(\mathfrak{S}_{z})^{\dagger}\mathfrak{S}_{0}$
reduce to
\be
\label{S*S+}
(\mathfrak{S}_{z})^{\dagger}\mathfrak{S}_{0}= \left\{ \begin{array}{lcc}
              P_{n_-}(L-2k)(a^-)^{|z|}\,, & -N<z<0\,,\\\vspace{-0.4cm}
             \\(a^+)^{z}P_{n_-}(L)\,,   &  0<z\leq n_+\,,\\\vspace{-0.4cm}
             \\ (-1)^{n_-}(a^+)^zT_{N-z}(L+2N)\,,  &  n_+<z<N\,,\\\vspace{-0.4cm}
             \\ (-1)^{n_-}(a^+)^z \,,  & N\leq z<2N\,.
             \end{array}
   \right.\\
\ee
Note that $\mathcal{G}_{\pm k}^{(-1)}$ and $\mathcal{G}_{\pm k}^{(+1)}$ with $k=|z|\leq n_-$ 
are two different matrix extensions of  the same operator $\mathfrak{A}_k^\pm$.

For a super-extended system based on  the positive scheme, we obtain     
\be 
\label{Q0'Qk'}
\{\mathcal{Q}_a^{'0},\mathcal{Q}_{b}^{'z}\}=
\delta_{ab}(\mathcal{G}^{'(2\theta(z)-1)}_{-z}+\mathcal{G}^{'(2\theta(z)-1)}_{+z})
-i\epsilon_{ab}(\mathcal{G}^{'(2\theta(z)-1)}_{-z}- \mathcal{G}^{'(2\theta(z)-1)}_{+z})\,,
\ee
where, again,  $z\in (-N,0)\cup(0,2N)$, and  $\mathcal{G}^{'(2\theta(z)-1)}_{\pm z}$
are given by  
\be
\label{superC'}
\mathcal{G}^{'(2\theta(z)-1)}_{-z}=
\left(
\begin{array}{cc}
 \mathfrak{S'}_{0}(\mathfrak{S'}_{z})^{\dagger} &  0  \\
 0&     (\mathfrak{S'}_{z})^{\dagger}\mathfrak{S'}_{0}   
\end{array}
\right)\,,\qquad
\mathcal{G}^{'(2\theta(z)-1)}_{+z}=(\mathcal{G}^{'(2\theta(z)-1)}_{-z})^\dagger\,.
\qquad 
\ee 
Now,  
$\mathfrak{S'}_{0}(\mathfrak{S'}_{z})^{\dagger}=\mathfrak{B}^{+}_{|z|}$ when $-N<z<0$,
while for positive index $z$ this operator reduces to  
$\mathfrak{B}^{-}_{z}$ 
 when $0<z\leq n_-,$ and to $\mathfrak{C}^{-}_{z}$
 when $n_-<z<2N$. For the other matrix element we have
 \begin{eqnarray}
(\mathfrak{S}_{z}')^{\dagger}\mathfrak{S}_{0}'= \left\{ \begin{array}{lcc}
(a^+)^{|z|}P_{n_+}(L)\,, & -N<z<0\,,\\\vspace{-0.4cm}
              \\(a^-)^{z}P_{n_+}(L)\,, &  0<z\leq n_-\,,\\\vspace{-0.4cm}
             \\ (-1)^{n_-}T_{N-k}(L)(a^-)^z\,, & n_-<z<N\,,\\\vspace{-0.4cm}
             \\ (-1)^{n_-}(a^-)^z\,, & N<z<2N\,.
             \end{array}
   \right. 
 \end{eqnarray} 
Here, again, there are two different matrix extensions of the 
operators of the $\mathfrak{B}$-family given by 
$\mathcal{G}_{\pm k}^{'(+1)}$ and $\mathcal{G}_{\pm k}^{'(-1)}$ when $k\leq n_-$. 

By comparing both schemes 
one can note two other special features.
It turns out  that $\mathcal{G}_{\pm k}^{(1)}=\mathcal{G}_{\pm k}^{'(1)}$ when 
$k\geq N$, and this corresponds to  the automorphism discussed 
in section~\ref{SecOSP22Conformal}. In the same way,
for $\max(n_-,n_+)<k<N$, operators $\mathcal{G}_{\pm k}^{(1)}$ and $\mathcal{G}^{'(1)}_{\pm k}$
are different matrix extensions of  $\mathfrak{C}^{\pm}_k$. 

{}From here and in what follows 
we do not specify whether we have the super-extended system 
corresponding to the negative or the positive scheme, and will just use, respectively, 
 the unprimed or primed 
notations for operators   of the alternative dual schemes. 
In particular, we have
\begin{equation}
[\mathcal{H},\mathcal{G}_{\pm k}^{(2\theta(z)-1)}]=\pm 2k \mathcal{G}_{\pm k}^{(2\theta(z)-1)}
\,,\qquad k\equiv|z|\,, \qquad z \in (-N,0)\cup(0,2N)\,,
\end{equation} 
that shows explicitly that our new bosonic operators have the nature of  ladder operators 
of the super-extended system $\mathcal{H}$.  
Commutators $[\mathcal{G}_{-k}^{(1)},\mathcal{G}_{+k}^{(1)}]$ and 
$[\mathcal{G}_{-k}^{(-1)},\mathcal{G}_{+k}^{(-1)}]$ 
produce polynomials in $\mathcal{H}$ and $\sigma_3$,
 which can be calculated by using the polynomials  
$\mathcal{P}_{\rho,j}$ defined in  (\ref{sl2rh}). 
The algebra generated by  $\mathcal{H}$, $\mathcal{G}_{\pm k}^{(2\theta(z)-1)}$
and $\sigma_3$ is  identified 
as  a deformation of $\mathfrak{sl}(2,\R)\oplus \mathfrak{u}(1)$,  
where a concrete form of deformation depends on 
the system, $\mathcal{H}$,  and  on $z$.
Each of these nonlinear 
bosonic algebras expands further  up to  
a certain closed nonlinear deformation of superconformal  $\mathfrak{osp}(2|2)$ algebra
generated by the  subset
of operators
\begin{equation}
\label{U1}
\mathcal{U}_{0,z}^{(2\theta(z)-1)}\equiv \{\mathcal{H},\sigma_3,
\mathbb{I},\mathcal{G}_{\pm |z|}^{(2\theta(z)-1)}, \mathcal{Q}_{a}^{0},
\mathcal{Q}_{a}^{z}\}\,,\qquad z \in (-N,0)\cup(0,2N)\,,
\end{equation} 
see Appendix \ref{apen-comm}.

The deficiency of  any of these nonlinear superalgebras 
is that none of them is a spectrum-generating algebra for the
super-extended system\,: application 
of operators from the set (\ref{U1}) and of their products 
does not allow one to connect two arbitrary eigenstates in the spectrum 
of $\mathcal{H}$. 
To find the spectrum-generating superalgebra  
for this kind of the super-extended systems, one can try 
 to include into the superalgebra simultaneously 
 the operators  $\mathcal{G}^{(1)}_{\pm N}$
 and, say, $\mathcal{G}^{(1)}_{\pm 1}$ or $\mathcal{G}^{(-1)}_{\pm 1}$.
 The operators  $\mathcal{G}^{(1)}_{\pm N}$ 
 provide us with matrix extension of the operators $\mathfrak{C}_{N}^\pm$
 being ladder operators for deformed subsystems $L_{(-)}$ or  $L_{(+)}$.
 Analogously, operators $\mathcal{G}^{(1)}_{\pm 1}$ or $\mathcal{G}^{(-1)}_{\pm 1}$
 supply us with matrix extensions  of the ladder operators 
$\mathfrak{A}_{ 1}^\pm$ or $\mathfrak{B}_{ 1}^\pm$ 
($\mathfrak{A}_{ 2}^\pm$ or $\mathfrak{B}_{ 2}^\pm$)
when systems $L_{(\pm)}$ are of the class $c=1$ or $c=2$ with even (odd) $N$.
Therefore, it is enough to unify  the sets of generators 
  $\mathcal{U}_{0,1}^{(1)}$ and  $\mathcal{U}_{0,N}^{(1)}$. 
 Having in mind the commutation relations 
 between operators of the three families  
$\mathfrak{A}$, $\mathfrak{B}$ and $\mathfrak{C}$, one can find, however, 
that the commutators of  the operators 
 $\mathcal{G}^{(1)}_{\pm N}$  with $\mathcal{G}^{(1)}_{\pm 1}$
 generate other bosonic matrix operators  $\mathcal{G}^{(1)}_{\pm k}$.
 The commutation of these operators with supercharges  $\mathcal{Q}_{a}^{0}$
 generates the rest of the fermionic operators we considered, see
 Appendix \ref{apen-comm}  for details.
The set of higher order generators 
is completed by considering all non-reducible bosonic and fermionic generators,
which do not decompose into the products of other generators.
In correspondence with that was noted above, we 
 arrive finally at two different extensions of the sets of 
operators with index less than $N$. By this reason it is convenient also 
to  introduce   the operators 
\begin{eqnarray}
\label{G_k}
&\mathcal{G}_{\pm k}^{(0)}\equiv \Pi_- (a^\pm)^k, \qquad
 k=1,\ldots,N-1, \qquad 
 \Pi_-=\frac{1}{2}(1-\sigma_3)\,,&
\end{eqnarray} 
which help us to fix in a unique way the bosonic set of generators. For our purposes we choose 
to write all the operators $\mathcal{G}_{\pm k}^{(-1)}$ in terms of $\mathcal{G}_{\pm k}^{(1)}$ and 
$\mathcal{G}_{\pm k}^{(0)}$ when $k\leq n_+$ in the negative scheme, 
and when $k\leq n_-$ in the extended system associated with the positive scheme.
For indexes outside the indicated scheme-dependent range, we neglect operators $\mathcal{G}_{\pm k}^{(-1)}$ 
 because they are not basic in correspondence with the discussion on reduction of ladder operators in the 
 previous  Sec. \ref{interladder}. 
As a result, we have to drop from  (\ref{U1}) all the operators $\mathcal{G}_{\pm |z|}^{(2\theta(z)-1)}$ with 
$z\in (-N,0)$.

By taking  anti-commutators  of fermionic operators $\mathcal{Q}_{a}^{N}$ 
with $\mathcal{Q}_{a}^{z}$, $z\neq 0$, we produce bosonic
dynamical integrals $\mathcal{J}_{\pm |z-N|}^{(1-2\theta(z-N))}$,
which  have exactly the same 
structure of  the even generators 
$\mathcal{G}_{\pm |z|}'^{(2\theta(z)-1)}$
in the extension associated with  the dual scheme.
In this way we obtain the subsets of 
operators
\begin{eqnarray}
\label{In11}
\mathcal{I}_{N,z}^{(1-2\theta(z-N))}\equiv \{\mathcal{H},\sigma_3,\mathbb{I},
\mathcal{J}_{\pm |z-N|}^{(1-2\theta(z-N))}, \mathcal{Q}_{a}^{N},\mathcal{Q}_{a}^{z}\}\,
\qquad z \in (-N,0)\cup(0,2N)\,,
\end{eqnarray}
which also generate closed nonlinear super-algerabraic structures. 
With the help of  (\ref{G_k}), we find similarly to the subsets (\ref{U1}),
that a part of the sets  (\ref{In11}) also can be reduced. 

Having  in mind the ordering relation between $n_-$ and $n_+$, 
the super-extended systems associated with 
the negative schemes 
can be   characterized finally by the following irreducible, 
in the sense of subection \ref{interladder},
subsets of symmetry generators\,:
\begin{table}[H]
\begin{center}
\begin{tabular}{|c| c|}
\hline
$n_-\leq n_+$   &$n_+<n_-$ \\
\hline
$  \mathcal{U}_{0,k}^{(1)}\,,\qquad 0<k<2N$ &  $\mathcal{U}_{0,k}^{(1)}\,,\qquad  k\in (0,n_+)\cup (n_-,2N) $\\
$   \mathcal{I}_{N,z}^{(1-2\theta(N-z))}\,,\qquad  z\in(-N,0)\cup(n_+,N)$ &$ \mathcal{I}_{N,z}^{(1-2\theta(N-z))}\,,\qquad  z\in(-N,0)\cup[n_+,N)$ \\
\hline
\end{tabular}
\caption{Symmetry generators subset.}
\label{Tabla 1}
\end{center}
\end{table}
\noindent

For more details, 
see  Appendix \ref{apen-red}. 
A similar result can be obtained for super-extended systems
associated with   positive schemes, where
the roles played by  families $\mathfrak{A}$ and $\mathfrak{B}$, 
and of numbers $n_-$ and $n_+$
are interchanged.

Finally, we arrive  at the following picture.
Any operator that can be generated via (anti)-commutation relations and which does 
not belong to the sub-sets appearing in Table \ref{Tabla 1}, can be written as a product of the basic generators.
For super-extensions of rationally deformed one-gap harmonic oscillator systems
we have considered, the spectrum-generating algebra is composed from  
the sets
 $\mathcal{U}_{0,k}^{(1)}$ and 
$\mathcal{I}_{N,z}^{(1-2\theta(N-z))}$
and from those operators generated by them via (anti)-commutation relations
which cannot be written as a product of the basic generators.
It is worth to stress that in this set of generators the unique 
true integrals of motion, in addition to $\mathcal{H}$ and $\sigma_3$,
 are the supercharges $\mathcal{Q}_a^{0}$, while the rest  has to
 be promoted to the dynamical integrals by unitary transforming them with the evolution operator.

For gapless rational extensions of the systems of class $c=2$, 
only the subset $\mathcal{U}_{0,2}^{(1)}$ has to be considered
instead of the family of sets $\mathcal{U}_{0,k}^{(1)}$. 
For super-extensions of rationally deformed systems of arbitrary form in the sense of the class $c$ 
and arbitrary number of gaps and their dimensions,  
the identification of their generalized super-Schr\"odinger or superconformal 
structures  is realized in a similar way. The procedure is based
on  the sets of operators (\ref{ladgen}) and 
(\ref{genA}), which include  the operators 
 (\ref{basicsubsetonegap})  and  (\ref{frakS})
 of the discussed one-gap case as subsets.
 As a result, 
for every irreducible pair of ladder operators (\ref{ladgen}) with index less than $N$
we have two super-extensions which are related  by operators of the form (\ref{G_k}). When we 
put together the 
subsets containing the spectrum-generating set of operators, 
we obtain all the other structures.

We would like to end this section highlighting 
some of the peculiarities of the simplest systems that can be treated with this machinery and these are

\emph{Peculiarities of one-gap deformations of the QHO}\,: 
The super-extended Hamiltonian constructed on the
base of the negative scheme with $n_-=1$ 
is characterized by unbroken  
$\mathcal{N}=2$ 
Poincar\'e supersymmetry, whose 
supercharges, being  the first 
order differential operators, generate
a  Lie superalgebra.  
The $\mathfrak{B}$ family of ladder operators in the sense of (\ref{basicsubsetonegap}) 
does not play any 
role in this scheme. 
On the other hand, the super-Hamiltonian provided by the positive scheme
possesses  $n_+$ 
singlet states while the ground state is a doublet. 
The $\mathcal{N}=2$ super-Poincar\'e 
algebra of such a system  is nonlinear
as its supercharges are of differential order 
$n_+=2\ell \geq 2$.  
\vskip0.1cm

\emph{Peculiarities of gapless deformations of $L_1$}\,:
The negative scheme produces a super-Hamiltonian with spontaneously broken 
supersymmetry, whose all energy levels are doubly degenerate;
its $\mathcal{N}=2$  super-Poincar\'e algebra  has linear nature. 
To construct the spectrum-generating algebra we only need a matrix extension
of the operators $\mathfrak{A}_2^\pm$. In a super-extended system produced by
the positive scheme,  $n_+>1$ physical and nonphysical states of $L_{0}$ of positive energy
(the latter being even eigenstates of harmonic oscillator)
are used as seed states for DCKA transformation.
Its supersymmetry is spontaneously broken, 
and the $\mathcal{N}=2$ super-Poincar\'e algebra is nonlinear. 
The nonlinearly deformed super-Poincar\'e symmetry cannot be expanded 
 to spectrum-generating superalgebra by combining it with matrix 
extension  of  the $\mathfrak{A}^\pm_2$, 
but this can be done by using matrix extensions of the 
$\mathfrak{B}_2^\pm$ or $\mathfrak{C}_2^\pm$ ladder operators,
see  (\ref{superC'}). 
The resulting  spectrum-generating superalgebra is a certain nonlinear 
deformation 
of the  $\mathfrak{osp}(2|2)$ superconformal symmetry.     
\vskip0.1cm
 

\section{Example 1: Gapless deformation of AFF model}
The example considered here corresponds to the same system 
analyzed in the previous chapter, in Sec. \ref{SecIsocase}. 
By construction, the super-Hamiltonian and its spectrum 
correspond to 
\begin{equation}
\label{hamilisodef}
\mathcal{H}=
\left(
\begin{array}{cc}
 H_1&    0 \\
0 &  H_0    
\end{array}\right),\qquad \mathcal{E}_{n}=4n+10\,,
\qquad n=0,1,\ldots\,,
\end{equation}
where $H_1=L_{(-)}+7$, with $L_{(-)}$ given in (\ref{reio3}), 
and $H_0=L_{0}+7$.
Due to complete isospectrality  of $ H_1$ and  $H_0$, all the energy levels
of the system (\ref{hamilisodef}) 
including the  lowest one $\mathcal{E}_{0}=10>0$ 
are  doubly  degenerate and we have here the case 
of spontaneously broken $\mathcal{N}=2$ super-Poincar\'e symmetry 
generated by Hamiltonian $\mathcal{H}$, the supercharges $\mathcal{Q}^0_a$
constructed in terms of $A_{(-)}^\pm$,
and by  $\Sigma=\frac{1}{2}\sigma_3$.

The generators that should be considered for the super-extension correspond to 
\be
\mathcal{U}_{0,2}^{(1)}=\{\mathcal{H},\mathbb{I},\mathcal{G}_{\pm2}^{(1)},\sigma_3,\mathcal{Q}_a^{0},\mathcal{Q}_a^{2} \}\,,
\ee where 
\begin{eqnarray}
&
\label{Q2Cpm2}
\mathcal{Q}^z_1=
\left(
\begin{array}{cc}
 0&    A^-_{(-)}(a^-)^z \\
(a^+)^zA^+_{(-)} &  0    
\end{array}\right),\,\, z=0,2\,, &\\&
\mathcal{G}_{-2}^{(1)}=
\left(
\begin{array}{cc}
 A_{(-)}^-(a^-)^2A^+_{(-)}&   0 \\
0 &  H_0(a^-)^2   
\end{array}\right),&\\&
\mathcal{Q}^z_2=i\sigma_3 \mathcal{Q}^z_1\,,\qquad\mathcal{G}_{+2}^{(1)}=(\mathcal{G}_{-2}^{(1)})^\dagger\,,
&
\end{eqnarray}
and the explisit form of $A_{(-)}^\pm$ is given in (\ref{A+-(-3)}).
The  complete set of superalgebraic relations they satisfy is 
\begin{eqnarray}
\label{nonlinear3}
&[\mathcal{H},\mathcal{Q}_a^{0}]=0\,,\qquad 
[\mathcal{H},\mathcal{Q}_a^{2}]=4i\epsilon_{ab}\mathcal{Q}_b^{2}\,,\qquad
[\sigma_3,\mathcal{Q}_a^{z}]=-2i\epsilon_{ab}\mathcal{Q}_b^{z}\,,\quad z=0,2\,,&\\
\label{nonlinear2}
&\{\mathcal{Q}_a^{0},\mathcal{Q}_a^{0}\}=2\delta_{ab}\mathcal{H}\,,\qquad
\{\mathcal{Q}_a^{0},\mathcal{Q}_b^{2}\}=\delta_{ab}(\mathcal{G}_{-2}^{(1)}+\mathcal{G}_{+2}^{(1)})+i
\epsilon_{ab}(\mathcal{G}_{-2}^{(1)}-\mathcal{G}_{+2}^{(1)})\,,&\\
\label{nonlinear1}
&[\mathcal{H},\mathcal{G}_{\pm 2}^{(1)}]=\pm4\mathcal{G}_{\pm2}^{(1)}\,,\qquad 
[\mathcal{G}_{\mp 2}^{(1)},\mathcal{Q}_a^{0}]=\pm 2(\mathcal{Q}_a^{2}\mp i\epsilon_{ab}\mathcal{Q}_b^{2})\,,&\\
\label{nonlinear4}
&[\mathcal{G}_{-2}^{(1)},\mathcal{G}_{+2}^{(1)}]=8(\mathcal{H}-4)(\mathcal{H}(2\mathcal{H}-9)+\Pi_-
(\mathcal{H}^2-4\mathcal{H}+24))\,,&\\
\label{nonlinear5}
&[\mathcal{G}_{\mp 2}^{(1)},\mathcal{Q}_a^{2}]= \pm 2(-80 + 4 \mathcal{H} + \mathcal{H}^2)(\mathcal{Q}_a^{0}\pm
i\epsilon_{ab}\mathcal{Q}_b^{0})\,,&\\
\label{nonlinear7}
&\{\mathcal{Q}_a^{2},\mathcal{Q}_b^{2}\}=2\delta_{ab}
(\eta+1)(\eta+3)(\eta+7)|_{\eta=\mathcal{H}+2\sigma_3-9}\,,&
\end{eqnarray}
where $\Pi_-=\frac{1}{2}(1-\sigma_3)$.   
The common eigenstates of  $\mathcal{H}$ and $\mathcal{Q}^0_1$ are  
\be
\label{states-3}
\Psi_{n}^{+}=
\left(
\begin{array}{c}
(\mathcal{E}_n)^{-1/2} A_{(-)}^-\psi_{2n+1}   \\
\psi_{2n+1}
\end{array}
\right),
\qquad 
\Psi_{n}^{-}=\sigma_3\Psi_{n}^+,
\ee
where $\mathcal{Q}^0_1\Psi^\pm_n=\pm\sqrt{\mathcal{E}_n}\Psi^\pm_n$,
and we have here the relations 
$\Psi_{n}^\pm=(\mathcal{G}_{+2}^{(1)})^n\Psi_0^{\pm}$ and $\mathcal{G}_{-2}^{(1)}\Psi_0^{\pm}=0$.
As a result one can generate all the complete set of eigenstates of the system 
by applying the generators of superalgebra to any of the two ground states 
$\Psi_0^+$ or $\Psi_0^-$, and therefore the restricted set of
generators we have chosen is the complete spectrum-generating 
set for the super-extended system (\ref{hamilisodef}).

\vskip0.1cm
 
The complete set of (anti)-commutation relations 
 (\ref{nonlinear1})-(\ref{nonlinear7}) corresponds to   a nonlinear deformation of superconformal 
algebra $\mathfrak{osp}(2|2)$. 
The first relation from   (\ref{nonlinear1})
and equation (\ref{nonlinear4})  represent
 a nonlinear  deformation of $\mathfrak{sl}(2,\mathbb{R})$ 
with commutator $[\mathcal{G}_{-2}^{(1)},\mathcal{G}_{+2}^{(1)}]$
 to be a cubic polynomial in $\mathcal{H}$.
{}From the superalgebraic relations it follows that like in the linear case 
of superconformal $\mathfrak{osp}(2|2)$ symmetry 
discussed in   Chap \ref{ChConformal}, Sec. \ref{SecOSP22Conformal},
here the extension of the 
set of generators $\mathcal{H}$, $\mathcal{Q}^0_a$ and $\Sigma$ of 
the $\mathcal{N}=2$ Poincar\'e super-symmetry 
by any one of the dynamical integrals $\mathcal{Q}^2_a$, $a=1,2$,
$\mathcal{G}_{+2}^{(1)}$ or $\mathcal{G}_{-2}^{(1)}$ recovers 
all the complete set of generators of the nonlinearly deformed 
superconformal $\mathfrak{osp}(2|2)$ symmetry.
\vskip0.2cm

Due to a gapless deformation of the AFF model,
here similarly to the case of the non-deformed superconformal $\mathfrak{osp}(2|2)$ symmetry, 
the super-extension
based on the positive scheme 
is characterized by essentially different physical properties.
The positive scheme of the system corresponds to the states$(1,2,3)$ 
and in this case we identify 
$\mathcal{H}'=\text{diag}\,(L_{(+)}-3,L_{0}-3)$
as the extended Hamiltonian. 
This $\mathcal{H}'$  is related to 
$\mathcal{H}$ defined by Eq. (\ref{hamilisodef})
by the equality $\mathcal{H}'=\mathcal{H}-6+4\sigma_3$.
For extended system $\mathcal{H}'$,
supercharges ${\mathcal{Q}'}_a^{0}$ 
have the form  similar to  $\mathcal{Q}_a^{0}$  in (\ref{Q2Cpm2}) 
but  with  $A^\pm_{(-)}$ changed for the third order  intertwining operators
$A^\pm_{(+)}$, constructed with the formula (\ref{generic-inter}). 
Being differential operators of the third order, 
they satisfy relations  
$[\mathcal{H}', {\mathcal{Q}'}_a^{0}]=0$  and 
$\{{\mathcal{Q}'}_a^{0},{\mathcal{Q}'}_b^{0}\}=2\delta_{ab}P_{n_+}(\mathcal{H}'+3)$
with $P_{n_+}(\mathcal{H}'+3)=\mathcal{H}'(\mathcal{H}'-2)(\mathcal{H}'-4)$.
The linear $\mathcal{N}=2$ super-Poincar\'e algebra
of the  system (\ref{hamilisodef}) is changed here for the nonlinearly 
deformed superalgebra with anti-commutator to be  polynomial 
of the third order in Hamiltonian. This system has two nondegenerate states
$(0,\psi_{1})^t$ and $(0,\psi_{3})^t$ of energies, respectively, $0$ and $4$,
and both them are annihilated by both supercharges  ${\mathcal{Q}'}_a^{0}$.
All higher energy levels $\mathcal{E}'_n=4n$ with $n=2,3,\ldots$
are doubly degenerate.
Thus, the nonlinearly deformed $\mathcal{N}=2$ super-Poincar\'e 
symmetry of this system can be identified as partially unbroken \textcolor{red}{[\cite{KliPly}]}
since the supercharges have differential order three but 
annihilate only two nondegenerate physical states. 
Here instead of the spectrum-generating set $\mathcal{U}_{0,2}^{(1)}$, formed by   true
and dynamical integrals, the same role is played by the set
of integrals 
$\mathcal{U}_{0,2}'^{(1)}=\{\mathcal{H}',\mathcal{G}'^{(1)}_{\pm 2}, \mathbb{I},\sigma_3, {\mathcal{Q}'}_a^{0},
{\mathcal{Q}'}_a^{2}\}$, where fermionic generators are 
${\mathcal{Q}'}_a^{z}=\mathcal{Q}_a^{4-z}$ with $z=0,2$ according with (\ref{frakS}) and (\ref{gencharge}).
Bosonic dynamical integrals  $\mathcal{G}'^{(1)}_{\pm 2}$ are given here by
\begin{eqnarray}
\label{Cpm2Hprime}
\mathcal{G}_{-2}'^{(1)}=
\left(
\begin{array}{cc}
 A_{(+)}^-(a^+)A^+_{(-)}&   0 \\
0 &  (L_{0}-1)(a^-)^2   
\end{array}\right),\qquad
\mathcal{G}_{+2}'^{(1)}=(\mathcal{G}_{-2}'^{(1)})^\dagger\,,
\end{eqnarray}
where equations in (\ref{superC'}) have been used for the case
of the present positive scheme. They are generated via anticommutation of
${\mathcal{Q}'}^0_a$ with ${\mathcal{Q}'}^2_b$.
The set of operators $\mathcal{U}_{0,2}'^{(1)}$ generates the nonlinearly deformed 
superconformal $\mathfrak{osp}(2|2)$ symmetry 
given by superalgebra of the form 
(\ref{nonlinear3})--(\ref{nonlinear7}),
but with coefficients to be polynomials of higher order in Hamiltonian
$\mathcal{H}'$
in comparison with the case of the system (\ref{hamilisodef}).

\section{Example 2: Rationally extended harmonic oscillator}\label{SecDefQHO}

The example we discuss in this subsection corresponds to the rational  extension  
of QHO based on the dual schemes $(1,2)\sim (-2)$,
for which  $N=3$. 
Different aspects of this system were extensively studied in  literature 
\textcolor{red}{[\cite{CarPly,CarInzPly}]}. Here, we investigate it in the 
light of the nonlinearly extended super-Schr\"odingerr symmetry.  

 The Hamiltonian produced via Darboux transformation based on the 
 negative scheme is
\begin{equation}
\label{H_-2}
L_{(-)}=-\frac{d^2}{dx^2}+x^2+8\frac{2 x^2-1 }{(1 + 2 x^2)^2}-2\,,
\end{equation}
whose spectrum is
$E_0=-5$, $E_{n+1}=2n+1$, $n=0,1,\ldots$.
In this system a gap of size 6 
separates the ground state energy  from the equidistant part of the spectrum,
 where levels are separated from each other by a distance $\Delta E=2$.
 The pair of ladder operators  of the $\mathfrak{C}$-family  
connects  here the isolated ground state  with the equidistant part of the spectrum,
and together with the ladder operators  $\mathfrak{A}^\pm_1$ they form 
the complete spectrum-generating set of operators for the system.
The  intertwining operators of the negative scheme are
\begin{equation}
A_{(-)}^-=\frac{d}{dx}-x-\frac{4x}{2x^2+1},\qquad
 A_{(-)}^+\equiv (A_{(-)}^{-})^{\dagger}\,.
\end{equation}
We also have the  intertwining operators $A_{(+)}^\pm$ 
constructed on the base of the  seed states of the positive scheme $(1,2)$. 
These four operators satisfy  their respective intertwining relations 
of the form (\ref{inter0}), and
 their alternate products (\ref{polyA})  reduce here to polynomials 
 $P_{n_-}(L_{(-)})=L_{(-)}+5\equiv  H_1$, $P_{n_-}(L)=L+5\equiv H_0$ and
$P_{n_+}(L_{(+)})=(L_{(+)}-3)(L_{(+)}-5)$, $P_{n_+}(L)=(L+3)(L+5)$, where $L=L_\text{os}$
is the Hamiltonian operator of the harmonic oscillator, and 
$L_{(+)}$ is the Hamiltonian 
produced by positive  scheme, which is related with  $L_{(-)}$, according to (\ref{L+L-}),
by $L_{(+)}-L_{(-)}=6$. 
Here, the eigenstate $A_{(-)}^{-}\widetilde{\psi_{-2}}=1/\psi_{-2}$  is the isolated  
ground state   of zero energy of the shifted Hamiltonian operator $H_1$.

The super-extended  Hamiltonian and its spectrum are 
\begin{equation}
\label{superdefHO}
\mathcal{H}=
\left(
\begin{array}{cc}
 H_1&    0 \\
0 &  H_0  
\end{array}\right),\qquad \mathcal{E}_{0}=0\,, \qquad \mathcal{E}_{n+1}=2n+6\,,\qquad n=0,1,\ldots\,.
\end{equation}
The ground state of zero energy is non-degenerate and corresponds to the ground state 
$(A_{(-2)}^- \widetilde{\psi_{-2}},0)^t$. Other energy levels 
are doubly degenerate
and correspond to eigenstates of the extended Hamiltonian (\ref{superdefHO}) 
and supercharge $\mathcal{Q}^0_1$, see below\,:
\begin{equation}
\Psi_{n+1}^{+}=
\left(
\begin{array}{c}
  (\mathcal{E}_{n+1})^{-1/2} A_{(-)}^-\psi_{n}   \\
 \psi_{n}
\end{array}
\right),\qquad
\Psi_{n+1}^{-}=\sigma_3\Psi_{n+1}^{+}\,.
\end{equation} 
The system (\ref{superdefHO}) 
is characterized by unbroken $\mathcal{N}=2$ Poincar\'e supersymmetry.
Now  we use the construction of Sec. \ref{susyextension} to produce generators
of the  extended nonlinearly deformed super-Schr\"odinger symmetry 
of the system. 
Following (\ref{gencharge}) and (\ref{superC}), 
we construct the odd operators 
$\mathcal{Q}_{a}^{z}$ with $z=-2,-1,0,\ldots,5$,  and 
matrix bosonic ladder operators $\mathcal{G}_{\pm k}^{(1)}$ with $k=1,\ldots,5$.  
Also we must consider the operators $\mathcal{G}_{\pm k}^{(0)}$ with $k=1,2$ 
defined in (\ref{G_k}).
To obtain all the
 ingredients, we have to use the version of relation  (\ref{reqgen1}) for this system 
translated to the supersymmetric extension of $\mathfrak{C}_{N+k}^\pm$ which is
\begin{equation}
\label{req}
\mathcal{G}_{\pm(3l+n)}^{(1)}=(-\mathcal{G}_{\pm 3}^{(1)})^l\mathcal{G}_{\pm n}^{(1)} \,,\qquad n=3,4,5\,,\qquad l=0,1,\ldots\,.
\end{equation} 
Then we generate the even part of the superalgebra\,: 
\begin{equation}
\label{slr1}
[\mathcal{H},\mathcal{G}_{\pm n}^{(1)}]=\pm 2n \mathcal{G}_{\pm n}^{(1)}\,,
\qquad
[\mathcal{H},\mathcal{G}_{\pm l}^{(0)}]=\pm 2l\mathcal{G}_{\pm l}^{(0)}\,,
\end{equation}
\begin{equation}
\label{slr3}
[\mathcal{G}_\alpha^{(1)},\mathcal{G}_\beta^{(1)}]=P_{\alpha,\beta} \mathcal{G}_{\alpha+\beta}^{(1)}+
M_{\alpha,\beta}\mathcal{G}_{\alpha+\beta}^{(0)}\,,
\quad 
\alpha,\beta=\pm1,\ldots,\pm5\,,
\end{equation}
\begin{equation}
\label{slr4}
[\mathcal{G}_\alpha^{(0)},\mathcal{G}_\beta^{(1)}]=\Pi_-(F_{\alpha,\beta}\mathcal{G}_{\alpha+\beta}^{(1)}+
N_{\alpha,\beta}\mathcal{G}_{\alpha+\beta}^{(0)})\,,
\quad 
\alpha=1,2\,,\quad\beta=\pm1,\ldots,\pm5\,,
\end{equation}
\begin{equation}
[\mathcal{G}_{-1}^{(0)},\mathcal{G}_{+1}^{(0)}]=2\Pi_-,\qquad [\mathcal{G}_{\pm 1}^{(0)},\mathcal{G}_{\mp 2}^{(0)}]=
\pm6\mathcal{G}_{\pm 1}^{(0)}\,,
\qquad 
[\mathcal{G}_{-2}^{(0)},\mathcal{G}_{+2}^{(0)}]=8\Pi_-(\mathcal{H}-5)\,,
\end{equation}
where we put $\mathcal{G}_0^{(1)}=\mathcal{G}_0^{(0)}=1$ and $P_{\alpha,\beta}$, $F_{\alpha,\beta}$, 
$M_{\alpha,\beta}$ and $N_{\alpha,\beta}$ are some polynomials in $\mathcal{H}$ and $\Pi_-=\frac{1}{2}(1-\sigma_3)$,
some of which are numerical coefficients,  whose explicit form is listed 
in Appendix \ref{list}.
We note that 
in Eqs. (\ref{slr3}) and (\ref{slr4}), the  operators
 $\mathcal{G}_{\pm n}^{(1)}$ with $1<n\leq 7$ can appear, 
where for $n>5$ we use relation (\ref{req}) (admitting  $\mathcal{G}_{\pm3}^{(0)}$ 
as coefficients  in the algebra).  
Additionally we note that  the operators 
$\mathcal{G}_{\pm m}^{(0)}$ with $m>2$ in both equations where they appear
are
 absorbed in generators $\mathcal{G}_{\pm m}^{(1)}$.

For eigenstates we have the relations 
 \begin{eqnarray}
 \label{C3psi}
&\Psi_{3j+k}^\pm= (\mathcal{G}_{+3}^{(1)})^j\Psi_k^\pm\,,
\qquad
\Psi_{0}=\mathcal{G}_{-3}^{(1)}\Psi_1^\pm,\qquad j=1,2,\ldots\,,\qquad k=1,2,3\,,&\\
 \label{C2psi}
&\Psi_{j}^\pm= (\mathcal{G}_{+ 1}^{(1)})^j\Psi_1^\pm\,,
\qquad
\mathcal{G}^{(1)}_{\pm 1}\Psi_{0}=\mathcal{G}_{-1}^{(1)}\Psi_{1}^\pm=0\,.&
\end{eqnarray}
Eq.  (\ref{C3psi}) shows that we can connect the
isolated  ground state with the equidistant part of the 
spectrum using $\mathcal{G}_{\pm 3}^{(1)}$, which  are not spectrum-generating operators. 
Eq.  (\ref{C2psi}) indicates  that the states in the  equidistant part of 
the spectrum  can be connected by  $\mathcal{G}_{\pm 1}^{(1)}$, but this part
of the spectrum cannot be connected by them  with  
the ground state. Thus 
we have to use a combination of both pairs of these operators. 
On the other hand, the odd operators $\mathcal{Q}_a^z$ satisfy relations (\ref{SUSY}),
where $\mathbb{P}_0=\mathcal{H}$, and, therefore, 
 we have again the linear $\mathcal{N}=2$ Poincar\'e  supersymmetry
as a sub-superalgebra  generated by $\mathcal{H}$, $\mathcal{Q}^0_a$ and $\Sigma$.
The general anti-commutation structure is given by 
\begin{equation}
\label{susy3}
\{\mathcal{Q}_a^{n},\mathcal{Q}_b^{m}\}=\delta_{ab}(\mathbb{C}_{nm}+(\mathbb{C}_{nm})^{\dagger})+
i\epsilon_{ab}(\mathbb{C}_{nm}-(\mathbb{C}_{nm})^{\dagger})\,,
\end{equation}
where $\mathbb{C}_{n,m}=\mathbb{C}_{n,m}(\mathcal{G}_{|n-m|}^{(1)},\mathcal{G}_{|n-m|}^{(0)})$
 in general are some linear combinations of the indicated 
 ladder operators with coefficients to be polynomials in 
$\mathcal{H}$, $\mathcal{G}_{\pm3}^{(0)}$ and $\sigma_3$. 
Some of these relations 
define  ladder operators, see Eq. 
(\ref{Cn+kQN}). 
For  $n=N=3$ and $m=-1,-2$  we can use 
(\ref{superC'})  knowing that $\mathcal{Q}_{a}'^{z}=\mathcal{Q}_{a}^{3-z}$,
see Sec. \ref{susyextension}. 
For structure of anti-commutation relations with 
other combinations of indexes,  see Appendix \ref{list}.  
To complete the description 
of the generated nonlinear supersymmetric  structure, 
we write down  the commutators between the independent lowering operators and 
supercharges\,:
\begin{equation}
\label{susy4}
[\mathcal{G}_{-m}^{(1)},\mathcal{Q}_a^{n}]=\mathbb{Q}_{m,n}^{1}(\mathcal{Q}_{a}^{n-m}+i
\epsilon_{ab}\mathcal{Q}_b^{n-m})+\mathbb{Q}_{m,n}^{2}(\mathcal{Q}_{a}^{m+n}-i
\epsilon_{ab}\mathcal{Q}_b^{m+n})\,,
\end{equation} 
\begin{equation}
\label{susy5}
[\mathcal{G}_{-m}^{(0)},\mathcal{Q}_a^{n}]=\mathbb{G}_{m,n}^{1}(\mathcal{Q}_{a}^{n-m}+i
\epsilon_{ab}\mathcal{Q}_b^{n-m})+\mathbb{G}_{m,n}^{2}(\mathcal{Q}_{a}^{m+n}-
i\epsilon_{ab}\mathcal{Q}_b^{m+n})\,.
\end{equation} 
Here $\mathbb{Q}_{m,n}^{j}$ and $\mathbb{G}_{m,n}^{j}$ with $j=1,2$ are 
polynomials in $\mathcal{H}$  or numerical coefficients, some of which are listed
 in  the sets of general commutation relations in  Appendix \ref{apen-comm}, 
while other are given explicitly in Appendix \ref{list}.
As the odd fermionic operators  are Hermitian, then 
$[\mathcal{G}_{+m}^{(1)},\mathcal{Q}_a^{z} ]=-([\mathcal{G}_{- m}^{(1)},
\mathcal{Q}_a^{z} ])^{\dagger}$, 
and we do not  write them explicitly. 
In matrix language, Eq. (\ref{susy4}) can be written as
\be
[\mathcal{G}_{-m}^{(1)},\mathcal{Q}_a^{n}]= 
\left(
\begin{array}{cc}
  0&  \mathfrak{S}_{n+m}^-  \\
 \mathfrak{S}_{n-m}^+ &  0     
\end{array}
\right)\,,
\ee    
and an important point here is that  the number $n-m$ could take values less than -2 and $n+m$ could 
be greater than 5,  but fermionic operators are defined with the index $z$ taking integer values
in the interval  $I=[-2,+5]$.  
It is necessary to  remember that we cut the series of $\mathfrak{S}_z^\pm$ 
because operators outside the defined interval are reduced to combinations (products) of other basic operators.
In this way, we formally apply the definition of $\mathfrak{S}_z^\pm$ outside of the indicated
 interval and 
 use the relation in Appendix \ref{apen-red} to show that these ``new'' generated operators reduce to combinations 
of operators with index values in the interval $I$  
and of the generators $\mathfrak{C}_{\pm 3}$.  

Finally, the subsets which produce closed sub-superalgebras here are those defined by $\mathcal{U}_{0,z}^{(1)}$
 in (\ref{U1}), with $z=1,\ldots,5$ in addition to $\mathcal{I}_{N,-k}^{(1)}$ given in (\ref{In11}) with $k=1,2$.
 
With respect to the positive scheme, the super-Hamiltonian is given by $\mathcal{H}'=\text{diag}\,
(L_{(+)}-3,L_0-3)$. 
It has two positive energy singlet states  of the form $(0,\psi_n)$ with $n=1,2$;  besides, 
there are  two ground states $\Psi_0^+=(\phi_0,\psi_0)$ and $\Psi_0^-=\sigma_3\Psi_0^+$ of energy $-2$.
According 
to the construction from the previous section, 
the fermionic operators here are $\mathcal{Q}^{'z}_a=\mathcal{Q}^{3-z}_a$, and
the basic subsets which generate 
closed sub-superalgebras are $\mathcal{U}_{0,k}'^{(1)}$  and $\mathcal{I}_{N,l}'^{(1-2\theta(l))}$ 
with $k=3,4,5$ and $l=-1,-2,4,5$.  

One can  note  that considering  $\mathcal{G}_{\pm 3}^{(1)}$ as coefficients, 
the subset $\{\mathcal{H}, \mathcal{G}_{\pm 3}^{(1)},\sigma_3,
 \mathcal{Q}_a^{-2},\mathcal{Q}_a^{1},\mathcal{Q}_a^{4}, \mathbb{I}  \}   $
also generates  a closed nonlinear superalgebraic structure.

\section{Remarks}
In fact, the construction in Sec. \Ref {susyextension} offers more possibilities: 
in principle, the choice of the constant $ \lambda_* $ in the Hamiltonian (\ref{Hlambda*})
 can be modified in such a way that another pair of fermionic operators in the scheme (\ref{gencharge})  
will be the true integrals of the motion. As a result, the super-extended system will have a different spectrum. 
We schematically discussed this picture in the original work \textcolor{red}{[\cite{InzPly2}]}. 
Another possibility is to choose $ L_0 = L_{(-)} $ and $ L_ {[n]} = L_{(+)} $ and,
 as a consequence, the intertwining operators will be the ladder operators in (\ref {ladgen}),
 and one can expect that the use of intermediate systems in the DCKA procedure will 
provide lower order intertwining operators, however this is still an open problem.

Finally, the discussion in these last two chapters involved AFF 
models with integer coupling constant $ m (m + 1) $, 
so the next natural step is to try to generalize for the case $ \nu (\nu + 1) $ with $ \nu $ 
real equal to or greater than $ -1 / 2 $. This is the objective of the next chapter.

\chapter{The Klein four-group and Darboux duality}
\label{ChKlein}

The invariance of the QHO eigenvalue problem to the discrete transformation 
$ (x, E) \rightarrow (ix, -E) $ was the basis of the construction 
presented in the last two chapters.
The presence of nonphysical eigenstates gives rise to the so-called Darboux duality, 
which was the key to building the spectrum-generating ladder operators for extended rational systems.
In this chapter we demonstrate that the Schr\"odinger equation for the AFF model 
with $\nu\geq-1/2$ 
has an even larger discrete symmetry group,
 which will be responsible for the generalization of Darboux duality for these systems. 
 Such a discrete group
has its particular consequences when it acts on eigenstates and (super) symmetry generators.

With the generalization of the Darboux duality at hand, constructing spectrum-generating 
ladder operators for rational deformations of the general AFF models,
 as well as their nonlinear algebras, is straightforward. 
It is interesting to recall that when $ \nu $ is a half-integer number, the Jordan states associated with confluent 
Darboux transformations naturally enter in the framework. 
In particular, some deformed systems undergo structural changes when we set 
$\nu = \ell - 1/2$ with $\ell=0,1\ldots\,$.
The results contained in this chapter were reported in our work \textcolor{red}{[\cite{InzPly3}]}.


\section{The Klein four-group in AFF model}
\label{secK4group}

Parameterizing the coupling constant in parabolic form 
$g=\nu(\nu+1)$, which is symmetric with respect to $\nu=-\frac{1}{2}$, 
we artificially induce the invariance of the equation 
\begin{equation}
\label{timedependent1}
\left(-\frac{\partial^2}{\partial x^2}+
x^2+
\frac{\nu(\nu+1)}{x^2}\right)\psi=i\frac{\partial}{\partial t}\psi
\end{equation}
 with respect to the transformation $\rho_1:\nu\rightarrow-\nu-1$.
Equation 
(\ref{timedependent1}) is also invariant 
with respect to the transformation 
$\rho_2:(x,t)\rightarrow(ix,-t)$.
These two transformations generate 
the Klein four-group as a symmetry of equation (\ref{timedependent1}):
$K_4\simeq \Z_2\times \Z_2=(1,\rho_1,\rho_2,\rho_1\rho_2=\rho_2\rho_1)$,
 where each element is its own inverse. 
 At the level of the stationary Schr\"odinger equation,
the action of $\rho_2$ reduces to the transformation
$\rho_2:(x,E_{\nu,n})\rightarrow(ix,-E_{\nu,n})$,
which  means that $\rho_2$
is a completely  broken $\Z_2$ symmetry,
for which  the transformed eigenstates $\rho_2(\psi_{\nu,n})=\psi_{\nu,n}(ix)$ 
 with eigenvalues $-E_{\nu,n}$ are nonphysical solutions. 
The transformation
 $\rho_1$ at the same level of the stationary Schr\"odinger equation 
 implies that the energy eigenvalues change as 
 $E_{\nu,n}\rightarrow\rho_1(E_{\nu,n})=E_{-\nu-1,n}=4n-2\nu+1$. 
 The difference between the original energy level and 
 the transformed one is $E_{\nu,n}-E_{-\nu-1,n}=\Delta E\cdot (\nu+1/2)$, 
 where $\Delta E=4$ is the distance between two consecutive levels.
 So, if we take   $\nu=\ell-1/2$ with 
 $\ell=0,1,\ldots$, we obtain 
 $\rho_1(E_{\ell-1/2,n})= E_{\ell-1/2,n-\ell}$,
 and find that physical energy levels with $n\geq\ell$
 are transformed into physical energy levels 
 but lowered by $4\ell$.
 Under  the action of $\rho_1$, the eigenstates in  (\ref{AFFless})
 are transformed into 
 the functions 
\begin{eqnarray}
\label{psi-nu-1}
 &\rho_1(\psi_{\nu,n})=
\sqrt{\frac{n!}{\Gamma(n-\nu+1/2)}}x^{-\nu}L_n^{(-\nu-1/2)}(x^2)e^{-x^2/2}:=\psi_{-\nu-1,n}
 \,.&
\end{eqnarray}
In the case of $\nu\neq \ell-1/2$,
functions (\ref{psi-nu-1}) do not satisfy boundary condition at $x=0$ 
because of the presence of the factor $x^{-\nu}$,
and they are nonphysical, formal eigenstates
of $\mathcal{H}_{\nu}$.
The case of $\nu= \ell-1/2$ requires, however,
a separate consideration.
 To analyze this case,  we  observe that
 \begin{eqnarray}
& 
 \rho_1(\psi_{\ell-1/2,n})=\sqrt{\frac{n!}{\Gamma(n-\ell+1)}}x^{-\ell+1/2}L_n^{(-\ell)}(x^2)e^{-x^2/2}\,.
&
 \end{eqnarray}
Due to the poles of Gamma function, this expression vanishes when $n<\ell$, i.e.,
$\rho_1$ annihilates the first $\ell$ eigenstates
of the system.  On the other hand,
the identity 
\be
\frac{(-\eta)^{m}}{m!}L_{n}^{(m-n)}(\eta)=\frac{(-\eta)^{n}}{n!}L_{m}^{(n-m)}(\eta)\,,
\ee 
with integer $m$ and $n$, which follows from (\ref{Laguerre}),
allows us to write $\rho_{1}(\psi_{\ell-1/2,n})=(-1)^{\ell}\psi_{\ell-1/2,n-\ell}$ when $n\geq\ell$, 
and 
this is coherent with the change of
the energy eigenvalues under application to them of 
transformation $\rho_1$.
 In conclusion, $\rho_1$ 
corresponds to a  symmetry which 
is just the identity operator when $\ell=0$,  
while for $\ell\geq 1$
this symmetry annihilates the $\ell$ lowest physical eigenstates, 
but restores them by acting on the higher eigenstates~\footnote{This is similar  to a 
picture of a 
Hilbert's  hotel under departure of clients from  first $\ell$  rooms with numbers $n=0,\ldots,\ell-1$ 
with simultaneous  translation of the clients from rooms with numbers $n=\ell,\ell+1,\ldots$, 
into the rooms with numbers $n-\ell$.
Note that the power $(\mathcal{C}_\nu^-)^{\ell}$ of lowering generator of
 conformal symmetry
with $\nu=\ell-\frac{1}{2}$ acts on physical eigenstates in a  way similar to 
$\rho_1$, but violating normalization of the states.}. 
{}From this point of view, in 
the case of half-integer $\nu$, 
transformation $\rho_1$ does not produce anything new.  Nevertheless, we can also 
construct a finite set of nonphysical solutions of the same nonphysical
nature as in  (\ref{psi-nu-1}) given by the functions
\begin{eqnarray}
\label{halfintegernonphysical}
& \psi_{-\ell-1/2,k}:=
\rho_1\left(\sqrt{\frac{\Gamma(k+l+1)}{k!}}\psi_{\ell-1/2,k}\right)=
x^{-\ell+1/2}L_n^{(-\ell)}(x^2)e^{-x^2/2}, \quad k=0,\ldots,\ell-1,
&
\end{eqnarray}
singular at $x=0$, 
whose corresponding eigenvalues are $E_{-\ell-1/2,n}=4n-2\ell+2$.  
\vskip0.1cm

We note that the combined transformation $\rho_1\rho_2(\psi_{\nu,n})$ 
always produces nonphysical solutions for all values of $\nu$
due to the presence of $\rho_2$. 
Wave eigenfunctions transformed by the $K_4$ generators $\rho_2$ 
 and $\rho_1\rho_2$ diverge exponentially at infinity,
 and for the  following consideration it is convenient to introduce 
 a special common notation for them: 
$\psi_{r(\nu),n}(x)$, with $r(\nu)=-\nu-1$ for functions that vanish at infinity 
and $\psi_{r(\nu),-n}(x)=\psi_{r(\nu),n}(ix)$ for functions that diverge when $x\rightarrow \infty$.
In the case of $\nu=\ell-1/2$, $\ell\geq 1$,  we  have
$E_{-\ell-1/2,\ell-n-1}=-E_{-\ell-1/2,n}$ for $n<\ell$, 
and one finds that (\ref{halfintegernonphysical}) and their partners 
in the sense of Eq. (\ref{tildepsi}) are related with 
nonphysical eigenstates  produced 
by $\rho_2$ and their partners,   
\be
\label{tilderelation}
\psi_{-\ell-1/2,\ell-1-n}\propto  \widetilde{\psi}_{-\ell-1/2,-n}\,,\qquad 
\widetilde{\psi}_{-\ell-1/2,n}\propto \psi_{-\ell-1/2,-\ell+1-n}\,.
\ee 

 Now, let us study the quantum conformal symmetry of
 the AFF model from the perspective of the 
discrete  Klein four-group.
 Keep in mind that under these transformations,
$\mathfrak{sl}(2,\R)$ ladder operators 
$\mathcal{C}_\nu^\pm$ introduced in (\ref{Ladderdimensionless})
change  as 
\be
\rho_1(\mathcal{C}_\nu^{\pm})=\mathcal{C}_\nu^{\pm}\,,
\qquad
\rho_2(\mathcal{C}_\nu^{\pm})=\rho_3(\mathcal{C}_\nu^{\pm})=-\mathcal{C}_\nu^{\mp}\,,
\ee
so what we have here is  a group of  automorphisms of the conformal algebra. 
Knowing that  $\mathcal{C}_\nu^-$ annihilates the ground state, we can use the $K_4$
group to obtain the kernels of $\mathcal{C}_\nu^\pm$ in the case $\nu\geq-1/2$, 
\begin{eqnarray}
\label{kerCnu}
\ker\,\mathcal{C}_{\nu}^-=\text{span}\,\{
\psi_{\nu,0},\psi_{-\nu-1,0}
\}\,,\qquad
\ker\,\mathcal{C}_{\nu}^+=\text{span}\,\{
\psi_{\nu,-0},\psi_{-\nu-1,-0}
\}\,.
\end{eqnarray}
For $\nu=-1/2$, the kernels of 
 $\mathcal{C}_{-1/2}^\pm$ 
 are similar to (\ref{kerCnu})  but 
 with  the states  $\psi_{-\nu-1,0}$ and $\psi_{-\nu-1,-0}$ are replaced, respectively, by  the Jordan states 
\begin{eqnarray}
\label{Jordan0}
&\Omega_{-1/2,0}=\left(a-\frac{1}{2}\ln x\right)\psi_{-1/2,0}\,,\qquad
\Omega_{-1/2,-0}=\left(b-\frac{1}{2}\ln x\right)\psi_{-1/2,-0}\,,&
\end{eqnarray} 
where  $a$ and $b$ are constants. 

In the context of the Darboux transformations, 
 the equations in (\ref{kerCnu}) indicate 
that the second order differential operators $-\mathcal{C}_\nu^\pm$ 
are generated by the choice of  the seed states $(\psi_{\nu,\mp0},\psi_{-\nu-1,\pm0})$, and
by means of Eq. (\ref{Darstates})  we can  write the equalities
\begin{eqnarray}
\label{WrC-}
\mathcal{C}_\nu^\mp\phi_{r(\nu),z}=-\frac{W(\psi_{\nu,\pm0},\psi_{-\nu-1,\pm0},
\phi_{r(\nu),z})}{W(\psi_{\nu,\pm0},\psi_{-\nu-1,\pm0})}
\,,
\end{eqnarray}
where $\phi_{r(\nu),z}$ with $z=\pm n$, $n\in \N$, corresponds to an eigenstate or a
Jordan state of $L_\nu$. 
The Wronskian form of these equalities is useful 
to find  the action of the ladder operators
on the states $\widetilde{\psi}_{r(\nu),\pm 0}$ and $\breve{\Omega}_{-1/2,0}$.  
Using  some Wronskian identities from the Appendix \ref{ApenWI}, specifically the
Eqs. (\ref{ide2}) and (\ref{tech1}), 
as wells as the relations 
\begin{equation}
\label{tools1}
W(\psi_{\nu,\pm0},\psi_{-\nu-1,\pm0})=-(2\nu+1) e^{\mp x^2}\,,\qquad
W(\psi_{-1/2,\pm0},\Omega_{-1/2,\pm0})=e^{\mp x^2}\,,
\end{equation}
one can find that 
\begin{eqnarray}
\label{tools2}
\mathcal{C}_\nu^-\widetilde{\psi}_{r(\nu),0}\propto\psi_{r(-\nu-1),-0}\,,\qquad
\mathcal{C}_\nu^+\widetilde{\psi}_{r(\nu),-0}\propto\psi_{r(-\nu-1),0}\,,\\
\label{tools3}
\mathcal{C}_{-1/2}^\mp\widetilde{\psi}_{-1/2,\pm0}\propto\Omega_{-1/2,\mp0}\,,\qquad
\mathcal{C}_{-1/2}^\mp\breve{\Omega}_{-1/2,\pm0}\propto\psi_{-1/2,\mp0}\,.
\end{eqnarray}
So far, we realize that the states of Jordan should play some role in the case of half-integer $\nu$,
however, let us first consider the general case.
For this, we use (\ref{Dar-jor}) and 
the $\mathfrak{sl}(2,\R)$ algebra 
 to prove the relations   
\be
\label{Jordann}
\Omega_{r(\nu),\pm n}\propto(\mathcal{C}_{\nu}^{\pm})^n\Omega_{r(\nu),\pm 0}\,,\qquad
\breve{\Omega}_{r(\nu),\pm n}\propto(\mathcal{C}_{\nu}^{\pm})^n\breve{\Omega}_{r(\nu),\pm 0}\,.
\ee
Thus, the ladder operators act in a similar  way as they act
 on  eigenstates of $L_\nu$, but with 
a difference when $n=0$.  
When $\nu\not=-1/2$, we obtain the relations
$
\mathcal{C}_{\nu}^{\pm}\Omega_{r(\nu),\mp 0}\propto \widetilde{\psi}_{r(-\nu-1),\pm 0}
$ and $\mathcal{C}_{\nu}^{\pm}\breve{\Omega}_{r(\nu),\mp 0}\propto \Omega_{r(-\nu-1),\pm 0}$.
 Due to  (\ref{tilderelation}) one can make the identification 
 $\breve{\Omega}_{-\ell-1/2,\pm0}=\Omega_{-\ell-1/2,\mp(\ell-1)}$, so in the half-integer case $\nu=\ell-1/2$ 
with $\ell\geq 1$ we obtain
\begin{equation}
\label{ConJordan}
\mathcal{C}^{\pm}_{\ell-1/2}\Omega_{\ell-1/2,\mp0}\propto\psi_{-\ell-1/2,\mp (\ell-1)}\,,\qquad
\mathcal{C}^{\pm}_{\ell-1/2}\Omega_{-\ell-1/2,\mp0}\propto\psi_{\ell-1/2,\mp (\ell-1)}\,.
\end{equation}
 Acting   on  these
 relations by  $(\mathcal{C}_{\ell-1/2}^\pm)^{\ell}$, we obtain zero, and conclude that 
\begin{eqnarray}
\label{spanChalf}
\begin{array}{ll}
\ker (\mathcal{C}_{\ell-1/2}^\pm)^{\ell+k}&=\text{span}\{\psi_{\ell-1/2,\mp0},\ldots,\psi_{\ell-1/2,\mp(\ell+k-1)},\psi_{-(\ell-1/2)-1,\mp0},\ldots,\\
&\qquad\qquad\psi_{-(\ell-1/2)-1,\mp(\ell-1)},\Omega_{\ell-1/2,\mp0},\ldots,\Omega_{\ell-1/2,\mp(k-1)}\}\,
\end{array}
\end{eqnarray} 
for $k=1,2,\ldots$.
The whole picture is summarized in 
Figure   \ref{figure0}.
\begin{figure}[H]
\begin{center}
\includegraphics[scale=0.5]{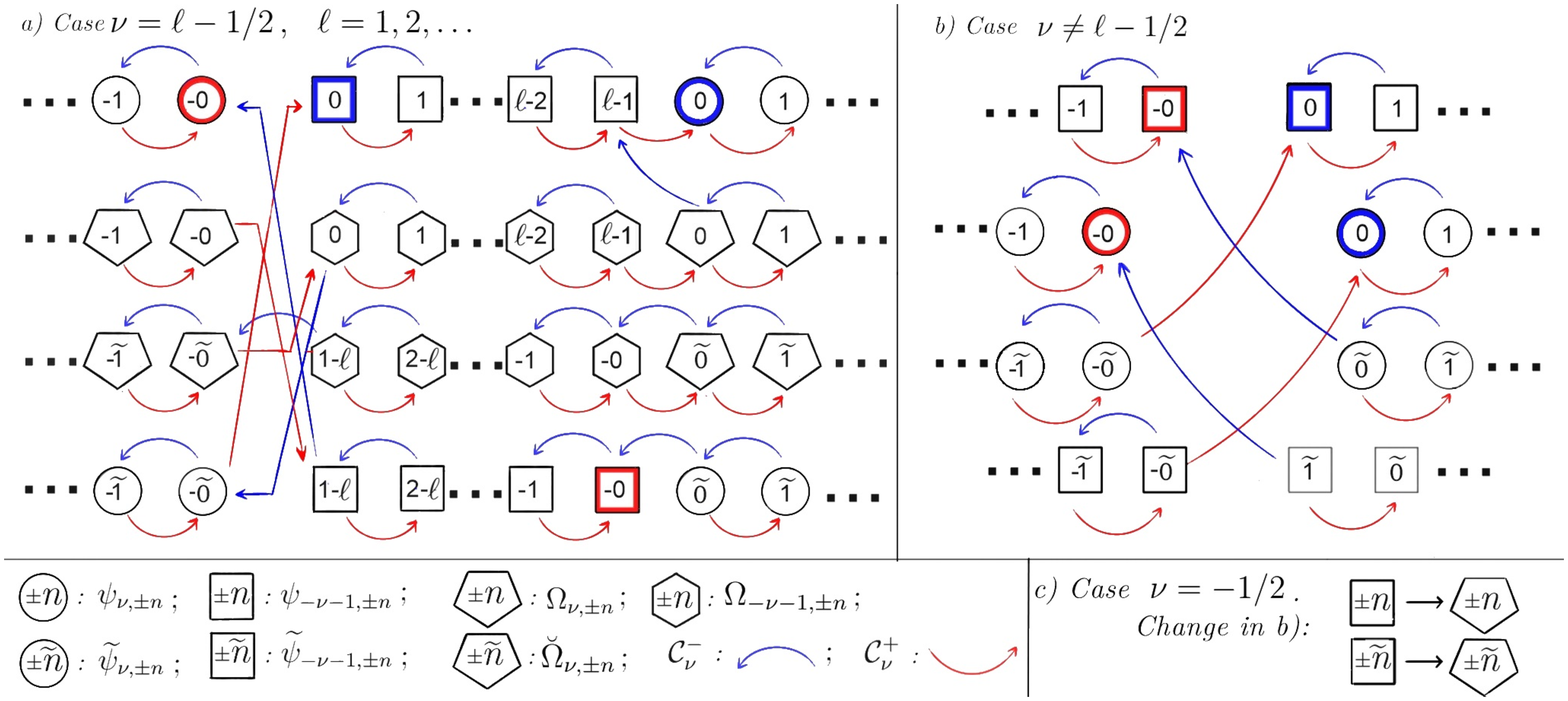} 
\caption[Ladder operators action on eigenstates and Jordan states, Sec. 8.1 ]{\small{The action of the ladder operators in dependence on the value of $\nu$.
Diagram   \emph{a)} illustrates the case of  half-integer 
$\nu=\ell-1/2$ with $\ell=1,\ldots,$ where it  is shown 
how  Jordan states can be related to eigenstates by the action 
of $\mathcal{C}_{\nu}^\pm$.
Diagram  \textit{b)} corresponds to 
non-half-integer values of  $\nu$.  In \textit{c)}, it is indicated how 
the case with $\nu=-1/2$ 
can be obtained from 
  \textit{b)} by 
changing the corresponding states. 
The shapes with borders highlighted in blue (red) represent  the states
annihilated 
by $\mathcal{C}_{\nu}^-$ ($\mathcal{C}_{\nu}^+$).}}
\label{figure0}
\end{center} 
\end{figure}


\section{Superconformal symmetry and the Klein four-group}
\label{section3.2}

Here, we inspect the action of the Klein four-group on a supersymmetric extension of the AFF model.
 To do so, we must pay attention to the intertwining operators $ A_\nu^\pm $ and $B_\nu^\pm $ 
introduced in Chap. \ref{ChConformal}, Eqs. (\ref{Anu}) and (\ref{Bnu}) 
(with  $ \omega = 1 $). Acting on them, the group produces
\begin{eqnarray}
\label{rho1}
\rho_1(A^\mp_{\nu})=-B^\pm_{\nu-1}\,,\qquad
\rho_1(B^\mp_{\nu})=-A^\pm_{\nu-1}\,,\\
\label{rho2}
\rho_2(A^\pm_\nu)=-iB^\pm_\nu\,,\qquad
\rho_2(B^\pm_\nu)=-iA^\pm_\nu \,.
\end{eqnarray}
These relations are valid for $\nu>-1/2$, while for  
 $\nu=-1/2$ the transformation $\rho_1$ reduces to the identity.
 
The symmetry generators of the super-extended AFF model, 
namely 
 $\{\mathcal{H}_{\nu}^{e},\mathcal{R}_\nu,\mathcal{C}_\nu^\pm,\mathcal{Q}_\nu^{a},\mathcal{S}_\nu^{b}\}$,
were 
defined in  Eqs. (\ref{Poincare1}), (\ref{Rnu}), (\ref{Snu})
 and (\ref{Gnu}). 
 The basic blocks to construct these objects are the intertwining operators
 $A_\nu^\pm$ and $B_\nu^\pm$, so the role of the Klein four-group
at the supersymmetric level is at hand. Nevertheless, before to apply 
the relations (\ref{rho1})-(\ref{rho2}) in the supersymmetric generators,
it is convenient to remember that 
the corresponding  superalgebra (\ref{HRQ0})-(\ref{QSGG}) has the automorphism 
$f=f^{-1}$,
which corresponds to  the transformations 
$\mathcal{H}_{\nu}^{e}\rightarrow \mathcal{H}_{\nu}^{e}-4\mathcal{R}_{\nu}=\mathcal{H}_{\nu}^b$,
$\mathcal{R}_{\nu}\rightarrow -\mathcal{R}_\nu$, 
$\mathcal{G}_{\nu}^\pm\rightarrow \mathcal{G}_{\nu}^{\pm}$,
$\mathcal{Q}_\nu^1\rightarrow -\mathcal{S}_{\nu}^{1}$, 
$\mathcal{Q}_\nu^2\rightarrow \mathcal{S}_{\nu}^{2}$,
$\mathcal{S}_\nu^1\rightarrow -\mathcal{Q}_{\nu}^{1}$
$\mathcal{S}_\nu^2\rightarrow \mathcal{Q}_{\nu}^{2}$. Then,
the action of $\rho_1$ gives us  
\begin{eqnarray}
\label{gentransformed1}
&\rho_1(\mathcal{H}_{\nu}^{e})=\sigma_1(\mathcal{H}_{\nu-1}^{e}-4\mathcal{R}_{\nu-1})\sigma_1\,,\qquad
\rho_1(\mathcal{G}_{\nu}^\pm)=\sigma_1(\mathcal{G}_{\nu-1}^\pm)\sigma_1\,,&\\
&\rho_1(\mathcal{R}_{\nu})=\sigma_{1}(-\mathcal{R}_{\nu-1})\sigma_{1}\,,&\label{gentransformed1+}\\
&\rho_1(\mathcal{Q}_{\nu}^1)=\sigma_1(-\mathcal{S}_{\nu-1}^1)\sigma_1\,,\qquad
\rho_1(\mathcal{Q}_{\nu}^2)=\sigma_1(\mathcal{S}_{\nu-1}^2)\sigma_1\,,\label{gentransformed1++}& \\
&\rho_1(\mathcal{S}_{\nu}^1)=\sigma_1(-\mathcal{Q}_{\nu-1}^1)\sigma_1\,,\qquad
\rho_1(\mathcal{S}_{\nu}^2)=\sigma_1(\mathcal{Q}_{\nu-1}^2)\sigma_1\,,\label{gentransformed2}&
\end{eqnarray}
which in fact is a combination of the shift $\nu\rightarrow \nu-1$, the action of $f$ and the unitary rotation. 
The transformed generators 
(\ref{gentransformed1})-(\ref{gentransformed2}) 
still satisfy the same superconformal algebra, 
i.e. $\rho_1$ is an automorphism of the $\mathfrak{osp}(2|2)$ symmetry,
however the new generators describe another
super-extended system: Unlike the  initial system $\mathcal{H}_\nu^{e}$,
in the transformed one 
the $\mathcal{N}=2$ Poincar\'e  supersymmetry 
is  spontaneously broken in the case of $\nu>-1/2$, 
see Chap. \ref{ChConformal}. The only  
exception from this rule corresponds to 
the case $\nu=-1/2$, where 
the transformed Hamiltonian
reduces to 
$\sigma_1\mathcal{H}_{-1/2}^{e}\sigma_1$, 
and represents a unitarily transformed 
super-Hamiltonian with the unbroken 
 $\mathcal{N}=2$ Poincar\'e  supersymmetry.

On the other hand, one can verify that when $\rho_1$
 acts on the Hamiltonian
$\mathcal{H}_\nu^{b}$, it produces $\sigma_1(\mathcal{H}_{\nu-1}^{e})\sigma_1$, 
and this time  the  $\mathcal{N}=2$ Poincar\'e  supersymmetry
 of the system is changed from the
spontaneously broken phase (in the case of $\nu>-1/2)$
to the phase of unbroken supersymmetry,
 with the only exception of the system $\mathcal{H}_{-1/2}^b$
 with unbroken supersymmetry,
which unitary transforms into  $\sigma_1\mathcal{H}_{-1/2}^{b}\sigma_1$.  
This action of transformation $\rho_1$ on super-extended 
systems can be compared with 
the case of the non-extended AFF system, where $\rho_1$
acts identically on its Hamiltonian  and generators of the conformal symmetry,
though, as we saw, it acts  nontrivially  on eigenstates of the system.   
  
On the other hand, the action of $\rho_2$ produces 
\begin{eqnarray}
\label{gentransformedrho2}
&\qquad \rho_2(\mathcal{H}_{\nu}^{e})= -\mathcal{H}_{\nu}^{b}\,,\quad
\rho_2(\mathcal{G}_{\nu}^\pm)= -\mathcal{G}_{\nu}^{\mp}\,,\quad
\rho_2(\mathcal{R}_{\nu})= \mathcal{R}_\nu\,,&\\
&\rho_2(\mathcal{Q}_\nu^1)= -i\mathcal{S}_{\nu}^{1}\,, \qquad
\rho_2(\mathcal{Q}_\nu^2)= -i\mathcal{S}_{\nu}^{2}\,,&\\
&\rho_2(\mathcal{S}_\nu^1)= -i\mathcal{Q}_{\nu}^{1}\,,\qquad
\rho_2(\mathcal{S}_\nu^2)= -i\mathcal{Q}_{\nu}^{2}\,.&
\end{eqnarray}
Transformed Hamiltonian operator is similar here to the Hamiltonian  
produced  by the automorphism $f$  but  multiplied by $-1$.
This correlates with the anti-Hermitian nature of 
the transformed fermion generators of superalgebra.
Accordingly,  the spectrum of the transformed matrix Hamiltonian
is  negative, not bounded from below, and each of its level is doubly degenerate 
for  $\nu\geq-1/2$.

In correspondence with the described picture, 
the application of the combined transformation $ \rho_2\rho_1$
is just another automorphism of the superconformal algebra
(\ref{HRQ0})-(\ref{anti2}), which produces 
anti-Hermitian  odd generators,  and 
$\rho_2\rho_1(\mathcal{H}_{\nu}^{e})=\sigma_1(-\mathcal{H}_{\nu-1}^{e})\sigma_1$.
The discrete spectrum of  the transformed Hamiltonian
is not restricted from below and is  given
by the numbers $\mathcal{E}_n=-4n$, $n=0,1,\ldots$,
where each negative energy level is doubly degenerate, 
while non-degenerate zero energy level corresponds to the 
state  $(\psi_{\nu,0},0)^t$.  



\section{Dual Darboux schemes}
\label{Mirror} 

With the new set of nonphysical solutions, in this section we extend the idea of dual schemes for the 
AFF model with $ \nu \geq-1/2 $. As we have shown in Sec. \ref{secK4group}, the case in which $ \nu $
 takes half-integer values is special, because the Jordan states take relevance through the properties of the conformal
  symmetry generators\footnote{
Operators $\mathcal{C}_\nu^\pm$ can be interpreted as the second
 order intertwining operators associated with the seed states $(\psi_{-\nu-1,0},\psi_{\nu,0})$ for $\nu>1/2$, 
and to the confluent scheme $(\Omega_{-1/2,0},\psi_{-1/2,0})$, when $\nu=1/2$.},
 which are simultaneously the ladder operators for corresponding AFF systems, see equation (\ref{ConJordan}). 
For this reason, we start first with the case where $ \nu $ is not a half-integer.
 Let us choose a generic set of physical and nonphysical eigenstates of $ L_{\nu} $
 as seed states,
\begin{eqnarray}
\label{unioncollection}
\{\alpha\}=(\psi_{\nu,k_1},\ldots,\psi_{\nu,k_{N_1}},\psi_{-\nu-1,l_1},\ldots,\psi_{-\nu-1,l_{N_2}})
\,,\qquad
k_{i},l_{j}= \pm0,\pm1,\ldots\,,
\end{eqnarray}
where $i=1,\ldots,N_1$ and $j=1,\ldots,N_2$, and, 
for simplicity,
we suppose that  
$|k_1|<\ldots<|k_{N_1}|$ and  $|l_1|<\ldots<|l_{N_2}|$.
Let us assume that in the scheme  
(\ref{unioncollection}) there are no repeated states 
 and   both $k_i$ and $l_j$ carry the same sign for all $i$ and $j$.
Also let us define the index number 
\be
n_N=\text{max}\,(|k_1|,\ldots,|k_{N_1}|,|l_1|,\ldots,|l_{N_2}|)\,.
\ee
which can correspond to a state with index $\nu$ or $-\nu-1$. 
By means of the algorithm described in Appendix \ref{DualAFF1}
one can show that 
\begin{eqnarray}
\label{eqschemes3}
&W(\{\alpha\})=e^{-(n_{N}+1)x^2}W(\{\Delta_-\})\,,&\\\nonumber
&\{\Delta_-\}:=(\psi_{-\nu-1,-0},\psi_{\nu,-0},\ldots,\check{\psi}_{-\nu-1,-r_i},
\check{\psi}_{\nu,-s_i},\ldots,\psi_{-\nu-1,-n_N},{\psi}_{\nu,-n_N}
)\,,&
\end{eqnarray}   
is satisfied, where the marked  states
 $\check{\psi}_{-\nu-1,-r_i}$ and $\check{\psi}_{\nu,-s_i}$, with 
 $r_i=n_{N}-k_i$ and $s_j=n_{N}-l_j$,
are omitted 
from the set $\{\Delta_-\}$.
On the contrary, if $k_i$ and $l_j$ carry the minus sign, we have the equality
\begin{eqnarray}
\label{eqschemes4}
&W(\{\alpha\})=e^{(n_{N}+1)x^2}W(\{\Delta_+\})\,,&\\\nonumber
&\{\Delta_+\}:=(\psi_{-\nu-1,0},\psi_{\nu,0},\ldots,\check{\psi}_{-\nu-1,r_i},
\check{\psi}_{\nu,s_j},\ldots,{\psi}_{-\nu-1,n_N},{\psi}_{\nu,n_N}
)\,,&
\end{eqnarray}   
where now $r_i=n_N-|k_i|$ and $s_j=n_N-|l_j|$.
These relations are also valid if one of the
numbers ${N_1}$ or ${N_2}$
is equal to zero, which means that in the corresponding scheme there are only 
states of the same kind
with respect to the first index,  $-\nu-1$ or $\nu$, respectively. 

When considering  $\nu=\ell-1/2$ with  $\ell=0,1,2,\ldots$, some repeated states
could appear  due to $\rho_{1}(\psi_{\ell-1/2,n})=(-1)^{\ell}\psi_{\ell-1/2,n-\ell}$.
This means that the Wronskian must vanish, however, that happens because, in the general case, 
this object takes the form
$ \Lambda (\nu) f (x; \nu) $, where $ \Lambda (\nu) $ disappears in these special cases
(see the example (\ref{W(nu2-nu-1,2)}) below).
To obtain a deformed AFF system with the potential modified
 by $ -2 \ln (f (x; \nu))'' $ for half-integer $ \nu $, 
as well as its dual scheme, we will have relations analogous to (\ref{eqschemes3}) 
and (\ref{eqschemes4}), but changing each state of the form 
$ \psi_{-\nu-1, \pm(\ell + k)} $ by $ \Omega_{\ell-1 / 2, \pm k} $,
which means that we are dealing with the confluent Darboux transformation, 
see Appendix \ref{AFFhalf} for a detailed derivation. 
The general  
rules of the Darboux duality  can be summarized and better understood with the 
examples presented diagrammatically  in Fig. \ref{Kleinfigure1}. 
\begin{figure}[H]
\begin{center}
\includegraphics[scale=0.28]{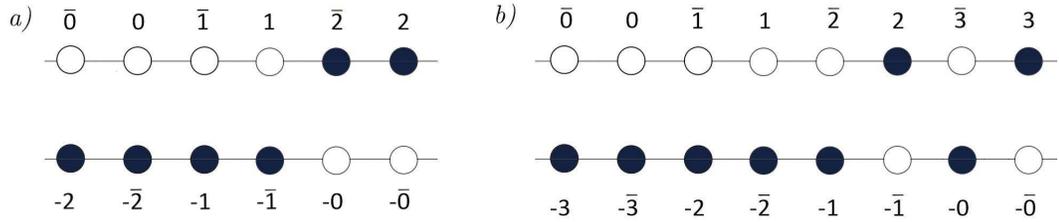} 
\caption[Two mirror diagrams, Sec. 8.3]{\small{Two ``mirror diagrams'' corresponding to dual schemes 
for the conformal mechanics model. 
The numbers $\pm n$ indicate the states $\psi_{\nu,\pm n}$, and symbols 
$\pm\bar{n}$ correspond 
to the  states $\psi_{-\nu-1,\pm n}$.} }
\label{Kleinfigure1}
\end{center} 
\end{figure}
\vskip-0.5cm
These types of diagrams are read in the same way as for the harmonic oscillator
 mirror diagram presented in Chap. \ref{ChRQHO} and in this case they 
correspond to  the following Wronskian relations:  
\begin{eqnarray}
\label{Interpoeg}
W(\psi_{-\nu-1,2},\psi_{\nu,2})=e^{-3x^2}W(\psi_{-\nu-1,-1},\psi_{\nu,-1},\psi_{-\nu-1,-2},\psi_{\nu,-2})\,,\\
\label{Adlerkreineg}
W(\psi_{\nu,2},\psi_{\nu,3})=e^{-4x^2}W(\psi_{\nu,-0},\psi_{\nu,-1},\psi_{-\nu-1,-2},\psi_{\nu,-2},,\psi_{-\nu-1,-3},\psi_{\nu,-3})\,,
\end{eqnarray}
whose  explicit forms are 
\begin{eqnarray}\label{W(nu2-nu-1,2)}
\begin{array}{ll}
W(\psi_{\nu,2},\psi_{-\nu-1,2})=&(2\nu+1)e^{-x^2}\big(45 - 72 \nu +16 (-4x^6 + x^8) \\
& +8x^4(15 - 4 \nu (1 + \nu)) + \nu^2 (-7 +2 \nu (2 + \nu))\big),
\end{array}\\
\label{W(nu2nu3)}
\begin{array}{ll}
W(\psi_{\nu,2},\psi_{\nu,3})=&e^{-x^2} x^{
 3 + 2 \nu} \big(16 x^8 - 32 x^6 (5 + 2 \nu) + 
   24 x^4 (5 + 2 \nu)^2 - \\
  & 8 x^2 (3 + 2 \nu) (5 + 2 \nu) (7 + 
      2 \nu) +  (3 + 2 \nu) (5 + 2 \nu)^2 (7 + 
      2 \nu)\big).
\end{array}
\end{eqnarray}
The transformation which relates the AFF systems 
described by  $L_\nu$ with $L_{\nu+m}$ 
can also be understood within  this picture. Furthermore, using 
a diagram similar to those  in Fig. \ref{Kleinfigure1}, one can 
show that the schemes 
$\{\Delta_+\}=(\psi_{r(\nu),0},\ldots,\psi_{r(\nu),m-1})$ and 
$\{\Delta_-\}=(\psi_{r(\nu),-0},\ldots,\psi_{r(\nu),-(m-1)})$ are dual.


\section{Rationally deformed AFF systems}
\label{Ladders}

A rational deformation of the AFF model can be generated 
by taking a set of the seed states
\begin{equation}\label{alpKA}
\{\alpha_{KA}\}=(\psi_{\nu,l_1},\psi_{\nu,l_1+1},\ldots,\psi_{\nu,l_m},\psi_{\nu,l_m+1})\,,
\end{equation}
composed from  $m$ pairs of neighbour physical states.
Krein-Adler theorem \textcolor{red}{[\cite{Krein,Adler}]} guarantees that the resulting  system 
described by the Hamiltonian operator of the form
\begin{eqnarray}
\label{deformed1}
&L^{KA}_{(\nu,m)}=L_{\nu+m}+4m+\frac{F_\nu(x)}{Q_\nu(x)}&
\end{eqnarray}
is  nonsingular on $\R^+$. Here
$F_\nu(x)$ and $Q_\nu(x)$ are real-valued  polynomials,   
$Q_\nu(x)$ has no zeroes on $\R^+$, its degree  is two more than that  
of $F_\nu(x)$, and so, the last rational term in (\ref{deformed1}) vanishes at infinity. 
The spectrum of the system (\ref{deformed1}) 
is the equidistant spectrum of the AFF model with  the removed energy levels
corresponding to the seed states. Consequently,  any gap 
in the resulting system has a size $12+8k$, where $k=0,1,\ldots$ 
correspond to  $k$  adjacent pairs in the set (\ref{alpKA}) 
which produce a given gap.
An example of this kind of systems is 
generated by the scheme $(\psi_{\nu,2},\psi_{\nu,3})$,
whose dual negative scheme is given by 
equation (\ref{Adlerkreineg}). 
\vskip0.1cm

Another class of rationally extended AFF systems 
is provided by  isospectral deformations generated by the
schemes  of  the form 
\begin{equation}\label{isoscheme}
\{\alpha_{iso}\}=(\psi_{\nu,-s_1},\ldots,\psi_{\nu,-s_m})\,,
\end{equation}
which contains the states of the form 
 $\rho_2(\psi_{\nu,n}(x))=\psi_{\nu,n}(ix)$.
As the functions used in this scheme are proportional to  $x^{\nu+1}$ 
and do not have real zeros other than $x=0$,  one obtains a regular 
on $\R^+$ system of the form
\begin{eqnarray}
\label{deformed2}
&L^{iso}_{(\nu,m)}=L_{\nu+m}+2m+f_\nu(x)\,,&
\end{eqnarray}
where $f_\nu(x)$ is a rational function disappearing at infinity  \textcolor{red}{[\cite{Grand}]},
and one can find that  potential of the system (\ref{deformed2}) is a convex on $\R^+$ function.
In this case the transformation does not remove or add energy levels, 
and,  consequently, the initial system $\mathcal{H}_\nu$  and the deformed system (\ref{deformed2}) 
are completely isospectral superpartners.
 Some concrete examples of the systems  (\ref{deformed2}) with 
 integer values of $\nu$ were  considered in the 
 two previous chapters, see also \textcolor{red}{[\cite{CarInzPly}]} . 
 \vskip0.1cm
 Consider yet another generalized Darboux scheme 
 which allows us to interpolate between different
 rationally deformed AFF systems.
 For this we assume that the initial AFF system
 is characterized by the parameter
 $\nu=\mu+m$, where 
 $-1/2<\mu\leq 1/2$ 
and $m$ can take any non-negative integer value. 
For these ranges of values of the parameter $\nu$,
real zeros of the functions $\psi_{\mu+m,n-m}$ are located between 
zeros of $\psi_{-(\mu+m)-1,n}$,
so that we can rethink the Krein-Adler theorem
 and consider the scheme 
\be
\label{Interpol}
\{\gamma_{\mu}\}=(\psi_{-(\mu+m)-1,n_1},\psi_{(\mu+m),n_1-m},\ldots,
\psi_{-(\mu+m)-1,n_{N}},\psi_{(\mu+m),n_{N}-m})\,,
\ee
which includes  $2N$ states and  where 
we suppose that $n_i-m\geq 0$
for all $i=1,\ldots, N$. 
The DCKA transformation based on the set (\ref{Interpol}) produces 
the system 
\be
\label{Intersys}
L_{\mu+m}^{def}:=L_{\mu+m}-2(\ln W(\gamma_\nu))''=L_{\mu+m}+4N+h_{\mu+m}(x)/q_{\mu+m}(x)\,,
\ee
where the constant $4N$ is provided by the Gaussian factor in the Wronskian, 
and the last term is a rational function vanishing at infinity 
and having  no zeros on the whole real line, including the origin, if an only if $-1/2<\mu\leq1/2$,
 see Appendix \ref{apWron}. 
 Let us analyze now some special values of $\mu$.
 
\textit{The case $\mu=0$}\,: by virtue of  relations between Laguerre and Hermite polynomials
mentioned in Chap \Ref{ChConformal}, see equation 
(\ref{hermiteLaguerre}),  in this case we obtain 
those systems which were generated in \textcolor{red}{[\cite{CarInzPly}]}  and discussed 
in Chap. \ref{ChRQHO}, we refer to systems  (\ref{REIO}).

\textit{The case $\mu=1/2$}\,:  we have here the relation    
\be\rho_1(\psi_{m+1/2,n_i})=\psi_{-m-3/2,n_i}=(-1)^{m+1}\psi_{m+1/2,n_i-m-1}\,,
\ee
 due to which 
the scheme (\ref{Interpol})
transforms into 
\be
\{\gamma_{{1}/{2}}\}=(\psi_{1/2+m,n_1-m-1},\psi_{1/2+m,n_1-m},\ldots,
\psi_{1/2+m,n_{N}-m-1},\psi_{1/2+m,n_{N}-m})\,,
\ee 
which corresponds to  (\ref{alpKA}) with
$l_i=n_i-m-1$. We additionally suppose 
that $n_i-m-1\not=n_{i-1}-m$, otherwise the Wronskian vanishes. Note that 
when $\mu\not=1/2$, the image of the states $\psi_{\mu+m,n_i-m-1}$
under Darboux mapping (\ref{Darstates})
is a physical state, but in the case $\mu=1/2$ such states are mapped into zero 
since  the argument  $\psi_{1/2+m,n_i-m-1}$  appears twice 
in the Wronskian of the numerator.

\textit{The case $\mu=-1/2$}\,: this case was not included
 in the range of $\mu$ from the beginning due to relation
$\rho_1(\psi_{m-1/2,n_i})=\psi_{-m-1/2,n_i}=(-1)^{m}\psi_{m-1/2,n_i-m}$
which 
would mean  the appearance of the repeated states in the scheme 
(\ref{Interpol}) and vanishing of the corresponding Wronskian.
However, in Appendix \ref{apWron} we show that  the  limit  relation 
$\lim_{\mu\to-1/2}{W(\{\gamma_\mu\})}/{(\mu+\frac{1}{2})^N}\propto
W(\{\gamma\})$ is valid, 
where the scheme $\{\gamma\}$ is
\be
\label{Interpol2}
\{\gamma\}=(\psi_{m-1/2,n_1-m},\Omega_{m-1/2,n_1-m},\ldots,\psi_{m-1/2,n_N-m},\Omega_{m-1/2,n_N-m})\,,
\ee
which 
corresponds to a non-singular  confluent Darboux transformation, 
\textcolor{red}{[\cite{Jordan1}]}. 

By considering this last comment, in conclusion we have that 
when $-1/2\leq\mu<1/2$, the states $\psi_{-(\mu+m)-1,n_i}$ 
(and $\Omega_{m-1/2,n_i-m}$ in the case of $\mu=-1/2$) are nonphysical states.
This  means that 
 only the physical states 
$\psi_{\nu+m,n_i-m}$ indicate the energy levels removed under
 the corresponding Darboux transformation, i.e., there are gaps 
of the minimum size $2\Delta E=8$, where $\Delta E=4$ is the 
distance between energy levels of the AFF model, which can merge 
to produce energy gaps of the size $8+4k$.
On the other hand, when $\mu=1/2$,
we have a typical Krein-Adler  scheme with gaps of the size $12+4k$.

 To give an example, we put $m=0$, that
  means $\nu=\mu$,  and consider the scheme $(\psi_{-\nu-1,2},\psi_{\nu,2})$ given in (\ref{W(nu2-nu-1,2)})
with $-1/2<\nu\leq1/2$, and in the case of $\nu=-1/2$ we have 
the scheme $(\psi_{-1/2,2},\Omega_{-1/2,2})$.
 The potential of the rationally deformed AFF system 
 generated by the corresponding Darboux transformation
is shown in Fig. \ref{potential1}
and Fig. \ref{potential12}.
\begin{figure}[H]
\begin{center}
\includegraphics[scale=0.78]{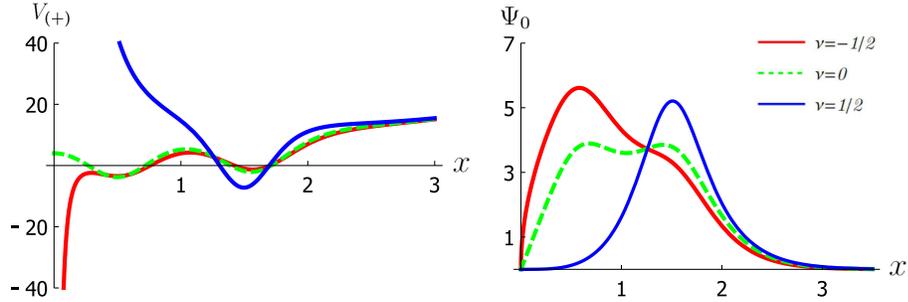} 
\caption[First picture of potential and ground state behavior with respect to $\nu$, Sec. 8.3]{\small{On the left, 
a graph of the corresponding potential is shown which is 
produced by the 
associated Darboux transformation applied to the AFF model
with three indicated values of the parameter $\nu$ 
versus  the dimensionless coordinate $x$.
For $\nu=-1/2$,  
the corresponding limit is taken, and the resulting system has an attractive potential with a (not shown) 
potential 
barrier at $x=0$.
For $\nu=0$, we obtain a rationally extended half-harmonic 
oscillator. The case $\nu=1/2$ corresponds to the
Krein-Adler scheme $(\psi_{1/2,1},\psi_{1/2,2})$ with a 
gap equal to $12$. On the right, 
the ground states 
of the corresponding generated systems are shown 
as functions of dimensionless coordinate $x$.}}
\label{potential1}
\end{center} 
\end{figure} 
\vskip-0.5cm
\begin{figure}[H]
\begin{center}
\includegraphics[scale=0.78]{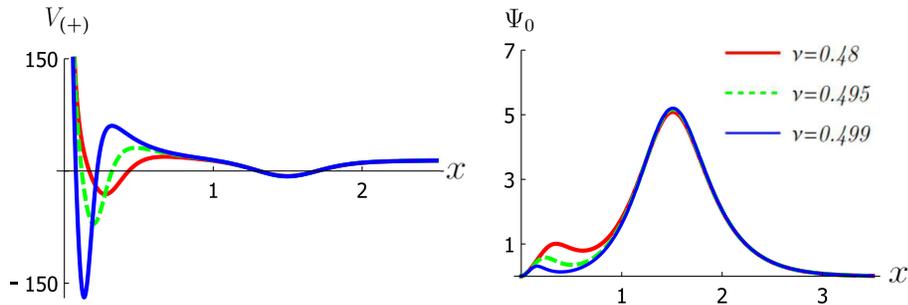} 
\caption[Second picture of potential and ground state behavior with respect to $\nu$, Sec. 8.3]{\small{On the left, the potential of deformed systems with $\nu$ close
 to $1/2$ is shown. On the right, the ground states of the corresponding 
 systems are displayed. }} 
\label{potential12}
\end{center} 
\end{figure} 
 As it is seen from the figures, 
 the first minimum of the potential grows in its absolute value, 
  its position moves to $0$,  and it disappears at $\nu=1/2$,
while  the  local maximum near zero also grows,
its position approaches zero,   and it goes to infinity  in the limit.
 Besides, the first maximum of the ground state 
 vanishes when $\nu$ approximates the limit value $1/2$. 
 Coherently with the described behavior of the potential, 
 the image of the  Darboux-transformed state $\psi_{\nu,1}$, 
 which is  the first excited state of the new system when $-1/2\leq \nu<1/2$,  vanishes when 
 $\nu\rightarrow 1/2$, the corresponding energy level disappears from the spectrum at $\nu=1/2$, 
 and the size of the gap increases from $8$ to $12$.
  
The described three possible selection rules to choose the
seed states correspond to  the negative scheme  (\ref{isoscheme}), which 
generates  isospectral deformations,  
the  positive Krein-Adler scheme (\ref{alpKA}),
 and the positive interpolating   scheme (\ref{Interpol}). 
Then we can apply the Darboux duality  to obtain 
the corresponding dual schemes for them. 
The positive and negative  dual schemes 
will be used in the next subsection
 to construct complete sets 
of the spectrum-generating ladder 
operators for the rationally deformed 
conformal mechanics systems.


\section{Intertwining and ladder operators}\label{SecIntLad}

In this paragraph we proceed to construct the intertwining and ladder operators 
of rational deformed system obtained by means of  the 
seed states selection rules detailed above. 
For simplicity we do not touch
here the schemes that contain Jordan states. However, 
we have relations (\ref{Polly2}) and  (\ref{spanChalf}), 
and relations (\ref{eqschemes3}) and (\ref{eqschemes4})
which
were extended to such cases with the corresponding substitutions. This means that 
the properties summarized below are also valid for the schemes containing Jordan states.
Suppose that the positive (negative) scheme possesses $n_+$ ($n_-$) seed states. 
Then the generated Hamiltonian $L_{(\pm)}$ satisfy the relation  

\be
\label{dualL}
L{(+)}-L_{(-)}=\Delta E(n_{n_+}+1)=2(n_++n_-)\,,\qquad \Delta E=4\,,
\ee
where $n_{n_+}$ is the bigest quantum number in the positive scheme.
Let us  denote by  $A_{(+)}^\pm$ and 
$A_{(-)}^\pm$ the intertwining operators  of the positive and negative schemes
being differential operators of  the orders $n_+$ and $n_-$, respectively.  
They satisfy  the intertwining relations 
\be
\label{inter-relation}
A_{(\pm)}^-L_{\nu}=L_{(\pm)}A_{(\pm)}^-\,,\qquad
 A_{(\pm)}^+L_{(\pm)}=
L_{\nu}A_{(\pm)}^+\,.
\ee
As the  states 
$\widetilde{\psi}_{r(\nu),\pm n}$
behave asymptotically as  $e^{\pm x^2/2}$, 
the states produced from them by application of
differential operators $A_{(\pm)}^-$   
will carry the same exponential factor. Having this asymptotic behavior in mind,
let us suppose that $\psi_{r(\nu),-l_*}$ and $\psi_{r(\nu),n_*}$ are some arbitrary states 
from the negative and positive scheme, respectively.   
By using  (\ref{inter-relation}), we obtain the relations
\begin{eqnarray}
\label{complement1}
&A_{(-)}^-\widetilde{\psi}_{r(\nu),-l_*}=A_{(+)}^-\rho_1(\psi_{r(\nu),n_{n+}-l_*})\,,\qquad
A_{(+)}^-\widetilde{\psi}_{r(\nu),n_*}=A_{(-)}^-\rho_1(\psi_{r(\nu),-(n_{n+}-n_*)})\,,\quad&
\end{eqnarray}
in both sides of which the functions satisfy the same 
second order differential equation and have the same behaviour at infinity. 
Note that in the dual schemes in 
(\ref{eqschemes3}) and (\ref{eqschemes4}),
the indexes $n_{n_+}-l_*$ and $-(n_{n_+}-n_*)$ are in correspondence with the indexes 
$r_i$, and $s_i$ of the states omitted from the positive and negative scheme, respectively.
 This helps us to find that 
\be
\ker \big(A_{(-)}^+A_{(+)}^-\big)=(\psi_{\nu,0},\psi_{-\nu-1,0},\ldots,
\psi_{\nu,n_{n_+}},\psi_{-\nu-1,n_{n_+}})=\ker\, (\mathcal{C}_{\nu}^-)^{n_{n_+}+1}\,,
\ee
from where we obtain  the identities 
\begin{equation}
\label{powerC}
A_{(-)}^+A_{(+)}^-=(-1)^{n_{n_+}+1-n_+}(\mathcal{C}_{\nu}^-)^{n_{n_+}+1}\,,
\qquad A_{(+)}^+A_{(-)}^-=(-1)^{n_{n_+}+1-n_+}(\mathcal{C}_{\nu}^+)^{n_{n_+}+1}\,.
\end{equation}

Finally, to have a complete picture we write   
the relations
\begin{eqnarray}
\label{complement3}
A_{(-)}^-\psi_{r(\nu),k}=A_{(+)}^-\psi_{r(\nu),n_{n+}+1+k'}\,,\qquad 
A_{(+)}^-\psi_{r(\nu),-k'}=A_{(-)}^-\psi_{r(\nu),-(n_{n+}+1+k')}\,.&
\end{eqnarray}  
Note that in the case $\nu=0$, first equation reduces to (\ref{relation-operators}). 

In the 
case of the dual schemes where $\nu=m-1/2$, similar relations 
are obtained but with $\psi_{-\mu-m-1,\pm n_i}$ and $\widetilde{\psi}_{-\mu-m-1,\pm n_i}$ replaced 
by $\Omega_{m-\frac{1}{2},\pm(n_i-m)}$ and 
$\breve{\Omega}_{m-\frac{1}{2},\pm(n_i-m)}$ when is required.

With the help of the described  intertwining operators, we can construct 
three  types  of  ladder operators 
for $L_{(\pm)}$ which are given by:
\begin{eqnarray}
\label{ladders}
&\mathcal{A}^{\pm}=A_{(-)}^-\mathcal{C}_{\nu}^\pm A_{(-)}^+\,, \quad
\mathcal{B}^{\pm}=A_{(+)}^-\mathcal{C}_{\nu}^\pm A_{(+)}^+\,,\quad
\mathcal{C}^{+}=A_{(-)}^-A_{(+)}^+\,,\quad \mathcal{C}^{-}=A_{(+)}^-A_{(-)}^+\,.&
\end{eqnarray}
Let us denote these operators in the compact form
$\mathcal{F}_a^{\pm}=(\mathcal{A}^\pm,
\mathcal{B}^\pm,\mathcal{C}^\pm)$,  $a=1,2,3$, and use 
(\ref{dualL}) and (\ref{inter-relation}) to obtain the commutation 
relations 
\begin{eqnarray}
\label{defsl2R}
&[L_{(\pm)},\mathcal{F}_a^{\pm}]=\pm R_a\mathcal{F}_a^{\pm}\,,\qquad
[\mathcal{F}_{a}^-,\mathcal{F}_a^{+}]=\mathcal{P}_a(L_{(\pm)})\,,&\\\nonumber&
\begin{array}{ll}
R_1=R_2=4\,,&\mathcal{P}_1=(\eta+2\nu+3)(\eta-2\nu+1)P_{n_-}(\eta)P_{n_-}(\eta+4)|_{\eta=L_{(-)}-4}^{\eta=L_{(-)}}\,,\\
&\mathcal{P}_2=(\eta+2\nu+3)(\eta+2\nu+1)P_{n_+}(\eta)P_{n_+}(\eta+4)|_{\eta=L_{(+)}-4}^{\eta=L_{(+)}}\,,\\
R_3=4(n_{n_+}+1)\,,&\mathcal{P}_3=P_{n_+}(\eta)P_{n_-}(\eta)|_{\eta=L_{(+)}-4}^{\eta=L_{(-)}}\,,
\end{array}&
\end{eqnarray}
where
\begin{eqnarray}
\label{Poly2}
P_{n_-}(y)=\prod_{i=1}^{n_-}(y-\lambda_{i}^-)\,,\qquad
P_{n_+}(y)=\prod_{i=1}^{n_+}(y-\lambda_{i}^+)\,,
\end{eqnarray} 
and  $\lambda_{i}^\pm$ are the corresponding eigenvalues 
of the seed states in the positive and negative schemes.
Equations (\ref{defsl2R}) are three different but related  copies of
the nonlinearly deformed conformal algebra 
$\mathfrak{sl}(2,\R)$. 
One can verify  the commutators between 
generators with different values of index $a$ do not vanish, and therefore 
the complete structure is rather complicated.

Similarly to the non-deformed case, be means of a unitary transformation 
produced by $U=e^{-itL_{(\pm)}}$ we obtain  the integrals of motion 
${}_H\mathcal{F}_a^\pm(t)=e^{\mp R_a}\mathcal{F}_a^\pm$, and by linear combinations of them 
 construct the Hermitian generators 
$\mathfrak{D}_a(t)=(\mathcal{F}_a^-(t)-\mathcal{F}_a^+(t))/(i2R_a)$ and $
\mathfrak{K}_a(t)=(\mathcal{F}_a^+(t)+\mathcal{F}_a^-(t)+2L_{(\pm)})/R_a^2$,
which generate three copies of a
 nonlinear deformation of the Newton-Hooke algebra, 
\begin{eqnarray}
[L_{(\pm)},\mathfrak{D}_a]=-i\left(L_{(\pm)}-\frac{(R_a)^2}{2}\mathfrak{K}_a\right)\,,\qquad
[L_{(\pm)},\mathfrak{K}_a]=-2i\mathfrak{D}_a\,,\\\nonumber
[\mathfrak{D}_a,\mathfrak{K}_a]=\frac{1}{iR_a^3}\left(\mathcal{P}_a(L_{(\pm)})-
2R_aL_{(\pm)}+R_a^3\mathfrak{K}_a\right)\,,
\end{eqnarray}
which are hidden symmetries of the system described by $L_{(\pm)}$.
 
In the isospectral case, the operators $\mathcal{A}^\pm$ are the spectrum generating
ladder operators, where   
their action on physical eigenstates of $L_{(\pm)}$ 
is similar  to that of  $\mathcal{C}_\nu^\pm$
in the AFF model. 
On the other hand, 
in rationally extended gapped systems obtained by   Darboux transfromations 
based on the  schemes  not containing  
Jordan states, the  separated  states 
have the form
$A_{(-)}^-\widetilde{\psi}_{-\nu-1,-l_j}=A_{(+)}^-\psi_{\nu,n_{n+}-l_j}$, where
the states  $\psi_{-\nu-1,-l_j}$
belong to the  negative scheme and $\psi_{\nu,n_{n+}-l_j}$ are the omitted 
states in the corresponding dual positive scheme. 
Since by construction the separated states
 belong to  the kernel of $A_{(-)}^+$, 
the operators $\mathcal{A}^\pm$ and $\mathcal{C}^-$ 
will always annihilate all them. 

In summary, the resulting picture is more or less the same as we had for the cases analyzed 
in the previous chapters.  We have three pairs of ladder operators; $ \mathcal{B}^\pm $ 
detect the upper and lower energy levels of each isolated valence band,
 $ \mathcal{A}^\pm $ operators annihilate all the isolated  states, and $ \mathcal{C}^\pm $
 operators connect isolated states with the equidistant part of the spectrum.


\section{An example}
\label{Examples}
In this section we will apply the machinery of the dual schemes and the construction 
of nonlinear deformations of the conformal algebra to  a nontrivial example 
 of rationally extended  systems with gaps.
  Remember that if we take 
 $\nu=\mu+m$, we replace 
 $\psi_{-(\mu+m)-1,\pm n}$ by 
$\Omega_{-(\mu+m)-1,\pm(n-m)}$ with $n>m$ when $\mu\rightarrow-1/2$ in each of the relations 
that we have in the following,  see Sec. \ref{Mirror}.

Consider a system generated on the base of the Darboux-dual schemes 
\begin{equation}
(\psi_{\nu,2},\psi_{\nu,3})\sim(\psi_{\nu,-0},
\psi_{\nu,-1},\psi_{\nu,-2},\psi_{-\nu-1,-2},\psi_{\nu,-3},\psi_{-\nu-1,-3})\,.
\end{equation}
Here,   $n_-=2$, $n_+=6$,
  $n_{n_{+}}=n_{n_{-}}=3$ and $n_-+n_+=2(n_{n_+}+1)=8=
  2\Delta E$.
The positive scheme, whose Wronskian is
 given explicitly in (\ref{W(nu2nu3)}), corresponds to the Krein-Adler  scheme that    
provides us with the system 
\begin{eqnarray}
\label{DeformedA-K}
&L_{(+)}=-\frac{d^2}{dx^2}+V_{(+)}(x)\,,&
\end{eqnarray}  
whose potential $V_{(+)}$ is plotted in 
Figure \ref{Potential2}. The spectrum of the system, 
$\mathcal{E}_{\nu,0}=2\nu+3$, $\mathcal{E}_{\nu,1}=2\nu+7$,
$\mathcal{E}_{\nu,n}=2\nu+4(n+2)+3$, $n=2,\ldots$, 
is characterized by the presence of  the gap of the size $3\Delta E=12$, 
which appears between the first and second excited states. 
The negative scheme generates the shifted Hamiltonian  operator
$L_{(-)}=L_{(+)}-4\Delta E$.  
In terms of the intertwining operators $A_{(+)}^\pm$ and $A_{(-)}^\pm$
of the respective positive and negative schemes,  
the physical eigenstates of (\ref{DeformedA-K})
are given by
\begin{eqnarray}
\Psi_j&=&A_{(+)}^-\psi_{\nu,j}=A_{(-)}^-\widetilde{\psi}_{-\nu-1,j-3}\,,\qquad j=0,1\,,\\
\Psi_j&=&A_{(+)}^-\psi_{\nu,j+2}=A_{(-)}^-\psi_{\nu,j-2}\,,\qquad j=2,3,\ldots\,.
\end{eqnarray}

\begin{figure}[H]
\begin{center}
\includegraphics[scale=0.6]{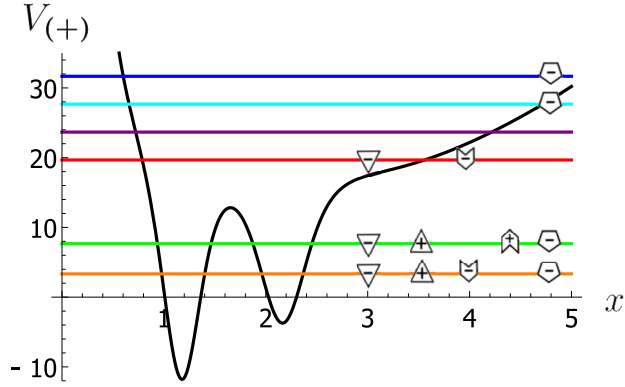} 
\caption[Behavior of the potential and physical states annihilated by 
the spectrum-generating ladder operators, Sec. 8.6 ]
{\small{The resulting potential with $\nu=1/3$ and energy levels of the system. 
The energy levels of the physical
states annihilated by the ladder operators $\mathcal{A}^-$, $\mathcal{A}^+$, $\mathcal{B}^-$, $\mathcal{B}^+$,
and  $\mathcal{C}^-$ are indicated from left to right.}} 
\label{Potential2}
\end{center} 
\end{figure} 
\noindent
The explicit form of the polynomials (\ref{Poly2}) for the system is
\begin{eqnarray}\label{Pn+y}
P_{n_+}(\eta)=(\eta-11-2\nu)(\eta-15-2\nu)\,,\\
P_{n_-}(\eta)=(\eta+9-2\nu)(\eta+13-2\nu)\prod_{i=0}^{3}(\eta+4n+3+2\nu)\,,\label{Pn-y}
\end{eqnarray}
and so,  $A_{(\pm)}^-A_{(\pm)}^+=P_{n_\pm}(\mathcal{H}_{\nu})$ and $A_{(\pm)}^-A_{(\pm)}^+=P_{n_\pm}(L_{(\pm)})$.

The spectrum-generating ladder operators are given by Eq.  (\ref{ladders}), 
and the nonlinearly deformed  conformal  algebras generated by each corresponding 
pair of the ladder operators and the Hamiltonian $L_{(+)}$  are obtained from 
(\ref{defsl2R})  by using polynomials (\ref{Pn+y}) and (\ref{Pn-y}). 
To clarify  physical nature of the ladder operators, one can 
 inspect their corresponding kernels by using  relations (\ref{tools2}) and (\ref{powerC}). 
 As a result, one gets that 
 the physical eigenstates annihilated by these operators are indicated in figure
\ref{Potential2} and all other functions in the respective kernels are nonphysical solutions.


\section{Remarks}

Note that the group $ K_4 $ can also be discussed in the context of Schr\"odinger equation 
\begin{equation}
\label{timedependent2}
\left(-\frac{\partial^2}{\partial x^2}+
\frac{\nu(\nu+1)}{x^2}\right)\psi=i\frac{\partial}{\partial t}\psi\,,
\end{equation}
the stationary solutions of which are 
$\Psi_\nu(x,t;\kappa)=\psi_\nu(x;\kappa)e^{-i\kappa^2 t}$, where $\psi_\nu(x;\kappa)=\sqrt{x}J_{\nu+\frac{1}{2}}(\kappa x)$ . The transformation
$\rho_2$ gives us the modified Bessel functions, besides  $\rho_1$ produces singular 
solutions when $\nu$ is not a half-integer number. In the case $\nu=\ell-1/2$ 
with $\ell=0,1,2\ldots\,,$ we have that 
$\rho_1(\psi_{\ell-1/2}(x,\kappa))=\sqrt{x}J_{-\ell}(\kappa x)=(-1)^{\ell}\psi_{\ell-1/2}(x,\kappa)$.


\chapter{Three-dimensional conformal mechanics  in a monopole background}
\label{Chapmono1}
The conformal algebra shown in  Chap. \ref{ChConformal} 
can be realized in higher-dimensional models. 
In the same sense, the conformal bridge is an algebraic construction,  
independent of the realization. This means that it also works for these higher-dimensional generalizations.

In this chapter, we will study a direct generalization of the AFF 
model in three dimensions, whose Hamiltonian corresponds to
\begin{equation}
\label{ClassicalH}
H=\frac{\vpi^2}{2m}+\frac{m\omega^2 r^2}{2}+\frac{\alpha}{2mr^2}\,, 
\end{equation}
where $\omega>0$, $\vpi=\vp-e\vA$,   $\vA$  
is a U(1) gauge potential  of a
Dirac magnetic monopole  at the origin 
with charge $g$, 
$\nabla\cross\vA=\vB=g\vr/r^3$, and the coupling
$\alpha$ should be chosen appropriately to prevent a fall to the center, see below.
We solve the Hamiltonian equations, 
study the  conformal Newton-Hooke symmetry
of the system, and investigate a hidden symmetry
which appears in  a special case   
 $\alpha=\nu^2$, $\nu=eg$. 
The results of this chapter 
are based on the article \textcolor{red}{[\cite{InzPlyWipf1}]} which was 
inspired by  the line of reasoning used in \textcolor{red}{[\cite{PlyWipf}]} to 
 identify the hidden symmetry
 and characterize the  particle's  trajectories. 
  \section{Classical case}  
\label{Classicalsection}

The particle's coordinates and  kinetic momenta obey the 
Poisson brackets  relations
$
\{r_i,\pi_j\}=\delta_{ij}\,,$ $\{r_i,r_j\}=0\,,$ and $\{\pi_i,\pi_j\}=e\epsilon_{ijk}B_{k}\,,$
which give rise to the equations of motion
\begin{equation}
\label{CanonEq}
\dot{\vr}=\frac{1}{m}\vpi\,,\qquad
\dot{\vpi}=\frac{1}{mr^3}(\alpha\vn-\nu\,\vr
\times\vpi)-m\omega^2\vr\,,
\end{equation}
where $\vn={\vr}/{r}$.
From  (\ref{CanonEq}) we  derive the  equations  
$
\frac{d r}{dt}=\frac{1}{m}\pi_r\,,$ and 
$\dot{\vn}=\frac{1}{mr^2}\,\vJ\cross\vn\,,$
where we denote $\pi_r=\vn\cdot\vpi$,  and 
\begin{equation}
\label{ClassicPoincare}
\vJ=\vr\cross\vpi-\nu\vn
\,
\end{equation}
is the conserved Poincar\'e vector  
identified as the 
angular momentum  of the system,
\begin{equation}
\{J_i,J_j\}=\epsilon_{ijk}J_k\,,\qquad
\{J_i,r_j\}=\epsilon_{ijk}r_k\,,\qquad
\{J_i,\pi_j\}=\epsilon_{ijk}\pi_k\,. 
\end{equation}
In terms of this conserved quantity the Hamiltonian can be presented in the form 
\begin{equation}\label{HAFF}
H=\frac{\pi_r^2}{2m}+\frac{\mathscr{L}^2}{2mr^2}+\frac{m\omega^2r^2}{2}\,,
\qquad\mathscr{L}^2
:=\vJ^2-\nu^2+\alpha\,,
\end{equation}
which reveals that the variables $r$ and $\pi_r$,  $\{r,\pi_r\}=1$, 
behave like $y$ and $p$ in the one-dimensional AFF model (\ref{mostgeneralH}). 
From (\ref{HAFF}) one also reads the following assertions: 
\begin{itemize}
\item There is no fall to the center
if $\mathscr{L}^2> 0$, i.e. $\alpha>0$,
 that we will assume from now on. 
 \item  The possible values of the angular  momentum  $J$ and energy 
 obey the relation
$
\frac{\mathscr{L}\omega}{H}\leq 1\,.$ 
\item The turning points 
for the radius are
\begin{equation}
\label{returning1}
r_{\pm}^2=\frac{H}{m\omega^2}(1\pm \rho)\,,\qquad 
0\leq \rho=\sqrt{1- \frac{\mathscr{L}^2\omega^2}{H^2}}<1\,,\qquad
r_+r_-=\frac{\mathscr{L}}{m\omega}\,.
\end{equation}
\end{itemize}
On the other hand,
 to solve the equations of motion it is worth parameterizing  $\vn$  as
\begin{eqnarray}
&
\label{n(phi)}
\vn(t)=
\vn_\parallel+\vn_{\bot}(t)
=-\nu\,\frac{\vJ}{J^2}
+\vn_{\bot}(t),\qquad \vJ\cdot\vn_{\bot}(t)=0\,,\qquad 
\vJ\cdot\vn=\vJ\cdot\vn_\parallel=-\nu. &\\&
\label{nparaort}
\vn_{\bot}(t)=\vn_{\bot}(0)\cos\varphi(t)
+\hvJ\cross\vn_{\bot}(0)\sin\varphi(t)\,.&
\end{eqnarray}  
From (\ref{nparaort}) and the equation of motion for $ \vn $ 
we get $ \dot{\varphi} = \frac{J}{mr^2} $. 
These relations involve a clockwise rotation of $ \vn_{\bot} $
from the perspective of  vector $ \vJ $. Thus, if $ \vJ $ is oriented along $ \ve_z $, and $
 \nu <0 $, $ 0 <\theta <\pi/2 $, where $ \theta = \arccos(- \nu / J ) $, the path of the particle
 is on the upper sheet of the cone and $ \vn_{\bot} $ rotates clockwise in the horizontal plane. 
If on the other hand  $ \vJ $ is oriented along $ - \ve_z $ and $ \nu> 0 $, $ \pi / 2 <\theta <\pi $,
 then the path is again on the upper sheet of the cone, but the vector $ \vn_{\bot} $ 
rotates counterclockwise in the $ (x, y) $ plane looking from $ \ve_z $.
We also note that when $J=\nu$, then $\theta=\pi$ so $\vn$ is co-linear 
to $\vJ$ and there is no rotation at all.  In the following we exclude that case. 

The corresponding solutions for the angular and radial variables are
\begin{eqnarray}
&
\label{r(t)1}
r^2(t)=\frac{H}{m\omega^2}(1- \rho\cos(2\omega t))\,, 
\qquad
\label{phi(t)}
\varphi(t)=\frac{J}{\mathscr{L}} \arctan(\frac{r_\mathrm{max}}{r_\mathrm{min}}\tan(\omega t))\,,&
\end{eqnarray} 
where the initial conditions $r(t=0)=r_-:=r_\text{min}$ and 
$\varphi(t=0)=0$ are assumed (also we redefine $r_+:=r_\text{max}$).  
By expressing time thought $\varphi(t)$ and introducing it 
in to  $r^2(t)$,
we obtain the trajectory equation 
\begin{eqnarray}
\frac{1}{r^2(\varphi)}=\frac{mH}{\mathscr{L}^2}\left[1+\gamma\cos(\frac{2\mathscr{L}}{J}\varphi)\right]\,,
\end{eqnarray}
which shows us that
the angular 
period is $\pi J/\mathscr{L}$. 
The condition for a periodic trajectory is
\begin{equation}
\frac{2\mathscr{L}}{J}2\pi l_r=2\pi l_a\quad
\Longleftrightarrow
\quad
\frac{2\mathscr{L}}{J}=\frac{l_a}{l_r},
\qquad l_r,l_a=1,2,\ldots\,.
\end{equation}
From the definition of $\mathscr{L}$  in (\ref{HAFF}) 
we find  that the trajectories are closed for arbitrary 
values of $J$ if  and only if  
$\alpha=\nu^2$. 
On the other hand, when $\alpha\neq \nu^2$, 
the trajectory will be closed only for 
special values of the angular momentum given by
the condition
\begin{equation}
\label{alpha nu J}
\alpha=\nu^2
+\left(\frac{1}{4}\frac{l_a^2}{l_r^2}-1\right)J^2\,,
\end{equation}
 and in this case the condition 
$\frac{\mathscr{L}\omega}{H}\leq 1\,$ takes the form 
$\frac{l_a}{l_r}\leq \frac{2H}{\omega J}$.
Figure \ref{figure1} illustrates several particular orbits
lying on the corresponding conical surface 
in a general case $\alpha\neq \nu^2$ and 
in the special case $\alpha= \nu^2$.
Trajectories  $\vr(\varphi)$ are shown there 
for fixed values of $H$, $\vJ$ and $\nu$, but for different 
 values of $\alpha$. 
\begin{figure}[H]
\begin{center}
\includegraphics[scale=0.350]{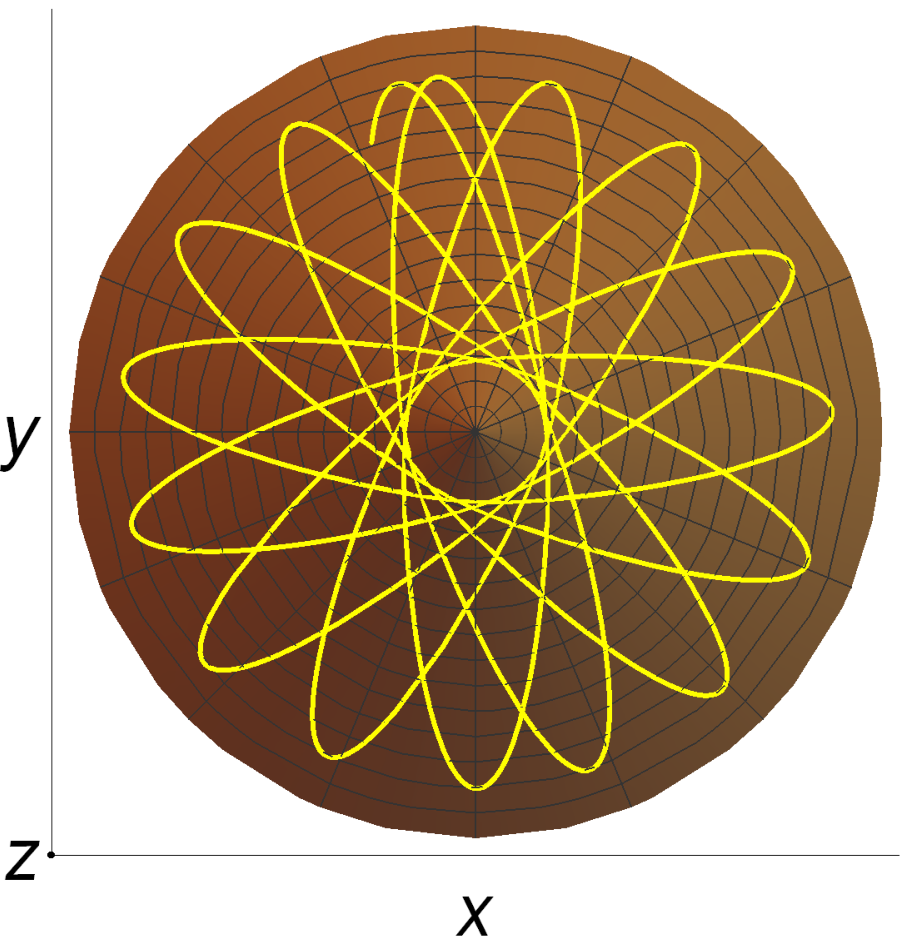}\hskip8mm
\includegraphics[scale=0.350]{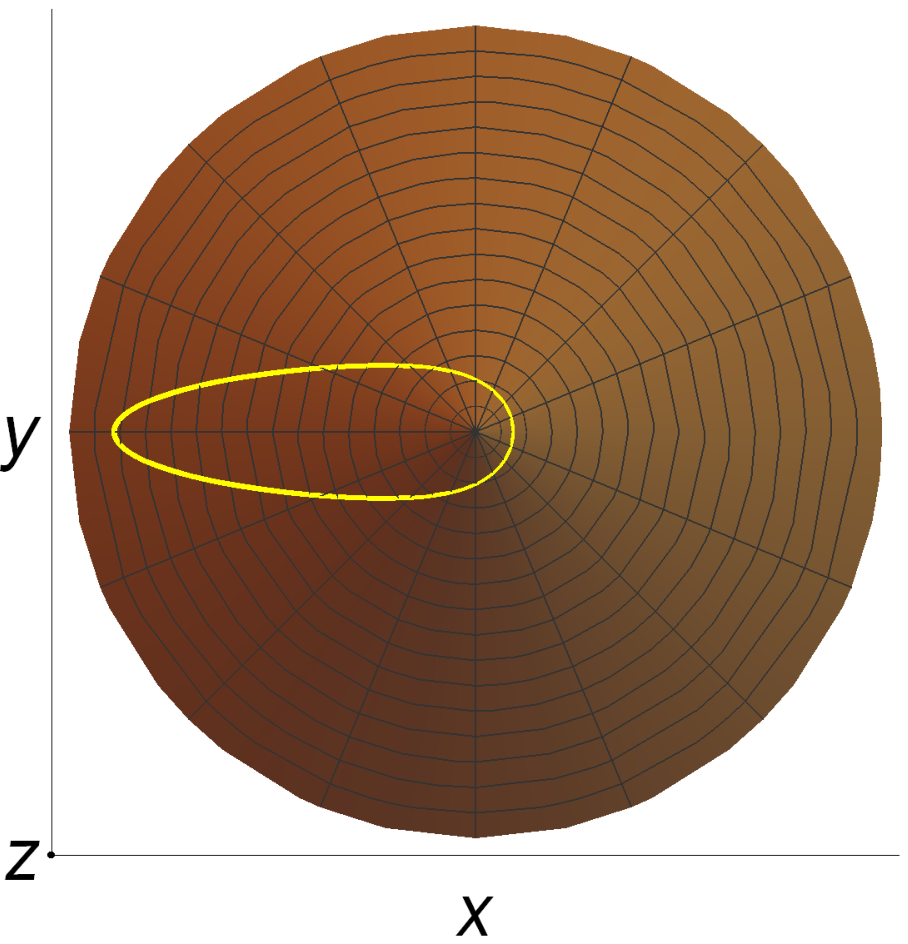}\hskip8mm
\includegraphics[scale=0.350]{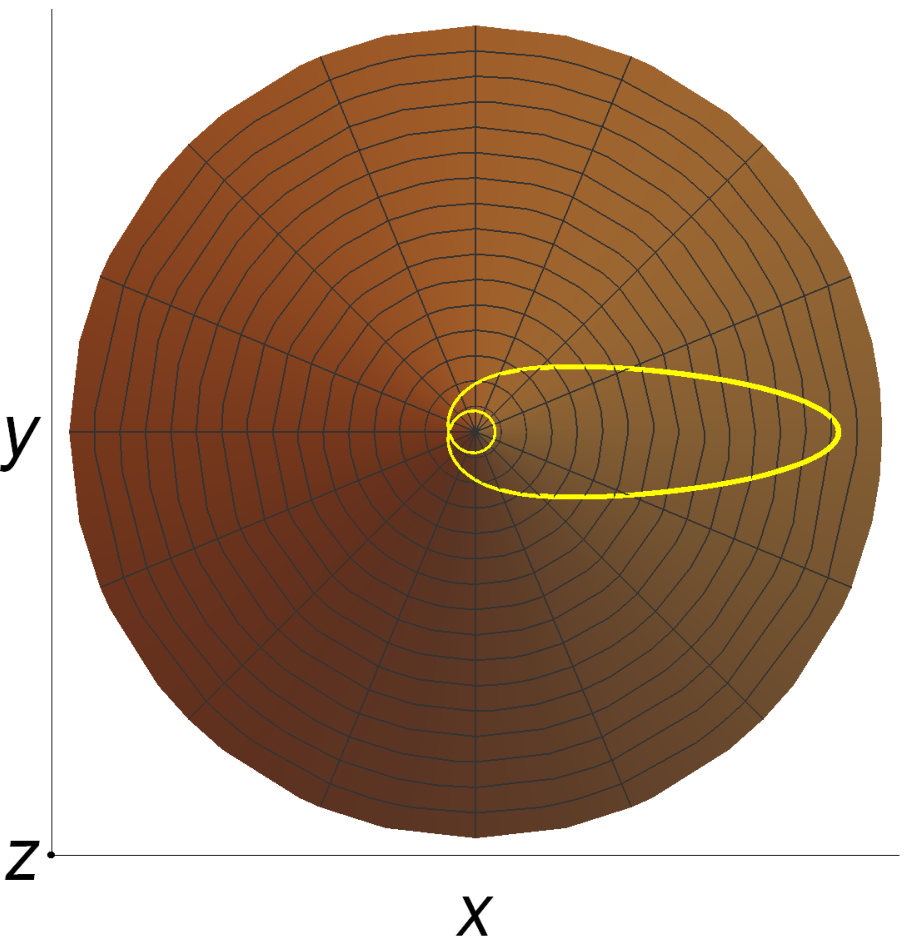}\hskip8mm\\
\includegraphics[scale=0.350]{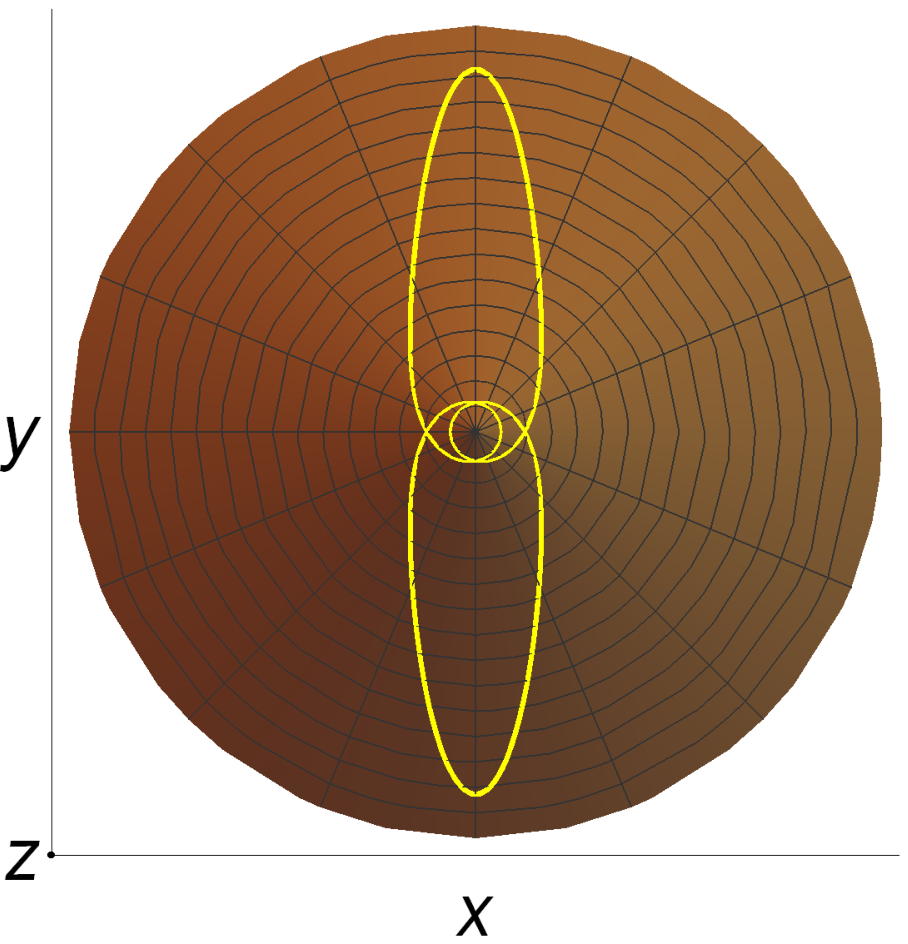}\hskip8mm
\includegraphics[scale=0.350]{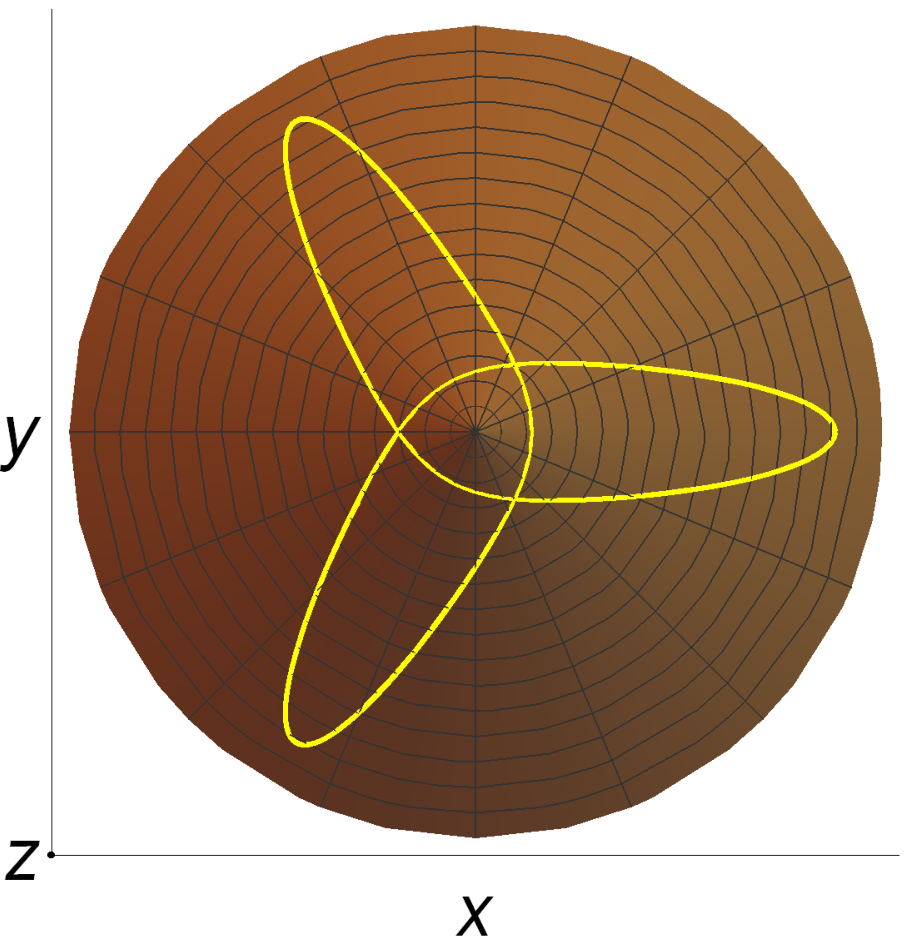}\hskip8mm
\includegraphics[scale=0.350]{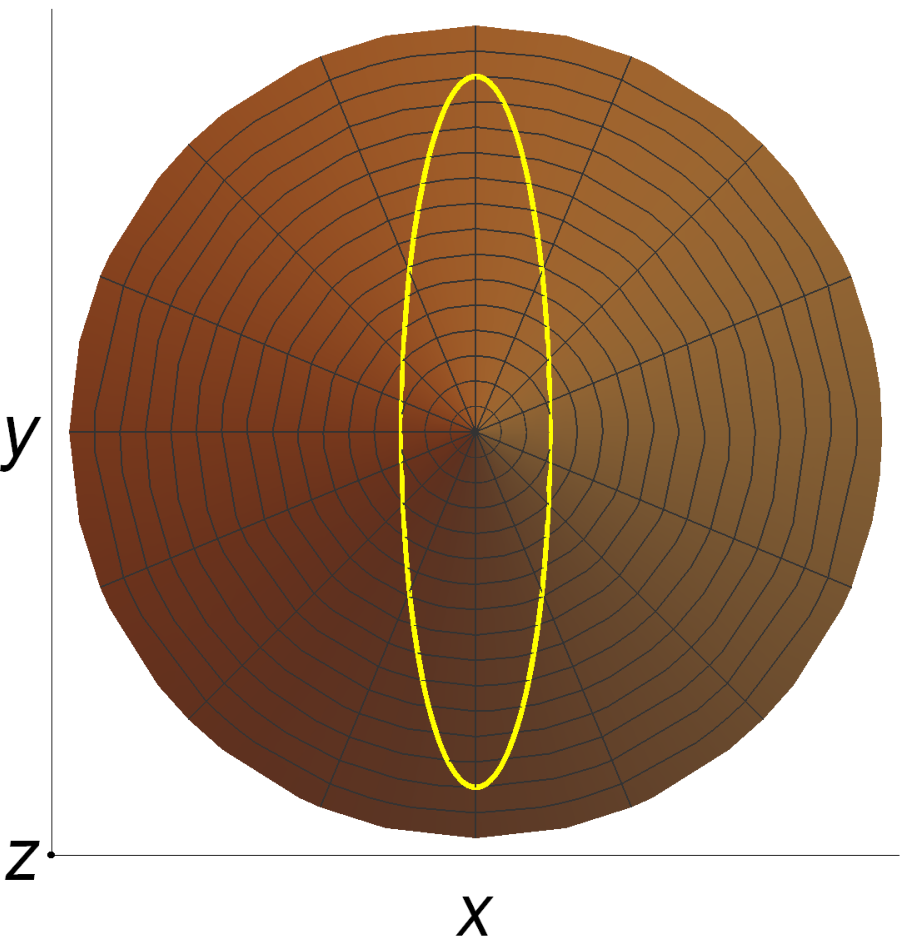}
\caption[Classical trajectories 1, Sec. 9.1]{\small{
The depicted trajectories  correspond to the vector 
$\vJ$ oriented along $\ve_z$.
The first figure in the top row represents the generic case with
non-closed trajectory. 
The other figures are examples of closed trajectories
with parameters satisfying  the relation  (\ref{alpha nu J}),
with quotients  $l_a/l_r=\{1,\,1/2,\,2/3,\,3/2,\,2\}$ are sequentially shown.  
The last}
relation  $l_a/l_r=2$ corresponds to  the special case $\alpha=\nu^2$. }
\label{figure1}
\end{center} 
\end{figure}
\vskip-0.7cm
Below we shall see that when  $\alpha=\nu^2$,
the projection to the plane  orthogonal to $\vJ$ 
 of the trajectory shown on the last 
plot  is an ellipse centered at 
the origin of the coordinate system 
similarly to the case of the three-dimensional 
isotropic harmonic oscillator.
This corresponds to a fundamental universal 
property   of the magnetic monopole background which
we discuss in the last section.
Since the center of the projected elliptical trajectory is
in the center of an ellipse, the angular period $P_a$ is twice 
the radial period $P_r$, $P_a/P_r=2$,  similarly to the
isotropic harmonic oscillator. This is different from the 
picture of the finite orbits in Kepler problem
where the force center is in one of the foci, and as a result
$P_a=P_r$. This similarity with the isotropic oscillator 
and contrast to the Kepler problem are also reflected
in the spectra of the systems at the quantum level.

As we have the AFF model form of the Hamiltonian in (\ref{HAFF}), 
we can intermediately write the rest of the Newton-Hooke 
conformal algebra generators. They 
are given by 
\begin{eqnarray}
\label{KyD1}
&\mathcal{D}=\frac{1}{2}\left(rp_r\cos(2\omega \tau)+\left(m \omega  r^2-
H{\omega}^{-1}\right)\sin(2\omega \tau)\right)\,,&\\
&\mathcal{K}= \cos(2\omega \tau)m\frac{r^2}{2}-\frac{H}{\omega^2}\sin^2(\omega\tau)
-\frac{\sin(2\omega\tau)}{2\omega}rp_r\,.\label{KuD2}&
\end{eqnarray}
Together with $H$ they satisfy the algebra 
(\ref{sl2RAFF}). The Casimir invariant corresponds to $\mathscr{F}=\frac{\mathscr{L}^2}{4}$. 

To conclude this part of the  analysis,
we comment on  the limit 
$\omega\rightarrow 0$. In this case the generators $H$, $D$ and $K$
take the form 
\begin{equation}
\label{freemotion}
H_0=\frac{\pi_r^2}{2m}+\frac{\mathscr{L}^2}{2mr^2}\,,\qquad
D_0=\frac{1}{2}r\pi_r-H_0t\,,\qquad
K_0=\frac{mr^2}{2}-Dt-H_0t^2\,,
\end{equation}
and  satisfy the conformal algebra. 

The case $\alpha=0$ of the system $H_0$ 
corresponds  to a geodesic motion on the dynamical cone \textcolor{red}{[\cite{Plymono1,Plymono2}]}.
The special case of $\alpha=\nu^2$, on the other hand,  was studied in
\textcolor{red}{[\cite{PlyWipf}]}. It was shown there that 
the trajectory of the particle, projected 
 to the plane orthogonal to
$\vJ$, is  a straight line along which the projected particle's motion 
takes place with constant velocity. Consistently with these
peculiar properties,  in the special case 
$\alpha=\nu^2$ the system with $H_0$ 
possesses a hidden symmetry  described by 
the integral of motion $\vV=\vpi\times\vJ$ being a sort of 
Laplace-Runge-Lenz vector, in the plane orthogonal to which 
and parallel to $\vJ$ 
the particle's trajectory lies  \textcolor{red}{[\cite{PlyWipf}]}.
In Fig. \ref{figure2Cono} some plots of the trajectories are shown 
for the 
system (\ref{freemotion}). 
\begin{figure}[H]
\begin{center}
\includegraphics[scale=0.350]{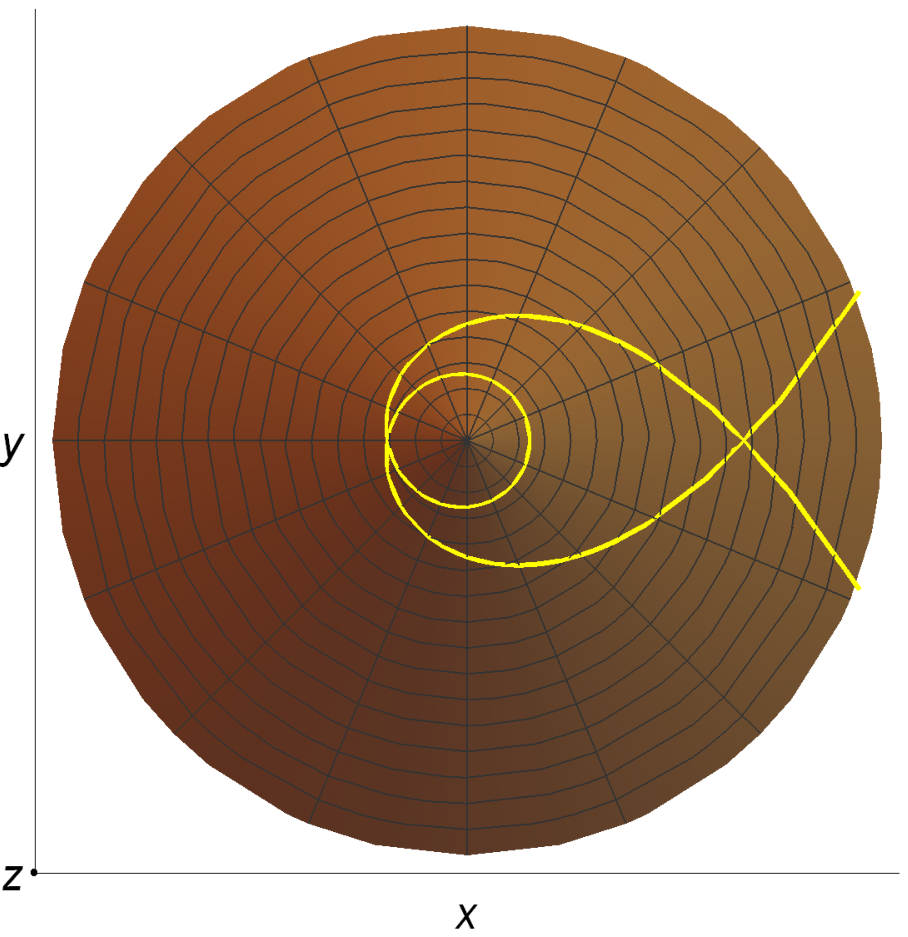}\hskip8mm
\includegraphics[scale=0.350]{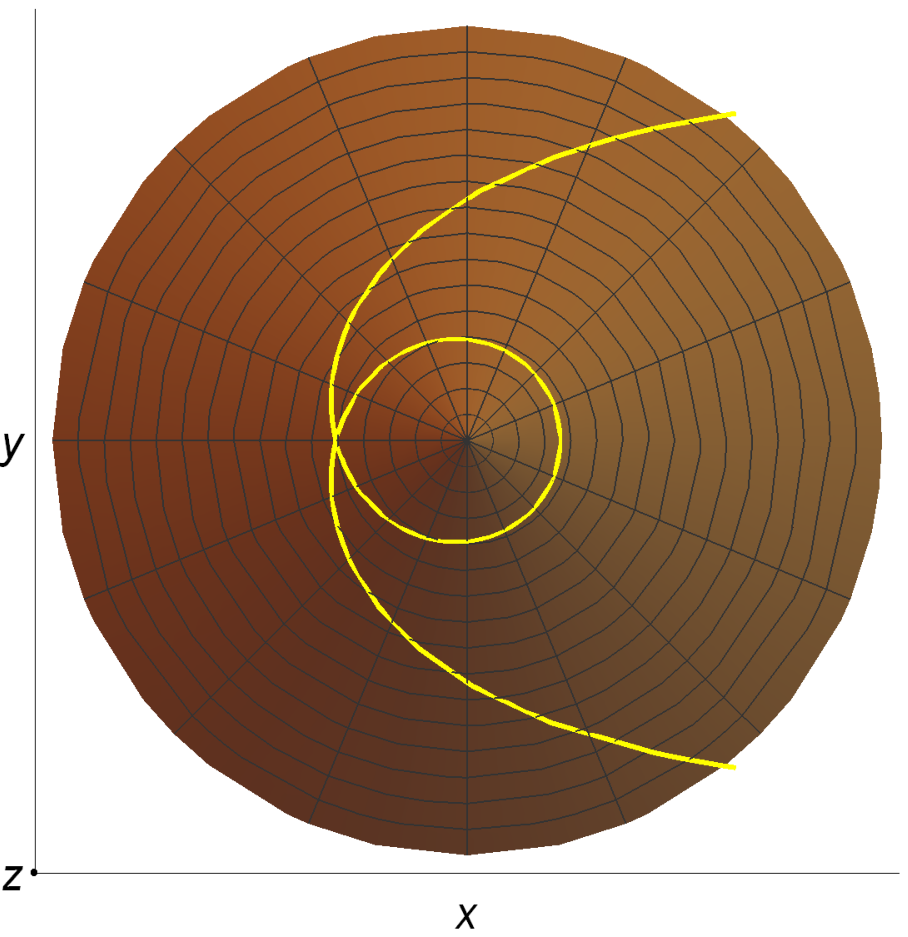}\hskip8mm
\includegraphics[scale=0.350]{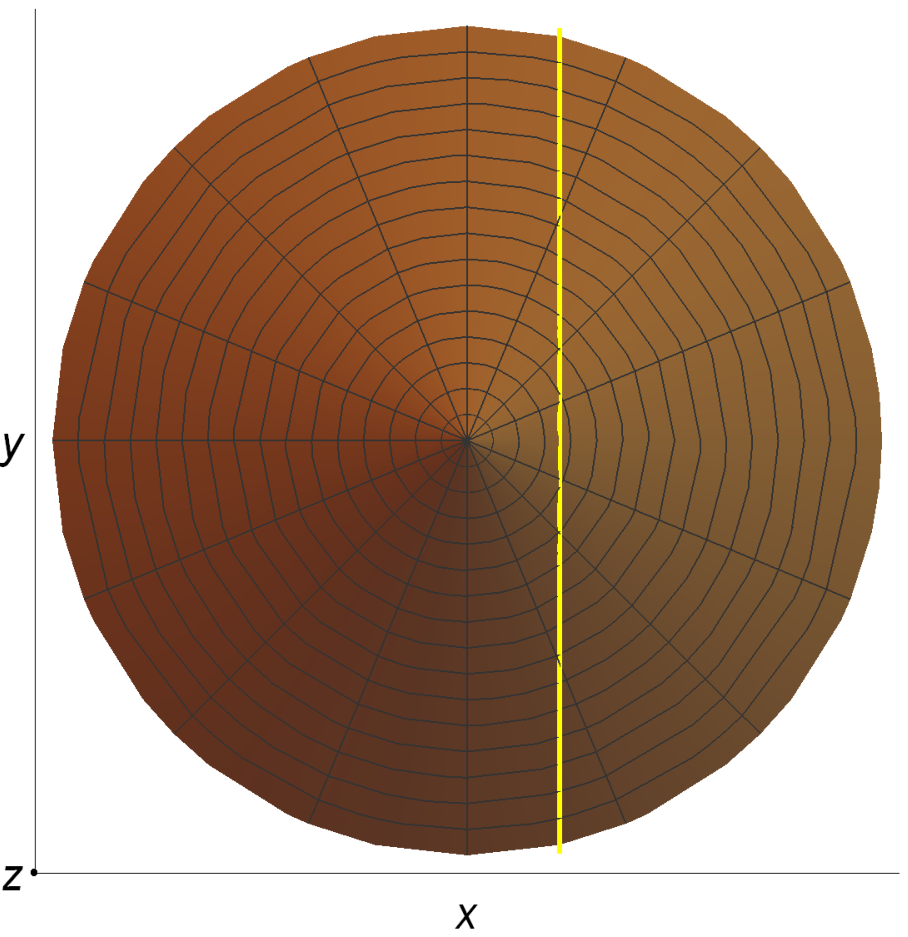}
\caption[Classical trajectories 2, Sec. 9.1]{\small{Each plot represents a trajectory for a specific  value of 
$\alpha$ chosen according to (\ref{alpha nu J}) 
with the vector $\vJ$ oriented along  $\ve_z$. 
From left to right the cases $l_a/l_r=\{3/2,\,1/2,\,2\}$ are shown, 
where the last plot 
corresponds to the special case  $\alpha=\nu^2$.} }
\label{figure2Cono}
\end{center} 
\end{figure}


\subsection{
The case $\alpha=\nu^2$\,: hidden symmetries}\label{hidclassym}

In the case $\alpha=\nu^2$ the particle described
by the Hamiltonian (\ref{ClassicalH}), which 
now is 
\begin{equation}
\label{ClassicHnu}
H=\frac{\vpi^2}{2m}+\frac{m\omega^2}{2}r^2+\frac{\nu^2}{2mr^2}\,,
\end{equation}
 admits
the vector  integrals  of motion 
responsible for the closed nature of the trajectories 
for arbitrary choice of initial conditions. 
The integrals  are derived by an algebraic
approach  as in Fradkin's construction 
for the isotropic  three-dimensional harmonic oscillator 
\textcolor{red}{[\cite{Frad}]}.  

Let us first introduce the vector quantities  
\begin{align}
\vI_1&=\vpi\cross\vJ\cos(\omega t)+\omega m \vr\cross\vJ\sin(\omega t)\,,
\label{I1}\\
\vI_2&=\vpi\cross\vJ\sin(\omega t)-\omega m \vr\cross\vJ\cos(\omega t)\,.
\label{I2}
\end{align}
Using the corresponding equations of motion for $\vr$ 
and $\vpi$  
is not difficult to show that $\dot{I_i}=0$ so they are 
dynamical integrals of motion.

The evaluation of these integrals 
in the initial conditions give us  
\begin{equation}\label{I(0)}
\vI_1(0)=\frac{J^2}{r_\mathrm{min}}
\vn_{\bot}(0)\,,\qquad
\vI_2(0)=m\omega r_\mathrm{min}
\vJ\cross\vn_{\bot}(0)\,,
\end{equation}
thus,  $\vI_1$ and $\vI_2$   are orthogonal to each other. 
On the other hand, the lengths of these vectors 
are also dynamical integrals which for the initial 
conditions take the form 
\begin{equation}
\vert\vI_{1}\vert
=m\omega\sqrt{J^2-\nu^2}\,r_\mathrm{max},\qquad
\vert\vI_{2}\vert
=m\omega\sqrt{J^2-\nu^2}\,r_\mathrm{min}\,,\label{lengthint}
\end{equation}
where we have taken into account Eqs. (\ref{I(0)}) and
the second equation in  (\ref{returning1}).
The sum of their squares, however, is a true integral of motion 
whose value is a function of $H$ and $J$,
\be\label{I12+I22}
\vI_1^2+\vI_2^2=2mH(J^2-\nu^2)\,.
\ee
These vectors point
in the direction of the semi-axes of the elliptic trajectory
in the plane orthogonal to $\vJ$.  
The  lengths of semi-major and semi-minor axes 
correspond to  those of the vectors $r\vn_{\bot}(0)$ and 
$r\hvJ\cross \vn_{\bot}(0)$,  and are equal to 
$r_\mathrm{max}\sqrt{1-\nu^2/J^2},$ and  
$r_\mathrm{min}\sqrt{1-\nu^2/J^2}$.
As it is shown in \textcolor{red}{[\cite{InzPlyWipf1}]},
in  a general case of 
$\alpha\neq \nu^2$, the periodic change  of the scalar
product of $\vI_1$ and $\vI_2$, which would not 
be integrals,  
implies a precession of the orbit, see 
Fig. \ref{figure1}.

Using the definition of  $\vI_1$ and $\vI_2$ in (\ref{I1})
and (\ref{I2}), 
we can express the position $\vr(t)$ of the particle
as follows,  
\begin{equation}
\label{r(t)2}
\vr(t)=\frac{1}{m\omega J^2}\left( \vJ\cross \vI_{1}\sin \omega t-\vJ\times\vI_{2}\cos \omega t
-\nu \frac{\sqrt{I_1^2\sin^2\omega t+I_2^2\cos^2\omega t}}{\sqrt{J^2-\nu^2}}\vJ\right)\,,
\end{equation} 
where we again see that $\vI_{1}=\vI_1(0)$ and  
$\vI_{2}=\vI_2(0)$
correspond to the orthogonal set that define the elliptic trajectory 
in the plane.
\vskip0.1cm

Alternatively, one can follow a more algebraic approach to 
extract information on the trajectories without explicitly 
solving the equations of motion.
It is well known from the seminal paper  \textcolor{red}{[\cite{Frad}]}
that for the three-dimensional isotropic harmonic oscillator all 
symmetries of the trajectories are encoded in a tensor integral 
of motion. During the rest of this 
subsection
we construct 
an analogous tensor for the system at hand to find 
the
trajectories by a
linear algebra techniques.
We begin with the  tensor integrals
\begin{equation}
T^{ij}=T^{(ij)}+T^{[ij]}\,,\qquad
T^{(ij)}=\frac{1}{2}(I_{1}^{i}I_1^j+I_{2}^i I_2^j)\,,\qquad
T^{[ij]}=\frac{1}{2}(I_{1}^{i}I_2^j-I_{1}^j I_2^i)\,.\label{symm_tensor}
\end{equation}
They,  unlike the vectors $\vI_1$ and $\vI_2$,  but like
the quadratic expression (\ref{I12+I22}) are the true,
not depending explicitly on time integrals of motion,
$\frac{d}{dt}T^{ij}=\{T^{ij},H\}=0$.

In accordance with (\ref{I12+I22}),
their  components satisfy relations
\begin{equation}
\text{tr}(T)= m(J^2-\nu^2)H\,,\qquad
\epsilon_{ijk}T^{[jk]}=m\omega(J^2-\nu^2)J_{i}\,.
\end{equation}
As the anti-symmetric part of $T^{ij}$ is related to the Poincar\'e integral, 
we only need to use the symmetric part $T^{(ij)}$, which 
is related but not identical to Fradkin's tensor.
Since the vectors (\ref{I1}), (\ref{I2}) are orthogonal to each other and
to $\vJ$,  we immediately conclude that
$\vJ,\vI_1$ and $\vI_2$ are eigenvectors of $T^{(ij)}$ with eigenvalues equal,
respectively,
 to zero and 
\begin{align}
\lambda_1=\vert\vI_1\vert^2&=\frac{1}{2}m^2\omega^2(J^2-\nu^2)r^2_\mathrm{max}\,,\\
\lambda_2=\vert\vI_2\vert^2&=\frac{1}{2}m^2\omega^2(J^2-\nu^2)r^2_\mathrm{min}\,,
\label{eigenvalues}
\end{align}
Also one can show 
that the quadratic form $\vr^T T\vr$
is time-independent,
\begin{equation}
2r_i T^{ij}r_j=(\vI_1\cdot\vr)^2+(\vI_2\cdot\vr)^2=(J^2-\nu^2)^2\,.
\label{quad_form}
\end{equation}
In a coordinate system with orthonormal base $\ve_x=\hat{\vI}_1,\ve_y=\hat{\vI}_2$ and
$\ve_z=\hat{\vJ}$, the quadratic form (\ref{quad_form})
simplifies to
\begin{equation}
\lambda_1 x^2+\lambda_2 y^2=(J^2-\nu^2)^2\,.
\end{equation}
With $r_\mathrm{max}r_\mathrm{min}=J/(m\omega)$
one ends up with the equation for an ellipse
in the plane orthogonal to $\vJ$:
\begin{equation}
\frac{x^2}{r^2_\mathrm{min}}+\frac{y^2}{r^2_\mathrm{max}}=\frac{J^2-\nu^2}
{J^2}
\,.
\end{equation}
The lengths of the semi-major axis  and semi-minor axis  of
the ellipse are 
$r_\mathrm{max}\sqrt{1-\nu^2/J^2}$ and 
$r_\mathrm{min}\sqrt{1-\nu^2/J^2}$,
in accordance with that was found above.

Finally, the symmetric tensor components integral $T_{(ij)}$ satisfy 
the Poisson bracket relations
\begin{eqnarray}
&
\{J_i,T_{(jk)}\}=\epsilon_{ijl}T_{(lk)}+\epsilon_{ikl}T_{(jl)}\,, &\\&
\{T_{(ij)},T_{(lk)}\}=m(\epsilon_{ils}\mathcal{F}_{jk}+\epsilon_{iks}\mathcal{F}_{jl}+
\epsilon_{jls}\mathcal{F}_{ik}+\epsilon_{jks}\mathcal{F}_{im})J_s\,,&
\end{eqnarray}
where $
\mathcal{F}_{ij}=\tfrac{1}{4}m\omega^2(J^2-\nu^2)^2\delta_{ij}-HT_{(ij)}\,.
$

In fact, the quantum version of the tensor $T_{(ij)}$ was already considered in 
\textcolor{red}{[\cite{Vinet}]}, but  this is the first time that it has been obtained and used at the classical level.


\section{Quantum theory of the model with  $\alpha=\nu^2$}
\label{Quantumsection}
The$\quad$ quantum$\quad$ theory$\quad$ of$\quad$ the$\quad$ system$\quad$ with$\quad$ Hamiltonian$\quad$
(\ref{ClassicHnu}) 
is discussed in details 
in \textcolor{red}{[\cite{Mcin,Vinet,InzPlyWipf1}]} and here we summarize the results.
We shall use the units in which $m=1$ and $\hbar=1$.
\vskip0.1cm

In coordinate representation the basic 
commutation relations  are 
\begin{equation}
[\hat{r}_i,\hat{r}_j]=0\,,\qquad [\hat{r}_i,\hat{\Pi}_j]=i\delta_{ij}\,,\qquad
[\hat{\Pi}_i,\hat{\Pi}_j]=i\nu\epsilon_{ijk}\frac{\hat{r}_k}{r^3}\,.
\end{equation}
In what follows we shall skip the hat symbol $\,\hat{{}}\,\,\,$
to simplify the notation.  
The Hamiltonian 
(\ref{ClassicHnu}) can be written as 
\begin{equation}
\label{Qm Hamiltonian}
H=\frac{1}{2}\left[-\frac{1}{r^2}\frac{\partial}{\partial r}\left(r^2\frac{\partial}{\partial r}\right)+
\frac{1}{r^2}\vJ\!\,^2+\omega^2 r^2\right]\,,
\end{equation} 
where $\vJ$ is just the quantum version of the Poincar\'e integral (\ref{ClassicPoincare}), 
the components of which generate the 
$\mathfrak{su}(2)$ 
symmetry. 
The Dirac quantization condition 
implies that $\nu=eg$ must take an integer or half integer 
value \textcolor{red}{[\cite{Plymono1,Plymono2}]}.
Using the angular momentum treatment we obtain  
\begin{equation}
\label{so(3) rep1}
{\vJ}\!\,^2\mathcal{Y}_{j}^{j_3}=j(j+1)\mathcal{Y}_{j}^{j_3}\,,\quad
 J_3\mathcal{Y}_{j}^{j_3}=j_3\mathcal{Y}_{j}^{j_3}\,,\quad 
J_\pm\mathcal{Y}_{j}^{j_3}=c_{jj_3}^\pm\mathcal{Y}_{j}^{j_3\pm1}\,, 
\end{equation}
with $J_\pm=J_1\pm i J_2$, 
and 
\begin{equation}
j=|\nu|,|\nu|+1,\ldots\,,\qquad 
 j_3=-j,\ldots,j\,,\qquad
c_{jj_3}^\pm=\sqrt{(j\pm j_3+1)(j\mp j_3)}\,,
\end{equation}
where the indicated values for $j$ correspond to a super-selection rule. 
The case $\nu=0$ corresponds just to the quantum harmonic  isotropic oscillator.
Excluding the zero value for $\nu$, i.e. implying that $\vert \nu\vert$ takes
any nonzero integer or half-integer value,
the first relation in (\ref{so(3) rep1}) automatically 
provides the  necessary inequality  $\vJ^2=j(j+1)>\nu^2$.
The functions $\mathcal{Y}_{j}^{j_3}=\mathcal{Y}_{j}^{j_3}(\theta,\varphi;\nu)$
are the (normalized) monopole harmonics \textcolor{red}{[\cite{monoharm,Lochak,Plymono1,Plymono2}]},
which are well defined functions if and only if the combination $j\pm\nu$ 
is in $\N_0=\{0, 
1,2,\ldots\}$.

Then, the eigenstates and the spectrum of $H$ are given by 
\begin{align}
\psi_{n,j}^{j_3}(\vr
)&=f_{n,j}(\sqrt{\omega}r)
\mathcal{Y}_j^{j_3}(\theta,\varphi 
)\,,\nonumber
\\
f_{n,j}(x)&=\bigg(\frac{2n!}{\Gamma(n+j+3/2)}\bigg)^{1/2}\omega^{3/4}
\,x^{j}L_{n}^{(j+1/2)}(x^2)\,e^{-x^2/2}\,,\label{wavefunction}\\
 E_{n,j}&=\Big(2n+j+\tfrac{3}{2}\Big)\,\omega\,,\nonumber
\end{align}
where $L_{n}^{(j+1/2)}(y)$ are the generalized Laguerre polynomials. 
The degeneracy of
each level depends on $\nu$ and corresponds to   
\begin{eqnarray}
\label{degeneracy}
\mathfrak{g}(\nu,N)=
\left\{
\begin{array}{ccc}
\tfrac{1}{2}(N+\nu+1)(N-\nu+2)\,, & j-\nu &\text{even 
} \\
\\
\tfrac{1}{2}(N-\nu+1)(N+\nu+2)\,, & j-\nu &\text{odd} 
\end{array}
\right.\,,\qquad N=2n+j\,.
\end{eqnarray}

It is remarkable that
the system possesses  
$2\vert \nu \vert +1$  
degenerate ground states.
The ground states here are not invariant
under the 
action of the
total angular momentum $\vJ$, although the
Hamiltonian  operator commutes with $\vJ$ and hence is
spherically symmetric. Thus we see some analog of
spontaneous breaking of rotational symmetry in
the  
magnetic monopole
background.
This is of course in contrast to the isotropic harmonic oscillator 
in three dimensions which has a unique spherically symmetric
ground state and symmetry algebra $\mathfrak{su}(3)$. 
According to \textcolor{red}{[\cite{Vinet}]} the symmetry algebra for
the system under investigation is
$\mathfrak{su}(2)\oplus\mathfrak{su}(2)$. We do not further
dwell on these interesting aspects of symmetry but 
rather turn to the construction of spectrum-generating ladder operators.

Note that the coefficients at radial, $n$,
and angular momentum, $j$, quantum numbers
in the energy eigenvalue  $E_{n,j}=(2n+j+\frac{3}{2})\,\omega$ corresponds to  
the ratio  $P_a/P_r=l_a/l_r=2$ 
between the classical angular and radial periods 
in the special case $\alpha=\nu^2$ under investigation.
This can be compared with the structure of the
principle quantum number $N=n_r+l+1$
defining the spectrum 
in the quantum model of the hydrogen atom,
where the corresponding classical periods are equal.

The explicit wave functions in (\ref{wavefunction}) are
specified by the discrete quantum numbers $n$, 
$j$ and $j_3$.
Our target now  is to identify the 
 ladder operators for radial, $n$, and 
 angular momentum, $j$, quantum numbers
  (we already have the ladders 
operators for $j_3$),  which are 
based on the conformal  and hidden symmetries of the system. 

In the algebraic approach we do not fix the representation
for the position and momentum operators and thus use
Dirac's ket notation  
for eigenstates.
\vskip0.2cm

\noindent
\emph{Ladder operators for $n$.} Let us first consider 
the quantum version of the $\mathfrak{sl}(2,\R)$ symmetry, 
\begin{equation}
\label{Qsl2r}
[H,\cC]=-2\omega\cC\,,\qquad 
[H,\cC^\dagger]=2\omega\cC^\dagger
\,,\qquad
[\cC,\cC^\dagger]=4\omega H\,,
\end{equation}
where the generators $\cC,\cC^\dagger$ 
are the quantum versions of combinations of Newton-Hooke
symmetry generators
in the Schr\"odinger picture at $t=0$, i.e.,
\begin{equation}
\label{ladern}
\cC=H-\omega^2r^2- \frac{i\omega}{2}(\vr\cdot\vpi+\vpi\cdot\vr)\,,
\end{equation} 
and their action on the eigenstates is 
\begin{eqnarray}
&
\cC\ket{n,j,j_3}=\omega\, d_{n,j}\ket{n-1,j,j_3}\quad,\quad
\cC^\dagger\ket{n,j,j_3}=\omega\, d_{n+1,j}\ket{n+1,j,j_3}\,,&\\&
d_{n,j}=\sqrt{2n(2n+2j+1)}\,.\label{d-coeff}&
\end{eqnarray}

\emph{Ladder operators for $j$.} We introduce the
complex vector operator 
\begin{equation}
\va=\frac{1}{2}(\vb\cross\vJ-\vJ\cross\vb)=(\vb\cross\vJ-i\vb)\,,\qquad
\vb=\frac{1}{\sqrt{2}}
(\vpi-i\omega m \vr)\,,
\label{b}
\end{equation}
together with its Hermitian conjugation. 
The vector operator $\va$ is the quantum version of the 
complex classical quantity $\frac{1}{\sqrt{2}}
(\vI_1+i\vI_2)$ in 
Schr\"odinger picture at $t=0$,
and its components satisfy the relations 
\begin{eqnarray}
&\label{comHaJa}
[H,a_{i}]=-\omega a_i\,,\qquad [J_i,a_{j}]=i\epsilon_{ijk}a_{k}\,,\qquad
[a_{i},a_j]=-i\epsilon_{ijk}\cC J_k\,,
&\\&
\label{comaa}
[a_{i}^\dagger,a_{j}]=-\omega[(2\vJ^2+1-\nu^2)\delta_{ij}-J_{i}J_j)]-iH\epsilon_{ijk}J_k\,,
&
\end{eqnarray} 
The action of these operators is computed algebraically in
\textcolor{red}{[\cite{InzPlyWipf1}]} and for us 
 is sufficient to consider  $a_3$ and $a_3^\dagger$
and their actions on the ket-states
\begin{align}
\label{eta on psi1}
a_3\ket{n,j,j_3}&=A_{n,j,j_3}\ket{n,j-1,j_3}+B_{n,j,j_3}\ket{n-1,j+1,j_3}\,,
\\
a_3^\dagger\ket{n,j,j_3}&=A_{n,j+1,j_3}\ket{n,j+1,j_3}+B_{n+1,j-1,j_3}\ket{n+1,j-1,j_3}\,,
\end{align}
where the squares of the positive coefficients are
\begin{align}
\label{Anlm1}
\big(A_{n,j,j_3}\big)^2&=\omega(2n+2j+1)\,
	\frac{(j^2-j^2_3)(j^2-\nu^2)}{(2j)^2-1}\,, &
	\big(B_{n,j,j_3}\big)^2&=\frac{2n}{2n+2j+3}\big(A_{n,j+1,j_3}\big)^2\,. 
\end{align}
We see that the operators $a_3$ and $a_3^\dagger$
change the quantum numbers $n$ and $j$, but the result is a 
superposition of the two 
eigenstate 
vectors. Their 
action
is depicted in Fig. \ref{figure2Esqueme}.
\begin{figure}[H]
\begin{center}\includegraphics[scale=0.3]{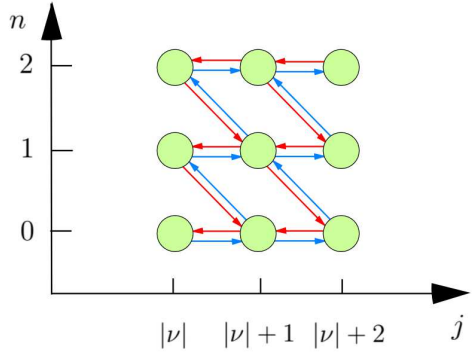}\hskip5mm
\caption[Angular ladder operators action, Sec. 9.2]{\small{The  
circles 
represent the  first two 
	quantum numbers of the eigenstates
	$\ket{n,j,j_3}$. Red arrows indicate the action of $a_3$ and blue 
	arrows correspond to the action of $a_3^\dagger$. Note that some circles have two emergent  
	arrows of the same color, which means that the action of the rising/lowering
 operator on that states produce a superposition of two states.} 
  }
\label{figure2Esqueme}
\end{center} 
\end{figure} 
Clearly, if we are working in a representation where $H$, $\vJ^2$ and $J_3$ are simultaneously 
diagonalized,
it would be rather natural to try to find 
ladder operators that map a given eigenstate
into just one eigenstate
with a  different quantum number $j$ and not a superposition
of eigenstates (having in mind the picture of the usual harmonic oscillator). To find such operators we introduce 
the nonlocal operator  
 \begin{eqnarray}
&
\mathscr{J}=\sqrt{\vJ^2+\frac{1}{4}}-\frac{1}{2}\,,\qquad
\mathscr{J}\ket{n,j,j_3}=j\ket{n,j,j_3}\,,
 \end{eqnarray}
and 
construct the operators
\begin{equation}
\mathscr{T}_{\pm}=\omega(\mathscr{J}+\ha)a_3\pm(H-\omega)a_3\mp a_3^\dagger\cC\,
\label{defLadder2}
\end{equation}
together with their Hermitean conjugate. Actually $\mathscr{T}_\pm$ 
and $\mathscr{T}_\pm^\dagger$
are the third components of the vector operators $\mathbfcal{T}_\pm$
and $\mathbfcal{T}_\pm^\dagger$ 
which are given by (\ref{defLadder2}) wherein $a_3$ and $a_3^\dagger$ are replaced by
$\va$ and $\va^\dagger$ on the right hand side.
But in what follows it suffices to consider $\mathscr{T}_\pm$ 
and $\mathscr{T}_\pm^\dagger$
which are ladder operators for the energy,
\be
[H,\mathscr{T}_\pm]=\omega\mathscr{T}_\pm\,,\qquad 
[H,\mathscr{T}_\pm^\dagger]=-\omega\mathscr{T}_\pm^\dagger\,.
\ee
They decrease and increase the angular momentum 
according to
\begin{align}
\label{ActionA-}
\mathscr{T}_+\ket{n,j,j_3}&=\omega(2j+1)A_{n,j,j_3}\ket{n,j-1,j_3}\,,\\
\label{ActionA+}
\mathscr{T}_-\ket{n,j,j_3}&=\omega(2j+3)B_{n,j,j_3}\ket{n-1,j+1,j_3}\,,
\end{align}
and the analogous Hermitian conjugate relations.
These nonlocal objects were inspired by a similar construction 
presented in 
\textcolor{red}{[\cite{DiracOs3}]}
 for the three-dimensional isotropic harmonic oscillator.
  
Now one can generate in a simple way all eigenstates of the
commuting observables $H,\vJ^2$ and $J_3$
by acting with the local ladder operators $\cC,\cC^\dagger, J_\pm$ and
with the nonlocal ladder operators $\mathscr{T}_+,\mathscr{T}_{+}^\dagger$ 
on just  one eigenstate. The same can 
be achieved with local ladder operators when
one uses $a,a^\dagger$ instead of 
$\mathscr{T}_+,\mathscr{T}_+^\dagger$, but then the recursive construction
gets more involved, since $a,a^\dagger$ map
into a superposition of eigenstates.


\subsection{The conformal bridge in monopole background}

\label{Conformal bridge in mono}
Here we show how the generators of the conformal symmetry as well as 
the hidden symmetry  of the quantum 
 system (\ref{Qm Hamiltonian}) 
 can be obtained from generators of the corresponding 
 symmetries of the quantum system studied in 
 \textcolor{red}{[\cite{PlyWipf}]}. This will be realized  by means 
the conformal bridge transformation introduced in Chap. \ref{ChBridge} .

Similarly to
the classical case, in the limit $\omega\rightarrow 0$ 
the quantum version of the generators (\ref{freemotion}) 
has the form
\begin{eqnarray}
&
\label{spin0monopole}
H_0=\frac{1}{2}\left(\vpi^2+\frac{\nu^2}{r^2}\right)=\frac{1}{2}
\left(-\frac{1}{r^2}\frac{\partial}{\partial r}\left(r^2\frac{\partial}{\partial r}\right)+
\frac{1}{r^2}\vJ\!\,^2\right)\,,&
\\&
D_0=\frac{1}{4}(\vr\cdot\vpi+\vpi\cdot\vr ) - H_0t\,,\qquad
K_0=\frac{1}{2}
r^2 -Dt-H_0t^2\,.&\label{K0r2}
\end{eqnarray}
They produce the quantum conformal algebra 
\begin{equation}
\label{quantumconformal}
[D_0,H_0]=iH_0\,,\qquad
[D_0,K_0]=-iK_0\,,\qquad
[K_0,H_0]=2iD_0\,. 
\end{equation}
The Hamiltonian $H_0$ is a non-compact generator of the 
conformal algebra $\mathfrak{sl}(2,\R)$ with a continuous spectrum $(0,\infty)$.
In the same limit and in the quantum version, the vector integrals $\vI_1$ and 
$\vI_2$ transform into vectors 
\begin{equation}
\label{LRL-G}
\vI_1\rightarrow\frac{1}{2}\left(\vpi\cross\vJ-\vJ\cross\vpi\right):=\vV\,,\qquad
\frac{\vI_2}{\omega}\rightarrow\frac{1}{2}\big(\vpi t-\vr)\cross\vJ-\vJ\cross(\vpi t-\vr)\big):=\vG\,,
\end{equation}
which we identify, respectively,  as the  
 Laplace-Runge-Lentz vector and the Galilei boost 
 generator 
for the system $H_0$ \textcolor{red}{[\cite{PlyWipf}]} in the Weyl-ordered form.
The commutator relations of the 
vectors $\vV$ and $\vG$ with the generators of the conformal algebra 
are
\begin{eqnarray}
&
[H_0,G_i]=-iV_i\,,\qquad
[K_0,V_i]=iG_i\,,\qquad
[H_0,V_i]=[K_0,G_i]=0\,,&\\&
[D_0,V_i]=\frac{i}{2}V_i\,,\qquad
[D_0,G_i]=-\frac{i}{2}G_i\,.&
\end{eqnarray}
In order to go in the opposite  direction, i.e., to 
recover our system $H$ and its symmetry generators starting from the generators 
(\ref{spin0monopole}), (\ref{K0r2}) and 
 (\ref{LRL-G}), we implement the conformal bridge transformation  \textcolor{red}{[\cite{InzPlyWipf2}]},
 \begin{equation}
\mathfrak{S}=e^{-\omega K_0}e^{\frac{1}{2\omega}H_0}e^{i\ln 2 D_0}\,,
\end{equation}
where generators are fixed at $t=0$. 
 A similarity transformation generated by  $\mathfrak{S}$ yields 
\begin{eqnarray}
\label{bridge 1}
&
\mathfrak{S}(\vJ)\mathfrak{S}^{-1}=\vJ\,,\qquad
\mathfrak{S}(\vV)\mathfrak{S}^{-1}=\va\,,\qquad
\mathfrak{S}(\omega\vG)\mathfrak{S}^{-1}=-i\va^\dagger\,,&\\
&
\label{bridge 2}
\mathfrak{S}(H_0)\mathfrak{S}^{-1}=\frac{1}{2}\cC\,\qquad
\mathfrak{S}(2i\omega D_0)\mathfrak{S}^{-1}=H
\,,\qquad
\mathfrak{S}(\omega^2 K_0)\mathfrak{S}^{-1}=-\frac{1}{2}\cC^\dagger\,,
&
\end{eqnarray}
where $H=H_0+\omega^2K_0$ is the quantum Hamiltonian (\ref{Qm Hamiltonian}). 
Then, as we know from Chap. \ref{ChBridge}, the eigenstates of $H$ are mapped from 
the rank $n$ Jordan states of zero energy of $H_0$, which also satisfy 
the equation 
$2i\omega D_0\chi_{n,j}^{j_3}=\omega(2n+j+3/2)\chi_{n,j}^{j_3}$. Besides, 
the coherent states are obtained from the wave-type eigenstates of $H_0$. 
On one hand, the mentioned Jordan states are 
\begin{equation}
\chi_{n,j}^{j_3}(r,\theta,\phi)=r^{j+2n}\mathcal{Y}_{j}^{j_3}(\theta,\phi)\,,
\end{equation} 
and after the transformation we get
\begin{eqnarray}
&\mathfrak{S}\chi_{n,j}^{j_3}=\frac{(-1)^n}{\sqrt{2}}\left(\frac{2}{\omega}\right)^{n+\frac{j}{2}+\frac{3}{4}}
\left[n! \Gamma(n+j+3/2)\right]^{\frac{1}{2}}\psi_{n,j}^{j_3}\,. &
\end{eqnarray}
On the other hand, the corresponding eigenstates of $H_0$ are 
\begin{eqnarray}
\label{Jstates}
&\phi_j^{j3}(\vr;\kappa)=\frac{1}{\sqrt{r}}J_{j+\frac{1}{2}}(\kappa r)\mathcal{Y}_{j}^{j_3}
=\sum_{n=0}^{\infty}\frac{(-1)^{n}(\kappa/2)^{2n+j+1/2}}{n!\Gamma(n+j+3/2)}\chi_{n,j}^{j_3}(\vr)\,,&
\end{eqnarray}
and the normalized coherent states of $H$ are 
\begin{eqnarray}
\label{coherent}
&\zeta^{j_3}_{j}(\vr;\kappa)=
N\mathfrak{S}\phi_j^{\,j3}(\vr;\frac{\kappa}{\sqrt{2}})=\sqrt{2}Ne^{-\frac{\omega 
x^2}{2}+\frac{\kappa^2}{4\omega}}\phi_j^{\,j3}(\vr;\kappa)\\\nonumber&=
\frac{N}{\omega^{1/2}}\sum_{n=0}^{\infty} \frac{1}
{\sqrt{n!\Gamma(n+j+\frac{3}{2})}}\left(\frac{\kappa}{2\sqrt{\omega}}\right)^{2n+j+
1/2}\psi_{n,j}^{j_3}(\vr)\,,
&
\end{eqnarray}
where  
$N^2=\sqrt{\omega}/(I_{j+\frac{1}{2}}(\frac{|\kappa|^2}{2\omega}))$, 
the term  $I_{j+\frac{1}{2}}(z)$ is the modified Bessel function of the first kind,
and we have put  the modulus in its argument 
because $\kappa$ admits an analytic extension for complex values, as is usual 
for coherent states.

\section{Remarks}

As we have shown, hidden symmetries appear only  when 
$ \alpha = \nu^2 $. In this case, one always has closed trajectories,
 the angular period is twice the radial period, and even more,
 the projected dynamics in the plane orthogonal to the Poincar\'e vector turns out to
 be similar to that of the three-dimensional isotropic harmonic oscillator trajectory.
 In fact, such an interesting ``coincidence'' is actually an universal property of the
 monopole background: Consider the system described by the Hamiltonian
\begin{equation}
\label{centralpotentialH}
H=\frac{\vpi^2}{2m}+\frac{\nu^{2}}{2mr^{2}}+ U(r)\,,
\end{equation}
where $ U (r) $ is an arbitrary central potential.  The dynamical variables
 $ \vr \cross \vJ $ and $ \vpi \cross \vJ $ satisfy the same equations of motion as the vector 
variables $ \vr \cross \vL $ and $ \vp \cross \vL $ when $ \nu = eg = 0 $, 
where $ \vL $ is the usual angular momentum:
\begin{table}[H]
\begin{center}
\begin{tabular}{|c| c|}
\hline
$\nu\not=0$ &$\nu=0 $\\\hline
$\frac{d}{dt}(\vr\cross\vJ)=\frac{1}{m}\vpi\cross\vJ $& 
$\frac{d}{dt}(\vr\cross\vL) =\frac{1}{m}\vp\cross\vL$\\\hline
$\frac{d}{dt}(\vpi\cross\vJ)=U'(r)\,\vn\cross \vJ $& 
$\frac{d}{dt}(\vp\cross\vL)=U'(r)\,\vn\cross \vL$\\
\hline
\end{tabular}
\caption{Comparison of dynamics in the presence and absence of the monopole charge.}
\label{Tabla 2}
\end{center}
\end{table}
\noindent
\vskip-0.5cm
Therefore, the movement in the plane orthogonal to 
$ \vJ $ is equivalent to the dynamics obtained in the absence of the magnetic monopole source, 
and if we know the solutions $ \vr = \vr (t) $ and $ \vp = \vp (t) $ in the case 
$ \nu = 0 $, the dynamic for $ \vpi \cross \vJ $ and $ \vr\cross \vJ $ is at hand,
\begin{equation}
\vr(t)=\frac{1}{J^2}\left(\vJ\cross(\vr(t)\cross\vJ)+\sqrt{\frac{|\vr(t)\cross\vJ|}{J^2-\nu^2}}\vJ\right)\,.
\end{equation} 
On the other hand, if we take the system 
$ \widetilde {H}_\nu = \frac{1}{2m}\vpi ^ 2 + \widetilde{U} (r) $
 with arbitrary central potential $ \widetilde {U} (r) $, the corresponding dynamical problem 
is reduced to that of the system (\ref{centralpotentialH}) with central potential 
$ U (r) = \widetilde{U} (r) - \nu^2 / 2mr^ 2 $. The indicated similarities and relations allow, 
in particular, to identify immediately the analog of the Laplace-Runge-Lenz vector 
(\ref{LRL-G}) for a particle in the monopole background in the cases   $ \widetilde{U} = 0 $,  $ U = 0 $
 and for the Kepler problem 
with $ U = q / r $. This was done previously in \textcolor{red}{[\cite{Plymono2,PlyWipf}]} and \textcolor{red}{[\cite{Vinet}]} 
using different approaches. 

In the next chapter we will study how to extend this picture to supersymmetric quantum mechanics.


\chapter{A charge-monopole superconformal model}
\label{Chapmono2}

In this chapter we extend our system by means of an additional contribution in the Hamiltonian (\ref{Qm Hamiltonian})
that involves spin degrees of freedom. 
The supplemented term describes a strong long-range spin-orbit coupling and one of its direct 
consequences is the appearance of two independent subsets of energy levels. 
In one of these subsets or towers, infinitely degenerate energy levels appear, 
while in the other, the levels have finite degeneration. The system is studied in detail in
 Sec. \ref{sectionSOC}. 

In the Sec. \ref{osp22 extension}, we show that thanks to this term, the system introduced earlier 
supports a factorization in terms of intertwining operators  that naturally leads 
us to a supersymmetric extension, which is nothing more than a three-dimensional
 realization of the superalgebra $ \mathfrak{osp} (2 \vert 2) $. Finally, in Sec. \ref{DimRed},
 it is shown that by means of certain dimensional reductions, it is possible to obtain
 supersymmetric AFF models in their exact  and spontaneously broken supersymmetric phase. 
Something special about the models obtained in this way is that the coupling constant 
in the potential is $j (j + 1)$, where $j$ can takes integer or half-integer values, starting from 
$\nu = (eg)^2$.

\section{Introducing spin degrees of freedom: Spin-orbit coupling}
\label{sectionSOC}
Let us consider the following two Hamiltonians with strong spin-orbit coupling  
\begin{equation}
\label{spinorbitH}
H_{\pm\omega}= \frac{1}{2}\left(\vpi^2 +\omega^2 r^2+\frac{\nu^2}{r^2}\right)\pm\omega\vsigma\cdot\vJ
=H\pm\omega\vsigma\cdot\vJ\,.
\end{equation}
The Hamiltonians $H_{\pm\omega}$ are similar to those which appear
as subsystems of the nonrelativistic limit of the supersymmetric 
Dirac oscillator discussed in \textcolor{red}{[\cite{DiracOs1,DiracOs2}]}. Thus
the eigenvalue problems can be solved similarly as in those 
references, but the usual spherical harmonics
are replaced by the monopole harmonics. Actually,
if we choose a spin-orbit coupling $\omega'\vsigma\cdot\vJ$
with $0\leq \omega<\omega'$, then the spectra of both
Hamiltonians would be unbounded 
from
below. 
On the other hand, for
$0\leq \omega'<\omega$ all energies will have finite
degeneracy. Only in the very particular case $\omega'=\omega$, which we 
consider here, the spectra are bounded from below and
half of the energies have a finite degeneracy whereas
the other half have infinite degeneracy. This 
reminds us  the BPS-limits in 
field theory, where different interactions balance and
supersymmetry is observed.

The operators $H$ and $\vsigma\cdot\vJ$ commute and as a consequence
$H_{\pm \omega}$ commute with the ``total angular momentum''
\begin{equation}
\vK=\vJ+\vs=\vJ+\tfrac{1}{2}\vsigma\,,\qquad
[K_i,K_j]=i\epsilon_{ijk}K_k\,.
\end{equation}
The possible eigenvalues of $\vK^2$ are $k(k+1)$.
It is well-known how to construct the simultaneous eigenstates
of $\vK^2$ and $K_3$:
\begin{equation}
\ket{n,k,k_3,\pm}
=\sum_{m_s}C^{kk_3}_{jj_3\has m_s} \ket{n,j,j_3}
\otimes\ket{\ha,m_s}_{k=j\pm\has}\,,
\label{eigenHpm}
\end{equation}
where the Clebsch-Gordan
 coefficients 
\begin{equation}
C^{kk_3}_{jj_3\has m_s}=\bra{j,j_3,\tfrac{1}{2},m_s}k,k_3\rangle\,
\end{equation}
on the right hand
side are nonzero only if $j_3+m_s=k_3$ and 
if the triangle-rule holds, which means that
the total angular momentum $k$ is either $j+\frac{1}{2}$ 
or $j-\frac{1}{2}$. In the first case the eigenstates
of the total angular momentum are
denoted by $\ket{\dots,k,k_3,+}$ and in the second case
by $\ket{\dots,k,k_3,-}$.
The sums (\ref{eigenHpm}) contain just two terms, since the eigenvalue $m_s$ 
of the third spin-component 
$s_3=\frac{1}{2}\sigma_3$ 
is either $\frac{1}{2}$ 
or $-\frac{1}{2}$.
In the coordinate representation  
the wavefunctions corresponding
to these kets are given by 
\vskip-0.5cm
\begin{eqnarray}
&
\label{Wspin+-}
\bra{\vr}\ket{n,k,k_3,\pm}=f_{n,j}(\sqrt{\omega}r)\bra{\vn}\ket{k,k_3,\pm}\,,&\\&
\label{Omega}
\bra{\vn}\ket{k,k_3,\pm}=
\frac{1}{\sqrt{2k+1\mp 1}}
\left(\begin{array}{cc}
\pm \sqrt{k\pm k_3+(1\mp 1)/2}\,\mathcal{Y}_{k\mp 1/2}^{k_3-1/2}(\theta,\varphi;\nu)\\
\sqrt{k\mp k_3 +(1\mp 1)/2}\,\mathcal{Y}_{k\mp 1/2}^{k_3+1/2}(\theta,\varphi;\nu)
\end{array}\right):=\Omega_{k}^{k_3\,\pm}\,.
&
\end{eqnarray}
If $\nu=eg$ is integer-valued then $j$ is a non-negative
integer and $k$ a positive half-integer. If $eg$ is half-integer, then
$j$ is a positive half-integer and $k$ is in $\N_0$.

The vector 
in (\ref{eigenHpm}) is a simultaneous eigenstate of $\vJ^2$ 
with eigenvalue $j(j+1)$, of $\vK^2$ with eigenvalue $k(k+1)$, of $H$ with eigenvalue
$(2n+j+\frac{3}{2})\omega$,  where $j=k\mp 1/2$, 
and finally of the operator $\vsigma\cdot\vJ$:
\begin{equation}
\vsigma\cdot\vJ\ket{n,k,k_3,\pm}=\big(\pm(k+\tfrac{1}{2})-1\big)
\ket{n,k,k_3,\pm}\,.
\end{equation}
As a consequence the action of the Hamiltonians in (\ref{spinorbitH}) 
on these states is 
\begin{align}
H_{+\omega}\ket{n,k,k_3,\pm}&=\omega\left(2n+k+\tfrac{1}{2} \pm k \right)
\ket{n,k,k_3,\pm}\,,\label{Hket+}\\
H_{-\omega}\ket{n,k,k_3,\pm}&=\omega\left(2n+k+\tfrac{5}{2} \mp(k+1)\right)
\ket{n,k,k_3,\pm}\,.
\label{Hket-}
\end{align} 
We see that the discrete eigenvalues of both
Hamiltonians $H_{\pm\omega}$ fall into two families:
in one
family all energies are  infinitely degenerate
and in the other family they all have finite degeneracy
(due to their dependence on the quantum number $k$). 
More explicitly, for $k=j\mp\frac{1}{2}$ the eigenvalues 
of $H_{\mp\omega}$ have infinite degeneracy
and for $k=j\pm\frac{1}{2}$ they  have finite 
degeneracy $\mathfrak{g}(N,\nu)=N^2-\nu^2$,  where $N=n+j+1$.
A similar peculiar behavior is observed in the Dirac 
oscillator spectrum \textcolor{red}{[\cite{DiracOs1}]}.

Operators $K_\pm=K_1\pm iK_2$ are the ladder operators
for the magnetic quantum number $k_3$. 
The ladder operators for the radial quantum number 
are 
given in (\ref{ladern}), 
and their 
action
on the simultaneous eigenstates 
reads
\begin{align}
\cC\ket{n,k,k_3,\pm}&=\omega d_{n,j}\ket{n-1,k,k_3,\pm}\,,\\
\cC^\dagger\ket{n,k,k_3,\pm}&=\omega d_{n+1,j}\ket{n+1,k,k_3,\pm}\,,
\end{align}
with coefficients defined in (\ref{d-coeff}).
Thus, as for the spin-zero particle system in monopole background, 
we can
easily construct local ladder operators for $n$ and $k_3$.
But again, finding ladder operators for $k$ is
more difficult.
One way to proceed is to follow the ideas employed
for the Dirac oscillator in \textcolor{red}{[\cite{DiracOs1,DiracOs2,DiracOs3}]}.
First we decompose the total Hilbert space in two subspaces,
$\mathscr{H}=\mathscr{H}^{(+)}\oplus \mathscr{H}^{(-)}$, where 
each
$\mathscr{H}^{(\pm)}$ is spanned by the states 
$\ket{n,k,k_3,\pm}$.
Actually we can construct nonlocal operators 
which project orthonormally onto these subspaces,
\begin{align}
\mathscr{P}_{+}&=\ha+\sqrt{\vK^2+\tfrac{1}{4}}-
\sqrt{\vJ^2+\tfrac{1}{4}}\,,\\
\mathscr{P}_{-}&=\ha-\sqrt{\vK^2+\tfrac{1}{4}}+
\sqrt{\vJ^2+\tfrac{1}{4}}\,,
\end{align}
and 
reproduce or annihilate the eigenstates,
\begin{equation}
\mathscr{P}_{(\pm)}\big\vert_{\mathscr{H}^{(\pm)}}
=\id\big\vert_{\mathscr{H}^{(\pm)}}\,,\qquad
\mathscr{P}_{(\pm)}\big\vert_{\mathscr{H}^{(\mp)}}=0\,.
\end{equation} 
In next step we introduce the operators  
\begin{equation}
\label{PAP+}
\mathcal{A}_{\pm}=\mathscr{P}_\pm \mathscr{T}_\pm\mathscr{P}_\pm\,,
\end{equation}
where the nonlocal $\mathscr{T}_\pm$ have been
defined in (\ref{defLadder2}). 
The presence of the projectors will ensure that 
$\mathcal{A}_\pm$ only
acts 
on eigenstates in $\mathscr{H}^{(\pm)}$, and 
its action on these eigenstates  
can be computed straightforwardly using the relations (\ref{ActionA-}) 
and (\ref{ActionA+}):
\begin{align}
\label{nonlocalAaction}
&\mathcal{A}_+ \ket{n,k,k_3,+}=
(k-1)\sqrt{n+k}\,\Lambda_{k,k_3,j}\, 
\ket{n,k-1,k_3,+}\,, 
\\
&\mathcal{A}_{-} \ket{n,k-1,k_3,-}=
(k+1)\sqrt{n}\,\Lambda_{k,k_3,j}
\, 
\ket{n-1,k,k_3,-}\,,
\end{align}
with 
\[
\Lambda_{k,k_3,j}=\frac{\omega^{3/2}}{k}\sqrt{2(k^2-k_3^2)(j^2-\nu^2)}\,.
\]
These relations mean that the operators $\mathcal{A}_\pm$ 
and their adjoint act as ladder operators for the quantum 
number $k$. Together with operators $K_\pm,\cC,\cC^\dagger$
they generate all eigenstates in the full Hilbert space 
from just two eigenstates, one from each subspace 
$\mathscr{H}^{(\pm)}$.

\section{The $\mathfrak{osp}(2\vert 2)$ superconformal extension}
\label{osp22 extension}
In this subsection we construct and analyze  supersymmetric partners
of the Hamiltonians $H_{\pm \omega}$ by introducing 
 factorizing operators. 
From these we obtain two $\mathcal{N}=2$ super-Poincar\'e quantum systems
which are related to each other by a common integral of motion
which generates an $R$-symmetry.
Supplementing the supercharges of one of these systems  by supercharges of another, 
we extend the $\mathcal{N}=2$ super-Poincar\'e  symmetry up to 
the $\mathfrak{osp}(2\vert 2)$ superconformal symmetry 
realized by a three-dimensional system of 
spin-1/2 particle in a monopole background.

Consider 
the  first-order scalar operators 
\begin{equation}
\label{intertwiningQ}
\Theta=i\vsigma\cdot\vb-\frac{1}{\sqrt{2}}\frac{\nu}{r}\,,\qquad 
\Xi=i\vsigma\cdot\vb^\dagger-\frac{1}{\sqrt{2}}\frac{\nu}{r}\,,
\end{equation}
and their adjoint $\Theta^\dagger$ and $\Xi^\dagger$.
The products of these operators with their adjoint are
\begin{equation}
H_{[1]}:=
\Theta\Theta^\dagger=H_{+\omega}+\tfrac{3}{2}\omega\,,\qquad
\breve{H}_{[1]}:=
\Xi\Xi^\dagger=H_{-\omega}-\tfrac{3}{2}\omega\,,
\label{H0}
\end{equation}
where $H_{\pm\omega}$ are given in (\ref{spinorbitH}).
The associated superpartners take the form
\begin{align}
\label{H0a}
H_{[0]}&:=
\Theta^\dagger \Theta=\breve{H}_{[1]}-\nu\left(\tfrac{1}{r^2}+2\omega \right)\sigma_r
\,,
\\
\label{H0b}
\breve{H}_{[0]}&:=\Xi^\dagger \Xi
=H_{[1]}-\nu\left(\tfrac{1}{r^2}-2\omega \right)\sigma_r \,,
\end{align} 
wherein the projection of $\vsigma$ to the normal
unit vector appears,
\begin{equation}
\label{sigman}
\sigma_r=\vn\cdot\vsigma=\left(\begin{array}{cc}
\cos\theta & e^{-i\varphi}\sin\theta\\
e^{i\varphi}\sin\theta & -\cos\theta
\end{array}\right)\,.
\end{equation}
The first order operators satisfy the intertwining relations 
\begin{eqnarray}
&\label{thetainter}
\Theta H_{[0]}=H_{[1]}\Theta\,,\qquad 
\Theta^\dagger H_{[1]}=H_{[0]}\Theta^\dagger\,,  
&\\&
\label{Xiinter}
\Xi\breve{H} _{[0]}=\breve{H}_{[0]}\Xi\,,\qquad
\Xi^\dagger \breve{H}_{[1]}=\breve{H}_{[0]}\Xi^\dagger\,. 
&
\end{eqnarray} 
To compute the action of the intertwining  operators $\Theta^\dagger$ and 
$\Xi^\dagger$ in eigenstates of $H_{\pm \omega}$
 is useful to express them in the form 
\begin{align}
\label{Qnu}
\Theta^\dagger
=\frac{\sigma_r}{\sqrt{2}}\left(-
\frac{1}{r}\frac{\partial}{\partial r} r+\omega r+
\frac{1+\vsigma\cdot\vJ}{r}\right)\,,
\\
\label{Wsigmaform}
\Xi^\dagger
=\frac{\sigma_r}{\sqrt{2}}\left(-
\frac{1}{r}\frac{\partial}{\partial r} r-\omega r+
\frac{1}{r}(1+\vsigma\cdot\vJ)
\right)\,.
\end{align}
Then the strategy is to apply directly this operators on the eigenstates of 
$H_\omega$ in their coordinate representation (\ref{Wspin+-}), 
obtaining in this way the eigenstates of systems 
 $H_{[0]}$ and  $\breve{H}_{[0]}$. The action of operators 
 $\Theta$ and 
$\Xi$ in these new eigenvectors follows from the intertwining relations 
 (\ref{thetainter})-(\ref{Xiinter}). 
The final result is 
\begin{align}
\Theta^\dagger\ket{n,k,k_3,\pm}&=\pm\sqrt{2\omega(n+1+\beta_{\pm}k)}
\,\Vert n+\beta_{\mp},k,k_3,\pm\rangle\,,\quad \beta_\pm=\tfrac{1}{2}(1\pm 1)\,,\label{qdonpsi}
\\
\Theta\Vert n,k,k_3,\pm\rangle&=\pm
\sqrt{2\omega(n+\beta_{\pm}(k+1))}\,\ket{n-\beta_{\mp},k,k_3,\pm}\,,
\label{Qonphi+}
\\
\Xi^\dagger\ket{n,k,k_3,\pm}&=\pm\sqrt{2\omega (n+\beta_{\mp}(k+1)}\,
\Vert n-\beta_\pm,k,k_3,\pm\rangle\,,
\label{W+onpsi1} \\
\Xi\,\Vert n,k,k_3,\pm\rangle
&=\pm\sqrt{2\omega (n+1+\beta_\mp k)}\,\,\ket{n+\beta_{\pm},k,k_3,\pm}\,.
\label{W+onphi2}
\end{align}
Where in coordinate representation the normalized spinors $\Vert n,k,k_3,\pm\rangle$ 
have the explicit form
\begin{align}
\langle\vr\Vert n,k,k_3,\pm\rangle&=f_{n,j\pm1}\sigma_r\Omega_{k}^{k_3\,\pm}\,,
\label{spinors2a}\end{align}
and  $\Omega_{k}^{k_3\,\pm}$ are given in (\ref{Omega}). 

From these equations it is easy to show that 
\begin{align}
H_{[0]}\Vert n,k,k_3,\pm\rangle&=2\omega(n+\beta_{\pm}(k+1))\Vert n,k,k_3,\pm\rangle\,,
\\
\breve{H}_{[0]}
\,\Vert n,k,k_3,\pm\rangle&=2\omega(n+1+k\beta_{\mp})\,\Vert n,k,k_3,\pm\rangle\,,
\label{eigenbreve}
\end{align} 
and note that in one hand, $\Vert 0,k,k_3,-\rangle$ are zero-modes of
$H_{[0]}$ since they are annihilated by $\Theta$,
and on the other hand 
 $\Xi^\dagger$ as well as $\breve{H}_{[1]}$ annihilate the set  
of states $\ket{0,k,k_3,+}$.

Having at hand the eigenstates $\Vert n,k,k_3,\pm\rangle$,
one may find spectrum-generating ladder operators. In 
this context  Eqs. (\ref{qdonpsi}), (\ref{Qonphi+}), (\ref{W+onpsi1}) and (\ref{W+onphi2}) 
can be used to construct such operators 
for the quantum number $n$. They read
\begin{align}
\tilde{\mathcal{C}}=\Xi^\dagger \Theta\,,\qquad
\tilde{\mathcal{C}}^\dagger=\Theta^\dagger \Xi\,,
\end{align} 
and act on the eigenvectors  $\Vert \dots\rangle$
as 
follows:
\begin{align}
\tilde{\mathcal{C}}^\dagger\,\Vert n,k,k_3,\pm\rangle&=2\omega d_{n+1,j\pm 1}
\,
\Vert n+1,k,k_3,\pm\rangle\,,\nonumber\\
\tilde{\mathcal{C}}\,\Vert n,k,k_3,\pm\rangle&=2\omega d_{n,j\pm 1}
\,
\Vert n-1,k,k_3,\pm\rangle\,.
\label{W+Q}
\end{align} 
Actually, the first order operators $\Theta$ and $\Xi^\dagger$ 
factorize the earlier considered second order 
ladder operator (\ref{ladern}) according to $\cC=\Theta\Xi^\dagger$.

Having constructed lowering and 
raising operators for $n$, 
we are still missing ladder operators for $k$ and $k_3$. For the 
latter we may of course use $K_\pm$, since
$\Theta$, $\Xi$ and their adjoint 
are scalar operators with respect to $\vK$. But once more,
for the angular momentum quantum number $k$ we can introduce
nonlocal 
``dressed'' operators
\begin{eqnarray}
\label{nonlocaldressed}
&\tilde{\mathcal{A}}_-=\Theta\sqrt{\frac{1}{H_{[1]}}}\mathcal{A}_-\sqrt{\frac{1}{H_{[1]}}}\Theta^\dagger\,,\qquad
\tilde{\mathcal{A}}_+=\Xi\sqrt{\frac{1}{\breve{H}_{[1]}}}\mathcal{A}_+\sqrt{\frac{1}{\breve{H}_{[1]}}}\Xi^\dagger\,,&\qquad
\end{eqnarray}
and their  
adjoint
operators, where $\mathcal{A}_\pm$ 
have been given in (\ref{PAP+}). The operators $\tilde{\mathcal{A}}_\pm$
are the analogs to $\mathcal{A}_\pm$ for the vectors $\Vert n,k,k_3,\pm\rangle$, 
as we can see from the equations 
\begin{align}
\label{nonlocalAaction2}
&\tilde{\mathcal{A}}_+ \Vert n,k,k_3,+\rangle=
(k-1)\sqrt{n+k}\,\Lambda_{k,k_3,j}\, \Vert n,k-1,k_3,+\rangle \,, 
\\
&\tilde{\mathcal{A}}_{-}\Vert n,k-1,k_3,-\rangle=
(k+1)\sqrt{n}\,\Lambda_{k,k_3,j}
\, \Vert n-1,k,k_3,-\rangle\,.
\end{align}

In a final step we combine the four $2\times 2$ matrix
Hamiltonians introduced above into two 
$4\times 4$ matrix super-Hamiltonians  
 as follows:
\begin{eqnarray}
\label{superH}
\mathcal{H}=\left(\begin{array}{cc}
H_{[1]} & 0\\
0 & H_{[0]}
\end{array}\right)\,,\qquad
\breve{\mathcal{H}}=\left(\begin{array}{cc}
\breve{H}_{[1]} & 0\\
0 & \breve{H}_{[0]}
\end{array}\right)\,.
\end{eqnarray}
In the limit $\nu\rightarrow 0$ they turn into
different versions  of the Dirac oscillator in the nonrelativistic limit, see \textcolor{red}{[\cite{DiracOs1}]}.
Both operators commute with 
the $\Z_2$-grading operator $\Gamma=\sigma_3\otimes \mathbb{I}_{2\times  2}$, 
$[\Gamma,\mathcal{H}]=[\Gamma,\breve{\mathcal{H}}]=0$,
and 
their difference is the (bosonic) integral of motion
\begin{equation}
\mathcal{R}=\frac{1}{2\omega}(\mathcal{H}-\breve{\mathcal{H}})=
(\vJ\cdot \vsigma +\tfrac{3}{2})\Gamma-2\nu\sigma_r\Pi_-= 
\left(\begin{array}{cc}
\vsigma\cdot\vJ+\frac{3}{2} & 0\\
0&-(\vsigma\cdot\vJ+2\nu\sigma_r+\frac{3}{2}) \\
\end{array}\right),
\label{rwisym}
\end{equation} 
where  $\Pi_-$ is a projector,  
\be\label{Pi+-}
\Pi_\pm=\tfrac{1}{2}(1\pm\Gamma)\,.
\ee
In the fermionic sectors of the systems $\mathcal{H}$ and $\breve{\mathcal{H}}$
we have the nilpotent operators 
\begin{eqnarray}
\label{QyW}
&
\cQ=\left(\begin{array}{cc}
0 & \Theta\\
0 & 0
\end{array}\right)\,,\qquad
\cW=\left(\begin{array}{cc}
0 & 0 \\
\Xi^\dagger & 0
\end{array}\right)\,,&
\end{eqnarray} 
$\{\Gamma,\mathcal{Q}\}=\{\Gamma,\mathcal{W}\}
=0$, and their adjoint operators.

The even integral $\mathcal{R}$ in (\ref{rwisym})
generates an $R$-symmetry
for both systems.
Having in mind 
that $\mathcal{H}$ and $\breve{\mathcal{H}}$
can be diagonalized 
simultaneously, from now on 
we treat  $\mathcal{H}$ as the Hamiltonian of the super-extended system
and $\breve{\mathcal{H}}=\mathcal{H}-2\omega\mathcal{R}$ as its integral.
Then,  
by anti-commuting $\cQ$ and $\cW$ we obtain the bosonic generator 
\begin{equation}
\mathcal{G}=\{\mathcal{W},\mathcal{Q}\}=\left(\begin{array}{cc}
\mathcal{C} & 0\\
0 & \tilde{\mathcal{C}}
\end{array}\right)\,, \qquad [\Gamma,\mathcal{G}]=0\,,
\end{equation}
together with its adjoint. They are  
composed from 
 the ladder operators of  
 sub-systems $H_{[1]}$ and $H_{[0]}$ 
 of our system $\mathcal{H}$. 

Taking together, these  scalar generators with respect to  
\begin{equation}
 \mathcal{K}_i=
 \left(\begin{array}{cc}
K_i & 0\\
0 & K_i
\end{array}\right)\,, \qquad i=1,2,3\,,
\end{equation}
obey the $\mathfrak{osp(2,2)}$ superalgebra mentioned in 
Chap. \ref{ChConformal}, Eqs. 
 (\ref{Ospnil1})-(\ref{Ospnilf}),
and  therefore this construction maybe
considered as generalization of the super-extended AFF model  to three dimensions.

The common eigenstates of $\mathcal{H}$, $\mathcal{R}$, $\Gamma$, $\mathcal{K}_3$ and 
$\mbfgr{\mathcal{K}}^2$ are given by      
\begin{eqnarray}
\label{supervectors}
\ket{n,k,k_3,\pm,1}=\left(\begin{array}{cc}
\ket{n,k,k_3,\pm}\\
0
\end{array}\right)
\,,\qquad
\ket{n,k,k_3,\pm,-1}=\left(\begin{array}{cc}
0\\
\Vert n,k,k_3,\pm\rangle
\end{array}\right)\,,
\end{eqnarray}
which satisfy the eigenvalue equations 
\begin{align}
\label{spectalEq}
\mathcal{H}\ket{n,k,k_3,\pm,\gamma}&=2\omega\big(n+\tfrac{1}{2}(1+\gamma)+\beta_\pm(k+
\tfrac{1}{2}(1-\gamma))\big)\ket{n,k,k_3,\pm, \gamma 
}\,, \\
\Gamma \ket{n,k,k_3,\pm,\gamma}&=
\gamma \ket{n,k,k_3,\pm,\gamma} \,,\qquad \gamma=\pm 1
\,, \\
\label{REq}
\mathcal{R} \ket{n,k,k_3,\pm,\gamma}&= 
[\pm(k+\tfrac{1}{2})+\tfrac{\gamma}{2}] \ket{n,k,k_3,\pm,\gamma}\,,
\\
\label{Konsusy}
\mbfgr{\mathcal{K}}^2\ket{n,k,k_3,\pm,\gamma}&= 
k(k+1) \ket{n,k,k_3,\pm,\gamma}\,,\\
\mathcal{K}_3\ket{n,k,k_3,\pm,\gamma}&=k_3\ket{n,k,k_3,\pm,\gamma}\,.
\end{align}  
The operators $\mathcal{Q}$ and $\mathcal{Q}^\dagger$ ($\mathcal{W}$ 
and $\mathcal{W}^\dagger$) defined in (\ref{QyW}),   
interchange the state vectors $\ket{n,k,k_3,\pm,\gamma}$ and 
$\ket{n,k,k_3,\pm,-\gamma}$ according to the rules in 
(\ref{qdonpsi}), (\ref{Qonphi+}) and (\ref{W+onpsi1}),
(\ref{W+onphi2}).
The
ground  
states of $\mathcal{H}$ ($\breve{\mathcal{H}}$) which 
are given by $\ket{n,k,k_3,-,-1}$ ($\ket{n,k,k_3,+,+1}$ ) 
are invariant under transformations generated 
by these fermionic operators, therefore the
quantum system $\mathcal{H}$ 
exhibits the unbroken
$\mathcal{N}=2$ Poincar\'e
supersymmetry. 
  
Finally, the spectrum-generating ladder operators for the
supersymmetric system correspond to  
operators $\mathcal{G}$ and $\mathcal{G}^\dagger$  (associated with  $n$),
$\mathcal{K}_\pm$ that change $k_3$, and 
the matrix nonlocal operators  
\begin{eqnarray}
\left(
\begin{array}{cc}
\mathcal{A}_\pm &0\\
0 & \tilde{\mathcal{A}}_\pm
\end{array}\right)\,,\qquad
\left(
\begin{array}{cc}
\mathcal{A}_\pm^\dagger &0\\
0 &  \tilde{\mathcal{A}}_\pm^\dagger
\end{array}\right)\,.\qquad
\end{eqnarray}
related to the angular quantum number $k$. 

\section{Dimensional reductions}
\label{DimRed}

The system studied in the last section and the one presented in 
Chap \ref{ChConformal}, Sec. \ref{SecOSP22Conformal} share the same symmetry,
 and in this paragraph we will show that they are related by a dimensional reduction.
 For the sake of simplicity, we put $ \omega = 1 $ here, and denote $ \sqrt {\omega} r = r $
 as $ x $. 

The first step is to note that the Hamiltonian $ \mathcal{H}$ 
 can be presented in the following form 
\begin{equation}
\label{HintermsRK}
\mathcal{H}=\frac{1}{2}\left[-\frac{1}{x^2}\frac{\partial}{\partial x}\left(x^2\frac{\partial}{\partial x}\right)
+ x^2\right]\mathbb{I}_{4\cross 4}+ \frac{1}{2x^2}(\mbfgr{\mathcal{K}}^2-\Gamma\mathcal{R}+\tfrac{3}{4})+\mathcal{R}\,.
\end{equation} 
Then, to do the reduction we introduce the set of equations 
\begin{eqnarray}
\label{reduction1}
&
\mbfgr{(\mathcal{K}}^2-k(k+1))\ket{\chi,\pm}=0\,,\qquad 
(\mathcal{K}_3-k_3)\ket{\chi,\pm}=0\,,\qquad 
&\\&
\label{reduction2}
\mathcal{P}_{\pm}\ket{\chi,\pm}=0\,,\qquad
\mathcal{P}_{\pm}=\frac{1}{2k+1}(\Pi_\pm+k\mp \mathcal{R})\,,
&
\end{eqnarray}
where 
$k=j\pm\tfrac{1}{2}\,,$ and  
$k_3=j_3\pm\tfrac{1}{2}\,$.
$k=j\pm\tfrac{1}{2}\,,$ and  
$k_3=j_3\pm\tfrac{1}{2}\,$. Here, 
the most general form of $\ket{\chi,\pm}$ is 
\begin{equation}\label{anbn}
\ket{\chi,\pm}=\sum_{n=0}^{\infty}a_n^\pm\ket{n,k,k_3,\pm,1}+b_n^\pm \ket{n,k,k_3,\pm,-1}=
\sum_{n=0}^{\infty}
\left(\begin{array}{cc}
a_n^\pm \ket{n,k,k_3,\pm}\\
b_n^\pm \Vert n,k,k_3,\pm\rangle
\end{array}\right)\,,
\end{equation}
and effectively, operators $\mathcal{P}_{\pm}$ are projectors onto 
the orthogonal subspaces $\ket{\chi,-}$ and $\ket{\chi,+}$. 
These states satisfy 
\be
\mathcal{H}\ket{\chi,-}= \frac{1}{x}\mathcal{H}_j^{-}x\otimes\I_{2\cross 2}\ket{\chi,-}\,,\qquad 
\mathcal{H}\ket{\chi,+}= \sigma_1(\frac{1}{x}\mathcal{H}_{j+1}^{+}x)\sigma_1\otimes\I_{2\cross 2}\ket{\chi,+}\,,
\ee
where $\mathcal{H}_j^{-}=\mathcal{H}_j^{e}$ 
and $\mathcal{H}_j^{+}=\mathcal{H}_j^{b}$ are the one-dimensional supersymmetric 
extension of the AFF model in exact and spontaneously broken phase, 
see Chap \ref{ChConformal}, Sec. \ref{SecOSP22Conformal}.
Moreover, if we call as $\mathcal{B}_{a}$ and 
 $\mathcal{F}_{b}$ 
(where index $\mathcal{B}_{1}$ is the Hamiltonian and so on)
 the bosonic and fermionic generators of the three-dimensional system, respectively, 
and in the same vein  $\mathscr{B}_{j,a}^{\pm}$ and 
 $\mathscr{F}_{j,b}^{\pm}$ are their analogs  for one-dimensional system 
in their respective supersymmetric phases, 
we get 
\begin{eqnarray}
&
\mathcal{B}_{a}\ket{\chi,-}=\frac{1}{x}\mathscr{B}_{j,a}^{-}x\otimes\mathbb{I}_{2\cross 2}\ket{\chi,-}\,,\qquad
\mathcal{F}_{b}\ket{\chi,-}=\frac{1}{x}\mathscr{F}_{j,b}^{-}x\otimes\sigma_r\ket{\chi,-}\,,
&
\\
&
\mathcal{B}_{a}\ket{\chi,+}=\sigma_1(\frac{1}{x}\mathscr{B}_{j+1,a}^{+}x)\sigma_1\otimes\mathbb{I}_{2\cross 2}\ket{\chi,+},
\quad
\mathcal{F}_{b}\ket{\chi,+}=\sigma_1(\frac{1}{x}\mathscr{F}_{j,b}^{+}x)\sigma_1\otimes\sigma_r\ket{\chi,+}.\quad
&
\end{eqnarray}
In these equations the generators take the form of a direct product 
of two  matrix operators: In  case of bosonic (fermionic) operators  one has 
 $x^{-1}B\otimes x \I_{2\cross2}$ ( $x^{-1}F\otimes x\sigma_r$), where $B$ ($F$) 
is  a particular  bosonic (fermionic) operator of the one-dimensional AFF model 
in its corresponding  supersymmetric phase. Note that in the odd 
  sector we still have angular dependence due to $\sigma_r$. 
 
 To complete the reduction we introduce the  operators
\begin{equation}
\mathcal{O}_{\pm}=\left(\begin{array}{cc}
\ket{v}\bra{k,k_3,\pm} & 0\\
0 & \ket{v}\bra{k,k_3,\pm}\sigma_r\ 
\end{array}\right)\,,\qquad
\ket{v}=\left(\begin{array}{c}
1\\
1
\end{array}\right),
\end{equation}
and their adjoint, as well as the unitary operator
 \begin{equation}
U=\left(\begin{array}{cccc}
1 & 0 & 0 &0 \\
0 & 0 & 1 & 0\\
0 & 1&0  &0 \\
0 & 0 & 0 & 1
\end{array}\right)\,,\qquad
UU^\dagger=1\,,\qquad
\text{det}\, U=-1\,.
\end{equation}  

Operators $\mathcal{O}_{\pm}$  effectively integrate the angular variables, 
so the bosonic generators do not change, but the
 fermionic generators are transformed into
\begin{eqnarray}
&
\mathcal{O}_{-}\mathcal{F}_{b}\mathcal{O}_{-}^\dagger\ket{\Psi,-}=
\frac{1}{x}\mathscr{F}_{j,b}^{-}x\otimes\sigma_1\,\ket{\Psi,-}\,,
&\\&
\mathcal{O}_{+}\mathcal{F}_{b}\mathcal{O}_{+}^\dagger\ket{\Psi,+}=
\sigma_1(\frac{1}{x}\mathscr{F}_{j+1,b}^{+}x)\sigma_1\otimes\sigma_1\,\ket{\Psi,+}\,,
&
\end{eqnarray}
where $\mathcal{O}_\pm\ket{\chi,\pm}=\ket{\Psi,\pm}$.
On the other hand, by means of the unitary transformation produced by $ U $,
 we are able to present the bosonic and fermionic generators, already transformed by $\mathcal{O}_\pm$,
 in the form $ \I_{2 \cross2} \otimes x^{- 1} Bx $ and $ \sigma_1 \otimes x^{- 1} F x $, respectively.
 From these expressions one simply extracts the one-dimensional generators by means 
 of the projectors $\Pi_\pm$, and it is also easy to show that the objects 
 $ \Pi_\pm U\ket{\Psi, \pm} $ take the form of  the eigenstates of the AFF supersymmetric model
divided by $x$.

In summary, we have two schemes of dimensional reductions 
made up by a projection on a subspace with fixed $ k $, the integration of the remaining angular variables,
and a unitary transformation. Let us denote these two schemes as
 $\delta_{\pm}=\{\mathcal{P_\pm},\mathcal{O}_\pm, U\}$. 
Then, by applying the scheme $ \delta_-$ ($ \delta_+ $) in 
our three-dimensional $\mathcal{N}=2$ $\mathfrak{osp}(2|2)$ superconformal system we obtain
a super-extension of the AFF model in the exact (spontaneously broken) supersymmetric phase,
 and there is a one-to-one correspondence between bosonic and fermionic
 generators of the three-dimensional model
with those associated with the one-dimensional model.


\section{Remarks}
We end this chapter with a comment related to supersymmetry and Dirac Hamiltonian.
Taking the nilpotent operators $ \mathcal{Q}^\pm $ given in (\ref{QyW}), 
a Hermitian  supercharge can be constructed, and this has the form
\be
\label{PB-Dirac Op}
\mathcal{Q}_{0}=-\sqrt{2}(\mathcal{Q}^++\mathcal{Q}^-)=\gamma^{i}(p_i-e\mathscr{A}_i)+e\gamma^{0}\mathscr{A}_0\,,
\end{equation}
where $
\mathscr{A}_{0}=\frac{g}{r}\,,$
$\mathscr{A}_{i}=A_{i}-i\frac{\omega}{e}\gamma^{5}\,r_i\,,
$
and 
 $\gamma^{5}=\Gamma$ is  our grading operator in Sec. \ref{osp22 extension}. 
Then the operator (\ref{PB-Dirac Op}) can be viewed
as a parity breaking Euclidean Dirac operator with
components of the gauge potential satisfying the relations 
$-\partial_i \mathscr{A}_{0}=\epsilon_{ijk}\partial_{j}\mathscr{A}_{k}
=gr_i/r^3$.
Hence we are dealing with a new type of parity breaking dyon
background. Actually, the $\gamma^{5}$ terms do not allow for 
an
$\mathcal{N}=4$ supersymmetric extension and we
only have  $\mathcal{N}=2$ supersymmetry, with the 
second supercharge given by $i\sqrt{2}(\mathcal{Q}^+-\mathcal{Q}^-)=i\gamma^5\mathcal{Q}_{0}$.
It is 
interesting  
to relate a parity-breaking Dirac operators with 
supersymmetric quantum mechanics.  
In this context it  is not clear whether  
a (pseudo)classical supersymmetric system exists 
whose quantization would produce our three-dimensional 
superconformal system,  
or we have here a kind of 
a classical anomaly \textcolor{red}{[\cite{clasanomaly}]}. 
Also, 
the fact that the ground state is infinitely degenerate is maybe due to this parity breaking term.

\chapter*{Conclusions and Outlook}
\label{Conclusion}

\addcontentsline{toc}{chapter}{Conclusions}

In conclusion, we recall the  problems  
\emph{a) to d)} that were originally listed in the introduction,
 but now in the light of the obtained results. This will also allow 
to point out interesting problems for further research.

\underline{\emph{a) Connection between different mechanical systems through symmetries}} \\
We addressed the problem of establishing a mapping between 
the two forms of dynamics (in the sense of Dirac \textcolor{red}{[\cite{Dirac}]})
 associated with conformal algebra.

The indicated mapping is  the conformal bridge transformation introduced in 
\textcolor{red}{[\cite{InzPlyWipf1}]} (Chap. \ref{ChBridge}), that relates an asymptotically free system 
with an harmonically confined one.  
The transformation maps  rank $n$ Jordan states of the zero energy (and eigenstates)
of the first system  to  eigenstates (coherent states) of the second.  
The conformal bridge also maps  symmetry generators 
 from one system to the another. From its general nature, this mapping 
provides a new approach to study higher dimensional (in the sense of degrees of freedom)
 conformal invariant systems, such as 
the  Calogero model \textcolor{red}{[\cite{Calogero1,Calogero2}]}.  Actually, we have already shown its applicability for 
 the Landau problem analyzed in Chap. \ref{ChBridge}, as well as for the monopole 
background model in Chap. \ref {Chapmono1}. A fairly natural question is whether there is any analog transformation at the level of supersymmetric quantum mechanics,
in such a way we could include in this mapping fermionic integrals of motion. 
There could also be some relationship between this transformation and the
 Riemann hypothesis, since Hamiltonians of the form $ xp $ have been used in this direction
\textcolor{red}{[\cite{Connes,Berry,Regniers,Sierra,Bender2017}]}.

\underline{\emph{b) Hidden and  bosonized supersymmetry}}\\
We wanted to establish  the origin of the hidden 
bosonized superconformal symmetry of the harmonic oscillator in one dimension  \textcolor{red}{[\cite{Hiden1,BalSchBar,CarPly2,Hiden3}]}.

It was shown that  such a bosonized  supersymmetry  originates from a nontrivial supersymmetric system,
via the nonlocal Foldy-Wouthuysen transformation
 \textcolor{red}{[\cite{ InzPly1}]} (Chap. \ref{ChHiddenboson}). 
The only fermionic true integrals of that system are the trivial 
Pauli matrices, and  other operators are dynamical integrals, in the sense of 
the total Heisenberg equation. 
In contrast to the usual super-harmonic oscillator,
 the system has spontaneously broken supersymmetry.
 We explain the nature of this system through confluent Darboux transformation
and in the scheme of free anomaly quantization for second order supersymmetry  \textcolor{red}{[\cite{PlyPara,KliPly,Plyushchay}]}.
The question about what happens in higher dimensional cases remains open,
 however we think that the conformal bridge transformation  could provide  us an answer.

\underline{\emph{c) Hidden symmetries in rationally extended conformal mechanics}}\\
 The objective was to find the 
spectrum generating ladder operators for rational deformations of the AFF model and 
its supersymmetric extensions.

We have used the DCKA transformation
 to produce a rational extension of the AFF model. The nature of the resulting 
Hamiltonians depends on the choice of the seed states: We can produce isospectral 
and non-isospectral rational deformations that have 
an arbitrary number of gaps of different sizes in their spectra. 
Starting from the harmonic oscillator 
\textcolor{red}{[\cite{CarInzPly}]} (Chap. \ref{ChRQHO}), we implemented 
an algorithmic procedure that takes a set of seed states for
 DCKA transformation (them could be physical or nonphysical, but not a mixture),
 and produces a new set of seed states 
 of a different nature.
 Both Darboux schemes essentially generate the same system, up to an additive constant.  
This is what we called a Darboux duality for the harmonic oscillator, and we have used it to construct
the spectrum-generating ladder operators for rational deformations of the AFF 
model with potential  $ x^2 + m (m + 1) / x^2 $ where $ m = 0,1, \ldots. $.
 These ladder operators fall into three categories; Operators of the type $ \mathcal{A} $
 that irreducibly act on the equidistant part of the spectrum but annihilate all separate states. 
Operators of type $\mathcal{B}$ that act similarly to $ \mathcal{A} $ on the
 equidistant part of the spectrum but annihilate only the upper (rising operator) 
 and lower (lowering operator) states in
 each separate  band. Finally, operators of type $ \mathcal{C} $, that 
 connect the separated part of the spectrum with its equidistant part. 
These results are analogous to what was obtained for rational extensions 
of the harmonic oscillator in \textcolor{red}{[\cite{CarPly}]}.

This phenomenon in which different possible options of  Darboux schemes produce the same system,
also appears in the context of deformations of the free particle, specifically,
 in the construction of the so-called reflectionless potentials, see \textcolor{red}{[\cite{MatSal}]} for a background on the subject. 
The main difference between these systems and the
rational deformations of the harmonic oscillator (as well as deformations of the AFF model),
 is that the Darboux
 schemes produce there the same potential without any additive constant.
 This implies that the Darboux dressing procedure provides there the true integrals of motion, 
which are the so-called Lax-Novikov integrals, see \textcolor{red}{[\cite{Boson3,Arancibia,Arancibia2,Arancibia2014,plyushchay2020exotic}]}
for more information.

The next  step was to study the complete nonlinear supersymmetry 
that characterizes the rational super-extensions of the AFF model and the 
harmonic oscillator \textcolor{red}{[\cite{InzPly2}]} (Chap. \ref{ChNonLinearSUSY}).
By means of a set of algebraic relations, we have obtained  a
 large chain of new higher-order dynamical integrals that act irreducibly in the system,
 in a similar way as powers of the first-order  ladder operators do in the 
case of the simplest harmonic oscillator. We stopped the generation of integrals
 when we realized  that certain objects  can be written in terms of more basic elements than they are, 
otherwise one would have an infinite-dimensional algebra of the $W$ type, see \textcolor{red}{[\cite{deBoer}]}
and references therein. 
With fermionic generators we have a similar picture. 
Despite having so many new operators, which we cannot avoid because they arise from the
commutation relations between operators of the type $\mathcal{A}$, 
$\mathcal{B}$ and $\mathcal{C}$,
 the role they play is not clear since the spectrum-generating set was already built.
Perhaps there is a more basic structure behind this construction,
 hidden in the virtual systems produced by the Darboux chain, but this is still an open question.
In this context, another interesting problem 
to investigate is whether these higher order generators can be obtained 
by means of a quantization prescription of a pseudo-classical system, however one must bear
 in mind that higher order supersymmetry presents a quantum anomaly 
\textcolor{red}{[\cite{KliPly,Plyushchay}]}.

In \textcolor{red}{[\cite{InzPly3}]} (Chap. \ref{ChKlein}) we extend the Darboux duality to the case of the 
AFF model with potential  $ x^2 + \nu (\nu + 1) / x^2 $, where $\nu\geq-1/2$. 
This is possible due to the Klein four-group
 associated to the   Schr\"odinger equation of the model.
 Having the Darboux duality for this system allows us to extend 
the notion of the three classes of ladder operators described above, now for any possible deformation 
of the AFF model. We have not considered spectrum-generating 
algebras and supersymmetric extensions for these cases, so this remains as an open problem.
 Within all this, the cases in which 
$\nu$ is a half-integer number are really special: When this happens, the confluent Darboux transformation
 is involved in some of the recipes for constructing rationally extended potentials, and some rational extensions
 undergo significant structural changes.  
Such changes are reflected both
 in the available energy levels, such as in the number of physical states, and also in the kernels of the
of spectrum-generating ladder operators, where now nonphysical states and Jordan states appear.

On the other hand, systems very similar to these, but without the harmonic term, appeared in a completely different context, through the so-called $ \mathcal{PT} $ regularization 
\textcolor{red}{[\cite{Correa2016,JM1, JM2}]}. These models are intimately related to the Korteweg-de Vries 
equation due to the Lax pair formalism \textcolor{red}{[\cite{MatSal}]} and help to provide new types
 of  solutions. It would be interesting to clarify if there is a generalization of the conformal
 bridge for  deformed systems, that could provide us a new knowledge
 related to integrable models.

\underline{\emph{d) Hidden symmetries in three-dimensional conformal mechanics}}\\
For this problem, we have considered a particle with electric charge $e$ in a Dirac monopole background,
i.e., a $U(1)$ external vector  potential $\vA$, the curl of which gives us 
the spherically symmetric magnetic field produced by a monopole source with charge $g$, see details in 
\textcolor{red}{[\cite{Sakurai,Mcin,Vinet,InzPlyWipf1}]} and in the references cited there. 
The particle  was also 
  subjected 
to a central potential of the form $ V (\vr) = \frac{\alpha}{2m \vr ^ 2} 
+\frac {m \omega \vr^2}{ 2} $.
We investigated the possibility of obtaining hidden integrals of motion for this system, and 
we also looked for a possible supersymmetric extension of this model. 

It was found that the system has hidden symmetries when $\alpha=(eg)^2$.
At the classical level,
 they control the periodic nature of the trajectory, besides  
in the quantum case, these integrals reveal the nature of
 spectrum degeneration of the system.

 To construct the hidden integrals at the classical level,
 we have used the fact that the projection of the particle's trajectory 
into the orthogonal plane  to the Poincar\'e vector integral (the modified 
 angular momentum of the system),  is analogous to the orbit 
of the three-dimensional harmonic oscillator. 
Actually, we demonstrated that this is a universal property of this background, i.e., 
if we change the harmonic trap for an arbitrary central potential, the dynamics 
in the mentioned plane will be the same that would occur in the absence of the monopole charge.

It is also necessary  to emphasize that the system has the
  $\mathfrak{sl} (2, \R) $ symmetry  and is connected with 
  an $\mathfrak{so}(2,1) $  invariant  system previously analyzed  in
\textcolor{red}{[\cite{PlyWipf}]}, 
   by means of the 
conformal bridge transformation. This brings us another way to get the integrals of the hidden symmetries. 

Inspired by the so-called ``Dirac oscillator''
proposed in \textcolor{red}{[\cite{DiracOs1, DiracOs2, DiracOs3}]}, we introduced a special spin-orbit coupling 
term into the Hamiltonian of our  system in the monopole background  (Chap. \ref{Chapmono2}), 
and this naturally leads us to the construction of a supersymmetric extension. The resulting model
is a three-dimensional realization of the $ \mathfrak{osp} (2 | 2) $ superconformal  symmetry, and 
some of its interesting proprieties appear in the following list: 
\begin{itemize}
\item In the limit $\nu\rightarrow 0$, the Hamiltonian of our system takes the form
$$
\mathcal{H}_{\text{DO}}=
\frac{1}{2}\left(\vp^2+\omega^2\vr^2\right)\I_{4\cross 4}+\omega\Gamma(
\vsigma\cdot  \vL + \frac{3}{2})\,,\qquad \Gamma=\sigma_3\otimes \mathbb{I}_{2\times  2}\,.
$$
which is identified with the mentioned 
 Dirac oscillator Hamiltonian 
in the non-relativistic limit.

\item In the limit $\omega\rightarrow 0$, our  Hamiltonian operator is transformed into 
$$
\mathcal{H}_{\text{dyon}}=
\frac{1}{2}\left((\vp-e\vA)^2+\frac{\nu^2}{r^2}\right)\I_{4\cross 4}+\frac{\nu}{r^2}\vsigma\cdot\vr \Pi_-\,,\qquad 
 \Pi_\pm=\frac{1}{2}(1\pm \Gamma)\,,
$$
which$\quad$ is$\quad$ interpreted$\quad$ as$\quad$ the$\quad$ Pauli$\quad$  Hamiltonian$\quad$ of$\quad$ a$\quad$ 
supersymmetric dyon $(c=1)$ \textcolor{red}{[\cite{PlyWipf}]}. 
This system has 
  the exceptional superconformal symmetry 
$ D (2; 1, \alpha) $ with $ \alpha = 1/2 $, which is larger than 
$ \mathfrak{osp} (2 | 2) $ superalgebra, so we believe that 
some important structures are still missing in our construction.

\item The system has two classes of energy levels organized in two independent towers. 
The eigenvalues associated with one of these towers are infinitely degenerate, 
while the energies in the other tower have finite degeneracy.

\item Through the application of two different dimensional reduction schemes,
 the system is transformed into the super-extended AFF model. 
One scheme gives us the extended system in the spontaneously broken
 supersymmetric phase, while the other scheme produces the system in the exact supersymmetric phase.

\end{itemize}

This type of system opens an interesting line of research, which consists 
in exploring the supersymmetric structure of a Dirac Hamiltonian that breaks parity symmetry
(since the Hermitian supercharges of our model can be interpreted in this way), 
 and searching for applications for systems with infinitely degenerate ground energy.


\bibliography{mibiblio}

\begin{thebibliography}{}

\bibitem[Adler, 1994]{Adler}
Adler, V.~E. (1994).
\newblock \emph{``A modification of Crum's method"}.
\newblock {\em Theor. Math. Phys.}, 101:1381.

\bibitem[Aizawa, 2011]{Aizawa}
Aizawa, N. (2011).
\newblock \emph{``Lowest weight representations of super Schr{\"o}dinger
  algebras in low dimensional spacetime"}.
\newblock In {\em Journal of Physics: Conference Series}, volume 284, page
  012007.
  \href{https://iopscience.iop.org/article/10.1088/1742-6596/284/1/012007}{IOP
  Publishing}.

\bibitem[Akulov and Pashnev, 1983]{SCM1}
Akulov, V. and Pashnev, A. (1983).
\newblock \emph{``Quantum superconformal model in the (1, 2) space"}.
\newblock {\em
  \href{https://link.springer.com/article/10.1007%2FBF01086252}{Teoreticheskaya
  i Matematicheskaya Fizika}}, 56(3):344--349.

\bibitem[Ammon and Erdmenger, 2015]{Ammon}
Ammon, M. and Erdmenger, J. (2015).
\newblock {\em Gauge/Gravity Duality: Foundations and Applications}.
\newblock Cambridge University Press.

\bibitem[Andrzejewski, 2014]{NH3}
Andrzejewski, K. (2014).
\newblock {\emph{``Conformal Newton--Hooke algebras, Niederer's transformation
  and Pais--Uhlenbeck oscillator"}}.
\newblock {\em
  \href{https://www.sciencedirect.com/science/article/pii/S037026931400731X?via%3Dihub}{Phys.
  Lett. B}}, 738:405--411.

\bibitem[Arancibia et~al., 2014]{Arancibia2014}
Arancibia, A., Correa, F., Jakubsk\'y, V., Mateos~Guilarte, J., and Plyushchay,
  M.~S. (2014).
\newblock {Soliton defects in one-gap periodic system and exotic
  supersymmetry}.
\newblock {\em
  \href{https://journals.aps.org/prd/abstract/10.1103/PhysRevD.90.125041}{Phys.
  Rev. D}}, 90(12):125041.

\bibitem[Arancibia et~al., 2013]{Arancibia}
Arancibia, A., Mateos~Guilarte, J., and Plyushchay, M.~S. (2013).
\newblock {\emph{``Effect of scalings and translations on the supersymmetric
  quantum mechanical structure of soliton systems"}}.
\newblock {\em
  \href{https://journals.aps.org/prd/abstract/10.1103/PhysRevD.87.045009}{Phys.
  Rev. D}}, 87(4):045009.

\bibitem[Arancibia and Plyushchay, 2014]{Arancibia2}
Arancibia, A. and Plyushchay, M.~S. (2014).
\newblock {\emph{``Transmutations of supersymmetry through soliton scattering,
  and self-consistent condensates"}}.
\newblock {\em
  \href{https://journals.aps.org/prd/abstract/10.1103/PhysRevD.90.025008}{Phys.
  Rev. D}}, 90(2):025008.

\bibitem[Arnold et~al., 1989]{Arnold}
Arnold, V.~I., Weinstein, A., and Vogtmann, K. (1989).
\newblock {\em Mathematical Methods Of Classical Mechanics}.
\newblock Graduate Texts in Mathematics. Springer.

\bibitem[Asorey et~al., 2007]{Aso}
Asorey, M., Carinena, J., Marmo, G., and Perelomov, A. (2007).
\newblock {\emph{``Isoperiodic classical systems and their quantum
  counterparts"}}.
\newblock {\em
  \href{https://www.sciencedirect.com/science/article/abs/pii/S0003491606001461?via%3Dihub}{Annals
  Phys.}}, 322:1444--1465.

\bibitem[Balantekin et~al., 1988]{BalSchBar}
Balantekin, A., Schmitt, H., and Barrett, B.~R. (1988).
\newblock \emph{``Coherent states for the harmonic oscillator representations
  of the orthosymplectic supergroup Osp (1/2 N, R)"}.
\newblock {\em \href{https://aip.scitation.org/doi/10.1063/1.528189}{J. Math.
  Phys.}}, 29(7):1634--1639.

\bibitem[Balasubramanian and McGreevy, 2008]{GAdS2}
Balasubramanian, K. and McGreevy, J. (2008).
\newblock {\emph{``Gravity duals for non-relativistic CFTs"}}.
\newblock {\em
  \href{https://journals.aps.org/prl/abstract/10.1103/PhysRevLett.101.061601}{Phys.
  Rev. Lett.}}, 101:061601.

\bibitem[Barbon and Fuertes, 2008]{BarFue}
Barbon, J.~L. and Fuertes, C.~A. (2008).
\newblock {\emph{``On the spectrum of nonrelativistic AdS/CFT"}}.
\newblock {\em
  \href{https://iopscience.iop.org/article/10.1088/1126-6708/2008/09/030}{JHEP}},
  09:030.

\bibitem[Beckers et~al., 1987]{beckers2}
Beckers, J., Dehin, D., and Hussin, V. (1987).
\newblock \emph{``Symmetries and supersymmetries of the quantum harmonic
  oscillator"}.
\newblock {\em
  \href{https://iopscience.iop.org/article/10.1088/0305-4470/20/5/024/meta}{J.
  Phys. A}}, 20(5):1137.

\bibitem[Beckers and Hussin, 1986]{beckers1}
Beckers, J. and Hussin, V. (1986).
\newblock \emph{``Dynamical supersymmetries of the harmonic oscillator"}.
\newblock {\em
  \href{https://www.sciencedirect.com/science/article/abs/pii/0375960186903166?via%3Dihub}{Phys.
  Lett. A}}, 118(7):319--321.

\bibitem[Belavin et~al., 1984]{Polyakov2}
Belavin, A., Polyakov, A., and Zamolodchikov, A. (1984).
\newblock \emph{``Infinite conformal symmetry in two-dimensional quantum field
  theory"}.
\newblock {\em
  \href{https://www.sciencedirect.com/science/article/pii/055032138490052X?via%3Dihub}{Nucl.
  Phys. B}}, 241:333--380.

\bibitem[Bellucci et~al., 2005]{HSUSY4}
Bellucci, S., Krivonos, S., and Nersessian, A. (2005).
\newblock {\emph{``N=8 supersymmetric mechanics on special Kahler manifolds"}}.
\newblock {\em
  \href{https://linkinghub.elsevier.com/retrieve/pii/S0370269304015795}{Phys.
  Lett. B}}, 605:181--184.

\bibitem[Bellucci et~al., 2006]{HSUSY5}
Bellucci, S., Nersessian, A., and Yeranyan, A. (2006).
\newblock {\emph{``Hamiltonian reduction and supersymmetric mechanics with
  Dirac monopole"}}.
\newblock {\em
  \href{https://journals.aps.org/prd/abstract/10.1103/PhysRevD.74.065022}{Phys.
  Rev. D}}, 74:065022.

\bibitem[Bender, 2007]{PT2}
Bender, C.~M. (2007).
\newblock {\emph{``Making sense of non-Hermitian Hamiltonians"}}.
\newblock {\em
  \href{https://iopscience.iop.org/article/10.1088/0034-4885/70/6/R03}{Rept.
  Prog. Phys.}}, 70:947.

\bibitem[Bender et~al., 2017]{Bender2017}
Bender, C.~M., Brody, D.~C., and M{\"u}ller, M.~P. (2017).
\newblock \emph{``Hamiltonian for the zeros of the Riemann zeta function"}.
\newblock {\em
  \href{https://journals.aps.org/prl/abstract/10.1103/PhysRevLett.118.130201}{Phys.
  Rev. Lett.}}, 118(13):130201.

\bibitem[Bentez et~al., 1990]{DiracOs2}
Bentez, J., y~Romero, R.~M., N{\'u}ez-Y{\'e}pez, H., and Salas-Brito, A.
  (1990).
\newblock \emph{``Solution and hidden supersymmetry of a Dirac oscillator"}.
\newblock {\em
  \href{https://journals.aps.org/prl/abstract/10.1103/PhysRevLett.65.2085}{Phys.
  Rev lett.}}, 64(14):1643.

\bibitem[Berezin and Marinov, 1975]{psedoclasical2}
Berezin, F. and Marinov, M. (1975).
\newblock \textit{``Classical spin and Grassman algebra"}.
\newblock {\em Pis' ma v Zhurnal Ehksperimental'noj i Teoreticheskoj Fiziki},
  21(11):678--680.

\bibitem[Berezin and Marinov, 1976]{psedoclasical1}
Berezin, F. and Marinov, M. (1976).
\newblock \emph{``Particle spin dynamics as the Grassmann variant of classical
  mechanics"}.
\newblock Technical report, Gosudarstvennyj Komitet po Ispol'zovaniyu Atomnoj
  Ehnergii SSSR.

\bibitem[Berry and Keating, 1999]{Berry}
Berry, M.~V. and Keating, J.~P. (1999).
\newblock \emph{``The Riemann zeros and eigenvalue asymptotics"}.
\newblock {\em \href{https://epubs.siam.org/doi/10.1137/S0036144598347497}{SIAM
  review}}, 41(2):236--266.

\bibitem[Bonatsos et~al., 1994]{Bonatsos}
Bonatsos, D., Daskaloyannis, C., Kolokotronis, P., and Lenis, D. (1994).
\newblock {\emph{``The Symmetry algebra of the N-dimensional anisotropic
  quantum harmonic oscillator with rational ratios of frequencies and the
  Nilsson model''}}.
\newblock {\em \href{https://arxiv.org/abs/hep-th/9411218}{arXiv:
  hep-th/9411218}}.

\bibitem[Bonezzi et~al., 2017]{Hiden3}
Bonezzi, R., Corradini, O., Latini, E., and Waldron, A. (2017).
\newblock {\emph{``Quantum Mechanics and Hidden Superconformal Symmetry"}}.
\newblock {\em
  \href{https://journals.aps.org/prd/abstract/10.1103/PhysRevD.96.126005}{Phys.
  Rev. D}}, 96(12):126005.

\bibitem[Brezhnev, 2008]{Brezh}
Brezhnev, Y.~V. (2008).
\newblock \emph{``What does integrability of finite-gap or soliton potentials
  mean?"}.
\newblock {\em
  \href{https://royalsocietypublishing.org/doi/10.1098/rsta.2007.2056}{Phyl.
  Trans. of the Royal Society A}}, 366(1867):923--945.

\bibitem[Britto-Pacumio et~al., 2000]{BlackHold5}
Britto-Pacumio, R., Michelson, J., Strominger, A., and Volovich, A. (2000).
\newblock {Lectures on Superconformal Quantum Mechanics and Multi-Black Hole
  Moduli Spaces}.
\newblock {\em
  \href{https://link.springer.com/chapter/10.1007%2F978-94-011-4303-5_6}{NATO
  Sci. Ser. C}}, 556:255--284.

\bibitem[Brodsky et~al., 2015]{App1}
Brodsky, S.~J., de~Teramond, G.~F., Dosch, H.~G., and Erlich, J. (2015).
\newblock {\emph{``Light-Front Holographic QCD and Emerging Confinement"}}.
\newblock {\em
  \href{https://www.sciencedirect.com/science/article/abs/pii/S0370157315002306}{Phys.
  Rept.}}, 584:1--105.

\bibitem[Calogero, 1969]{Calogero1}
Calogero, F. (1969).
\newblock \emph{``Solution of a three-body problem in one-dimension,''}.
\newblock {\em \href{http://aip.scitation.org/doi/10.1063/1.1664820} {J. Math.
  Phys.}}, 10:2191.

\bibitem[Calogero, 1972]{Calogero2}
Calogero, F. (1972).
\newblock \emph{``Solution of the one-dimensional N body problems with
  quadratic and/or inversely quadratic pair potentials,''}.
\newblock {\em \href{http://aip.scitation.org/doi/10.1063/1.1665604}{J. Math.
  Phys.}}, 12.

\bibitem[Cari\~nena et~al., 2018]{CarInzPly}
Cari\~nena, J.~F., Inzunza, L., and Plyushchay, M.~S. (2018).
\newblock {\emph{``Rational deformations of conformal mechanics"}}.
\newblock {\em
  \href{https://journals.aps.org/prd/abstract/10.1103/PhysRevD.98.026017}{Phys.
  Rev. D}}, 98(2):026017.

\bibitem[Cari\~nena and Plyushchay, 2016a]{CarPly2}
Cari\~nena, J.~F. and Plyushchay, M.~S. (2016a).
\newblock {\emph{``Ground-state isolation and discrete flows in a rationally
  extended quantum harmonic oscillator"}}.
\newblock {\em
  \href{https://journals.aps.org/prd/abstract/10.1103/PhysRevD.94.105022}{Phys.
  Rev. D}}, 94(10):105022.

\bibitem[Cari\~nena and Plyushchay, 2016b]{Car2}
Cari\~nena, J.~F. and Plyushchay, M.~S. (2016b).
\newblock {\emph{``Ground-state isolation and discrete flows in a rationally
  extended quantum harmonic oscillator"}}.
\newblock {\em
  \href{https://journals.aps.org/prd/abstract/10.1103/PhysRevD.94.105022}{Phys.
  Rev. D}}, 94(10):105022.

\bibitem[Cari\~nena and Plyushchay, 2017]{CarPly}
Cari\~nena, J.~F. and Plyushchay, M.~S. (2017).
\newblock {\emph{``ABC of ladder operators for rationally extended quantum
  harmonic oscillator systems"}}.
\newblock {\em
  \href{https://iopscience.iop.org/article/10.1088/1751-8121/aa739b}{J. Phys.
  A}}, 50(27):275202.

\bibitem[Cariglia, 2014]{Cariglia}
Cariglia, M. (2014).
\newblock \emph{``Hidden symmetries of dynamics in classical and quantum
  physics''}.
\newblock {\em
  \href{https://journals.aps.org/rmp/abstract/10.1103/RevModPhys.86.1283}{Rev.
  Mod. Phys. }}, 86:1283.

\bibitem[Cariglia et~al., 2018]{Cariglia2}
Cariglia, M., Galajinsky, A., Gibbons, G., and Horvathy, P. (2018).
\newblock {\emph{``Cosmological aspects of the Eisenhart--Duval lift"}}.
\newblock {\em
  \href{https://link.springer.com/article/10.1140%2Fepjc%2Fs10052-018-5789-x}{Eur.
  Phys. J. C}}, 78(4):314.

\bibitem[Cari{\~n}ena and De~Lucas, 2011]{ermakov4}
Cari{\~n}ena, J.~F. and De~Lucas, J. (2011).
\newblock \emph{``Lie systems: theory, generalisations, and applications"}.
\newblock {\em
  \href{https://www.impan.pl/pl/wydawnictwa/czasopisma-i-serie-wydawnicze/dissertationes-mathematicae/all/479//88196/lie-systems-theory-generalisations-and-applications}{Dissertationes
  Math.}}, 479.

\bibitem[Casalbuoni, 1976]{Casalbuoni}
Casalbuoni, R. (1976).
\newblock \emph{``The classical mechanics for Bose-Fermi systems"}.
\newblock {\em Il Nuovo Cimento A (1965-1970)}, 33(3):389--431.

\bibitem[Cen et~al., 2020]{Cen}
Cen, J., Correa, F., and Fring, A. (2020).
\newblock \emph{``Nonlocal gauge equivalence: Hirota versus extended continuous
  Heisenberg and Landau--Lifschitz equation"}.
\newblock {\em
  \href{https://iopscience.iop.org/article/10.1088/1751-8121/ab81d9}{J. Phys.
  A}}, 53(19):195201.

\bibitem[Chamon et~al., 2011]{Jack}
Chamon, C., Jackiw, R., Pi, S.-Y., and Santos, L. (2011).
\newblock {\emph{``Conformal quantum mechanics as the CFT$_1$ dual to
  AdS$_2$"}}.
\newblock {\em
  \href{https://www.sciencedirect.com/science/article/pii/S0370269311006447?via%3Dihub}{Phys.
  Lett. B}}, 701:503--507.

\bibitem[Connes, 1999]{Connes}
Connes, A. (1999).
\newblock \emph{``Trace formula in noncommutative geometry and the zeros of the
  Riemann zeta function"}.
\newblock {\em
  \href{https://link.springer.com/article/10.1007/s000290050042}{Selecta
  Mathematica}}, 5(1):29.

\bibitem[Contreras-Astorga and Schulze-Halberg, 2015]{Jordan3}
Contreras-Astorga, A. and Schulze-Halberg, A. (2015).
\newblock \emph{``On integral and differential representations of Jordan chains
  and the confluent supersymmetry algorithm"}.
\newblock {\em
  \href{https://iopscience.iop.org/article/10.1088/1751-8113/48/31/315202}{J.
  Phys. A}}, 48(31):315202.

\bibitem[Cooper et~al., 1995]{Cooper}
Cooper, F., Khare, A., and Sukhatme, U. (1995).
\newblock \emph{``Supersymmetry and quantum mechanics''}.
\newblock {\em
  \href{https://www.sciencedirect.com/science/article/pii/037015739400080M?via\%3Dihub}{Phys.
  Rept.}}, 251:267.

\bibitem[Correa and Fring, 2016]{Correa2016}
Correa, F. and Fring, A. (2016).
\newblock ``\emph{{Regularized degenerate multi-solitons}"}.
\newblock {\em
  \href{https://link.springer.com/article/10.1007/JHEP09(2016)008}{JHEP}},
  09:008.

\bibitem[Correa et~al., 2008]{Boson3}
Correa, F., Jakubsky, V., Nieto, L.~M., and Plyushchay, M.~S. (2008).
\newblock {\emph{``Self-isospectrality, special supersymmetry, and their effect
  on the band structure"}}.
\newblock {\em
  \href{https://journals.aps.org/prl/abstract/10.1103/PhysRevLett.101.030403}{Phys.
  Rev. Lett.}}, 101:030403.

\bibitem[Correa et~al., 2015]{Jordan1}
Correa, F., Jakubsky, V., and Plyushchay, M.~S. (2015).
\newblock {\emph{``$PT$-symmetric invisible defects and confluent Darboux-Crum
  transformations"}}.
\newblock {\em
  \href{https://journals.aps.org/pra/abstract/10.1103/PhysRevA.92.023839}{Phys.
  Rev. A}}, 92(2):023839.

\bibitem[Correa et~al., 2014]{CorLefPly}
Correa, F., Lechtenfeld, O., and Plyushchay, M. (2014).
\newblock {\emph{``Nonlinear supersymmetry in the quantum Calogero model"}}.
\newblock {\em
  \href{https://link.springer.com/article/10.1007/JHEP04(2014)151}{JHEP}},
  04:151.

\bibitem[Correa et~al., 2007]{CorNiePly}
Correa, F., Nieto, L.~M., and Plyushchay, M.~S. (2007).
\newblock {\emph{``Hidden nonlinear supersymmetry of finite-gap Lame
  equation"}}.
\newblock {\em
  \href{https://www.sciencedirect.com/science/article/abs/pii/S0370269306014274?via%3Dihub}{Phys.
  Lett. B}}, 644:94--98.

\bibitem[Correa and Plyushchay, 2007]{Boson2}
Correa, F. and Plyushchay, M.~S. (2007).
\newblock {\emph{``Hidden supersymmetry in quantum bosonic systems"}}.
\newblock {\em
  \href{https://www.sciencedirect.com/science/article/abs/pii/S0003491606002831?via%3Dihub}{Annals
  Phys.}}, 322:2493--2500.

\bibitem[Crum, 1955]{Crum}
Crum, M. (1955).
\newblock \emph{``Associated Sturm-Liouville systems"}.
\newblock {\em Quart. J. Math. Oxford}, 6:121.

\bibitem[Darboux, 1882]{Darboux}
Darboux, G. (1882).
\newblock \emph{``Sur une proposition relative aux \'equations lin\'eaires"}.
\newblock {\em C. R. Acad. Sci Paris}, 94:1456.

\bibitem[de~Alfaro et~al., 1976]{AFF}
de~Alfaro, V., Fubini, S., and Furlan, G.
  (\href{https://link.springer.com/article/10.1007%2FBF02785666}{1976}).
\newblock \emph{``Conformal invariance in quantum mechanics"}.
\newblock {\em
  \href{https://link.springer.com/article/10.1007/BF02785666}{Nuovo Cim. A}},
  {\bf 34}:569--612.

\bibitem[de~Azcarraga et~al., 1999]{BlackHold3}
de~Azcarraga, J., Izquierdo, J., Perez~Bueno, J., and Townsend, P. (1999).
\newblock {\emph{``Superconformal mechanics and nonlinear realizations"}}.
\newblock {\em
  \href{https://journals.aps.org/prd/abstract/10.1103/PhysRevD.59.084015}{Phys.
  Rev. D}}, 59:084015.

\bibitem[de~Boer et~al., 1996]{deBoer}
de~Boer, J., Harmsze, F., and Tjin, T. (1996).
\newblock {\emph{``Nonlinear finite W symmetries and applications in elementary
  systems''}}.
\newblock {\em
  \href{https://www.sciencedirect.com/science/article/abs/pii/0370157395000755}{Phys.
  Rept.}}, 272:139--214.

\bibitem[de~Crombrugghe and Rittenberg, 1983]{Hiden1}
de~Crombrugghe, M. and Rittenberg, V. (1983).
\newblock \emph{Supersymmetric quantum mechanics}.
\newblock {\em
  \href{https://www.sciencedirect.com/science/article/abs/pii/0003491683903160}{Annals
  Phys.}}, 151(1):99--126.

\bibitem[Deur et~al., 2015]{Brod2}
Deur, A., Brodsky, S.~J., and de~Teramond, G.~F. (2015).
\newblock {\emph{``Connecting the Hadron Mass Scale to the Fundamental Mass
  Scale of Quantum Chromodynamics"}}.
\newblock {\em
  \href{https://www.sciencedirect.com/science/article/pii/S0370269315007388}{Phys.
  Lett. B}}, 750:528--532.

\bibitem[Dirac, 1949]{Dirac}
Dirac, P.~A. (1949).
\newblock {\emph{``Forms of Relativistic Dynamics"}}.
\newblock {\em
  \href{https://journals.aps.org/rmp/abstract/10.1103/RevModPhys.21.392}{Rev.
  Mod. Phys.}}, 21:392--399.

\bibitem[Dodonov et~al., 1974]{Dodonov}
Dodonov, V., Malkin, I., and Man'Ko, V. (1974).
\newblock \emph{``Even and odd coherent states and excitations of a singular
  oscillator"}.
\newblock {\em Physica}, 72(3):597--615.

\bibitem[Donets et~al., 2000]{SCM4}
Donets, E., Pashnev, A., Rivelles, V.~O., Sorokin, D.~P., and Tsulaia, M.
  (2000).
\newblock {\emph{``N=4 superconformal mechanics and the potential structure of
  AdS spaces"}}.
\newblock {\em
  \href{https://www.sciencedirect.com/science/article/abs/pii/S0370269300006705}{Phys.
  Lett. B}}, 484:337--346.

\bibitem[Dorey et~al., 2001]{PT1}
Dorey, P., Dunning, C., and Tateo, R. (2001).
\newblock {\emph{``Supersymmetry and the spontaneous breakdown of PT
  symmetry}"}.
\newblock {\em
  \href{https://iopscience.iop.org/article/10.1088/0305-4470/34/28/102https://iopscience.iop.org/article/10.1088/0305-4470/34/28/102}{J.
  Phys. A}}, 34:L391.

\bibitem[Dubov et~al., 1994]{Dubov}
Dubov, S.~Y., Eleonskii, V., and Kulagin, N. (1994).
\newblock \emph{``Equidistant spectra of anharmonic oscillators"}.
\newblock {\em
  \href{https://aip.scitation.org/doi/abs/10.1063/1.166056}{Chaos}},
  4(1):47--53.

\bibitem[Duval et~al., 1991]{DGH}
Duval, C., Gibbons, G.~W., and Horvathy, P. (1991).
\newblock {\emph{``Celestial mechanics, conformal structures and gravitational
  waves"}}.
\newblock {\em
  \href{https://journals.aps.org/prd/abstract/10.1103/PhysRevD.43.3907}{Phys.
  Rev. D}}, 43:3907--3922.

\bibitem[Duval and Horvathy, 1994]{Duval}
Duval, C. and Horvathy, P. (1994).
\newblock \emph{``On Schr{\"o}dinger superalgebras"}.
\newblock {\em \href{https://aip.scitation.org/doi/10.1063/1.530521}{J. Math.
  Phys}}, 35(5):2516--2538.

\bibitem[El-Ganainy et~al., 2018]{PT3}
El-Ganainy, R., Makris, K.~G., Khajavikhan, M., Musslimani, Z.~H., Rotter, S.,
  and Christodoulides, D.~N. (2018).
\newblock \emph{``Non-Hermitian physics and PT symmetry"}.
\newblock {\em \href{https://www.nature.com/articles/nphys4323}{Nature
  Physics}}, 14(1):11--19.

\bibitem[Ermakov, 1880]{ermakov1}
Ermakov, V. (1880).
\newblock \emph{``Transformation of differential equations"}.
\newblock {\em Univ. Izv. Kiev}, 20(1).

\bibitem[Evnin and Rongvoram, 2017]{Evnin}
Evnin, O. and Rongvoram, N. (2017).
\newblock {\emph{``Hidden symmetries of the Higgs oscillator and the conformal
  algebra"}}.
\newblock {\em
  \href{https://iopscience.iop.org/article/10.1088/1751-8113/50/1/015202/meta}{J.
  Phys. A}}, 50(1):015202.

\bibitem[Falomir and Pisani, 2005]{Falomir2}
Falomir, H. and Pisani, P. (2005).
\newblock {\emph{``Self-adjoint extensions and SUSY breaking in supersymmetric
  quantum mechanics"}}.
\newblock {\em
  \href{https://iopscience.iop.org/article/10.1088/0305-4470/38/21/011}{J.
  Phys. A}}, 38:4665--4683.

\bibitem[Falomir et~al., 2002]{Falomir1}
Falomir, H., Pisani, P., and Wipf, A. (2002).
\newblock {\emph{``Pole structure of the Hamiltonian zeta function for a
  singular potential"}}.
\newblock {\em
  \href{https://iopscience.iop.org/article/10.1088/0305-4470/35/26/306}{J.
  Phys. A}}, 35:5427--5444.

\bibitem[Fedoruk et~al., 2012]{SCM5}
Fedoruk, S., Ivanov, E., and Lechtenfeld, O. (2012).
\newblock {\emph{``Superconformal Mechanics"}}.
\newblock {\em
  \href{https://iopscience.iop.org/article/10.1088/1751-8113/45/17/173001/meta}{J.
  Phys. A}}, 45:173001.

\bibitem[Foldy and Wouthuysen, 1950]{FW}
Foldy, L.~L. and Wouthuysen, S.~A. (1950).
\newblock \textit{``On the Dirac theory of spin 1/2 particles and its
  non-relativistic limit"}.
\newblock {\em
  \href{https://journals.aps.org/pr/abstract/10.1103/PhysRev.78.29}{Phys.
  Rev.}}, 78(1):29.

\bibitem[Fradkin, 1965]{Frad}
Fradkin, D.~M. (1965).
\newblock \emph{``Three-Dimensional Isotropic Harmonic Oscillator and SU3''}.
\newblock {\em \href{https://aapt.scitation.org/doi/10.1119/1.1971373}{Am. J.
  Phys.}}, 33:207.

\bibitem[Francesco et~al., 1997]{Francesco}
Francesco, P., Mathieu, P., and Senechal, D. (1997).
\newblock {\em Conformal Field theory}.
\newblock Graduate texts in contemporary physics. Springer.

\bibitem[Fubini and Rabinovici, 1984]{SCM2}
Fubini, S.~P. and Rabinovici, E. (1984).
\newblock \emph{``Superconformal quantum mechanics"}.
\newblock {\em
  \href{https://www.sciencedirect.com/science/article/pii/055032138490422X?via%3Dihub}{Nucl.
  Phys. B}}, 245(CERN-TH-3825):17--44.

\bibitem[Galajinsky, 2010]{NH2}
Galajinsky, A. (2010).
\newblock {Conformal mechanics in Newton-Hooke spacetime}.
\newblock {\em
  \href{https://www.sciencedirect.com/science/article/pii/S0550321310001173?via%3Dihub}{Nucl.
  Phys. B}}, 832:586--604.

\bibitem[Galajinsky, 2015]{Galaj}
Galajinsky, A. (2015).
\newblock {\emph{``$\mathcal N =4$ superconformal mechanics from the $su(2)$
  perspective"}}.
\newblock {\em
  \href{https://link.springer.com/article/10.1007/JHEP02(2015)091}{JHEP}},
  02:091.

\bibitem[Galajinsky, 2018]{NH4}
Galajinsky, A. (2018).
\newblock {\emph{``Geometry of the isotropic oscillator driven by the conformal
  mode"}}.
\newblock {\em
  \href{https://link.springer.com/article/10.1140%2Fepjc%2Fs10052-018-5568-8}{Eur.
  Phys. J. C}}, 78(1):72.

\bibitem[Gamboa and Plyushchay, 1998]{clasanomaly}
Gamboa, J. and Plyushchay, M. (1998).
\newblock {\emph{``Classical anomalies for spinning particles"}}.
\newblock {\em
  \href{https://www.sciencedirect.com/science/article/pii/S055032139700792X?via%3Dihub}{Nucl.
  Phys. B}}, 512:485--504.

\bibitem[Gamboa et~al., 1999]{Gamboa2}
Gamboa, J., Plyushchay, M., and Zanelli, J. (1999).
\newblock {\emph{``Three aspects of bosonized supersymmetry and linear
  differential field equation with reflection"}}.
\newblock {\em
  \href{https://www.sciencedirect.com/science/article/pii/S0550321398008323?via%3Dihub}{Nucl.
  Phys. B}}, 543:447--465.

\bibitem[Gazeau, 2009]{Gazeau}
Gazeau, J.-P. (2009).
\newblock {\em Coherent states in quantum physics}.
\newblock Wiley.

\bibitem[Ghanmi, 2012]{Hermite}
Ghanmi, A. (2012).
\newblock \emph{``Operational formulae for the complex Hermite polynomials"}.
\newblock {\em \href{https://arxiv.org/abs/1211.5746}{arXiv:1211.5746}}.

\bibitem[Gibbons and Townsend, 1999]{BlackHold1}
Gibbons, G. and Townsend, P. (1999).
\newblock {\emph{``Black holes and Calogero models"}}.
\newblock {\em
  \href{https://www.sciencedirect.com/science/article/abs/pii/S037026939900266X?via%3Dihub}{Phys.
  Lett. B}}, 454:187--192.

\bibitem[Gilmore, 2006]{Gilmore}
Gilmore, R. (2006).
\newblock {\em Lie groups, Lie algebras, and some of their applications}.
\newblock Dover Publications.

\bibitem[Ginsparg, 1988]{Ginsparg}
Ginsparg, P.~H. (1988).
\newblock {\emph{``Applied Conformal Field Theory"}}.
\newblock In {\em {Les Houches Summer School in Theoretical Physics: Fields,
  Strings, Critical Phenomena}}, pages 1--168.

\bibitem[G{\'o}mez-Ullate et~al., 2013]{Gomez2}
G{\'o}mez-Ullate, D., Grandati, Y., and Milson, R. (2013).
\newblock \emph{``Rational extensions of the quantum harmonic oscillator and
  exceptional Hermite polynomials"}.
\newblock {\em
  \href{https://iopscience.iop.org/article/10.1088/1751-8113/47/1/015203/meta}{J.
  Phys. A}}, 47(1):015203.

\bibitem[G{\'o}mez-Ullate and Milson, 2019]{gomez2019}
G{\'o}mez-Ullate, D. and Milson, R. (2019).
\newblock \emph{``Lectures on exceptional orthogonal polynomials and rational
  solutions to Painlev$\backslash$'e equations"}.
\newblock {\em \href{https://arxiv.org/abs/1912.07597}{arXiv:1912.07597}}.

\bibitem[Grandati, 2012]{Grand}
Grandati, Y. (2012).
\newblock {Multistep DBT and regular rational extensions of the isotonic
  oscillator}.
\newblock {\em
  \href{https://www.sciencedirect.com/science/article/abs/pii/S0003491612001054}{Annals
  Phys.}}, 327:2411--2431.

\bibitem[Henkel and Unterberger, 2003]{Henkel}
Henkel, M. and Unterberger, J. (2003).
\newblock \emph{``Schr{\"o}dinger invariance and spacetime symmetries"}.
\newblock {\em
  \href{https://www.sciencedirect.com/science/article/pii/S0550321303002529?via%3Dihub}{Nucl.
  Phys. B}}, 660(3):407--435.

\bibitem[Infeld and Hull, 1951]{Infeld}
Infeld, L. and Hull, T. (1951).
\newblock \emph{``The factorization method"}.
\newblock {\em
  \href{https://journals.aps.org/rmp/abstract/10.1103/RevModPhys.23.21#:~:text=The%20factorization%20method%20is%20an,are%20of%20importance%20to%20physicists.}{Rev.
  Mod. Phys.}}, 23(1):21.

\bibitem[Inzunza and Plyushchay, 2018]{InzPly1}
Inzunza, L. and Plyushchay, M.~S. (2018).
\newblock {\emph{``Hidden superconformal symmetry: Where does it come from?"}}.
\newblock {\em
  \href{https://journals.aps.org/prd/abstract/10.1103/PhysRevD.97.045002}{Phys.
  Rev. D}}, 97(4):045002.

\bibitem[Inzunza and Plyushchay, 2019a]{InzPly2}
Inzunza, L. and Plyushchay, M.~S. (2019a).
\newblock {\emph{``Hidden symmetries of rationally deformed superconformal
  mechanics"}}.
\newblock {\em
  \href{https://journals.aps.org/prd/abstract/10.1103/PhysRevD.99.025001}{Phys.
  Rev. D}}, 99(2):025001.

\bibitem[Inzunza and Plyushchay, 2019b]{InzPly3}
Inzunza, L. and Plyushchay, M.~S. (2019b).
\newblock {\emph{``Klein four-group and Darboux duality in conformal
  mechanics"}}.
\newblock {\em
  \href{https://journals.aps.org/prd/abstract/10.1103/PhysRevD.99.125016}{Phys.
  Rev. D}}, 99(12):125016.

\bibitem[Inzunza et~al., 2020a]{InzPlyWipf1}
Inzunza, L., Plyushchay, M.~S., and Wipf, A. (2020a).
\newblock {\emph{``Conformal bridge between asymptotic freedom and
  confinement"}}.
\newblock {\em
  \href{https://journals.aps.org/prd/abstract/10.1103/PhysRevD.101.105019}{Phys.
  Rev. D}}, 101(10):105019.

\bibitem[Inzunza et~al., 2020b]{InzPlyWipf2}
Inzunza, L., Plyushchay, M.~S., and Wipf, A. (2020b).
\newblock {\emph{``Hidden symmetry and (super)conformal mechanics in a monopole
  background"}}.
\newblock {\em
  \href{https://link.springer.com/article/10.1007%2FJHEP04%282020%29028}{JHEP}},
  04:028.

\bibitem[Ivanov et~al., 2003]{HSUSY2}
Ivanov, E., Krivonos, S., and Lechtenfeld, O. (2003).
\newblock {\emph{``New variant of N=4 superconformal mechanics"}}.
\newblock {\em
  \href{https://iopscience.iop.org/article/10.1088/1126-6708/2003/03/014}{JHEP}},
  03:014.

\bibitem[Ivanov et~al., 1989]{SCM3}
Ivanov, E., Krivonos, S., and Leviant, V. (1989).
\newblock \emph{``Geometric superfield approach to superconformal mechanics"}.
\newblock {\em
  \href{https://iopscience.iop.org/article/10.1088/0305-4470/22/19/015/pdf}{J.
  Phys. A}}, 22(19):4201.

\bibitem[Jackiw and Pi, 2011]{Jackiw}
Jackiw, R. and Pi, S.-Y. (2011).
\newblock {\emph{``Tutorial on scale and conformal symmetries in diverse
  dimensions}"}.
\newblock {\em
  \href{https://iopscience.iop.org/article/10.1088/1751-8113/44/22/223001}{J.
  Phys. A}}, 44:223001.

\bibitem[Jakubsk{\`y} et~al., 2010]{jakubsky}
Jakubsk{\`y}, V., Nieto, L.~M., and Plyushchay, M.~S. (2010).
\newblock \emph{``The origin of the hidden supersymmetry"}.
\newblock {\em
  \href{https://www.sciencedirect.com/science/article/pii/S0370269310008270}{Physics
  Letters B}}, 692(1):51--56.

\bibitem[Jauch and Hill, 1940]{Jauch}
Jauch, J.~M. and Hill, E.~L. (1940).
\newblock \emph{``On the problem of degeneracy in quantum mechanics''}.
\newblock {\em
  \href{https://journals.aps.org/pr/abstract/10.1103/PhysRev.57.641}{Phys.
  Rev.}}, 57:641.

\bibitem[Kirchberg et~al., 2003]{HSUSY1}
Kirchberg, A., Lange, J., Pisani, P., and Wipf, A. (2003).
\newblock {\emph{``Algebraic solution of the supersymmetric hydrogen atom in
  d-dimensions"}}.
\newblock {\em
  \href{https://www.sciencedirect.com/science/article/abs/pii/S0003491603000034}{Annals
  Phys.}}, 303:359--388.

\bibitem[Kirchberg et~al., 2005]{HSUSY3}
Kirchberg, A., Lange, J., and Wipf, A. (2005).
\newblock {\emph{``Extended supersymmetries and the Dirac operator"}}.
\newblock {\em
  \href{https://www.sciencedirect.com/science/article/abs/pii/S0003491604001496?via%3Dihub}{Annals
  Phys.}}, 315:467--487.

\bibitem[Kirsten and Loya, 2010]{kirsten}
Kirsten, K. and Loya, P. (2010).
\newblock \emph{``Spectral functions for the Schr{\"o}dinger operator on $\R+$
  with a singular potential"}.
\newblock {\em \href{https://aip.scitation.org/doi/10.1063/1.3263937}{J. Math.
  Phys.}}, 51(5):053512.

\bibitem[Klauder and Skagerstam, 1985]{Klauder}
Klauder, J.~R. and Skagerstam, B.-S. (1985).
\newblock {\em Coherent states: applications in physics and mathematical
  physics}.
\newblock World scientific.

\bibitem[Klishevich and Plyushchay, 2001]{KliPly}
Klishevich, S.~M. and Plyushchay, M.~S. (2001).
\newblock {\emph{``Nonlinear supersymmetry, quantum anomaly and quasiexactly
  solvable systems"}}.
\newblock {\em
  \href{https://www.sciencedirect.com/science/article/pii/S0550321301001973?via%3Dihub}{Nucl.
  Phys. B}}, 606:583--612.

\bibitem[Kozyrev et~al., 2017]{HSUSY6}
Kozyrev, N., Krivonos, S., Lechtenfeld, O., Nersessian, A., and Sutulin, A.
  (2017).
\newblock {\emph{``Curved Witten-Dijkgraaf-Verlinde-Verlinde equation and
  ${\cal N}{=}\,4$ mechanics"}}.
\newblock {\em
  \href{https://journals.aps.org/prd/abstract/10.1103/PhysRevD.96.101702}{Phys.
  Rev. D}}, 96(10):101702.

\bibitem[Krein, 1957]{Krein}
Krein, M.~G. (1957).
\newblock \emph{``On a continuous analogue of a Christoffel formula from the
  theory of orthogonal polynomials"}.
\newblock {\em Dokl. Akad. Nauk SSSR}, 113:970.

\bibitem[Labelle et~al., 1991]{Vinet}
Labelle, S., Mayrand, M., and Vinet, L. (1991).
\newblock \emph{``Symmetries and degeneracies of a charged oscillator in the
  field of a magnetic monopole"}.
\newblock {\em \href{https://aip.scitation.org/doi/10.1063/1.529259}{J. Math.
  Phys.}}, 32(6):1516--1521.

\bibitem[Landau and Lifshitz, 1965]{Landau}
Landau, L.~D. and Lifshitz, E. (1965).
\newblock {\em Quantum mechanics (volume 3 of a course of theoretical
  physics)}.
\newblock Pergamon Press Oxford, UK.

\bibitem[Leinaas and Myrheim, 1977]{leinaas1}
Leinaas, J.~M. and Myrheim, J. (1977).
\newblock \emph{``On the theory of identical particles"}.
\newblock {\em \href{https://link.springer.com/article/10.1007/BF02727953}{Il
  Nuovo Cimento B}}, 37(1):1--23.

\bibitem[Leinaas and Myrheim, 1988]{leinaas2}
Leinaas, J.~M. and Myrheim, J. (1988).
\newblock \emph{``Intermediate statistics for vortices in superfluid films"}.
\newblock {\em
  \href{https://journals.aps.org/prb/abstract/10.1103/PhysRevB.37.9286}{Phys.
  Rev. B}}, 37(16):9286.

\bibitem[Leiva and Plyushchay, 2003]{Leiva}
Leiva, C. and Plyushchay, M.~S. (2003).
\newblock {\emph{``Superconformal mechanics and nonlinear supersymmetry"}}.
\newblock {\em
  \href{https://iopscience.iop.org/article/10.1088/1126-6708/2003/10/069}{JHEP}},
  10:069.

\bibitem[Lochak, 1985]{Lochak}
Lochak, G. (1985).
\newblock \emph{``Wave equation for a magnetic monopole"}.
\newblock {\em
  \href{https://link.springer.com/article/10.1007/BF00670815}{International
  journal of theoretical Physics}}, 24(10):1019--1050.

\bibitem[Mackenzie and Wilczek, 1988]{mackenzie}
Mackenzie, R. and Wilczek, F. (1988).
\newblock \emph{``Peculiar spin and statistics in two space dimensions"}.
\newblock {\em
  \href{https://www.worldscientific.com/doi/abs/10.1142/S0217751X88001181}{J.
  Phys. A}}, 3(12):2827--2853.

\bibitem[Maldacena, 1999]{Maldacena}
Maldacena, J. (1999).
\newblock \emph{``The Large N limit of superconformal field theories and
  supergravity"}.
\newblock {\em
  \href{https://link.springer.com/article/10.1023/A:1026654312961}{Int. J.
  Theor. Phys.}}, 38:1113--1133.

\bibitem[Mateos~Guilarte and Plyushchay, 2017]{JM1}
Mateos~Guilarte, J. and Plyushchay, M.~S. (2017).
\newblock {\emph{``Perfectly invisible $\mathcal{PT}$-symmetric zero-gap
  systems, conformal field theoretical kinks, and exotic nonlinear
  supersymmetry"}}.
\newblock {\em
  \href{https://link.springer.com/article/10.1007%2FJHEP12%282017%29061}{JHEP}},
  12:061.

\bibitem[Mateos~Guilarte and Plyushchay, 2019]{JM2}
Mateos~Guilarte, J. and Plyushchay, M.~S. (2019).
\newblock {\emph{``Nonlinear symmetries of perfectly invisible $PT$-regularized
  conformal and superconformal mechanics systems"}}.
\newblock {\em
  \href{https://link.springer.com/article/10.1007%2FJHEP01%282019%29194}{JHEP}},
  01:194.

\bibitem[Matveev and Salle, 1991]{MatSal}
Matveev, V.~B. and Salle, M.~A. (1991).
\newblock {\em Darboux Transformations and Solitons}.
\newblock Springer series in nonlinear dynamics. Springer-Verlag.

\bibitem[McIntosh and Cisneros, 1970]{Mcin}
McIntosh, H.~V. and Cisneros, A. (1970).
\newblock \emph{``Degeneracy in the presence of a magnetic monopole"}.
\newblock {\em \href{https://aip.scitation.org/doi/10.1063/1.1665227}{J. Math.
  Phys.}}, 11(3):896--916.

\bibitem[Michelson and Strominger, 1999]{BlackHold2}
Michelson, J. and Strominger, A. (1999).
\newblock {\emph{``Superconformal multiblack hole quantum mechanics"}}.
\newblock {\em
  \href{https://iopscience.iop.org/article/10.1088/1126-6708/1999/09/005}{JHEP}},
  09:005.

\bibitem[Milne, 1930]{ermakov2}
Milne, W. (1930).
\newblock \emph{``The numerical determination of characteristic numbers"}.
\newblock {\em
  \href{https://journals.aps.org/pr/abstract/10.1103/PhysRev.35.863}{Phys.
  Rev.}}, 35(7):863.

\bibitem[Moshinsky and Szczepaniak, 1989]{DiracOs1}
Moshinsky, M. and Szczepaniak, A. (1989).
\newblock \emph{``The Dirac oscillator"}.
\newblock {\em
  \href{https://iopscience.iop.org/article/10.1088/0305-4470/22/17/002/meta}{J.
  Phys. A}}, 22(17):L817.

\bibitem[Moutard, 1875]{Moutard2}
Moutard, T. (1875).
\newblock \emph{``Note sur les {\'e}quations diff{\'e}rentielles lin{\'e}aires
  du second ordre"}.
\newblock {\em CR Acad. Sci. Paris}, 80:729--733.

\bibitem[Moutard, 1878]{Moutard1}
Moutard, T. (1878).
\newblock \emph{``Sur la construction des {\'e}quations de la forme
  $\frac{1}{z}\frac{\partial^2 z}{\partial x \partial y}=\lambda(x, y)$ qui
  admettenent une int{\'e}grale g{\'e}n{\'e}rale explicite"}.
\newblock {\em J. Ec. Pol}, 45:1--11.

\bibitem[Nakahara, 2003]{Nakahara}
Nakahara, M. (2003).
\newblock {\em Geometry, Topology and Physics}.
\newblock Graduate student series in physics. Institute of Physics Publishing.

\bibitem[Niederer, 1972]{Niedfree}
Niederer, U. (1972).
\newblock \emph{``The maximal kinematical invariance group of the free
  Schr{\"o}dinger equation"}.
\newblock Technical report.

\bibitem[Niederer, 1973]{NH1}
Niederer, U. (1973).
\newblock \emph{``Maximal kinematical invariance group of the harmonic
  oscillator"}.
\newblock {\em Helv. Phys. Acta}, 46:191.

\bibitem[Ohashi et~al., 2017]{App3}
Ohashi, K., Fujimori, T., and Nitta, M. (2017).
\newblock {\emph{``Conformal symmetry of trapped Bose-Einstein condensates and
  massive Nambu-Goldstone modes"}}.
\newblock {\em
  \href{https://journals.aps.org/pra/abstract/10.1103/PhysRevA.96.051601}{Phys.
  Rev. A}}, 96(5):051601.

\bibitem[Pauli, 1926]{Pauli}
Pauli, W. (1926).
\newblock \emph{``{\"U}ber das Wasserstoffspektrum vom Standpunkt der neuen
  Quantenmechanik"}.
\newblock {\em Zeitschrift f{\"u}r Physik}, 36(5):336--363.

\bibitem[Pelc and Horwitz, 1997]{Pelc}
Pelc, O. and Horwitz, L. (1997).
\newblock {\emph{``Generalization of the Coleman-Mandula theorem to higher
  dimension"}}.
\newblock {\em \href{https://aip.scitation.org/doi/10.1063/1.531846}{J. Math.
  Phys.}}, 38:139--172.

\bibitem[Perelomov, 2012]{Perelomov}
Perelomov, A. (2012).
\newblock {\em Generalized coherent states and their applications}.
\newblock Springer Science \& Business Media.

\bibitem[Pinney, 1950]{ermakov3}
Pinney, E. (1950).
\newblock \emph{``The nonlinear differential equation $y''+ p(x)y+ cy^{-3}=
  0$"}.
\newblock {\em Proceedings of the American Mathematical Society}, 1(5):681.

\bibitem[Pioline and Waldron, 2003]{PioWal}
Pioline, B. and Waldron, A. (2003).
\newblock {\emph{``Quantum cosmology and conformal invariance"}}.
\newblock {\em
  \href{https://journals.aps.org/prl/abstract/10.1103/PhysRevLett.90.031302}{Phys.
  Rev. Lett.}}, 90:031302.

\bibitem[Plyushchay, 1996]{Boson1}
Plyushchay, M. (1996).
\newblock {\emph{``Deformed Heisenberg algebra, fractional spin fields and
  supersymmetry without fermions"}}.
\newblock {\em
  \href{https://www.sciencedirect.com/science/article/abs/pii/S0003491696900123}{Annals
  Phys.}}, 245:339--360.

\bibitem[Plyushchay, 2000a]{PlyPara}
Plyushchay, M. (2000a).
\newblock {\emph{``Hidden nonlinear supersymmetries in pure parabosonic
  systems"}}.
\newblock {\em
  \href{https://www.worldscientific.com/doi/abs/10.1142/S0217751X00001981}{Int.
  J. Mod. Phys. A}}, 15:3679--3698.

\bibitem[Plyushchay, 1993]{Mikisl2R}
Plyushchay, M.~S. (1993).
\newblock \emph{``Quantization of the classical $\overline{SL(2,R)}$ system and
  representations of $\overline{SL(2,R)}$ group"}.
\newblock {\em \href{https://aip.scitation.org/doi/10.1063/1.530016}{J. Math.
  Phys.}}, 34(9):3954--3963.

\bibitem[Plyushchay, 2000b]{Plymono1}
Plyushchay, M.~S. (2000b).
\newblock {\emph{``Monopole Chern-Simons term: Charge monopole system as a
  particle with spin"}}.
\newblock {\em
  \href{https://www.sciencedirect.com/science/article/pii/S0550321300005307?via%3Dihub}{Nucl.
  Phys. B}}, 589:413--439.

\bibitem[Plyushchay, 2001]{Plymono2}
Plyushchay, M.~S. (2001).
\newblock {\emph{``Free conical dynamics: Charge-monopole as a particle with
  spin, anyon and nonlinear fermion-monopole supersymmetry"}}.
\newblock {\em
  \href{https://www.sciencedirect.com/science/article/abs/pii/S0920563201015638}{Nucl.
  Phys. B Proc. Suppl.}}, 102:248--255.

\bibitem[Plyushchay, 2017]{Plyushchay}
Plyushchay, M.~S. (2017).
\newblock {\emph{``Schwarzian derivative treatment of the quantum second-order
  supersymmetry anomaly, and coupling-constant metamorphosis"}}.
\newblock {\em
  \href{https://www.sciencedirect.com/science/article/abs/pii/S0003491616302743}{Annals
  Phys.}}, 377:164--179.

\bibitem[Plyushchay, 2019]{Plyunonlinear}
Plyushchay, M.~S. (2019).
\newblock \emph{``Nonlinear supersymmetry as a hidden symmetry"}.
\newblock In {\em Integrability, Supersymmetry and Coherent States. S{\"E}
  Kuru, J. Negro and L.M Nieto}, pages 163--186.
  \href{https://link.springer.com/chapter/10.1007%2F978-3-030-20087-9_6}{CRM
  Series in Mathematical Physics. Springer (Charm)}.

\bibitem[Plyushchay, 2020]{plyushchay2020exotic}
Plyushchay, M.~S. (2020).
\newblock \emph{``Exotic nonlinear supersymmetry and integrable systems"}.
\newblock {\em
  \href{https://link.springer.com/article/10.1134%2FS1063779620040589}{Physics
  of Particles and Nuclei}}, 51(4):583--588.

\bibitem[Plyushchay and Wipf, 2014]{PlyWipf}
Plyushchay, M.~S. and Wipf, A. (2014).
\newblock {\emph{``Particle in a self-dual dyon background: hidden free nature,
  and exotic superconformal symmetry"}}.
\newblock {\em
  \href{https://journals.aps.org/prd/abstract/10.1103/PhysRevD.89.045017}{Phys.
  Rev. D}}, 89(4):045017.

\bibitem[Polyakov, 1974]{Polyakov1}
Polyakov, A. (1974).
\newblock \emph{``Nonhamiltonian approach to conformal quantum field theory"}.
\newblock {\em Zh. Eksp. Teor. Fiz.}, \textbf{66}:23--42.

\bibitem[Prain et~al., 2010]{App2}
Prain, A., Fagnocchi, S., and Liberati, S. (2010).
\newblock {\emph{``Analogue cosmological particle creation: quantum
  correlations in expanding Bose Einstein condensates"}}.
\newblock {\em
  \href{https://journals.aps.org/prd/abstract/10.1103/PhysRevD.82.105018}{Phys.
  Rev. D}}, 82:105018.

\bibitem[Quesne, 2012]{Quesne2012}
Quesne, C. (2012).
\newblock \emph{``Novel enlarged shape invariance property and exactly solvable
  rational extensions of the Rosen-Morse II and Eckart potentials"}.
\newblock {\em \href{https://www.emis.de/journals/SIGMA/2012/080/}{SIGMA.
  Symmetry, Integrability and Geometry: Methods and Applications}}, 8:080.

\bibitem[Quesne and Moshinsky, 1990]{DiracOs3}
Quesne, C. and Moshinsky, M. (1990).
\newblock \emph{``Symmetry Lie algebra of the Dirac oscillator"}.
\newblock {\em
  \href{https://iopscience.iop.org/article/10.1088/0305-4470/23/12/011}{J.
  Phys. A}}, 23(12):2263.

\bibitem[Regniers and Van~der Jeugt, 2010]{Regniers}
Regniers, G. and Van~der Jeugt, J. (2010).
\newblock \emph{``The Hamiltonian $H= xp$ and classification of $osp (1| 2)$
  representations"}.
\newblock In {\em
  \href{https://aip.scitation.org/doi/abs/10.1063/1.3460159}{AIP Conference
  Proceedings}}, volume 1243, pages 138--147. American Institute of Physics.

\bibitem[Sakurai, 1994]{Sakurai}
Sakurai, J. (1994).
\newblock {\em Modern quantum mechanics}.
\newblock Addison-Wesley Pub. Co.

\bibitem[Schr{\"o}dinger, 1926]{Schrodinger}
Schr{\"o}dinger, E. (1926).
\newblock \emph{``Der stetige {\"U}bergang von der Mikro-zur Makromechanik"}.
\newblock {\em Naturwissenschaften}, 14(28):664--666.

\bibitem[Schulze-Halberg, 2013]{Jordan2}
Schulze-Halberg, A. (2013).
\newblock \emph{``Wronskian representation for confluent supersymmetric
  transformation chains of arbitrary order"}.
\newblock {\em
  \href{https://www.researchgate.net/publication/257868267_Wronskian_representation_for_confluent_supersymmetric_transformation_chains_of_arbitrary_order}{The
  European Physical Journal Plus}}, 128(6):68.

\bibitem[Sierra and Rodriguez-Laguna, 2011]{Sierra}
Sierra, G. and Rodriguez-Laguna, J. (2011).
\newblock {\emph{``The H=xp model revisited and the Riemann zeros"}}.
\newblock {\em
  \href{https://journals.aps.org/prl/abstract/10.1103/PhysRevLett.106.200201}{Phys.
  Rev. Lett.}}, 106:200201.

\bibitem[Son, 2008]{GAdS1}
Son, D. (2008).
\newblock {\emph{``Toward an AdS/cold atoms correspondence: A Geometric
  realization of the Schrodinger symmetry"}}.
\newblock {\em
  \href{https://journals.aps.org/prd/abstract/10.1103/PhysRevD.78.046003}{Phys.
  Rev. D}}, 78:046003.

\bibitem[Sundermeyer, 2014]{Sundermayer}
Sundermeyer, K. (2014).
\newblock {\em Symmetries in Fundamental Physics}.
\newblock Fundamental Theories of Physics 176. Springer Netherlands.

\bibitem[Takhtadzhian, 2008]{Tak}
Takhtadzhian, L.~A. (2008).
\newblock {\em Quantum mechanics for mathematicians}, volume~95.
\newblock American Mathematical Soc.

\bibitem[Weinberg, 1995]{Waimberg1}
Weinberg, S. (1995).
\newblock {\em Quantum theory of fields. Foundations}.
\newblock Cambridge University Press.

\bibitem[Weinberg, 2000]{Waimberg3}
Weinberg, S. (2000).
\newblock {\em The quantum theory of fields. Supersymmetry}.
\newblock Cambridge University Press.

\bibitem[Weinberg, 2012]{Waimberg2}
Weinberg, S. (2012).
\newblock {\em Lectures on Quantum Mechanics}.
\newblock Cambridge University Press.

\bibitem[Wigner, 1931]{Wigner0}
Wigner, E. (1931).
\newblock \emph{``Gruppentheorie und ihre Anwendung auf die Quantenmechanik der
  Atomspektrum"}.
\newblock {\em Fredrick Vieweg und Sohn, Braunschweig, Germany}, pages
  251--254.

\bibitem[Wigner, 2012]{Wigner}
Wigner, E. (2012).
\newblock {\em \emph{``Group theory: and its application to the quantum
  mechanics of atomic spectra"}}, volume~5.
\newblock Elsevier.

\bibitem[Witten, 1981]{Witten1}
Witten, E. (1981).
\newblock \emph{``Dynamical breaking of supersymmetry''}.
\newblock {\em
  \href{http://www.sciencedirect.com/science/article/pii/0550321381900067?via%3Dihub}{Nucl.
  Phys. B }}, 188:513.

\bibitem[Witten, 1982]{Witten2}
Witten, E. (1982).
\newblock \emph{``Constraints on supersymmetry breaking''}.
\newblock {\em
  \href{http://www.sciencedirect.com/science/article/pii/0550321382900712?via%3Dihub}{Nucl.
  Phys. B }}, 202:253.

\bibitem[Wu and Yang, 1976]{monoharm}
Wu, T.~T. and Yang, C.~N. (1976).
\newblock \emph{``Dirac monopole without strings: monopole harmonics"}.
\newblock {\em
  \href{https://www.sciencedirect.com/science/article/pii/0550321376901437}{Nucl.
  Phys. B}}, 107(3):365--380.

\bibitem[Zamolodchikov, 1985]{Zamolodchikov}
Zamolodchikov, A. (1985).
\newblock {\emph{``Infinite Additional Symmetries in Two-Dimensional Conformal
  Quantum Field Theory"}}.
\newblock {\em
  \href{https://link.springer.com/article/10.1007/BF01036128}{Theor. Math.
  Phys.}}, 65:1205--1213.

\bibitem[Zhedanov, 1992]{Zhedanov}
Zhedanov, A. (1992).
\newblock \emph{``The ``Higgs Algebra'' as a `quantum' deformation of su(2)''}.
\newblock {\em
  \href{https://www.worldscientific.com/doi/10.1142/S021773239200046X}{Modern
  Physics Letters A}}, 7(06):507--512.

\end{thebibliography}

\appendix

\chapter{Wronskian identities}
\label{ApenWI}

Here we consider the equalities between wavefunctions and 
Wronskians in the sense of ``up to a multiplicative constant'' 
when the corresponding constant is not essential.

\section{Wronskian relations due to DCKA transformation}
 
Suppose that we have two collections of (formal) eigenstates
 of (\ref{Sch}), $\{\phi_n\}=(\phi_1,\ldots,\phi_n)$ and 
$\{\varphi_l\}=(\varphi_1,\ldots,\varphi_l)$. 
In the first step, we generate a Darboux transformation
by taking the first collection as the set of the 
seed states, and obtain 
 the  intermediate Hamiltonian operator with potential 
$V_1=V(x)-2(\ln W(\{\phi_n\}) )''$.
In this way,  the states of the second collection $\{\varphi_l\}$ 
will be mapped into  the set of (formal in general case) eigenstates 
$\{\A_n\varphi_l\}=(\A_n\varphi_1,\ldots,\A_n\varphi_l)$.
Then, employing these states as the seed states 
for  a second Darboux transformation, 
we finally obtain a Schr\"odinger  operator with a potential 
$V_2=V_1(x)-2(\ln W(\{\A_n\varphi_l\}))''$.
Having in mind that  
the same result will be produced by a one-step generalized 
Darboux transformation 
based on the whole set of the chosen eigenstates of the system $L$,
we obtain  the equality 
\begin{equation}
\label{id1}
W(\{\phi_n\})W(\{\A_n\varphi_l\})=W(\phi_1,\ldots,\phi_n,\varphi_1,\ldots,\varphi_l)\,.
\end{equation}

Consider now the set of two states corresponding to a same eigenvalue 
${\lambda_j}$,   $\{\phi_{2}\}=(\phi_{1}=\psi_{j},\phi_{2}=\widetilde{\psi}_{j})$.  
In this case $W(\psi_{j},\widetilde{\psi}_{j})=1$,  and the corresponding intertwining operator 
reduces to  $\A_2=-(L-\lambda_j)$.
Using this observation and  Eq. (\ref{id1}), we derive the equality
$W(\psi_j,\widetilde{\psi}_j,\varphi_1,\ldots,\varphi_l)=W(\{\varphi_l\})$, 
which is generalized for the relation
\be
\label{ide2}
W(\psi_{1},\widetilde{\psi}_{1},\ldots,\psi_{s},\widetilde{\psi}_{s},\varphi_1,\ldots,\varphi_l)=W(\{\varphi_l\})\,.
\ee
In the case when  functions  $\varphi_1,\ldots,\varphi_l$ are not obligatorily to be eigenstates 
of  the operator $L$, the 
last relation changes  for
\begin{eqnarray}
\label{ide2+}
&W(\psi_{1},\widetilde{\psi}_{1},\ldots,\psi_{s},\widetilde{\psi}_{s},\varphi_1,\ldots,\varphi_l)=
W\left(\{\prod_{k=1}^s(-L+\lambda_k)\varphi_l\}\right).&
\end{eqnarray}
In the context of generalized Darboux transformations based 
on a mixture of 
eigenstates and Jordan states,  
a useful relation
\begin{equation}
\label{tech1}
W(\psi_{*},\widetilde{\psi}_{*},\Omega_*,\breve{\Omega}_*,\varphi_1,\ldots,\varphi_l)=
W(\varphi_1,\ldots,\varphi_l)\, 
\end{equation}
can be obtained by employing Eq. (\ref{ide2+})
with $s=1$, and  Eqs. (\ref{Omega12}) and (\ref{ide2}),
Here  we  imply that
$\varphi_i$ with $i=1,\ldots,l$ is the set of solutions of equation (\ref{Sch})
with $\lambda_i\neq \lambda_*$.

\section{Jordan states and Wronskian relations}
\label{apWron}
We show   here that the Wronskian  (\ref{Interpol}) 
takes non-zero values and that it reduces to  (\ref{Interpol2}) 
in the limit $\mu\to-1/2$. For this, consider first a generic system (\ref{Sch}) which has a set of the seed 
states $(\phi_1,\phi_2,\ldots,\phi_{2l-1},\phi_{2l})$ 
 with eigenvalues $\lambda_1<\lambda_2<\ldots<\lambda_{2l-1}<\lambda_{2l}$. Then 
 the following relation 
\begin{equation}
\label{producto1}
W(\phi_1,\phi_2,\ldots,\phi_{2l-1},\phi_{2l})=\prod_{i=0}^{l-1}W(\A_{2i}\phi_{2i+1},\A_{2i}\phi_{2i+2})\,,
\end{equation}
can be proved by induction, where $\A_{0}=1$, and $\A_{2i}$  with $i\geq 1$ corresponds to the intertwining 
operator associated with the scheme $(\phi_1,\ldots,\phi_{2i})$. 
{}From (\ref{producto1})
it follows that if  each factor 
$W(\A_{2i}\phi_{2i+1},\A_{2i}\phi_{2i+2})$ does not have zeros, 
then the complete  Wronskian neither has. 
To inspect the properties of the   Wronskian factors, we use the relation
\begin{equation}
\label{Wronskianderivative}
W'(\A_{2i}\phi_{2i+1},\A_{2i}\phi_{2i+2})=(\lambda_{2i+2}-\lambda_{2i+1})\A_{2i}\phi_{2i+1}\A_{2i}\phi_{2i+2}\,,
\end{equation} 
and integrate it from $a$ to $x$, 
\begin{equation}
\label{WronskianAphi}
W(\A_{2i}\phi_{2i+1},\A_{2i}\phi_{2i+2})=(\lambda_{2i+2}-\lambda_{2i+1})\int_a^x\A_{2i}
\phi_{2i+1}\A_{2i}\phi_{2i+2}d\zeta+\omega\,,
\end{equation}
where $ \omega=W(\A_{2i}\phi_{2i+1},\A_{2i}\phi_{2i+2})|_{x=a}$. In the case when  
functions $\A_{2i}\phi_{2i+1}$, $\A_{2i}\phi_{2i+2}$ and their first derivatives vanish 
in $b$,  we find  $\omega=-(\lambda_{2i+2}-\lambda_{2i+1})\int_a^b\A_{2i}\phi_{2i+1}\A_{2i}\phi_{2i+2}d\zeta$,
and then
\begin{equation}
\label{WronskianAphi2}
W(\A_{2i}\phi_{2i+1},\A_{2i}\phi_{2i+2})=-(\lambda_{2i+2}-\lambda_{2i+1})\int_x^b\A_{2i}\phi_{2i+1}
\A_{2i}\phi_{2i+2}d\zeta\,.
\end{equation}
Relation (\ref{producto1}) takes then the form
\begin{equation}
\label{Wronskianintegral}
W(\phi_1,\phi_2,\ldots,\phi_{2l-1},\phi_{2l})=\prod_{i=0}^{l-1}(\lambda_{2i+1}-\lambda_{2i+2})\int_x^b\A_{2i}\phi_{2i+1}\A_{2i}\phi_{2i+2}d\zeta_i\,.
\end{equation}

Analogously, one can consider a generic system,  choose $l$ solutions $\varphi_i$
of Eq. (\ref{Sch}), and construct $l$ corresponding Jordan states  $\Omega_i$ using Eq.
(\ref{omega1}). Assuming also that these states satisfy relations (\ref{omega2}), 
one can find that
\begin{equation}
\label{producto2}
W(\varphi_1,\Omega_1,\ldots,\varphi_{l},\Omega_l)=\prod_{i=0}^{l-1}W(\A_{2i}^{\Omega}\varphi_{i+1},\A_{2i}^{\Omega}\Omega_{i+1})=
\prod_{i=0}^{l-1}\int_{x}^{b}(\A_{2i}^{\Omega}\varphi_{i+1})^2d\zeta_i\,,\qquad 
\A_{0}^{\Omega}=1
\end{equation}
and $\A_{2i}^{\Omega}$ is the intertwining operator 
associated with the scheme
$
(\varphi_1, \Omega_1 \ldots, \varphi_{l}, \Omega_l)\,.
$
 Relation  (\ref{producto2}) can be proved in a  way similar to that 
 for (\ref{Wronskianintegral}). 

Let us turn now to the AFF model, where $a=0$, $b=\infty$, 
and choose the seed states in (\ref{producto1}) in correspondence with our 
picture: 
for $i=0,\ldots,l-1$
we fix $\phi_{2i+1}=\psi_{-\mu-m-1,n_{i+1}}$ and $\phi_{2i+2}=\psi_{\mu+m,n_{i+1}-m}$. 
This identification implies that 
$\lambda_{2i+1}=E_{-\mu-m-1,n_{i+1}}$, $\lambda_{2i+2}=E_{\mu+m,n_{i+1}-m}$, and  
$\lambda_{2i+2}-\lambda_{2i+1}=4(\mu+1/2)$. These both functions
and their first derivatives  
behave for large values of  $x$  as $e^{-x^2/2}$, and  vanish at $x=\infty$. 
This behavior is not changed by application of any differential operator 
with which we work.
 On the other hand,  near zero we have 
 $\A_{2i}\psi_{-\mu+m+1,n_{i+1}}\sim
x^{-\mu-m-i}$ and $\A_{2i}\psi_{\mu+m,n_{i+1}-m}\sim
x^{\mu+m+1+i}$. Therefore, for small values of $x$, 
$\A_{2i}\psi_{-\mu+m+1,n_{i+1}}\A_{2i}\psi_{\mu+m,n_{i+1}-m}\sim x$,
and $W(\A_{2i}\psi_{-\mu+m+1,n_{i+1}},\A_{2i}\psi_{\mu+m,n_{i+1}-m})$ 
takes a finite value when $x\rightarrow 0^+$. 
Knowing this and  Eq. (\ref{Wronskianderivative}),   we employ
the Adler method \textcolor{red}{[\cite{Adler}]}, and use the theorem on nodes of 
wave functions  to show that 
zeros and the minima and maxima of the functions 
$\A_{2i}\psi_{-\mu+m+1,n_{i+1}}$ and $\A_{2i}\psi_{\mu+m,n_{i+1}-m}$
do not coincide,  and that their corresponding Wronskian 
is non-vanishing.

 In the case $\mu=-1/2$, 
we put $\varphi_{j}=\psi_{m-1/2,n_{j+1}-n}$ with $j=0,\ldots,l-1$,  and 
then we arrive at  the relations 
\begin{eqnarray}
\label{WronskianAphi3}
&\frac{W(\{\gamma_\mu\})}{(4\mu+2)^N}
=(-1)^{l}
\prod\limits_{i=0}^{l-1}\int_x^\infty\A_{2i}\psi_{-\mu-m-1,n_{i+1}}
\A_{2i}\psi_{\mu+m,n_{n_{i+1}}-m}d\zeta_i\,,&\\&
W(\{\gamma\})=\prod\limits_{j=0}^{l-1}\int_x^\infty(\A_{2j}^{\Omega}\psi_{m-1/2,n_{j+1}-m})^2d\zeta_j\,,&
\end{eqnarray}
where the sets $\{\gamma_\mu\}$ and $\{\gamma\}$
are defined in (\ref{Interpol}) and (\ref{Interpol2}). We note that 
both equations are pretty similar each other, and 
 if we suppose that $\A_{2i}\to\A_{2i}^{\Omega}$ when $\mu\to-1/2$, and 
 take into account  the relation
$\psi_{m-1/2,n_{j}-m}\propto\psi_{-(m-1/2)-1,n_{j}}$, one proves by induction that 
\begin{equation}
\lim_{\mu\to-1/2}\frac{W(\{\gamma_\mu\})}{(4\mu+2)^N}
\propto W(\{\gamma\})\,.
\end{equation}

\chapter{The mirror diagram}
\label{Mirror appendix}
In this paragraph we prove the relations involved with 
mirror diagrams and Darboux duality using the Wronskian identities of 
Appendix \ref{ApenWI}
\section{Harmonic oscillator case}
\label{MirrorHarmonic}
To start, we consider a 
positive scheme 
$\{\Delta_+\}=(l_1^+,\ldots,l_{n_+}^+)$, where 
$l_i^+$ with  $i=1,\ldots, n_+$ are certain positive numbers
ordered  from low to high, and we supose that $l_1^+\not=0$ . By using the Wronskian 
identity (\ref{ide2}) we get the relation 
\begin{equation}
\label{trick}
W(\{\Delta_+\})=W(0,\widetilde{0},\{\Delta_+\})=e^{-x^2/2}
W(-1,\{a^-\Delta_+\})\,,
\end{equation}
where in the last step we have used the identity (\ref{id1})\footnote{For us, the Wronskian of a single function is the function itself, and $a^- \widetilde{\psi_{0}}=\psi_{0}(ix)$, which in our notations is $-1$. }, 
and $\{a^-\Delta_+\}$ means that $a^-$ acts in each state in the scheme.
Let us  repeat the trick a second time, obtaining 
$W(\{\Delta_+\})=e^{-x^2}W(-2,-1,\{(a^-)^2\Delta_+\})$. 
After $l_1^+$ times we get
\begin{align*}
W(\{\alpha\})& = e^{-\frac{l_1^+}{2}x^2}
W(-l_1^+,\ldots,-1,0,(l_2^+-l_1^+),\ldots,(l_{n_+}^+-l_1^+))\\
&=e^{-\frac{l_1^++1}{2}x^2}
W(\underbrace{-(l_1^++1),\ldots,-2}_{\text{negative states}},\underbrace{(l_2^+-l_1^+-1),\ldots,(l_{n_+}^+-l_1^+-1)}_{\text{positive states}})\,,
\end{align*}
where we have used the identity (\ref{id1})
with the ground state denoted by zero. 
So now, we have to answer the question: Is $l_2^+-l_1^+-1$ equal to $0$?. 
If the answer is negative, then we continue with the trick described in (\ref{trick}) another 
$l_2^+-l_1^+-1$ times. On the other hand, if 
 the answer is 
affirmative, we use again the identity (\ref{id1}) in order to do
 not have a ground state in the Wronskian of the right hand side. This step  is the responsible of 
the ``missing'' states  in the negative scheme constructed in this way. 
 We repeat the algorithm 
until positive eigenstates disappear in the right hand side, obtaining 
the relation  
\begin{equation}
W(\{\Delta_+\})=e^{(l_{n_+}^++1)x^2/2}W(\Delta_-)\,,\qquad
\Delta_-=(-\check{0},\ldots,-\check{n}_i^-,
\ldots,-l_{n_+}^+)\,,
\end{equation}
If one would like to start with the negative scheme, the algorithm is the same but instead of
$0$ and $\widetilde{0}$, is necessary to use  
the nonphysical states $ -0 $ and $ -\widetilde{0} $  in equation (\ref{trick}).

\section{The case of AFF model with $\nu\not=\ell-1/2$}
\label{DualAFF1}
To show the mirror diagram for this case, we follow 
the same spirit of last subsection, but in this case we 
have to use second order ladder operators. 
As a starting point,  consider the Wronskian of the set 
$\{\alpha\}$  
 defined in (\ref{unioncollection}).  
If the states 
 $\psi_{\nu,\pm 0}$ and $\psi_{-\nu-1,\pm 0}$ do not belong to (\ref{unioncollection}),
 we can replace the Wronskian $W(\{\alpha\})$ by 
\begin{eqnarray}
\label{updown1}
&W(\psi_{\nu,\pm0},\psi_{-\nu-1,\pm0},\widetilde{\psi}_{\nu,\pm0},\widetilde{\psi}_{-\nu-1,\pm0},\{\alpha\})=
e^{\mp x^2}W(\psi_{\nu,\mp0},\psi_{-\nu-1,\mp0},\{\mathcal{C}_\nu^\mp \alpha\})\,,&
\end{eqnarray}
where we used relations (\ref{id1}), (\ref{ide2}), (\ref{tools1}) and
(\ref{tools2}), and  $\{\mathcal{C}_\nu^\mp\alpha\}$ means 
that the ladder operators are applied to  all the states in the set. 
On the other hand, if $\psi_{r(\nu),\pm 0}$ belong to (\ref{unioncollection}),
we can replace the Wronskian of the initial set of the seed states by 
\begin{eqnarray}
\label{updown2}
W(\psi_{r(-\nu-1),\pm0},\widetilde{\psi}_{r(-\nu-1),\pm0},\{\alpha\})=
e^{\mp x^2}W(\psi_{r(-\nu-1),\mp0},\{\mathcal{C}_\nu^\mp\beta_1\})\,,
\end{eqnarray}
where $\{\beta_1\}$ is the scheme $\{\alpha\}$ with the omitted state $\psi_{r(\nu),\pm0}$. 
Finally, if  $\psi_{\nu,\pm 0}$ and $\psi_{-\nu-1,\pm 0}$ belong to (\ref{unioncollection}),
we have
\begin{eqnarray}
\label{updown3}
&W(\{\alpha\})=e^{\mp x^2}W(\{\mathcal{C}_\nu^\mp\beta_2\})\,,&
\end{eqnarray}
where $\{\beta_2\}$ is the scheme $\{\alpha\}$ with the omitted states 
$\psi_{\nu,\pm0}$ and $\psi_{-\nu-1,\pm0}$.  
Note that in all these three relations  we have lowered or raised the index 
of the states in $\{\alpha\}$, 
and also in the case of Eqs.  (\ref{updown1}) and (\ref{updown2})
we have included additional states which do not belong to the initial set. 
Also, we  note that an exponential factor has appeared. 
These identities can be applied to the Wronskians  on the right hand side of  Eqs. 
(\ref{updown1})-(\ref{updown3}), 
which will contribute with new exponential factors in new Wronskians, and so on.
For this reason,  if we restrict the initial set $\{\alpha\}$ by the conditions 
described above (that every state in the set has the second index of the same sign), 
and we repeat this procedure $n_N+1$ times
with positive (negative) sign of the indexes 
in (\ref{updown1})-(\ref{updown3}), 
we finally obtain  equation (\ref{eqschemes3}) or (\ref{eqschemes4}).

\section{The case of AFF model with $\nu=\ell-1/2$}
\label{AFFhalf}

To obtain the dual schemes in the half-integer case, 
we analyze first the relations
that exist  
between of $\mathcal{H}_{-1/2}$ and $\mathcal{H}_{-1/2+\ell}$.
The latter are given by the dual schemes  
$(\psi_{-1/2,\pm0},\ldots,\psi_{-1/2,\pm(\ell-1)})$,
whose Wronskians are  
\begin{eqnarray}
\label{Wrondual}
&W(\psi_{-1/2,\pm0},\ldots,\psi_{-1/2,\pm(\ell-1)})=x^{\ell^2/2}e^{\mp \ell x^2/2}\,.&
\end{eqnarray}
The corresponding intertwiners map eigen- and Jordan states of 
$\mathcal{H}_{-1/2}$ to those of $\mathcal{H}_{-1/2+\ell}$.
If we choose the scheme with positive indexes, some of these mappings
useful for the following are given by 
\begin{eqnarray}
\label{interLnu1}
\A_\ell^-\psi_{-1/2,n}=\psi_{-1/2+\ell,n-\ell}\,, \qquad 
\A_\ell^-\Omega_{\nu,-1/2}=\Omega_{-1/2+\ell,n-\ell}\,, \qquad n\geq\ell\,,\\
\A_\ell^-\Omega_{-1/2,l}=\psi_{-(-1/2+\ell)-1,l}\,,\qquad\l<\ell\,,
\end{eqnarray}
where $\A_\ell^-$  and its  Hermitian conjugate $\A_\ell^+$ 
are the  intertwining operators of the chosen Darboux transformation. 
On the other hand if we take the scheme with negative sign in indices, we obtain
another intertwining operators $\B_{\ell}^\pm$, which satisfy 
the relation $\B_{\ell}^\pm=(i)^\ell\rho_2(\A_{\ell}^\pm)$, i.e, 
their action on eigenstates and Jordan states can be obtained
by application of $\rho_2$ to  the relations that correspond to the 
 action of $\A_{m}^\pm$.  

Now, to derive the dual schemes let us assume that we have a collection
 of non-repeated seed states of the form 
$(\psi_{-1/2,0},\ldots,\psi_{-1/2,\ell-1},\{\vartheta_{-1/2}\})$, 
where  $\{\vartheta_{-1/2}\}$ contains $N_1$ arbitrary 
physical states 
$\psi_{-1/2,k_i}$ with
 $k_i>\ell-1$ for  $i=1,\ldots,N_1$, and  $N_2$ arbitrary Jordan states of 
 the form $\Omega_{-1/2,l_j}$
 with $j=1,\ldots,N_1$. 
 In the same way as we did in Sec. \ref{Mirror}, 
 we define $n_N$ as the largest of the numbers  $n_{N_1}$ and $n_{N_2}$, 
and also we suppose for simplicity that the signs of both $k_i$ and $k_j$ 
are positive.
Then we use (\ref{id1}) and (\ref{Wrondual}) to  write 
 $W(\psi_{-1/2,0},\ldots,\psi_{1/2,\ell-1},\{ \vartheta_{-1/2}\})=
 x^{\ell^2/2}e^{-\ell x^2/2}W(\{ \A_\ell^-\vartheta_{-1/2}\})$.
  The next step is to use 
the extension of the dual schemes for $\nu=-1/2$, i.e,
we change each function of the form 
$\psi_{-\nu-1,n}$ by $\Omega_{-1/2,n}$ in equation 
(\ref{eqschemes3}),  and 
use it to rewrite this last Wronskian relation as 
\begin{eqnarray}
\label{SchemeJor1}
&
W(\A_\ell^-\{ \vartheta_{-1/2}\})=x^{-\ell^2/2}e^{-(n_N+1-\ell/2)x^2}W(\{\Delta^{(-1/2)}_-\})\,,
&
\end{eqnarray}
where $\Delta_{-}^{(-1/2)}$ is the dual scheme of 
$(\psi_{-1/2,0},\ldots,\psi_{-1/2,\ell-1},\{\vartheta_{-1/2}\})$ given by 

\be
\{\Delta_{-}^{(-1/2)}\}=(\psi_{-1/2,-0},\ldots,\psi_{-1/2,-(\ell-1)},\{\vartheta_{-1/2}^-\})\,,\ee
and 
\begin{eqnarray}
\{\vartheta_{-1/2}^-\}=(\psi_{-1/2,-\ell},\Omega_{-1/2,-0},\ldots,\check{\psi}_{-1/2,-s_j},
\check{\Omega}_{-1/2,-r_i},\ldots,\psi_{-1/2,-n_{N}},\Omega_{-1/2,-n_{N}})\,.
\end{eqnarray}
Here, as well as in the non-half-integer case, 
the marked functions $\check{\psi}_{-1/2,-s_j}$ and  
$\check{\Omega}_{-1/2,-r_i}$  indicate  the omitted states with $s_j=n_{N}-l_j$ 
and $r_i=n_{N}-k_i$.
In the last step, we use  Eqs. (\ref{id1}) and (\ref{Wrondual}) with the 
negative sign to write the equality $
W(\{\Delta_{-}^{(-1/2)}\})=x^{\ell^2/2}e^{\ell x^2/2}W(\mathbb{B}_\ell^-\{\vartheta_{-1/2}^-\})
$ and as analog of  (\ref{SchemeJor1}) we obtain 
\begin{eqnarray}
\label{DualSchemeJor1}
W(\A_\ell^-\{\vartheta_{-1/2}\})=e^{-(n'_N+1)x^2}W(\mathbb{B}_\ell^-\{\vartheta_{-1/2}^-\})\,,
\qquad n'_N=n_N-\ell\,.
\end{eqnarray}
This relation is the dual scheme equation for the case $\nu=\ell-1/2$. By means of  
(\ref{interLnu1}) and its analogs for $\B_\ell^-$ obtained by the application of $\rho_2$, 
we conclude that in the scheme of the left hand side of the equation 
there are $N_1$ physical states
of the form
$\A_\ell^-\psi_{-1/2,k_i}=\psi_{\ell-1/2,k_i-\ell}$, and a mixture of $N_2$ Jordan states
and formal states produced by $\rho_2$ 
distributed in the following way:
we have Jordan states 
$\A_\ell^-\Omega_{-1/2,l_i}=\psi_{\ell-1/2,l_i}$ when $l_i<\ell-1$, and
formal states 
$\A_\ell^-\Omega_{-1/2,l_i}=\psi_{\ell-1/2,l_i-\ell}$  when $l_i\geq \ell$.
The omitted  states in the scheme on the right hand side are  
$\B_{\ell}^-\check{\psi}_{-1/2,-s_j}=\check{\psi}_{-1/2+\ell,-(s_j-\ell)}$
and  
$\B_{\ell}^-\check{\Omega}_{-1/2,-r_j}=\check{\psi}_{-\ell-1/2,-r_j}$ 
($\B_{\ell}^-\check{\Omega}_{-1/2,-r_j}=\check{\psi}_{-\ell-1/2,-(r_j-\ell)}$ )
when $r_j\leq \ell-1$  ($r_j>\ell$). 
Note that the largest index in both sides of the equation 
 is now given by $n_N'=n_N-\ell$. 
In comparison with the non-half-integer case, 
this is the same result that we would obtain 
if we consider equation (\ref{eqschemes3}) in the non-half-integer case, 
and then formally change the states 
of the form $\psi_{-\nu-1,l_i}$ by $\Omega_{-\ell-1/2,l_i-\ell}$  when 
$l_i\geq\ell$ in the limit 
$\nu\rightarrow \ell-1/2$. 

Relation analogous to  (\ref{eqschemes4})
would be obtained if we start from the case $\nu=-1/2$ with a scheme composed from
the eigenstates and Jordan states produced by $\rho_2$, and then 
apply the same arguments employed for the case analyzed above.

\chapter{Details for rationally extended systems}
Here we show some operator identities, 
as well as the explicit form of some polynomials in nonlinear algebras considered in the Chap. \ref{ChNonLinearSUSY}.
 
\section{Operator identities  (\ref{ide})}
\label{show}

We have
equalities $\ker\, (A_{(-)}^+A_{(+)}^-) = \Delta_+ \cup \tilde{\delta}=\{0,1,\ldots,n\}$,
where $\tilde{\delta}=\{A_{(-)}^+A_{(+)}^-
\widetilde{\psi_{-l_1}},\ldots, \\
A_{(-)}^+A_{(+)}^-\widetilde{\psi_{-l_{n_-}}} \}$,
and relation $\A_{(+)}^+\A_{(-)}^-\varphi_{n}\propto (a^+)^N\varphi_{n}$
following from  (\ref{relation-operators}) is used. The  first identity from (\ref{ide})
follows then from equality $\ker A_{(-)}^+A_{(+)}^-$ =$\ker(a^-)^N$ \textcolor{red}{[\cite{CarPly}]}.

In the second relation in (\ref{ide}), 
functions  $f(\eta)$ and $h(\eta)$ are the polynomials
\begin{equation}
\label{fyh}
f(\eta)\equiv \prod_{l^-_{k}-n_+<0}(\eta+2l^-_{k}+1),\qquad
h(\eta)\equiv \prod_{\check{n}^-_{k}-n_+\geq 0}(\eta+2\check{n}^-_{k}+1)\,,
\end{equation}
where $l^-_{k}\in \Delta_-$ and $\check{n}^-_{k}$ are 
the absent states in $\Delta_-$. Using the mirror diagram technique \textcolor{red}{[\cite{CarInzPly}]}, 
we obtain the equality    
$\ker f(L_{(-)})A_{(+)}^-(a^+)^{n_-}=\ker h(L_{(-)})A_{(-)}^-(a^-)^{n_+}$, where 
\begin{eqnarray}
\label{kerA}
\small{\begin{array}{rr}
\ker f(L_{(-)})A_{(+)}^-(a^+)^{n_-}=&\text{span}\{0,\ldots,(n_+-1),-0,\ldots,-(n_--1),\\
                                    &\qquad\qquad\{\widetilde{(\check{n}_i^{+}-n_+)}\},
                                       \{-\widetilde{(\check{n}_j^{-}-n_-)}\}\
                                       \}\,.
\end{array}}
\end{eqnarray} 
Indexes $i$ and $j$ are running here over the absent states of both schemes, provided the conditions 
$\check{n}_{i}^{+}-n_+\geq 0$ and 
$\check{n}_{j}^{-}-n_-\geq 0$ are met. A special case corresponds to the 
positive scheme 
 $\Delta_+=(l_1^+,l_1^++1,\ldots,l_1^++q)$, for which  the dual negative 
scheme is   $\Delta_-=(-(q+1),\ldots,-(q+l_1^+))$. Here  
$n_+=1+q$ and $n_-=l_{1}^+$,
there are no states to construct polynomials (\ref{fyh}),
and  we just put
$f(\eta)=h(\eta)=1$. 
Analogously, there are no tilted eigenstates in  (\ref{kerA}) in this case. In particular,
if $n_+$ is an even number, then the DCKA transformation will produce 
a deformed harmonic oscillator 
with one-gap of size $2(l_1^{-}+q+1)=2N$ in its spectrum, while 
if $q$ is an odd number and 
$l_{1}^+=1$, then we generate a gapless deformation of  $L_{1}^{iso}$ (by introducing the potential 
barrier at $x=0$).

\section{Relations between symmetry generators}\label{apen-red}
We first show explicitly how the three families appear by considering the 
commutators
\begin{eqnarray}
\small{\begin{array}{l}
[\mathfrak{C}_{N+l}^-,\mathfrak{A}_{k}^-]=P_{n_-}(\eta)|_{\eta=L_{(+)}}^{\eta=L_{(+)}+2k}\mathfrak{C}_{N+k+l}^-\,,\qquad
[\mathfrak{C}_{N+l}^+,\mathfrak{B}_{k}^+]=P_{n_-}(\eta)|_{\eta=L_{(+)}}^{\eta=L_{(-)}-2l}\mathfrak{C}_{N+k+l}^+\,,\\\,
[\mathfrak{A}_{k}^+,\mathfrak{C}_{N+l}^-]=(-1)^{n_-}T_k(L_{(-)})\mathfrak{A}_{N+l-k}^--P_{n_-}(L_{(+)})T_{l}(L_{(+)}+2l)\mathfrak{C}_{N+l-k}^-\,,  
\\{} 
 [\mathfrak{B}_{k}^-,\mathfrak{C}_{N+l}^+]=(-1)^{n_-}T_{k}(L_{(+)}+2k)\mathfrak{B}_{N+l-k}^+-
T_{l}(L_{(-)})P_{n_+}(L_{(-)}-2l)\mathfrak{C}_{N+l-k}^+\,,\\ {}
[\mathfrak{C}_{N+k}^+,\mathfrak{C}_{N}^-]=P_{n_+}(L_{(-)}-2k)\mathfrak{A}_{k}^--P_{n_-}(L_{(-)}+2N)\mathfrak{B}_{k}^-\,,
\\{}
[\mathfrak{C}_{N\pm k}^{\pm},\mathfrak{C}_{N\pm l}^{\pm}]=0\,,\qquad\l \geq 0\,,
\end{array}}
\end{eqnarray}
where polynomials $P_{n_\pm}(\eta)$ and $T_k(\eta)$ are defined by Eqs. (\ref{polyA}) and (\ref{Tk}).
These 
commutators should be interpreted as recursive relations which generate the elements of the 
three families of the ladder operators proceeding  from the spectrum-generating set 
of operators  with   $l=r(N,c)$ and $k=c$. 
On the other hand, the commutators of the  ladder operators with their own 
conjugate counterparts are
\begin{eqnarray}\small{
\begin{array}{c}
[\mathfrak{A}_k^-,\mathfrak{A}_{k}^+]=P_{n_-}(\eta-2k)P_{n_-}(\eta)T_k(\eta)|_{\eta=L_{(-)}}^{\eta=L_{(-)}+2k}\,,\\\,
[\mathfrak{B}_k^-,\mathfrak{B}_{k}^+]=P_{n_+}(\eta-2k)P_{n_+}(\eta)T_k(\eta)|_{\eta=L_{(+)}}^{\eta=L_{(+)}+2k}\,,\\\,
[\mathfrak{C}_{N\pm k}^-,\mathfrak{C}_{N\pm k}^+]=P_{n_-}(\eta)P_{n_+}(x-2k)T_k(\eta)|_{\eta=L_{(-)}}^{\eta=L_{(+)} \pm 2k}\,.
\end{array}}
\end{eqnarray}
In this way, we obtain a deformation of $\mathfrak{sl}(2,\R)$ in (\ref{sl2rh}). 

Below we present some relations between lowering ladder operators, from which analogous relations
for  raising operators can be obtained via Hermitian conjugation. 

The definitions of the three families  automatically  provide the following relations: 
\begin{eqnarray}
\label{a-c,b-c}
&\mathfrak{A}_{N+k}^-=(-1)^{n_-}P_{n_-}(L_{(-)})\mathfrak{C}_{N+k}^-\,,\qquad
\mathfrak{B}_{N+k}^-=(-1)^{n_-}\mathfrak{C}_{N+k}^-P_{n_+}(L_{(+)}),&
\\
\label{C-}
&\mathfrak{C}_{N-(N+k)}^-\equiv\mathfrak{C}_{-k}^-=(-1)^{n_-}P_{n_+}(L_{(-)}+2N)\mathfrak{A}_{k}^+\,,&\\
\label{reqgen1} 
&\mathfrak{C}_{2N+l+k}^-=(-1)^{n_-}\mathfrak{C}_{N+l}^-\mathfrak{C}_{N+k}^-\,, & \\
\label{reqgen2} 
&(\mathfrak{C}_{N+k}^-)^2=(-1)^{n_-}\mathfrak{C}_{2N+2k}^-\,,&
\end{eqnarray}
where $k,l=0,1,\ldots$. 
Eq. (\ref{a-c,b-c}) means that operators of families $\mathfrak{A}$ and $\mathfrak{B}$ with index $k\geq N$ 
are essentially the operators of the $\mathfrak{C}$ family.  
Eq. (\ref{C-}) shows that operators of the form $\mathfrak{C}_{-k}^\pm$ are not basic. 
If in  (\ref{reqgen1}) one fixes $l=r(N,c)$, then all the operators with index equal or
greater than  $N+r(N,c)$  reduce to the products  of the basic elements. 
Finally, Eq. (\ref{reqgen2})  
  means that the square of an operator of $\mathfrak{C}$-family with odd index 
 $N+k$ is a physical operator, but not basic. The unique special case is when $c=2$, $N$ is odd, and $k=0$
 since there is no product of physical operators of lower order which
 could make the same job. {}From here we conclude that the basic operators 
 are given by (\ref{genlad}). 
 
For one-gap systems we can use the  second equation in (\ref{ide})
(where $f(\eta)=h(\eta)=1$) to find some relations between operators with indexes less than $N$\,:
\begin{eqnarray}
\label{ABTOC}
\small{
\begin{array}{ll}
\mathfrak{C}^-_{n_+-k}=(-1)^{n_-}\mathfrak{A}_{n_+-k}^-T_{k}(L_{(-)})\,,&
\mathfrak{C}^-_{n_--k'}=(-1)^{n_-}\mathfrak{B}_{n_--k'}^-T_{k'}({L_{(-)}}+2(n_+-k'))\,,\\
\mathfrak{A}_{n_++k'}^-=(-1)^{n_-}\mathfrak{C}_{n_++k'}^-T_{k'}(L_{(-)})\,,&
\mathfrak{B}_{n_-+k}^-=(-1)^{n_-}\mathfrak{C}_{n_-+k}^-T_{k}(L_{(-)}+2n_+)\,,
\end{array}}
\end{eqnarray}
where $k=0,\ldots,n_+$ and $k'=0,\ldots,n_-$. 
By considering the ordering relation between $n_-$ and $n_+$, we can combine relations (\ref{ABTOC})
to represent operators of the $\mathfrak{A}$ family in terms of $\mathfrak{B}$ family or vice-versa. 
For the case $n_-<n_+$  we have    
\begin{equation}
\label{BcomoA}
\mathfrak{B}_{n_+-k}^-=T_{(n_+-n_--k)}(L_{(-)}+4n_+-2k)T_{k}(L_{(-)}+2n_+)\mathfrak{A}_{n_+-k}^-\,,
\end{equation}
where $k=0,\ldots, n_+-n_-$. In other words, only first  
$n_--1$ operators are basic.
In the case  $n_-=1$, there exist  no basic elements in the  $\mathfrak{B}$-family.
As  examples  corresponding to this observation we  have all the deformations produced 
by a unique nonphysical state of the form $\psi_{-n}(x)$.  
On the other hand, in the case $n_+<n_-$ we have
\begin{equation}
\label{AcomoB}
\mathfrak{A}_{n_--k}^-=T_{k}(L_{(-)}+2N)T_{(n_--n_+-k)}(L_{(-)}+2(n_--k))\mathfrak{B}_{n_--k}^-\,,
\end{equation}
where  $k=0,\ldots, n_--n_+$. 
According to this, only first  $n_+-1$ elements cannot be written in terms of the operators of 
 $\mathfrak{B}$ family. 
 The unique case in which there exist no basic elements of the families $\mathfrak{A}$ or $\mathfrak{B}$ 
 is when $n_-=n_+=1$, 
 which corresponds to the shape invariance of the harmonic oscillator. As a final result, the basic elements 
 of the three families are given by (\ref{basicsubsetonegap}).

We consider now the relations  between Darboux generators $A_{(\pm)}^-(a^\pm)^n$
and $A_{(\pm)}^-(a^\mp)^n$.  
Using the first relation in (\ref{ide}) and the definition of operators $\mathfrak{C}_{N+l}^\pm$, we obtain relations 
\begin{eqnarray}
&A_{(-)}^-(a^{-})^{N+l}=(-1)^{n_-}P_{n_-}(L_{(-)})A_{(+)}^-(a^{-})^{l}\,,\quad
A_{(+)}^-(a^{+})^{N+l}= (-1)^{n_-}P_{n_+}(L_{(+)})A_{(-)}^-(a^+)^{l}\,,&\nonumber\\
&A_{(+)}^-(a^-)^{N+l+k}=(-1)^{n_-}\mathfrak{C}_{N+l}^-A_{(+)}^-(a^-)^{k}\,,\quad
A_{(-)}^-(a^+)^{N+l+k}=(-1)^{n_-}\mathfrak{C}_{N+l}^+A_{(-)}^-(a^+)^{k}\,,&\nonumber
\end{eqnarray}
where $k,l=0,1,2,\ldots$. If we fix  $l=r(N,c)$, 
then one finds that the basic elements are just (\ref{genA}).
On the other hand for one-gap systems, with the help of (\ref{ide}) one can obtain relations 
\begin{eqnarray}
\label{Q->S+}
A_{(-)}^-(a^-)^{n_++k}=(-1)^{n_-}A_{(+)}(a^+)^{n_--k}T_{k}(L)\,,\\
\label{Q->S+2}
A_{(+)}^{-}(a^{+})^{n_-+k'}=(-1)^{n_-}A_{(-)}^-(a^-)^{n_+-k'}T_{k'}(L+2k')\,,
\end{eqnarray}
with $k=0,\ldots, n_-$ and $k'=0,\ldots,n_+$. These relations reduce 
the basic subsets of Darboux generators 
to (\ref{frakS}).

\section{(Anti)-Commutation relations for one-gap systems}
\label{apen-comm}

In this Appendix we summarize some (anti)commutation relations  
for one-gap deformations of harmonic oscillator systems. 

For the anticommutator of two fermionic operators
in (\ref{SUSY}) we have
\begin{eqnarray}
\label{hbP}
\mathbb{P}_z=\small{ \left\{ \begin{array}{lcc}
   P_{n_-}(\eta)T_{|z|}(\eta)\big|_{\eta=\mathcal{H}+2|z|\Pi_- + \lambda_{-}} & -N<z\leq0\\\vspace{-0.4cm}
   \\ P_{n_-}(\eta)T_z(\eta+2z)\big|_{\eta=\mathcal{H}-2|z|\Pi_- + \lambda_{-}} & 0<z\leq n_+\\                                                                       \end{array}
   \right. \,,}
 \end{eqnarray} 
 and for the positive scheme  
\begin{eqnarray}
\label{h2bP}
\mathbb{P}'_z= \small{\left\{ \begin{array}{lcc}
  P_{n_+}(\eta)T_{|z|}(\eta-2z)\big|_{\eta=\mathcal{H}'-2|z|\Pi_- + \lambda_{+}} & -N<z\leq 0\\\vspace{-0.4cm}
   \\ P_{n_+}(\eta)T_{z}(\eta)\big|_{\eta=\mathcal{H}'+2|z|\Pi_- + \lambda_{+}}  & 0<z\leq n_-\\                                                                       \end{array}
   \right. \,.}
 \end{eqnarray} 
 By virtue of the relation between dual  schemes, the expression $\mathbb{P}'_{z}(\mathcal{H}',\sigma_3)=\mathbb{P}_{N-z}(\mathcal{H}'+N(1+\sigma_3)-\lambda_-+\lambda_+,\sigma_3)$ helps to complete the set of polynomials.  
 
 For the negative scheme we also have
\begin{equation}
\label{CkQ0}
[\mathcal{G}_{-k}^{(2\theta(z)-1)},\mathcal{Q}_a^0]=\frac{1}{2}P_{n_-}(\eta)|_{\eta=\mathcal{H}+\lambda_{-}}^{\eta=\mathcal{H}+
\lambda_{-}+2|z|}(\mathcal{Q}_a^{z}-(2\theta(z)-1)i\epsilon_{ab}\mathcal{Q}_b^{z})\,,\qquad
\end{equation}
where $ z \in (-N,0)\cup(0,2N)$,
while for the positive scheme, where $\mathcal{Q}_{a}^{'z}=\mathcal{Q}_a^{N-z}$ and $\mathcal{G}_{\pm k}^{'(1)}=\mathcal{G}_{\pm k}^{(1)}$ 
when $k\geq N$,  we have
\begin{equation}
\label{B3}
[\mathcal{G}_{-z}^{'(2\theta(z)-1)},\mathcal{Q}_{a}^{'0}]=\frac{1}{2}P_{n_+}(x)|_{x=\mathcal{H}'+\lambda_{+}}^{x=\mathcal{H}'+
\lambda_{+}+2|z|}(\mathcal{Q}_a^{'z}+(2\theta(z)-1)i\epsilon_{ab}\mathcal{Q}_b^{'z})\,,
\end{equation}
where $ z \in (-N,0)\cup(0,2N)$. On the other hand, for the negative scheme the relation
$[\mathcal{G}_{z}^{(1)},\mathcal{Q}_a^z]=\frac{\mathcal{V}_{z}
(\mathcal{H})}{2}(\mathcal{Q}_a^{0}+i(2\theta(z)-1)\epsilon_{ab}\mathcal{Q}_b^{0})\,$
is valid,
where

\begin{eqnarray}
\label{C-kQk}
\mathcal{V}_{z}= \left\{ \begin{array}{lcc}
   P_{n_-}(\eta)T_z(\eta)\big|^{\eta=\mathcal{H}+\lambda_{-}-2z}_{\eta=\mathcal{H}+\lambda_{-}}\,,& -N<z<0\,,\\\vspace{-0.4cm}
   \\ P_{n_-}(\eta)T_z(\eta+2z)\big|^{\eta=\mathcal{H}+\lambda_{-}}_{\eta=\mathcal{H}+\lambda_{-}-2z} \,,& 0<z\leq n_+\,,\\\vspace{-0.4cm}
\\ P_{n_+}(\eta)T_{N-z}(\eta)\big|^{\eta=\mathcal{H}+\lambda_{-}+2N}_{\eta=\mathcal{H}+\lambda_{-}+2(N-z)}\,,& n_+<z\leq N\,,\\\vspace{-0.4cm}
\\ P_{n_+}(\eta)T_{z}(\eta+2z)\big|^{\eta=\mathcal{H}+\lambda_{-}+2N}_{\eta=\mathcal{H}+\lambda_{-}-2z}\,, & N<z<2N\,.
                                                                       \end{array}
   \right. 
    \end{eqnarray} 
In the positive scheme we have
$[\mathcal{G}_{z}^{'(1)},\mathcal{Q}_a^{'z}]=\frac{\mathcal{V}_{z}'(\mathcal{H}')}{2}(\mathcal{Q}_a^{'0}-i(2\theta(z)-1)\epsilon_{ab}\mathcal{Q}_b^{'0})$, 
where $\mathcal{V}_{z}'(\mathcal{H}')$ are given by
\begin{eqnarray}
\label{C-kQk2}
\mathcal{V}_{z}'= \left\{ \begin{array}{lcc}
   P_{n_+}(\eta)T_{z}(\eta+2z)\big|^{\eta=\mathcal{H}'+\lambda_{+}}_{\eta=\mathcal{H'}+\lambda_{+}-2z}\,,  & -N<z<0\,,\\\vspace{-0.4cm}
   \\P_{n_+}(\eta)T_z(\eta)\big|^{\eta=\mathcal{H}'+\lambda_{+}}_{\eta=\mathcal{H}'+\lambda_{+}}\,, & 0<z\leq n_-\,,\\\vspace{-0.4cm}
\\  P_{n_-}(\eta-2N)T_{N-z}(\eta-2z)\big|^{\eta=\mathcal{H}'+\lambda_{+}+2z}_{\eta=\mathcal{H}'+\lambda_{+}}\,,& n_-<z\leq N\,,\\\vspace{-0.4cm}
\\P_{n_-}(\eta)T_z(\eta)\big|^{\eta=\mathcal{H}'+\lambda_{+}+2z}_{\eta=\mathcal{H}'+\lambda_{+}-2N}\,, & N<z<2N\,.
                                                                       \end{array}
   \right. 
 \end{eqnarray} 
These are the missing relations which prove that the subsets $\mathcal{U}_{0,z}^{(2\theta-1)}$ defined in (\ref{U1}),
 satisfy closed superalgebras independently of choosing  the  scheme.
On the other hand, we can use them
to prove that the subsets  $\mathcal{I}_{N,z}^{(2\theta-1)}$ given in (\ref{In11}) also produce  closed superalgebras. 
Other useful relations are 
\be
\label{Cn+lQN+k}
[\mathcal{G}_{-(N+l)}^{(1)},\mathcal{Q}_a^{N+k}]=\frac{1}{2}T_k(\eta)P_{n_+}(\eta+2k)|_{\eta=\mathcal{H}+ \lambda_{-}}^{\eta=
\mathcal{H}+ \lambda_{-}+2(N-k)}(\mathcal{Q}_a^{l-k}+i\epsilon_{ab}\mathcal{Q}_b^{l-k})\,,
\ee
where $l>k$ and $l-k\leq n_+$. For $l<k$ we have
\be
\label{Cn+lQN+k2}
[\mathcal{G}_{-(N+l)}^{(1)},\mathcal{Q}_a^{N+k}]=P_{n_-}(\mathcal{H}+ 
\lambda_-)(\mathcal{Q}_a^{k-l}+i\epsilon_{ab}\mathcal{Q}_b^{k-l})\,,
\ee
and also we can write $[\mathcal{G}_{\pm(N\pm k)}^{(1)},\mathcal{G}_{\pm (N\pm l)}^{(1)}]=0$ for any values of $k$ and $l$.

\section{List of polynomial functions for Sec. \ref{SecDefQHO}}
\label{list}
\underline{Eq. (\ref{slr3})}\,: $P_{\alpha,\beta}(\mathcal{H})=-P_{-\alpha,-\beta}(\mathcal{H}-2(\alpha+\beta))$, 
and
\begin{eqnarray}
\nonumber\small{
\begin{array}{lll}
P_{-1,1}=\mathcal{H}(6\mathcal{H}-20)-8\Pi_-(\mathcal{H}-3)\,, &
P_{-1,-2}=-2\mathcal{H}+12(1-\Pi_-)\,, &
P_{-1,+2}=10(\mathcal{H}-4)-12\Pi_-\,,\\
M_{-1,+2}=12\,,&P_{-1,-3}=P_{-2,-3}=-6\,, & P_{-1,+3}=-12\,,\\
M_{-1,+3}=24\,, & P_{-1,-4}=P_{-2,-4}=-8\,, & P_{-1,+4}=16(\mathcal{H}-5-\Pi_-)\,,\\
P_{-1,-5}=P_{-2,-5}=-10\,, & P_{-1,+5}=20(\mathcal{H}-6-\Pi_-)\,,&
\end{array}}
\end{eqnarray}
\vskip-0.51cm
\begin{eqnarray}
\nonumber\small{
\begin{array}{ll}
P_{-2,+2}=(\mathcal{H}-4)[8\mathcal{H}(2\mathcal{H}-7)+\Pi_-(4\mathcal{H}^2-44\mathcal{H}+192)]\,,&
P_{-2,+3}=-18\mathcal{H}+96-4(\mathcal{H}-30)\Pi_-\,,\\
 M_{-2,+3}=M_{-2,+4}=-96\,, &P_{-2,+4}=-2(11\mathcal{H}-\Pi_-)+136\,, \\
P_{-2,+5}=1104-340\mathcal{H}+26\mathcal{H}^2+\Pi_-(576-104\mathcal{H}+10\mathcal{H}^2)\,, &
P_{-3,+4}=24\mathcal{H}-144+\Pi_-(2\mathcal{H}-76)\,,\\ M_{-3,+4}=848\,, &
P_{-3,+5}=30\mathcal{H}-180+\Pi_-(8\mathcal{H}-180)\,,\\ M_{-3,+5}=960\,, \\
P_{-4,+5}=40(32-10\mathcal{H}+\mathcal{H}^2-\Pi_-(7\mathcal{H}-32))\,,& M_{-4,+5}=-5760\,, \\
\end{array}}
\end{eqnarray}
\vskip-0.51cm
\begin{eqnarray}
\nonumber\small{
\begin{array}{l}
P_{-4,+4}=4[7\mathcal{H}^3 - 56\mathcal{H}^2+116\mathcal{H}+32+\Pi_-(\mathcal{H}^3 - 64\mathcal{H}^2+572\mathcal{H}+1472)]\,,\\
P_{-5,+5}=2(320-2848\mathcal{H}+1268\mathcal{H}^2-248\mathcal{H}^3+23\mathcal{H}^4)+
\\\qquad\qquad\qquad 8\Pi_- (10000-6212\mathcal{H}+1492\mathcal{H}^2-187\mathcal{H}^3+12\mathcal{H}^4)\,.
\end{array}}\\
\nonumber\end{eqnarray}
\vskip-0.51cm
\noindent 
\underline{Eq.  (\ref{slr4})}\,: $F_{\alpha,\beta}(\mathcal{H})=-F_{-\alpha,-\beta}(\mathcal{H}-2(\alpha+\beta))$, and 
\vskip-0.51cm
\begin{eqnarray}
\nonumber\small{
\begin{array}{lll}
F_{+1,-1}=F_{-1,+3}=0\,,& N_{+1,-1}=-F_{-1,-2}=-N_{-2,+1}=2\,, &
F_{-2,-1}=F_{-2,-2}=-4\,,\\
 F_{-1,+2}=-N_{-1,+3}=6\,, &F_{-1,+4}= F_{-2,+1}=8\,, & F_{-1,+5}=10\,,\\
N_{-1,+2}=F_{-2,+3}=-12\,, & F_{-2,+4}=-16& N_{-2,+3}=N_{-2,+4}=48\,,\\
F_{-1,+1}=4(\mathcal{H}-3)\,, & F_{-2,+5}=24(\mathcal{H}-7)\,,& F_{-2,+2}=12(\mathcal{H}-4)^2\,,
\end{array}}
\end{eqnarray}
while other elements are zero. 
\noindent 
\underline{Eq.  (\ref{susy3})}\,: 
$\mathbb{C}_{\alpha,\beta}=\mathbb{C}_{\beta,\alpha}$\,, and 
\vskip-0.51cm
\begin{eqnarray}
\nonumber\small{
 \begin{array}{ll}
 \mathbb{C}_{-2,-1}=\mathcal{G}_{+1}^{(1)}(\mathcal{H}-6)+8\mathcal{G}_{+1}^{(0)}\,, &
\mathbb{C}_{-2,0}=\mathcal{G}_{-2}^{(1)}+4\mathcal{G}_{-2}^{(0)}\,,\\
\mathbb{C}_{-2,1}=-(\mathcal{H}-4\Pi_-)\mathcal{G}_{-3}^{(1)}\,,&
\mathbb{C}_{-2,2}=(\mathcal{H}+4\Pi_-)\mathcal{G}_{-4}^{(1)}\,,\\
\mathbb{C}_{-2,5}=\mathcal{G}_{-7}^{(1)}=-\mathcal{G}_{-3}^{(1)}\mathcal{G}_{-4}^{(1)}\,,&
\mathbb{C}_{-1,0}= \mathcal{G}_{-1}^{(1)}+\mathcal{G}_{-1}^{(0)}\,,\\
\mathbb{C}_{-1,1}= -\mathcal{G}_{-2}^{(1)}+2\mathcal{G}_{-2}^{(0)}\,,&
\mathbb{C}_{-1,4}= \mathcal{G}_{-5}^{(1)}\,,\\
 \mathbb{C}_{-1,5}= \mathcal{G}_{-6}^{(1)}=-(\mathcal{G}_{-2}^{(1)})^2\,,&
\mathbb{C}_{1,2}= -(\mathcal{H}-2)(\mathcal{G}_{-1}^{(1)}+6\mathcal{G}_{-1}^{(0)})\,,\\
\mathbb{C}_{1,3}=-\mathcal{G}_{-2}^{(1)}+6\mathcal{G}_{-2}^{(0)}\,,&
\mathbb{C}_{1,4}=(\mathcal{H}-4\sigma_3)\mathcal{G}_{-3}^{(1)}\,,\\
\mathbb{C}_{1,5}=(\mathcal{H}-4-10\Pi_-)\mathcal{G}_{-4}^{(1)}\,,&
\mathbb{C}_{2,3}=(\mathcal{H}-2)\mathcal{G}_{-1}^{(1)}-12(\mathcal{H}-4)\mathcal{G}_{-1}^{(0)}\,,\\
\mathbb{C}_{2,4}=(\mathcal{H}-2)\mathcal{G}_{-2}^{(1)}-16(\mathcal{H}+3)\mathcal{G}_{-2}^{(0)}\,,&
\mathbb{C}_{2,5}=((\mathcal{H}-2)(\mathcal{H}-4)-8(\mathcal{H}-5)\Pi_-)\mathcal{G}_{-3}^{(1)}\,,\\
\mathbb{C}_{3,4}=(\mathcal{H}-2)\mathcal{G}_{-1}^{(1)}-16(\mathcal{H}-5)\mathcal{G}_{-1}^{(0)}\,,&
\mathbb{C}_{3,5}=-(\mathcal{H}+2)\mathcal{G}_{-2}^{(1)}-20(\mathcal{H}-4)\mathcal{G}_{-2}^{(0)}\,,\\
\end{array}}\\\nonumber
\small{\begin{array}{l}
\mathbb{C}_{4,5}=[(\mathcal{H}-2)(\mathcal{H}+6)-2\Pi_-(12\mathcal{H}-117))\mathcal{G}_{-1}^{(1)}-720\mathcal{G}_{-1}^{(0)}\,. 
\end{array}}
\end{eqnarray}

\vskip5cm
\noindent
\underline{Eq.  (\ref{susy4})}\,:
\vskip-0.51cm
\begin{eqnarray}
\nonumber\small{
\begin{array}{llll}
\mathbb{Q}_{1,-2}^{1}=2\,, &  \mathbb{Q}_{1,-2}^{2}=5\,, &
\mathbb{Q}_{2,3}^{1}=7(40-\mathcal{H})\,,&\mathbb{Q}_{2,3}^{2}=-3\,,\\
\mathbb{Q}_{1,-2}^{1}=2\,,& \mathbb{Q}_{1,-2}^{2}=5(\mathcal{H}-4)\,,&\mathbb{Q}_{2,4}^{1}=-10(\mathcal{H}-6)\,,
 & \mathbb{Q}_{2,4}^{2}=-4\,,\\
\mathbb{Q}_{1,-1}^{1}=-1\,, & \mathbb{Q}_{1,-1}^{2}=3\mathcal{H}-10\,,&
\mathbb{Q}_{2,5}^{1}=(336-118\mathcal{H}+11\mathcal{H}^2)\,, & \mathbb{Q}_{2,5}^{2}=5\,,\\
\mathbb{Q}_{1,2}^{1}=5(\mathcal{H}-4)\,, & \mathbb{Q}_{1,2}^{2}=\mathcal{H}\,,&
\mathbb{Q}_{3,1}^{1}=3\,, & \mathbb{Q}_{3,1}^{2}=1\,, \\
\mathbb{Q}_{1,3}^{1}=-10\,, & \mathbb{Q}_{1,3}^{2}=-6\,, & \mathbb{Q}_{3,2}^{1}=-8(\mathcal{H}-3)\,,
& \mathbb{Q}_{3,2}^{2}=4\,,\\
\mathbb{Q}_{1,4}^{1}=7(\mathcal{H}-36)\,, & \mathbb{Q}_{1,4}^{2}=4\,, &
\mathbb{Q}_{4,1}^{1}=2\,, & \mathbb{Q}_{4,1}^{2}=-1\,,\\
\mathbb{Q}_{1,5}^{1}=9(\mathcal{H}-56)\,, & \mathbb{Q}_{1,5}^{2}=5\,,&
\mathbb{Q}_{4,2}^{1}=10(3-\mathcal{H})\,,& \mathbb{Q}_{4,2}^{2}=-5\,,\\
\mathbb{Q}_{2,-2}^{1}=-2\,, & \mathbb{Q}_{2,-2}^{2}=4(\mathcal{H}-4)(2\mathcal{H}-7)\,, &
\mathbb{Q}_{5,1}^{1}=5\,, & \mathbb{Q}_{5,1}^{2}=3\,,\\
\mathbb{Q}_{2,-1}^{1}=-1\,, & \mathbb{Q}_{2,-1}^{2}=5(\mathcal{H}+2)\,,&
\mathbb{Q}_{5,2}^{1}=6(\mathcal{H}-1)\,, & \mathbb{Q}_{5,2}^{2}=3\,.\\
\end{array}}
\end{eqnarray}
\vskip-0.3cm
\noindent
\underline{Eq.  (\ref{susy5})}\,:
\vskip-0.7cm
\begin{eqnarray}
\nonumber\small{
\begin{array}{llll}
\mathbb{G}_{1,-2}^1=-1\,, & \mathbb{G}_{1,-2}^2=(8-\mathcal{H})\,, & \mathbb{G}_{2,-2}^1=-1\,,
 & \mathbb{G}_{2,-2}^2=(\mathcal{H}+8)(\mathcal{H}-6)\,,\\
\mathbb{G}_{1,-1}^1=1\,,  & \mathbb{G}_{1,-1}^2=(4-\mathcal{H})\,, & \mathbb{G}_{2,-1}^1=1\,,
  & \mathbb{G}_{2,-1}^2=4-\mathcal{H}\,,\\
\mathbb{G}_{1,0}^1=\mathbb{G}_{1,0}^2=1\,,& \mathbb{G}_{2,0}^1=-\mathbb{G}_{2,0}^2=1\,, &
\mathbb{G}_{1,1}^1=\mathcal{H}-4\,, & \mathbb{G}_{1,1}^2=-1\,,\\
\mathbb{G}_{2,1}^1=\mathcal{H}-2\,, & \mathbb{G}_{2,1}^2=\mathcal{H}\,, &\mathbb{G}_{1,2}^1=\mathcal{H}-4\,,
&\mathbb{G}_{1,2}^2=\mathcal{H}\,,\\
\mathbb{G}_{2,2}^1=-1\,,&\mathbb{G}_{2,2}^2=\mathcal{H}\,,&\mathbb{G}_{1,3}^1=\mathbb{G}_{1,3}^2=-1\,,
& \mathbb{G}_{2,3}^1=4-\mathcal{H}\,,\\
\mathbb{G}_{2,3}^2=-1\,,&\mathbb{G}_{1,4}^1=\mathcal{H}-4\,,&\mathbb{G}_{1,4}^2=-1\,,&
\mathbb{G}_{2,4}^1=2-\mathcal{H}\,,\\
\mathbb{G}_{2,4}^2=1\,,&\mathbb{G}_{1,5}^1=\mathcal{H}-4\,,
 & \mathbb{G}_{1,5}^2=-1\,,
 &\mathbb{G}_{2,5}^1=(\mathcal{H}-2)(\mathcal{H}-4)\,,\\
\mathbb{G}_{2,5}^2=-1\,.
\end{array}}
\end{eqnarray}

\end{document}